\pdfoutput=1 

\documentclass[11pt]{article}
\usepackage{vruler}
\usepackage{amssymb}

\usepackage{amsthm} %for \theoremstyle{definition}

\usepackage{tikz}
\usetikzlibrary{trees}
\usepackage{qtree}

\usetikzlibrary{decorations.pathmorphing}
\usetikzlibrary{decorations.markings}
\usepackage{verbatim}

\usepackage{amsmath}
\usepackage{amsthm} %for \theoremstyle{definition}
\usepackage{amssymb,amsfonts,textcomp}
\usepackage{array}
\usepackage{hhline}
\usepackage{graphicx}

\usepackage[T3,T1]{fontenc} %for \textquotedbl
\usepackage{txfonts} %/pxfonts for \circleright

\usepackage{latexsym} %for \Box
\usepackage{mathrsfs} %for \mathscr

\usepackage{MnSymbol} %for more curved arrow symbol and many others are possible \lcurvearrowright

\usepackage{tikz}
\usetikzlibrary{arrows,shapes,chains,matrix,positioning,scopes}% shapes for circle split
\usetikzlibrary{decorations.pathmorphing}
\usetikzlibrary{decorations.markings}
\usepgflibrary{plotmarks}%for squares and other shapes on graphs lines , 
\usetikzlibrary{patterns}
\usepackage{pdfpages}

\usetikzlibrary{mindmap}%for circle connection bar

\usepackage{graphicx}

\usepackage{multirow}
\usepackage{multicol}

\usepackage{subfig} 

\usepackage{float}
\usepackage{wrapfig}
\restylefloat{figure}
\restylefloat{table}
\usepackage{color, colortbl} %to color rows
\usepackage{float}
\restylefloat{figure}

\makeatletter
\def\section{\@startsection {section}{1}{\z@}{-2.5ex plus
     -1ex minus-.2ex}{2ex}{\reset@font\large\bf}}
\makeatother

\theoremstyle{definition}

\newcommand{\STRUT}{\rule{0mm}{4ex}}

\newcommand{\cat}{^\vee \!}
\newcommand{\pitch}{^\circ \!}
\newcommand{\pitcher}{^\wedge \!}

\newcommand{\pot}{o->}

\newcommand{\poti}{o-} %the start for line pots
\newcommand{\pote}{->} %the end for line pots

\newcommand{\potpitch}{^\circ \!}
\newcommand{\potonep}{^{\circ_1 \!}}
\newcommand{\pottwop}{^{\circ_2 \!}}
\newcommand{\potone}{^{\wedge_1 \!}}
\newcommand{\pottwo}{^{\wedge_2 \!}}

\newcommand{\catloaded}{\!\stackrel{\medbullet}{\cat}}

\newcommand{\catblocked}{\!\stackrel{-}{\cat}}

\newcommand{\jump}{right hook->}
\definecolor{currentcolor}{rgb}{0.8 0.4 0.2}%orange currentcolor is to be set to path color and then made lighter

\tikzstyle{stochasticjumpstyle}=[diamond,draw,fill=white,>=latex,>->,dashed]
\tikzstyle{stochasticPathstyle}=[>=latex,>->,dashed]
\tikzstyle{stochasticNodestyle}=[ellipse,inner sep=1pt,text=.,fill=.!20]%[fill=white,inner sep=1pt]%[ellipse,inner sep=1pt,draw,fill=white]
\tikzstyle{blankstyle}=[fill=white,inner sep=1pt]

\def\SnakeSegLen{0.6em}%defines snake segment length for signal jumps  in graphs
\def\SnakeAmp{0.11em}%defines snake amplitude for signal jumps  in graphs
\def\PrePostLen{5mm}

\tikzstyle{sendstyle}=[dashed,line width=1.1pt]%[dotted,ultra thick]

\tikzstyle{splitstyle}=[circle,draw]%not used

\newcommand{\sigjump}{*-to}
\newcommand{\sendsig}{*-latex}%{o->}%{triangle 60-*}%{triangle 60 reversed-*}%{*-*}%o-*

\tikzstyle{receivestyle}=[>->,line width=1.1pt,decorate, decoration={zigzag,segment length=\SnakeSegLen, amplitude=\SnakeAmp, pre length=\PrePostLen, post=curveto, post length=\PrePostLen},text=black]

\tikzstyle{receivesigstyle}=[draw,inner sep=2pt,fill=pink!20]
\tikzstyle{receivesigstyle3}=[draw,inner sep=2pt, fill=white]
\tikzstyle{receivesigstyle2}=[ellipse,shade, draw,double,fill=red!10]

\tikzstyle{sendsigstyle}=[diamond,draw,inner sep=1pt, text=black, fill=yellow!80]
\tikzstyle{sendsigstyle3}=[circle,draw, ball color=white]
\tikzstyle{sendsigstyle2}=[diamond,draw,double, inner sep=1pt, fill=white]

\tikzstyle{snakesendstyle}=[*->, decorate, decoration={snake, segment length=\SnakeSegLen, amplitude=\SnakeAmp,  pre length=\PrePostLen, post=curveto, post length=\PrePostLen}]

\tikzstyle{snakesendstyle1}=[line width=1.1pt, decorate, decoration={snake,segment length=\SnakeSegLen, amplitude=\SnakeAmp}]

\tikzstyle{snakesendstyle3}=[decorate, decoration={markings, mark=at position .75 with {\arrow[red,line width=5mm]{>}}, snake, segment length=\SnakeSegLen, amplitude=\SnakeAmp,  pre length=\PrePostLen, post=curveto, post length=\PrePostLen}]

\tikzstyle{snakesendstyle2}=[decorate, decoration={ zigzag,segment length=\SnakeSegLen, amplitude=\SnakeAmp, line around/.style={decoration={pre length=\PrePostLen,post length=\PrePostLen}}}]
\newcounter{foo}
\newcommand{\Cells}{\mbox{Cells}}

\usepackage{hyperref}

\makeatletter
\AtBeginDocument{\let\nameref\HyPsd@nameref}

\makeatother

\colorlet{anglecolor}{green!50!black}
\definecolor{darkgreen}{rgb}{0 0.6  0}

\definecolor{turquoise}{rgb}{0 0.41 0.41}
\definecolor{rouge}{rgb}{0.79 0.0 0.1}
\definecolor{vert}{rgb}{0.15 0.4 0.1}
\definecolor{mauve}{rgb}{0.6 0.4 0.8}
\definecolor{violet}{rgb}{0.58 0. 0.41}
\definecolor{orange}{rgb}{0.8 0.4 0.2}
\definecolor{bleu}{rgb}{0.39, 0.58, 0.93}

\definecolor{darkross}{rgb}{0.008,0.412,0.471}
\definecolor{middleross}{rgb}{0.012,0.580,0.663}
\definecolor{lightross}{rgb}{0.016,0.749,0.855}
\definecolor{darkblue}{rgb}{0.067,0.008,0.471}
\definecolor{middleblue}{rgb}{0.094,0.012,0.663}
\definecolor{lightblue}{rgb}{0.122,0.016,0.855}
\definecolor{darkpurple}{rgb}{0.471,0.008,0.412}
\definecolor{middlepurple}{rgb}{0.663,0.012,0.580}
\definecolor{lightpurple}{rgb}{0.855,0.016,0.749}
\definecolor{darkbrown}{rgb}{0.471,0.067,0.008}
\definecolor{middlebrown}{rgb}{0.663,0.094,0.012}
\definecolor{lightbrown}{rgb}{0.855,0.122,0.016}
\definecolor{darkolive}{rgb}{0.412,0.471,0.008}
\definecolor{middleolive}{rgb}{0.580,0.663,0.012}
\definecolor{lightolive}{rgb}{0.749,0.855,0.016}
\definecolor{darkgreen}{rgb}{0.008,0.417,0.067}
\definecolor{middlegreen}{rgb}{0.012,0.663,0.094}
\definecolor{lightgreen}{rgb}{0.016,0.855,0.122}
\definecolor{darkocre}{rgb}{0.471,0.298,0.008}
\definecolor{middleocre}{rgb}{0.663,0.420,0.012}
\definecolor{lightocre}{rgb}{0.855,0.541,0.016}

    \definecolor{lightblue}{rgb}{0,0,.7}
    \definecolor{orange}{rgb}{1,.7,0}
    \definecolor{darkorange}{rgb}{1,.4,0}
    \definecolor{darkgreen}{rgb}{0,.5,0}
    \definecolor{darkblue}{rgb}{0,0,.4}
    \definecolor{darkred}{rgb}{.4,0,0}
    \definecolor{gray}{rgb}{.2,.2,.2}
    \definecolor{darkgray}{rgb}{.2,.2,.2}
    \definecolor{shadecolor}{gray}{0.925}
    
\definecolor{darkred}{rgb}{0.65,0,0}
\definecolor{darkblue}{rgb}{0,0,.65}
\definecolor{darkgreen}{rgb}{0,0.5,0}
\definecolor{orange}{rgb}{1,.75,.25}
\definecolor{aqua}{rgb}{0,.25,.75}
\definecolor{grey}{rgb}{.5,.5,.5}
\definecolor{brown}{rgb}{.51,.35,.18}

\definecolor{lightblue}{rgb}{.3,.5,1}
\definecolor{orange}{rgb}{1,.7,0}
\definecolor{darkorange}{rgb}{1,.4,0}
\definecolor{darkgreen}{rgb}{0,.4,0}
\definecolor{darkblue}{rgb}{0,0,.4}
\definecolor{darkred}{rgb}{.56,0,0}
\definecolor{gray}{rgb}{.3,.3,.3}
\definecolor{darkgray}{rgb}{.2,.2,.2}

\definecolor{blue}{rgb}{0,0,1}
\definecolor{red}{rgb}{1,0,0}
\definecolor{pink}{rgb}{.933,0,.933}
\definecolor{green}{rgb}{0.133,0.545,0.133}
\definecolor{shadecolor}{gray}{0.925}

\definecolor{DarkBlue}{rgb}{0.000,0.000,0.545}
\definecolor{DarkChocolate}{rgb}{0.400,0.200,0.000}
\definecolor{DarkCyan}{rgb}{0.000,0.545,0.545}
\definecolor{DarkGoldenrod}{rgb}{0.720,0.525,0.044}
\definecolor{DarkGray}{rgb}{0.664,0.664,0.664}
\definecolor{DarkGreen}{rgb}{0.000,0.392,0.000}
\definecolor{DarkGrey}{rgb}{0.664,0.664,0.664}
\definecolor{DarkKhaki}{rgb}{0.740,0.716,0.420}
\definecolor{DarkLavender}{rgb}{0.400,0.200,0.600}
\definecolor{DarkMagenta}{rgb}{0.545,0.000,0.545}
\definecolor{DarkOliveGreen}{rgb}{0.332,0.420,0.185}
\definecolor{DarkOrange}{rgb}{1.000,0.550,0.000}
\definecolor{DarkOrchid}{rgb}{0.600,0.196,0.800}
\definecolor{DarkPeriwinkle}{rgb}{0.400,0.400,1.000}
\definecolor{DarkPurpleBlue}{rgb}{0.400,0.000,0.800}
\definecolor{DarkRed}{rgb}{0.545,0.000,0.000}
\definecolor{DarkRoyalBlue}{rgb}{0.000,0.200,0.800}
\definecolor{DarkSalmon}{rgb}{0.912,0.590,0.480}
\definecolor{DarkSeaGreen}{rgb}{0.560,0.736,0.560}
\definecolor{DarkSlateBlue}{rgb}{0.284,0.240,0.545}
\definecolor{DarkSlateGray}{rgb}{0.185,0.310,0.310}
\definecolor{DarkSlateGrey}{rgb}{0.185,0.310,0.310}
\definecolor{DarkSmoke}{rgb}{0.920,0.920,0.920}
\definecolor{DarkTurquoise}{rgb}{0.000,0.808,0.820}
\definecolor{DarkViolet}{rgb}{0.580,0.000,0.828}
\definecolor{DeepPink}{rgb}{1.000,0.080,0.576}
\definecolor{DeepSkyBlue}{rgb}{0.000,0.750,1.000}

\newenvironment{narrow}[2]{%
  \begin{list}{}{%
    \setlength{\topsep}{0pt}%
    \setlength{\leftmargin}{#1}%
    \setlength{\rightmargin}{#2}%
    \setlength{\listparindent}{\parindent}%
    \setlength{\itemindent}{\parindent}%
    \setlength{\parsep}{\parskip}%
  }%
  \item[]
}{\end{list}}

\tikzstyle{mystyle}=[scale= \PicSize,  %[****Crit. PicSize is not defined*****]
baseline=(current bounding box.center),
nodedescr/.style={fill=white,inner sep=2.5pt},
nodedescr2/.style={draw,circle,fill=white},
description/.style={fill=white,inner sep=2pt},
cancer loop/.style={preaction={draw=white, -,line width=6pt}, line width=1.1pt},
pot1 loop/.style={preaction={draw=white, -,line width=6pt}, red, line width=1.1pt},
pot2 loop/.style={preaction={draw=white, -,line width=6pt}, green, line width=1.1pt},
pot1/.style={preaction={draw=white, -,line width=6pt}, gray, solid, line width=1pt}, 
pot2/.style={preaction={draw=white, -,line width=6pt}, gray, solid, line width=1pt},
inPot/.style={ out=45,in=90,distance=3cm, min distance=2cm, line width=1.1pt, black!75},
inPot1/.style={ out=45,in=120,distance=3cm, min distance=2cm, line width=1.1pt, black!75},
inPot2/.style={ out=-45,in=-120,distance=3cm, min distance=2cm, line width=1.1pt, black!75},
in100Pot1/.style={ out=80,in=90,distance=3cm, min distance=2cm, line width=1.1pt, darkgreen},
in100Pot2/.style={ out=-80,in=-90,distance=3cm, min distance=2cm, line width=1.1pt, brown},
in2Pot1/.style={ out=45,in=90,distance=3cm, min distance=2cm, line width=1.1pt, black!75},
in2Pot2/.style={ out=-65,in=-90,distance=3cm, min distance=2cm, line width=1.1pt, black!75},
selfloop1/.style={ out=45,in=120,distance=6cm, min distance=4cm, line width=1.1pt, red},
selfloop2/.style={ out=-45,in=-120,distance=6cm, min distance=4cm, line width=1.1pt, blue},
loop1red/.style={ out=45,in=120,distance=6cm, min distance=4cm, line width=1.1pt, red},
loop2red/.style={ out=-45,in=-80,distance=6cm, min distance=4cm, line width=1.1pt, red},
cross line/.style={preaction={draw=white, -,	line width=6pt}},
jump1/.style={ out=45,in=135,distance=4cm, min distance=2cm, line width=1.1pt, red, dashed},
jump2/.style={ out=-45,in=-135,distance=4cm, min distance=2cm, line width=1.1pt, blue, dashed},
jumpover/.style={ out=80,in=90,distance=4cm, min distance=3cm, line width=1.1pt, black, dashed},
stochastic jump/.style={>=latex,>->,dashed, line width=1.1pt, green}
]

\def\PicSize{ 0.5} % 0.5 defines constant PicSize for uniform scale of TikZ pictures

\def\ColSepThree{3em}

\def\ColSepNarrow{2.7em}

\def\ColSepTight{1.7em} %1.9 works for all but one

\def\CellWallThickness{ 0.50mm}

\newcommand{\matrixsep}{0.75cm} %treat row and column separation same size

\numberwithin{equation}{section}
\usepackage[margin=1.4in]{geometry} %***set page dimensions easily. See geometryPageSize.pdf***

\usepackage[parfill]{parskip}    % Activate to begin paragraphs with an empty line rather than an indent

\begin{document}

%\setvruler %for line numbers
%\setvruler[scale][initial_count][step][digits][mode][odd_hshift][even_hshift][vshift] [height]
%\setvruler[10pt][1][1][4][1][0pt][0pt][0pt][\textheight] is the default

\title{Cancer Networks \\A general theoretical and computational framework for understanding cancer}

\author{Eric Werner \thanks{Balliol Graduate Centre, Oxford Advanced Research Foundation (http://oarf.org), Cellnomica, Inc. (http://cellnomica.com). We gratefully acknowledge the use of Cellnomica's Software Suite to construct the cancer and stem cell networks used to model and simulate all the multicellular processes that generated the {\em in silico} cancers described and illustrated in this paper.  \copyright Werner 2006-2011.  All rights
reserved. }\\
University of Oxford\\
Department of Physiology, Anatomy and Genetics, \\
and Department of Computer Science, \\
Le Gros Clark Building, 
South Parks Road, 
Oxford OX1 3QX  \\
email:  eric.werner@dpag.ox.ac.uk\\
}

\date{ } %This is to suppress the printing out of the date.

\maketitle

\thispagestyle{empty}

\begin{center}
\textbf{Abstract}

\begin{quote}
\it
We present a general computational theory of cancer and its developmental dynamics.  The theory is based on a theory of the architecture and function of developmental control networks which guide the formation of multicellular organisms. Cancer networks are special cases of developmental control networks. Cancer results from transformations of normal developmental networks. Our theory generates a natural classification of all possible cancers based on their network architecture. Each cancer network has a unique topology and semantics and developmental dynamics that result in distinct clinical tumor phenotypes.   We apply this new theory with a series of proof of concept cases for all the basic cancer types. These cases have been computationally modeled, their behavior simulated and mathematically described using a multicellular systems biology approach.  There are fascinating correspondences between the dynamic developmental phenotype of computationally modeled {\em in silico} cancers and natural {\em in vivo} cancers. The theory lays the foundation for a new research paradigm for understanding and  investigating cancer.  The theory of cancer networks implies that new diagnostic methods and new treatments to cure cancer will become possible. 
\end{quote}
\end{center}

\pagenumbering{roman}
\setcounter{page}{1}
\tableofcontents
%\listoffigures
\newpage
\pagenumbering{arabic}

%\listoftables

\section{Introduction}

We present a unified computational theory of cancer based on a theory of developmental control networks that, starting from a single cell, control the ontogenesis of multicellular organisms.  We investigate the space or range of all possible cancer networks. We model and simulate the major types of cancer networks.  We illustrate and discuss the result of simulations of instances of the major cancer network categories.  We relate the network types to their dynamic, developmental tumor phenotype.  The relevance to diagnosis, treatment and cure for the different cancer classes is indicated.  The goal is to present a new paradigm for cancer research that gives a deeper understanding of all cancers and will ultimately lead to a cure for cancer. 

While the research on cancer is enormous in scope and financing, a cure for all cancers is still nowhere in sight. The nature of cancer is still elusive. There is no real understanding of how cancer works, of its etiology, ontogeny and developmental dynamics. Models of cancer often display just so many pictures giving no details.  Up to now, mathematical models have been very limited in scope and offer limited  insight into the nature of cancer. Furthermore, the models are piecemeal, varying with each cancer. There seem to be as many molecular models of cancer as there are types of cancer.   

At present there is no unifying theory of cancer that relates the many different types of cancer.  At fault is the current paradigm about the nature of cancer. It is a gene-centered theory where so called cancer genes are responsible for cancer. 

The dominant  paradigm for understanding cancer is that cancer is caused by mutated, misbehaving genes.  When cancer genes are muted, and there is a failure of cell death, it leads to uncontrolled cell growth.  Cancer is seen as progressive, where multiple mutations are necessary to convert a normal cell into a cancerous cell.  What this account does not explain is how cancer is controlled. It does not explain how the mutation in a gene results in the development of a particular tumor.  It does not explain the way cancer cells and tumors develop, differentiate and de-differentiate.  

We propose instead that pathological developmental control networks cause cancer.  Therefore, we further propose that to understand cancer fully, we must understand the architecture of control networks that govern cancer development.  These genomic and epigenomic control networks are interpreted and executed by the cell.  From this perspective, cancer is a special case of multicellular developmental processes.  The following table~\autoref{CompareGN} compares and sums up the basic differences between the gene-centered theory of cancer and the control network theory of cancer. 

\begin{table}[H]
\begin{center}
\begin{footnotesize}%0.38\textwidth
\begin{tabular}{|p{25mm}|p{54mm}|p{54mm}|}
%\begin{tabular}{|p{0.2\textwidth}|p{0.4\textwidth}|p{0.4\textwidth}|}
\hline
\multicolumn{3}{|c|}{\begin{large}\bf Comparing Traditional Theories and Our Network Theory of Cancer\end{large} \STRUT } \\[1.5ex]
\hline \STRUT
 & \textbf{Traditional Theories of Cancer} & \textbf{Control Network Theory of Cancer} \\ \hline 
	\textbf{Cause} \STRUT &	Mutations in genes cause cancer & Mutations in developmental networks cause cancer  \\ \hline
	\textbf{Process} \STRUT &	Cancer is uncontrolled growth & Cancer is highly regulated process, controlled by developmental networks  \\ \hline
	\textbf{Likelihood} \STRUT  & Multiple mutations make cancers more likely & Multiple mutations in specific networks make cancers more likely\\ \hline
	\textbf{Severity} \STRUT  & Multiple mutations may make cancers more severe & Multiple mutations in specific networks determine the   architecture and {\bf metastatic potential} of cancer networks,  and, thereby, the potential severity of cancers\\ \hline
	\textbf{Networks}\STRUT &  \begin{center} {\bf \ldots} \end{center}  & Networks, their topology and locality are key to understanding cancer \\ \hline
	\textbf{Control States}\STRUT &  Not defined or confused with cell phenotype & Control states linked in networks specify development and differ from phenotype \\ \hline
	\multirow{3}{*}{\textbf{Classification}}\STRUT & Based on cell phenotype not theory & Based on network architecture and topology as related to cell phenotype \\ \cline{2-3}\STRUT
  & \textbf{Gene centred classification}: Classification is based on which gene is mutated  & \textbf{Networked 	centred classification}:  Classification is based on which area is mutated and how that transforms the network \\  \cline{2-3} \STRUT
  & Gene expression / protein levels & Cell phenotype is only indirectly related to network activation. Protein and RNA profiles are supportive evidence for network activation states \\ \hline
	\textbf{Dynamics} \STRUT & \begin{center} {\bf \ldots} \end{center} & {\bf A systematic relationship exists between cancer dynamics and genome network architecture}\\ \hline
%	\textbf{Dedifferentiation} \STRUT & Not explained & Results from re-activation of early stage developmental networks\\ \hline
	\textbf{Metastases} \STRUT & Caused by cell invasive properties. Metastatic types and dynamics not in the conceptual framework  & Different cancer networks have different metastatic potential, dynamics and distinct hierarchies of metastatic tumors \\ \hline
%	\textbf{Stage} \STRUT & {\bf Early, Middle, Late}: Based on cell type, tumor properties, cancer location, number of genetic mutations & Based on the {\bf location of network links} between the cancer network and the global developmental network,  network execution time, and genetic cellular strategies \\ \hline
	\textbf{Diagnostics} \STRUT & Diagnostics is based on cell and tissue phenotype, gene expression levels, gene mutations & Calls for new diagnostics methods and technologies to determine local genome network architecture. Traditional diagnostics still apply and can point to cancer network types. \\ \hline 
\end{tabular}
\end{footnotesize}
\end{center}
\caption{Comparison of the gene centered and network view of cancer}
\label{CompareGN}
\end{table}

While research into the molecular and genetic mechanisms underlying cancers is vast in scope, its focus has been so detailed as to miss the most fundamental unifying properties underlying all cancers.  If cancer were a forest then research has been looking not even at just single trees, but only at the bark of those trees.  No matter how detailed our knowledge of the bark of individual trees, it will give no real insight into the organization of the forest.  Without a deeper understanding of cancer, we cannot expect to adequately treat or cure cancer.  Hence, it should be no surprise that current treatment of cancer is piecemeal, arbitrary with many unwanted side effects.  It includes carcinogenic agents and procedures some of which can lead to future cancers in the same patient.  To cure cancer we need to understand cancer.  We need a comprehensive, unifying theory of cancer. 

The object of this work is to present a theory of cancer that unifies all cancers showing their commonalities and how they differ.  Our theory has fundamental implications for diagnostic methods,  drug discovery and approaches to treatment.   Most importantly, the theory implies that it is in principle possible to change any cancer cell into a normal cell.  This can be achieved by inverse operators that reverse the antecedent cancerous network transformations, thereby changing the cancer network into its normal precursor.  Therefore, the theory provides the foundation for a general strategy of how to cure cancer. 

The theory presented here may be considered by some as revolutionary in that it views cancer in a very different way from the dominant paradigm.  Contrary to the current view of cancer as uncontrolled cell growth (Hanahan~\cite{Hanahan2000}), we view cancer as a highly regulated developmental, multicellular process (Werner~\cite{Werner2003b,Werner2005}).   Our theory envisions cancer at higher levels of organization than the molecular level. At the same time our theory relates cancer to the cellular, epigenetic, genomic and molecular levels.  

Cancer is a multilevel phenomenon. At its core are cytogenic networks that generate and control cell growth and proliferation.  

If correct, this paradigm shift will generate whole new methodologies and approaches for understanding and treating cancer.  Importantly, unlike many other models, our theory is testable, and leads to precise predictions on how to control and cure cancers.  It is a sustainable new paradigm that offers scientists  a rich new framework for posing novel questions, hypotheses, explanations, and mechanisms, as well as guiding and opening new vistas for their research.

\textbf{Historical note}: The theory was motivated by the discovery of cancers in computationally simulated embryonic, multicellular systems.   Starting from a single cell, with an artificial genome, these embryo-like systems normally generated simple artificial multicellular organisms (MCOs) on the computer.  The surprise was that with certain types of mutations these multicellular systems developed what appeared to be tumors.   The static and dynamic phenomenological properties of these computationally simulated \textit{in silico} cancers were strikingly similar to those exhibited by naturally occurring \textit{in vivo} cancers. This led the author to search for a common theoretical framework that would explain both the computational \textit{in silico} cancers and the natural \textit{in vivo} cancers.  The first \textit{in silico} cancer was discovered in a Pensione in Rome, Italy in 1992 while the author was a visiting scientist at  the Italian National Research Council (CNR) in Rome, Italy in March of 1992. 

\subsection{Plan}

We hypothesize that the development of embryos and, indeed, the generation of all multicellular life from a single cell is a highly controlled process using interlinked control information in the genome in the form of developmental control networks.  We call such developmental control networks \emph{cenes} for control genes.  The global developmental control network underlying the ontogeny of multicellular organisms we call it the \emph{cenome}.  It is contained in the genome (the cell's DNA) but it is at a higher level of control than genes.  We distinguish cenes from genes. Protein coding genes are largely parts-genes used to generate the parts of the cell, its structural and functional units.  Multicellular development requires additional information to that contained in genes.  Cenes control genes and cenes control the cell's actions.  

We view cancer as the result of mutated or transformed normal developmental networks.  After introducing the theory of normal developmental control networks, we apply the theory to cancer.  We then describe all possible cancer networks. We start with slow growing {\bf linear cancer networks}, then develop the theory of {\bf exponential cancer networks} that result in extremely fast growth. Next we look at {\bf geometric cancer networks} lying in between the linear and exponential cancers. Linear cancer networks are a special instance of geometric cancer networks. We show that geometric cancer networks have growth properties related to the coefficients of Pascal's Triangle and these coefficients are related to the geometric numbers. Hence, the name geometric cancer networks. 

After that we develop a  {\bf theory of stem cell networks} and related them to geometric cancer networks. The theory of stem cell networks helps explain both the properties of stem cells and the properties of cancer stem cells.  

Next, we introduce {\bf stochastic cancer networks} which allow for probabilistic alternative paths in the multicellular development.  We look at the properties of stochastic cancer networks for linear, exponential and geometric networks showing the deterministic cancer networks can be approximated by stochastic cancer networks by adjusting probabilities.  

We then look at {\bf reactive communication cancer networks} that react to cellular and environmental signals but do not send signals.  Next we look at interactive communication cancer networks that receive and send signals to form communicating social systems of cells.  We show that communication can both exacerbate and ameliorate the effects of a cancer network. For instance, we show that, under certain conditions inter-cellular communication, an exponential cancer network need not lead to exponential growth. 

Next we describe particularly interesting cases of cancer that can be explained by our theory, these include {\bf bilateral cancers} such as those that occur in rare forms of breast cancer where the same tumor forms symmetrically in both breasts.  There is also the case of vacuous cancer networks that do not result in any cancer but lie hidden in the genome until activated by some circumstance.  

We develop a {\bf theory of metastases} (cancers that spread to other parts of the body) showing that different cancer networks have distinct metastatic phenotypes.  We show that geometric cancer networks generate a {\bf hierarchy of types of metastases} with distinct dynamic and proliferative properties. While some cancers can form dangerous metastases, other cancer networks form relatively harmless metastases.  We explore all the major types of possible metastases.  

We then discuss important factors that influence growth such as cell physics.  We discuss the relationship to treatment.  We relate our network theory of cancer with the previous gene centered theories of cancer. 

\subsection{What is cancer? } 

A careful review of the scientific literature leaves the deeper aspects of this question unanswered.   We know there are many types of cancers with differences in ontogeny and phenotype.  We know too that genes, so called oncogenes or cancer genes, are involved.  We know too that mutations of oncogenes can lead to cancer.  In addition, we know that in the standard multiple hit model of cancer, more than one mutation appears to be necessary for cancer to develop  (Knudson~\cite{Knudson1971}).  While this model indicates that genes play a causal role in the development of cancer, it is not an explanation of the functional cause of cancer. It gives no account of the functional, developmental dynamics of cancer.  We have no explanation as to why cancer develops as it does, why some cells in a tumor are cancerous and others not, why one cancer is dynamically different in phenotype and ontogeny than another, why some cancers are fast growing and others not. 

\subsection{A theoretical framework for understanding cancer}
We propose a general theoretical framework for understanding the dynamics of development of cancers.  Our theory is surprisingly simple and at the same time has extensive explanatory power.  In our view cancer is a special case of developmental multicellular processes generally.  The better we understand the development of organisms the better we will understand cancer. To model dynamics of cancer we need to relate the genome and cell architecture of cancerous cells with the dynamic development and differentiation of cancerous multicellular systems.  So too, since cell signaling is inherent to some cancers, we need to account for the relationship between cells, signals, genomes and cancer dynamics. Our theory will leave many open questions both theoretical and experimental. But that is as it should be.  Any new paradigm should propose new questions and problems with new approaches to answering and solving them.  It is hoped, therefore, that this paper will be a stimulus to new research based on this new paradigm. 

\subsection{Developmental networks and cancer}

We propose that pathological developmental control networks cause cancer. By a developmental control network we mean a control system of interlinked commands that changes the state of a cell or that of its offsprings. The network topology underlying all cancers contains at least one loop that leads to the original state that starts the cancer cascade. We can best illustrate the theory by a series of examples of typical cancers or cases.  For each case we provide the general structure of the genomic control network that underlies the cancer, we model what happens as the cancer develops, we make predictions of the expected clinical phenotype and we propose methods for differential diagnosis that distinguish the different types. 

\subsection{Implementation of developmental control networks}
Each of these cases of cancer networks are abstract and will require instantiation to particular {\em in vivo} cases.  The instantiation ultimately will also involve a molecular implementation of these networks.  However, the general principles presented here that govern the dynamics and development of a general  cancer type based on its network properties will apply to the special cases as well.  

For any abstract network many implementations are possible. For example, the networks may be implemented using protein transcription factors, but the same network could also be implemented using RNA.  The abstract properties of the developmental control network will be the same regardless of its implementation.  Thus nature had many different options for evolving developmental control networks. Indeed, we hypothesize that there was a switch in the implementation of genome control necessary for multicellular life to evolve. Even after the first primitive multicellular life evolved there may again have been new implementations of developmental networks necessary to achieve the increasing complexity of multicellular organisms.  

\subsection{Organizational information is not reducible to its parts}
The information in an organization such as a network is not, in general, contained in the parts used to implement or construct it.  This is because the properties of the components of the instantiation, be it using protein transcription factors or RNA or some other molecular mechanism, do not carry the information contained in the network composed out of those components.  The properties of the parts of a structure do not in general carry the information that is contained in that organized structure.  For example, the letters A, C, G, T do not carry the information contained in a sequence of these letters.  The interaction of the nucleic acids represented by A, C, G, T can generate a random sequence of such units, but the information in the parts is not sufficient to generate any particular sequence. Organization is not, in general, reducible to the information in its parts nor to the interaction of those parts.  

\subsection{Information and control of multicellular development}
%was {Control information} 

Essential to understanding cancer, and the development of organisms generally, is to understand how information functions to control what cells do.  We must understand not only how information can control a single cell, but also how it controls a dynamically developing, growing and dividing multicellular system of cells.  Thus, we assume that there is control information within a system of cells that controls the behavior and development of those cells.  Where is this control information located? I have argued elsewhere that the primary source and locus of this control information is in the genome and not primarily in the non-genomic portion of the cell (Werner~\cite{Werner2007a}).  While  control information is distributed throughout a cell, its primary source is the genome.  However, a great deal of this control information can be transferred to the cell from the genome.  Furthermore, signals from the cell itself, other cells and the environment interact with the genome's control networks leading to activation of new control networks and altered control states. In effect, the genome control architecture and the cell control architecture cooperate in process of development. 

\section{Developmental Control Networks (Cenes), the Cenome, and IES }
% was {The IES} 
We use the words {\em cene}, {\em developmental network}, {\em generative network}, {\em cytogenic network}, or {\em proliferative network} equivalently for a control network  that contains directive, control information that when interpreted and executed by the cell induces, controls and directs one or more cellular divisions resulting in a multicellular system.  We call developmental control networks {\em cenes} (for control genes) to distinguish them from genes that code for proteins.  Cenes are more like Mendel's original notion of a gene as the unit of inheritance that specifies some phenotypic property of the developing and mature multicellular organism.  'Gene', in its contemporary use, codes for protein parts of the cell.  To emphasize their role as parts, we call such genes {\em parts genes}.  Many organism share parts genes.  Many parts genes are evolutionarily conserved between distantly related species.   It has turned out that the genes in humans and chimps are 99\% identical, yet the organisms are patently very different. Parts genes cannot account for the vast  differences between organisms.  Just as a house and bridge may be built of the same parts yet have very different architectures, so the parts genes of two organisms may be the same yet they undergo different embryonic development resulting in very different phenotypes. 

We hypothesize that the difference in the development and morphology of multicellular organisms lies in their developmental control networks or cenes and not mainly in their parts genes.  Cenes can be combined and linked in many ways to form larger, more complex cenes. Indeed, we hypothesize that the rapid evolution of multicellular organisms involved the duplication, modification and combination of cenes to produce more complex cenes.   We define {\em cenome} to be the complete developmental control network encoded in the genome of an organism.  Thus, the cenome is just another cene that consists of numerous linked developmental subnetworks or sub-cenes.   Cenes and the cenome are the bases of the evolution and ontogeny of embryos and multicellular organisms. 

How can information in the genome control the actions of the cell? The cell and the genome form a cooperative system.  Since the information in the genome is primarily passive, the cell has to interpret and execute the control information in the genome.  We call this complex system the  {\em Interpretive Executive System} or \emph{IES}.   The IES of the cell interprets and executes the directives encoded in cenes.  Cell communication also is handled and interpreted by the IES. The IES interprets and executes both the  internal and the external signals the cell receives from its external environment, including other cells.  Such signaling systems interact by way of the IES with the genome to activate yet further control networks in the genome. 

In addition, genome control is self reflexive. The genome controls itself with the cooperation of the IES.   The genome interacts by way of the IES with itself producing a network of cascades of controlling information.  Implicit, therefore, in genomes are developmental control networks or cenes.   

\subsection{Self-sustaining cellular control networks versus developmental control networks}

It is essential not to confuse {\em developmental cenes} (developmental control networks) with {\em self-sustaining networks} such as metabolic gene networks and other non-developmental cellular molecular networks and pathways. The proteins produced by parts genes are not passive, but active molecules with agent like properties.  These molecular agents interact and self-assemble to produce complex cellular organization and networks of interacting molecular pathways that perform the essential life sustaining functions of the cell. This cellular organization including its networks of pathways is also the basis of sophisticated cellular behavior and strategic action vis a vis the environment.  Furthermore, the cell and the networks of proteins produced by parts genes form and constitute the system (the IES) that interprets and executes developmental control networks (cenes).  Thus, developmental control networks or cenes are a distinct level of control information in addition to the self-sustaining, intra-generational cellular control networks of the cell.  

\subsection{Pitchers, pots and catchers}

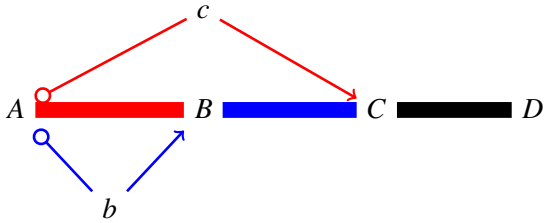
\begin{figure}[H]
\begin{tikzpicture}[style=mystyle]
\matrix (m) [matrix of math nodes, 
row sep=2em,
column sep=2em, 
text height=1.5ex, text depth=0.25ex]
{ \vphantom{a} & \vphantom{b}  &  c  &  \vphantom{c}  & \vphantom{c}\\
 A  &&  B  &&  C  && D \\
  \vphantom{a} &  b & \vphantom{b} &  \vphantom{d}  & \\ };
 \path[\poti]
 (m-2-1) edge [pot2, blue](m-3-2);
 \path[\pote]
(m-3-2) edge [pot2, blue] (m-2-3);
\path[\poti]
(m-2-1) edge [pot1, red](m-1-3);
 \path[\pote]
(m-1-3) edge [pot1, red] (m-2-5);
\path[solid, red, line width=6pt]
(m-2-1) edge (m-2-3);
\path[solid, blue, line width=6pt]
(m-2-3) edge (m-2-5);
\path[solid, black, line width=6pt]
(m-2-5) edge (m-2-7);
\end{tikzpicture} 
\caption{
    {\bf Network NDiv: A basic cell division network.} The developmental control network of the parent cell type $A$ divides into two terminal daughter cells $B$ and $C$. Two potential pitches pot1 with address $c$  activates control state $C$ in the first daughter cell while in parallel pot2 with address $b$ activates control state $B$ in the second daughter cell.  
  }
\label{fig:NDiv}
 \end{figure}

When a parent cell divides into two daughter cells that each differentiate into a different control state than that of the parent, something about the control state inherited from the parent cell has to change in the daughter cells.  This change we hypothesize is mediated by transcription factors or something like them that activate different areas of DNA in the daughter cells than were active in the parent cell. Each daughter cell gets a different activation unit from the parent.  In baseball a pitcher throws a ball to the catcher.  The event of throwing the ball is the pitch.  We use the pitcher-pitch-ball-catcher metaphor from baseball to describe the process of intergenerational control of cell states:  

A {\em potential pitcher} is an area of DNA in a parent cell that when executed throws a {\em potential pitch} or {\em pot} to a {\em catcher} or  {\em cat} in the daughter cell. A  {\em catcher} or {\em cat}  is an area of DNA in the daughter cell that catches the pot that was thrown by the parent cell. When the pot is caught by a catcher in the daughter cell, the pot binds to the corresponding area of DNA in the daughter cell.  These pots are inter-generational in that they pass from one generation to the next generation of cells.  We therefore call them {\em potential pitches} or {\em pots} because they are not active until after cell division when they are eventually caught by a catcher on the daughter cell DNA.  Thus, we distinguish {\em intra-generational pitches}, called {\em jumps} and which are caught within the lifetime of a cell, from {\em inter-generational potential pitches}  ({\em pots}) which are only caught by a daughter cells after the parent cell divides (\autoref{fig:NDiv}). 

We use the ball-arrow notation $A \circleright B$ to indicate a potential pitch or pot is thrown from a potential pitcher at area $A$ in the parent cell to a catcher at area $B$ in a daughter cell.  When the potential pitch is caught by a corresponding catcher at $B$ in a daughter cell, that daughter cell enters control state $B$.  When it is clear from the context, we use the term {\em pot} to refer either to the potential pitcher or to the potential pitch itself.  

When two daughter cells enter distinct control states (and/or phenotypic differentiation states mediated by control states), the parent cell must have thrown or sent each daughter cell a different pot.  By convention, we refer to theses possibly distinct pots as {\em pot1} and {\em pot2}, where it is specified prior to cell division which daughter cell receives pot1 and which daughter cell receives pot2. Two daughter cells enter identical control states when they receive identical pots, i.e., when pot1 = pot2. 

\subsection{Notation}  

By convention, we use upper case letters $A, B, C, \ldots$ for DNA potentially active control areas, lower case letters $a, b, c, \ldots$ for their respective addresses.  $a\pitcher$ is a {\em potential pitcher} that throws a potential pitch $a\potpitch$ with address $a$. $a\cat$ is a corresponding {\em catcher} with matching address $a$. Thus, $a\pitch$ is the {\em pot} thrown by pitcher $a\pitcher$ in the parent cell and caught by $a\cat$ in the daughter cell. We say a catcher $a\cat$ is \emph{loaded} if the pot has been caught, in symbols $a\catloaded$.  A catcher is {\em blocked} if it cannot be loaded, in symbols $a\catblocked$. 
%${\!\stackrel{-}{a\cat}}$ does not look good

In the graphical representation of developmental control networks (cenes), we usually leave out the details of the addressing structure of pitchers, pots and catchers since these are evident in the visual representation of the network links.  We show only the links, such as $A \circleright B$ (or at most $\stackrel{b}{A \circleright B}$), to indicate that an area $A$ throws a potential pitch to area $B$.  In more detail, $\stackrel{b}{A \circleright B}$ means that area $A$ has a pot pitcher $b\pitcher$ that when activated throws a potential pitch $b\potpitch$ with address $b$ to an area $B$ with catcher $b\cat$ with matching address $b$.  By convention, pot1 links are usually drawn on top of the genome and pot2 on the bottom (see \autoref{fig:NDiv}). 

\subsection{Simple cell division into two distinct cell types}

If we need more detail, $c\potone$ is a pot1 pitcher that throws a potential pot1 pitch $c\potonep$.  $b\pottwo$ is a pot2 pitcher that throws a pot2  pitch $b\pottwo$.  Combining pot1 and pot2 into synchronized throws to two daughter cells, an area $A^{c\potone}_{b\pottwo}$ that throws a pot1 pitch $c\potonep$ with address $c$ to the first daughter cell $C$ with catcher $c\cat$ and a pot2 pitch $b\pottwop$ with address $b$ to the second daughter cell $B$ with catcher $b\cat$.  Once the catchers are loaded, this activates the first daughter cell to state $C$ and the second daughter cell to state $B$.   

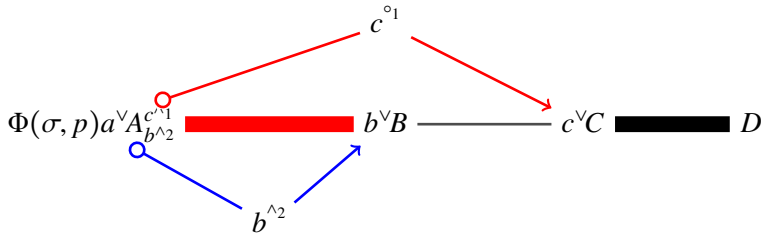
\begin{figure}[H]
\begin{tikzpicture}[style=mystyle]
\matrix (m) [matrix of math nodes, 
row sep=2em,
column sep=2em, 
text height=1.5ex, text depth=0.25ex]
{ \vphantom{a} & \vphantom{b}  &  c^{\potonep}  &  \vphantom{c}  & \vphantom{c}\\
 \Phi(\sigma, p) a\cat A^{c\potone}_{b\pottwo}  &&  b\cat B  &&  c\cat C  && D \\
  \vphantom{a} &  b^{\pottwo} & \vphantom{b} &  \vphantom{d}  & \\ };
 \path[\poti]
 (m-2-1) edge [pot2, blue](m-3-2);
 \path[\pote]
(m-3-2) edge [pot2, blue] (m-2-3);
\path[\poti]
(m-2-1) edge [pot1, red](m-1-3);
 \path[\pote]
(m-1-3) edge [pot1, red] (m-2-5);
\path[solid, red, line width=6pt]
(m-2-1) edge (m-2-3);
\path[solid, blue, line width=6pt]
(m-2-3) edge [pot1] (m-2-5);
\path[solid, black, line width=6pt]
(m-2-5) edge (m-2-7);
\end{tikzpicture} 
\caption{
    {\bf Network NDiv: Cell division network.} The developmental control network of the parent cell type $A$ divides into two terminal daughter cells $B$ and $C$. The \emph{ball-arrow notation} $\circleright$ links a parent cell with one of its daughter cells.  It indicates that when the parent cell's execution of the network reaches the control area on DNA that is next to the ball at beginning of the arrow then the parent cell divides and the linked resulting daughter cell activates the control area pointed to by the end of arrow.  In the above network, the cell in control state $A$ divides into daughter cells that differentiate into control states $B$ and $C$.  $C$ is terminal. The notation $A^{c\potone}_{b\pottwo} $ means that the area $A$ encodes pot1 with address $c$ and pot2 with address $b$.  The notation $b\cat B$ means that $B$ encodes a catcher with address $b$ that will catch the pitch $b\potonep$ thrown by pot2 ( $b\pottwo$ ) of $A$. The prefix $\Phi(\sigma, p)$ before the catcher $a\cat$ denotes further possible pre-activation conditions such as cell signaling, environmental, and other conditions (more on this below). 
  }
\label{fig:NDiv1}
 \end{figure}

Thus, each pot pitcher $\alpha\pitcher$ and its corresponding thrown pot $\alpha\pitch$ contains an address $\alpha$ that is matched by its corresponding catcher $\alpha\cat$.  We leave open how these addressing systems are implemented in DNA and we leave open how the intra and  inter-generational transfer mechanisms are molecularly implemented. 

An illustration of one possible scenario of how the pots might be transferred to their respective daughter cells.  While it may be helpful in understanding the process, this is just one of several possible implementations of the cell control inheritance.  Here the pots might attach to the opposite Centrioles or the opposite cell walls. In another scenario, the pots might directly attach to the chromosomes to be carried over to their intended daughter cells. Yet another scenario would have the replicated DNA itself marked (transformed by a viral like pot-vector) prior to division to be carried over and become active.  The point here is that any mechanism must in some way insure that different control pointers or pots are transferred to different daughter cells if they are to differentiate and develop differently. 

\begin{figure}[H]
\centering
\ifpdf
  \setlength{\unitlength}{1bp}%
  \begin{picture}(304.26, 170.08)(0,0)
  \put(0,0){\includegraphics{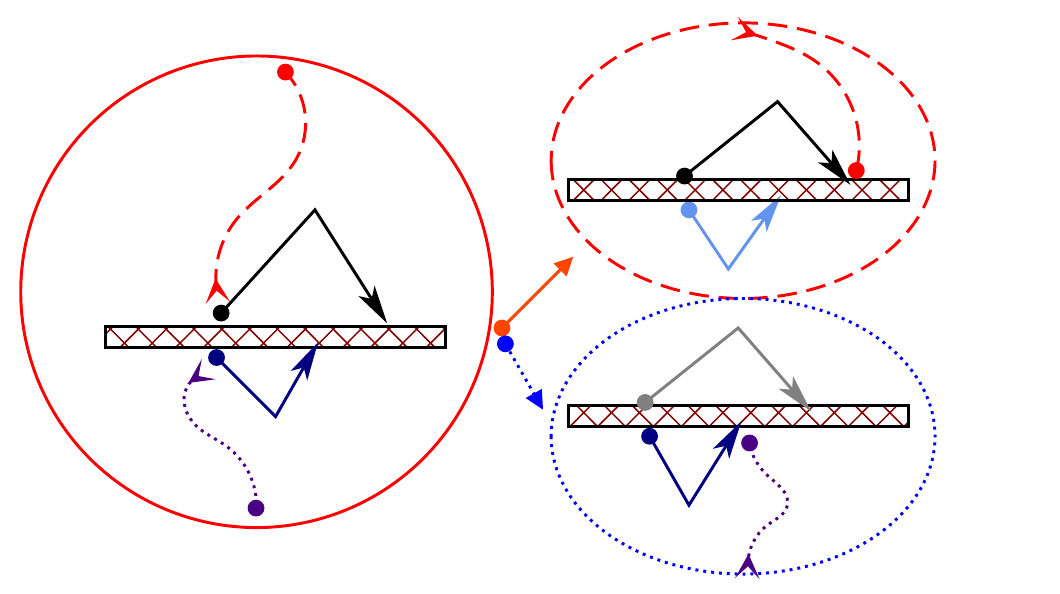}}
  \put(96.38,110.55){\fontsize{8.54}{10.24}\selectfont \makebox[0pt]{$b$}}
  \put(5.67,28.35){\fontsize{14.23}{17.07}\selectfont $A$}
  \put(272.13,133.23){\fontsize{14.23}{17.07}\selectfont $B$}
  \put(274.96,39.69){\fontsize{14.23}{17.07}\selectfont $C$}%B
  \put(85.04,45.35){\fontsize{8.54}{10.24}\selectfont \makebox[0pt]{c}}
  \put(252.28,124.72){\fontsize{8.54}{10.24}\selectfont $b^{\circ_{1}}$}
  \put(221.10,39.69){\fontsize{8.54}{10.24}\selectfont $c^{\circ_{2}}$}
  \put(62.36,28.35){\fontsize{8.54}{10.24}\selectfont $c^{\circ_{2}}$}
  \put(73.70,141.73){\fontsize{8.54}{10.24}\selectfont \makebox[0pt]{$b^{\circ_{1}}$}}
  \put(195.59,17.01){\fontsize{8.54}{10.24}\selectfont \makebox[0pt]{c}}
  \put(215.43,141.73){\fontsize{8.54}{10.24}\selectfont \makebox[0pt]{$b$}}
  \end{picture}%
\else
\begin{tikzpicture}[x=1.00mm, y=1.00mm, inner xsep=0pt, inner ysep=0pt, outer xsep=0pt, outer ysep=0pt]
\path[line width=0mm] (51.00,41.00) rectangle +(107.34,60.00);
\definecolor{L}{rgb}{1,0,0}
\path[line width=0.30mm, draw=L] (77.06,71.68) circle (23.96mm);
\definecolor{L}{rgb}{0.502,0,0}
\path[line width=0.15mm, draw=L]  (64.12,65.99) -- (61.98,68.12) (66.94,65.99) -- (64.81,68.12) (69.77,65.99) -- (67.64,68.12) (72.60,65.99) -- (70.47,68.12) (75.43,65.99) -- (73.30,68.12) (78.26,65.99) -- (76.13,68.12) (81.09,65.99) -- (78.95,68.12) (83.92,65.99) -- (81.78,68.12) (86.74,65.99) -- (84.61,68.12) (89.57,65.99) -- (87.44,68.12) (92.40,65.99) -- (90.27,68.12) (95.23,65.99) -- (93.10,68.12) (96.27,67.78) -- (95.93,68.12);
\path[line width=0.15mm, draw=L]  (61.73,67.38) -- (62.47,68.12) (63.16,65.99) -- (65.30,68.12) (65.99,65.99) -- (68.12,68.12) (68.82,65.99) -- (70.95,68.12) (71.65,65.99) -- (73.78,68.12) (74.48,65.99) -- (76.61,68.12) (77.31,65.99) -- (79.44,68.12) (80.13,65.99) -- (82.27,68.12) (82.96,65.99) -- (85.09,68.12) (85.79,65.99) -- (87.92,68.12) (88.62,65.99) -- (90.75,68.12) (91.45,65.99) -- (93.58,68.12) (94.28,65.99) -- (96.27,67.98);
\definecolor{L}{rgb}{0,0,0}
\path[line width=0.30mm, draw=L] (61.73,65.99) rectangle +(34.54,2.13);
\definecolor{L}{rgb}{0.392,0.584,0.929}
\path[line width=0.30mm, draw=L] (121.00,80.00) -- (125.00,74.00) -- (130.00,81.00);
\definecolor{F}{rgb}{0.392,0.584,0.929}
\path[line width=0.30mm, draw=L, fill=F] (121.00,80.00) circle (0.70mm);
\path[line width=0.30mm, draw=L, fill=F] (130.00,81.00) -- (127.80,79.13) -- (128.78,79.29) -- (128.94,78.31) -- (130.00,81.00) -- cycle;
\draw(85.00,80.00) node[anchor=base]{\fontsize{8.54}{10.24}\selectfont $a$};
\draw(53.00,51.00) node[anchor=base west]{\fontsize{14.23}{17.07}\selectfont $A$};
\definecolor{L}{rgb}{0.294,0,0.51}
\path[line width=0.30mm, draw=L, dash pattern=on 0.30mm off 0.50mm] (71.00,63.00) .. controls (70.57,62.78) and (70.22,62.43) .. (70.00,62.00) .. controls (69.30,60.66) and (69.86,59.06) .. (71.00,58.00) .. controls (71.88,57.18) and (73.03,56.71) .. (74.00,56.00) .. controls (76.20,54.39) and (77.35,51.71) .. (77.00,49.00);
\definecolor{F}{rgb}{0.294,0,0.51}
\path[line width=0.30mm, draw=L, fill=F] (71.00,63.00) -- (71.18,63.97) -- (70.42,62.61) -- (71.97,62.82) -- (71.00,63.00) -- cycle;
\path[line width=0.30mm, draw=L, fill=F] (77.02,49.70) circle (0.70mm);
\definecolor{L}{rgb}{1,0,0}
\path[line width=0.30mm, draw=L, dash pattern=on 2.00mm off 1.00mm] (73.00,72.00) .. controls (72.70,74.51) and (73.42,77.03) .. (75.00,79.00) .. controls (76.79,81.22) and (79.54,82.54) .. (81.00,85.00) .. controls (82.72,87.89) and (82.31,91.56) .. (80.00,94.00);
\definecolor{F}{rgb}{1,0,0}
\path[line width=0.30mm, draw=L, fill=F] (73.00,72.00) -- (73.76,71.37) -- (72.94,72.70) -- (72.37,71.24) -- (73.00,72.00) -- cycle;
\path[line width=0.30mm, draw=L, fill=F] (80.00,94.00) circle (0.70mm);
\path[line width=0.30mm, draw=L, dash pattern=on 2.00mm off 1.00mm] (126.50,85.00) [rotate around={180:(126.50,85.00)}] ellipse (19.50mm and 14.00mm);
\definecolor{L}{rgb}{0,0,1}
\path[line width=0.30mm, draw=L, dash pattern=on 0.30mm off 0.50mm] (126.50,57.00) [rotate around={180:(126.50,57.00)}] ellipse (19.50mm and 14.00mm);
\definecolor{L}{rgb}{0.502,0,0}
\path[line width=0.15mm, draw=L]  (111.34,80.99) -- (109.21,83.12) (114.17,80.99) -- (112.04,83.12) (117.00,80.99) -- (114.87,83.12) (119.83,80.99) -- (117.69,83.12) (122.66,80.99) -- (120.52,83.12) (125.48,80.99) -- (123.35,83.12) (128.31,80.99) -- (126.18,83.12) (131.14,80.99) -- (129.01,83.12) (133.97,80.99) -- (131.84,83.12) (136.80,80.99) -- (134.67,83.12) (139.63,80.99) -- (137.49,83.12) (142.45,80.99) -- (140.32,83.12) (143.27,83.01) -- (143.15,83.12);
\path[line width=0.15mm, draw=L]  (109.28,80.99) -- (111.41,83.12) (112.10,80.99) -- (114.24,83.12) (114.93,80.99) -- (117.06,83.12) (117.76,80.99) -- (119.89,83.12) (120.59,80.99) -- (122.72,83.12) (123.42,80.99) -- (125.55,83.12) (126.25,80.99) -- (128.38,83.12) (129.07,80.99) -- (131.21,83.12) (131.90,80.99) -- (134.04,83.12) (134.73,80.99) -- (136.86,83.12) (137.56,80.99) -- (139.69,83.12) (140.39,80.99) -- (142.52,83.12) (143.22,80.99) -- (143.27,81.04);
\definecolor{L}{rgb}{0,0,0}
\path[line width=0.30mm, draw=L] (108.73,80.99) rectangle +(34.54,2.13);
\definecolor{L}{rgb}{0.502,0,0}
\path[line width=0.15mm, draw=L]  (108.89,57.99) -- (108.73,58.15) (111.71,57.99) -- (109.58,60.12) (114.54,57.99) -- (112.41,60.12) (117.37,57.99) -- (115.24,60.12) (120.20,57.99) -- (118.07,60.12) (123.03,57.99) -- (120.90,60.12) (125.86,57.99) -- (123.72,60.12) (128.68,57.99) -- (126.55,60.12) (131.51,57.99) -- (129.38,60.12) (134.34,57.99) -- (132.21,60.12) (137.17,57.99) -- (135.04,60.12) (140.00,57.99) -- (137.87,60.12) (142.83,57.99) -- (140.69,60.12);
\path[line width=0.15mm, draw=L]  (108.90,57.99) -- (111.04,60.12) (111.73,57.99) -- (113.86,60.12) (114.56,57.99) -- (116.69,60.12) (117.39,57.99) -- (119.52,60.12) (120.22,57.99) -- (122.35,60.12) (123.05,57.99) -- (125.18,60.12) (125.87,57.99) -- (128.01,60.12) (128.70,57.99) -- (130.83,60.12) (131.53,57.99) -- (133.66,60.12) (134.36,57.99) -- (136.49,60.12) (137.19,57.99) -- (139.32,60.12) (140.02,57.99) -- (142.15,60.12) (142.84,57.99) -- (143.27,58.42);
\definecolor{L}{rgb}{0,0,0}
\path[line width=0.30mm, draw=L] (108.73,57.99) rectangle +(34.54,2.13);
\definecolor{L}{rgb}{1,0,0}
\path[line width=0.30mm, draw=L, dash pattern=on 2.00mm off 1.00mm] (127.00,98.00) .. controls (128.01,97.70) and (129.01,97.37) .. (130.00,97.00) .. controls (131.40,96.48) and (132.79,95.88) .. (134.00,95.00) .. controls (137.44,92.49) and (139.03,88.14) .. (138.00,84.00);
\path[line width=0.30mm, draw=L, fill=F] (127.00,98.00) -- (126.13,97.53) -- (127.67,97.80) -- (126.53,98.87) -- (127.00,98.00) -- cycle;
\path[line width=0.30mm, draw=L, fill=F] (138.00,84.00) circle (0.70mm);
\definecolor{L}{rgb}{0.294,0,0.51}
\path[line width=0.30mm, draw=L, dash pattern=on 0.30mm off 0.50mm] (127.00,44.00) .. controls (126.99,45.58) and (127.73,47.06) .. (129.00,48.00) .. controls (129.79,48.59) and (130.81,49.03) .. (131.00,50.00) .. controls (131.25,51.27) and (129.97,52.11) .. (129.00,53.00) .. controls (127.86,54.04) and (127.15,55.47) .. (127.00,57.00);
\definecolor{F}{rgb}{0.294,0,0.51}
\path[line width=0.30mm, draw=L, fill=F] (127.00,44.00) -- (127.66,43.27) -- (127.03,44.70) -- (126.27,43.34) -- (127.00,44.00) -- cycle;
\path[line width=0.30mm, draw=L, fill=F] (127.16,56.32) circle (0.70mm);
\definecolor{L}{rgb}{0,0,0.502}
\path[line width=0.30mm, draw=L] (117.00,57.00) -- (121.00,50.00) -- (126.00,58.00);
\definecolor{F}{rgb}{0,0,0.502}
\path[line width=0.30mm, draw=L, fill=F] (117.00,57.00) circle (0.70mm);
\path[line width=0.30mm, draw=L, fill=F] (126.00,58.00) -- (123.92,56.00) -- (124.89,56.22) -- (125.11,55.25) -- (126.00,58.00) -- cycle;
\path[line width=0.30mm, draw=L] (73.00,65.00) -- (79.00,59.00) -- (83.00,66.00);
\path[line width=0.30mm, draw=L, fill=F] (73.00,65.00) circle (0.70mm);
\path[line width=0.30mm, draw=L, fill=F] (83.00,66.00) -- (81.00,63.92) -- (81.96,64.18) -- (82.22,63.22) -- (83.00,66.00) -- cycle;
\definecolor{L}{rgb}{1,0.271,0}
\path[line width=0.30mm, draw=L] (102.00,68.00) -- (109.00,75.00);
\definecolor{F}{rgb}{1,0.271,0}
\path[line width=0.30mm, draw=L, fill=F] (102.00,68.00) circle (0.70mm);
\path[line width=0.30mm, draw=L, fill=F] (109.00,75.00) -- (107.52,74.51) -- (108.51,73.52) -- (109.00,75.00) -- cycle;
\definecolor{L}{rgb}{0,0,1}
\path[line width=0.30mm, draw=L, dash pattern=on 0.30mm off 0.50mm] (102.00,67.00) -- (106.00,60.00);
\definecolor{F}{rgb}{0,0,1}
\path[line width=0.30mm, draw=L, fill=F] (102.35,66.39) circle (0.70mm);
\path[line width=0.30mm, draw=L, fill=F] (106.00,60.00) -- (105.91,61.56) -- (104.70,60.87) -- (106.00,60.00) -- cycle;
\draw(147.00,88.00) node[anchor=base west]{\fontsize{14.23}{17.07}\selectfont $A$};
\draw(148.00,55.00) node[anchor=base west]{\fontsize{14.23}{17.07}\selectfont $B$};
\draw(81.00,57.00) node[anchor=base]{\fontsize{8.54}{10.24}\selectfont b};
\definecolor{L}{rgb}{0,0,0}
\path[line width=0.30mm, draw=L] (73.00,69.00) -- (83.00,80.00) -- (90.00,69.00);
\definecolor{F}{rgb}{0,0,0}
\path[line width=0.30mm, draw=L, fill=F] (73.47,69.52) circle (0.70mm);
\path[line width=0.30mm, draw=L, fill=F] (90.00,69.00) -- (89.09,71.74) -- (88.87,70.77) -- (87.91,70.99) -- (90.00,69.00) -- cycle;
\path[line width=0.30mm, draw=L] (120.00,83.00) -- (130.00,91.00) -- (137.00,83.00);
\path[line width=0.30mm, draw=L, fill=F] (120.55,83.44) circle (0.70mm);
\path[line width=0.30mm, draw=L, fill=F] (137.00,83.00) -- (135.68,85.57) -- (135.62,84.58) -- (134.63,84.65) -- (137.00,83.00) -- cycle;
\definecolor{L}{rgb}{0.502,0.502,0.502}
\path[line width=0.30mm, draw=L] (116.00,60.00) -- (126.00,68.00) -- (133.00,60.00);
\definecolor{F}{rgb}{0.502,0.502,0.502}
\path[line width=0.30mm, draw=L, fill=F] (116.55,60.44) circle (0.70mm);
\path[line width=0.30mm, draw=L, fill=F] (133.00,60.00) -- (131.68,62.57) -- (131.62,61.58) -- (130.63,61.65) -- (133.00,60.00) -- cycle;
\draw(140.00,85.00) node[anchor=base west]{\fontsize{8.54}{10.24}\selectfont $a^{\circ}$};
\draw(129.00,55.00) node[anchor=base west]{\fontsize{8.54}{10.24}\selectfont $b^{\circ}$};
\draw(73.00,51.00) node[anchor=base west]{\fontsize{8.54}{10.24}\selectfont $b^{\circ}$};
\draw(77.00,91.00) node[anchor=base]{\fontsize{8.54}{10.24}\selectfont $a^{\circ}$};
\draw(120.00,47.00) node[anchor=base]{\fontsize{8.54}{10.24}\selectfont b};
\draw(127.00,91.00) node[anchor=base]{\fontsize{8.54}{10.24}\selectfont $a$};
\end{tikzpicture}%
\fi
\caption{
{\bf A cell $A$ divides into two daughter cells $B$ and $C$.} This developmental control network regulates the division of a cell $A$ into two daughter cells $B$ and $C$.  At control state $A$ a two pitchers $b\pitcher$ and $c\pitcher$ throws two pot pitches $pot1 = b\potonep$ and $pot2 = c\pottwop$, respectively.  Prior to division these two pot types, pot1 and pot2, are separated  so that they will be transferred in opposite directions along the axis of division to different daughter cells, $B$ and $C$. After division these pots are transferred and caught by their catchers with matching addresses. $pot1 = b\potonep$ is caught by the catcher $b\cat$ and $pot2 = c\pottwop$ is caught by $c\cat$.  Once the catchers are loaded the daughter cells enters new control states, $B$ and $C$.  We leave open how prior and post cell division separation and transfer systems are implemented in the cell.
}
\end{figure}
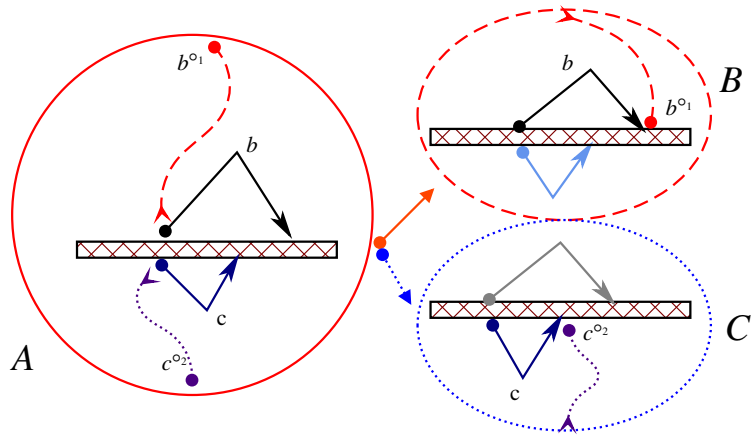

\subsection{Molecular implementation of developmental networks}

We use this special terminology of pots and their catchers for several reasons: First, we want to avoid confusion with normal transcription factors which are often intra-generational jumps. Second, we purposely abstract away from the details of the biological mechanisms that control cell division and differentiation.  Third, as a consequence, we leave open how pots are biologically instantiated or molecularly implemented. Fourth, the theory is a guide to developing and designing experiments to discover the actual biological implementations in multicellular animals and plants. Fifth, in particular, pots may or may not, in some or all cases, be implemented molecularly by protein transcription factor mechanisms.   

Thus, we want to leave the specific molecular implementation open because pots may be implemented by RNA in combination with protein transcription factors or other mechanisms as yet undiscovered.  Ultimately, we want to discover the syntax, semantics and pragmatics of the hidden source code of life underlying and controlling multicellular development.  By abstracting from the details of molecular implementation, we can focus on the universal properties of developing multicellular systems, and the corresponding universal properties of the control code and its meaning. 

\section{Architecture of developmental control networks}

By definition, \emph{development} will refer throughout the text as the generation of a multicellular system from one or more founder cells. We hypothesize that cancer is a form of development.  

A \emph{developmental control network} or \emph{cene} controls the development and ontogeny a multicellular system or organism. Developmental control networks have an architecture of interlinked control areas in genomes that control the states of cells. The architecture of cenes is based on the architecture of their addressing systems (Werner~\cite{Werner2011}). Developmental networks (cenes) can be interlinked to form larger developmental networks (cenes).   The nodes of developmental networks are built out of and linked to further more basic networks that control cell processes. Thus we distinguish \emph{cell differentiation networks} that control cell type from cenes which control multicellular development. Developmental networks or cenes are executed in parallel in a vast multicellular developing system.  

\subsection{Notation and Definitions}  

A {\bf cene} is a developmental control network.  \emph{Cene} stands for \emph{control gene} because they are more like Mendel's original concept than protein coding genes (Werner~\cite{Werner2011a}). The \emph{cenome} is the global developmental control network that controls the multicellular development, i.e., the embryogenesis and ontogeny of an multicellular organism.  An MCO is a multicellular organism or system. An MCS is a multicellular system.  Every MCO is an MCS. We propose that when an MCS develops from a single cell it does so by means of a developmental control network or cene.   

We use the phrases cenes, developmental control networks and developmental networks interchangeably.  However, because of the lack of familiarity with the term 'cene', we will often refer to cenes as developmental control networks, or simply developmental networks.  There is a danger of confusing developmental networks with metabolic networks, cell cycle networks, or other cellular networks that control cell differentiation. Cell cycle networks and cell differentiation networks are subsumed under (meta-controlled by) developmental control networks (cenes). 

Development of a multicellular system is \emph{mosaic} if that development is independent of outside communication.  It develops the same way, in an inflexible context independent way.  A network is \emph{mosaic} if the development it generates is mosaic.  Development that is not mosaic, but dependent on cell signaling or environmental conditions is called \emph{regulatory development}.  A network is \emph{regulatory} if the development it controls is regulatory.  Most networks will not be purely mosaic but instead are regulatory containing regulatory elements that respond conditionally to stimuli and signals.  Development is \emph{terminal} if cell proliferation eventually stops under all (normal) conditions. A network is {\em terminal} if the development it generates is terminal. This occurs if all possible developmental paths in the network terminate in a terminal node that halts with no further network controlled cell division.  Networks that contain antecedent activation conditions on nodes are called {\em conditional networks}. Regulatory networks are conditional on signal inputs.  A developmental path in a network is  \emph{nonterminal} if it contains a cycle or loop.  A network is {\em conditionally nonterminal} if all nonterminal paths have antecedent conditions for their activation.

\subsection{A small terminal developmental network}

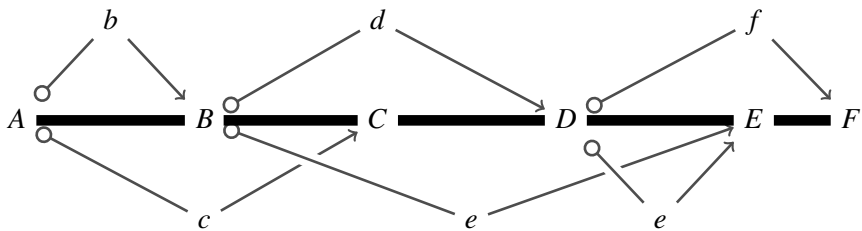
\begin{figure}[H]
\begin{tikzpicture}[style=mystyle]
%\matrix (m) [matrix of math nodes, fill=red!20, row sep=3em, %Note fill color
\matrix (m) [matrix of math nodes, 
row sep=2em, %\matrixrowsep, %0.75cm, %3em,
column sep=2em,%\matrixcolumnsep, %0.75cm, %3em, 
text height=1.5ex, text depth=0.25ex]
{ \vphantom{a} & b &  \vphantom{c}  & \vphantom{c}  & d  & \vphantom{d} &  \vphantom{d} & \vphantom{e} & f & \vphantom{e}\\
 A  && B && C  && D &&  E &  F\\
  \vphantom{a} &  \vphantom{b} & c &  \vphantom{c}  & \vphantom{b} &  e & \vphantom{e} & e\\ };
  \path[\poti]
(m-2-1) edge [pot1]( m-1-2);
\path[\pote]
(m-1-2) edge  [cross line, pot1] (m-2-3); %B = 3
  \path[\poti]
(m-2-1) edge [pot2 ](m-3-3);%yshift=-0.5cm
\path[\pote]
(m-3-3) edge [pot2] (m-2-5);%C = 5
%\path[\pot]
%(m-2-3) edge [inPot1,black, cross line,max distance=4cm, min distance=2cm] node[nodedescr] {$d$} (m-2-7); %D pot1 to A
  \path[\poti]
(m-2-3) edge  [pot1] (m-1-5);%B = 3 pot1 to D = 7
\path[\pote]
(m-1-5) edge [pot1] (m-2-7);%E = 9
  \path[\poti]
(m-2-3) edge [pot2] (m-3-6);
\path[\pote]
(m-3-6) edge [pot2] (m-2-9);
  \path[\poti]
(m-2-7) edge  [pot1] (m-1-9);%B = 3 pot1 to D = 7
\path[\pote]
(m-1-9) edge [pot1] (m-2-10);%E = 9
  \path[\poti]
(m-2-7) edge [pot2] (m-3-8);
\path[\pote]
(m-3-8) edge [pot2] (m-2-9);
\path[solid, line width=4pt]
(m-2-1) edge (m-2-3)
(m-2-3) edge (m-2-5)
(m-2-5) edge (m-2-7)
(m-2-7) edge (m-2-9)
(m-2-9) edge (m-2-10);
\end{tikzpicture}
\caption{ {\bf NBMC: A normal bounded multicellular network.} In the above network, the cell in control state $A$ divides into daughter cells that differentiate into control states $B$ and $C$.  $C$ is terminal.  $B$ divides into $D$ and $E$.  $D$ then divides into $E$ and $F$ both of which are terminal control states. The {\bf phenotype} of the multicellular system produced by this network starting at $A$ consists of five terminal cells: One each of type $C$, $D$, $F$ and two of type $E$.  
}
\label{fig:NBMC}
\end{figure}

Some normal developmental networks are terminal where all developmental paths result in terminal cells that no longer divide as seen in \autoref{fig:NBMC}.

%\section{Developmental networks}
  
\subsection{Normal Networks of type NN} 

\begin{figure}[H]
\begin{tikzpicture}[scale= \PicSize, 
nodedescr/.style={fill=white,inner sep=2.5pt},
description/.style={fill=white,inner sep=2pt},
cancer loop/.style={preaction={draw=white, -,line width=6pt}, line width=1.1pt},
pot1 loop/.style={preaction={draw=white, -,line width=6pt}, red, line width=1.1pt},
pot2 loop/.style={preaction={draw=white, -,line width=6pt}, green, line width=1.1pt},
pot1/.style={preaction={draw=white, -,line width=6pt}, gray, solid, line width=1pt}, 
pot2/.style={preaction={draw=white, -,line width=6pt}, gray, solid, line width=1pt},
inPot/.style={ out=45,in=90,distance=3cm, min distance=2cm, line width=1.1pt, black!75},
selfloop1/.style={ out=45,in=120,distance=6cm, min distance=4cm, line width=1.1pt, red},
selfloop2/.style={ out=-45,in=-135,distance=6cm, min distance=4cm, line width=1.1pt, blue},
loop1red/.style={ out=45,in=120,distance=6cm, min distance=4cm, line width=1.1pt, red},
loop2red/.style={ out=-45,in=-135,distance=6cm, min distance=4cm, line width=1.1pt, red},
cross line/.style={preaction={draw=white, -,	line width=6pt}}]
\matrix (m) [matrix of math nodes, row sep=3em,
column sep=3em, text height=1.5ex, text depth=0.25ex]
{ \vphantom{a} & \vphantom{b} &  \vphantom{c}  & \vphantom{c}  & &&\\
 A  &&  N_x  &&  N_y && D \\
 \vphantom{a} &  n_x & \vphantom{B} &  \vphantom{c}  & \vphantom{C} & \vphantom{d} & \vphantom{D}\\ };
\path[\poti]
%(m-2-1) edge [pot1]( m-1-2)
%(m-1-2) edge  [cross line, pot1] (m-2-5)
(m-2-1) edge [pot2, blue](m-3-2);%to B
\path[\pote]
(m-3-2) edge [pot2, blue] (m-2-3)
(m-2-1) edge [inPot,red, cross line,\pot] node[nodedescr] {$n_y$} (m-2-5); %A to Nx
%(m-2-7) edge [loop2red, cross line] node[nodedescr] {$ a $} (m-2-1); %end of Nx = D to A
%(m-2-3) edge [selfloop2, cross line] node[nodedescr] {$ \alpha_2 $} (m-2-1);
\path[solid,black!20, line width=6pt]
(m-2-1) edge (m-2-3);
\path[solid,blue!90, line width=10pt]
(m-2-3) edge (m-2-5);
\path[dotted,red!70, line width=10pt]
(m-2-5) edge (m-2-7);
%\draw[decorate,decoration={crosses,segment length=4pt}] (m-2-5) edge (m-2-7);
\end{tikzpicture}
\caption{
    {\bf Network NDev: Normal Developmental Network}  A divides into a cell controlled by a bounded network $N_x$ and a cell controlled by network $N_y$.  The daughter cell's developmental network $N_x$ may or may not generate further multicellular development and structure with possibly multiple cell types. 
  }
  \label{fig:NDev}
\end{figure}
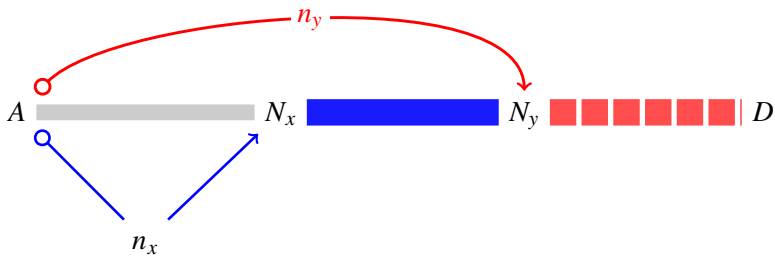

A {\em normal network} NN (\autoref{fig:NDev}) is a noncancerous developmental genomic network with possible signal inputs.  It is a non-cancerous network.  It may contain loops that do housekeeping chores that are continually active; however these do not generate abnormal cells.  It may contain conditional loops that induce some rounds of cellular development but only if there are signals that indicate some special situation such as a wound or other circumstance.  We will see that any normal developmental sub-network can become cancerous given the right mutations. 

In exponential cell division a cell divides into two daughter cells that both divide.  Normal developmental processes may be exponential for a time in that the daughter cells of a cell also generate two daughter cells that divide.  However, as we will see,  geometric networks show that this does not always lead to exponential growth.

\subsection{Networks that produce identical cell types}
\begin{figure}[H]
\begin{tikzpicture}[scale= \PicSize, 
nodedescr/.style={fill=white,inner sep=2.5pt},
description/.style={fill=white,inner sep=2pt},
cancer loop/.style={preaction={draw=white, -,line width=6pt}, line width=1.1pt},
pot1 loop/.style={preaction={draw=white, -,line width=6pt}, red, line width=1.1pt},
pot2 loop/.style={preaction={draw=white, -,line width=6pt}, green, line width=1.1pt},
pot1/.style={preaction={draw=white, -,line width=6pt}, gray, solid, line width=1pt}, 
pot2/.style={preaction={draw=white, -,line width=6pt}, gray, solid, line width=1pt},
inPot1/.style={ out=60,in=120,distance=3cm, min distance=2cm, line width=1.1pt, black!75},
inPot2/.style={ out=-60,in=-120,distance=3cm, min distance=2cm, line width=1.1pt, black!75},
selfloop1/.style={ out=45,in=120,distance=6cm, min distance=4cm, line width=1.1pt, red},
selfloop2/.style={ out=-45,in=-135,distance=6cm, min distance=4cm, line width=1.1pt, blue},
loop1red/.style={ out=45,in=120,distance=6cm, min distance=4cm, line width=1.1pt, red},
loop2red/.style={ out=-45,in=-135,distance=6cm, min distance=4cm, line width=1.1pt, red},
cross line/.style={preaction={draw=white, -,	line width=6pt}}]
\matrix (m) [matrix of math nodes, row sep=3em,
column sep=3em, text height=1.5ex, text depth=0.25ex]
{ \vphantom{a} & \vphantom{b} &  \vphantom{c}  & \vphantom{c}  & &&\\
 A  &&  N_x  &&  N_y  && D \\
 \vphantom{a} &   \vphantom{b} & \vphantom{B} &  \vphantom{c}  & \vphantom{C} & \vphantom{d} & \vphantom{D}\\ };
 \path[\pot]
(m-2-1) edge [inPot1,red, cross line] node[nodedescr] {$n_x$} (m-2-3)%A to Nx
(m-2-1) edge [inPot2,blue, cross line] node[nodedescr] {$n_x$} (m-2-3); %A to Nx
\path[solid,black!20, line width=6pt]
(m-2-1) edge (m-2-3);
\path[solid,blue!90, line width=10pt]
(m-2-3) edge (m-2-5);
\path[solid,red!70, line width=10pt]
(m-2-5) edge (m-2-7);
\end{tikzpicture}
\caption{
    {\bf Network NI$^1$: Normal Identical Cell Developmental Network}  A divides into two cells both controlled by a bounded network $N_x$.  Both daughter cell networks may or may not generate further multicellular development and structure with possibly multiple cell types. 
  }
  \label{fig:NI1}
\end{figure}
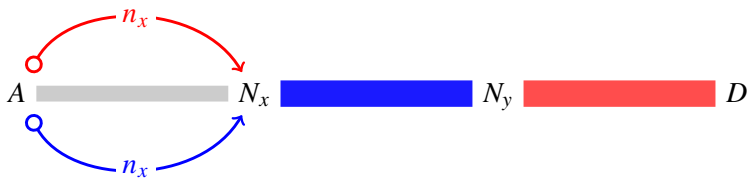

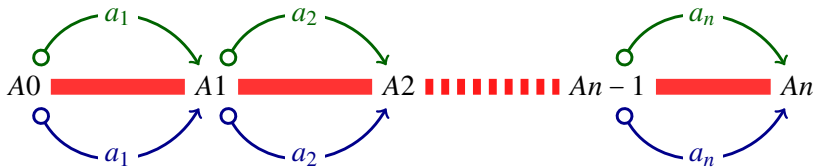
\begin{figure}[H]
\begin{tikzpicture}[scale= \PicSize, 
nodedescr/.style={fill=white,inner sep=2.5pt},
description/.style={fill=white,inner sep=2pt},
cancer loop/.style={preaction={draw=white, -,line width=6pt}, line width=1.1pt},
pot1 loop/.style={preaction={draw=white, -,line width=6pt}, red, line width=1.1pt},
pot2 loop/.style={preaction={draw=white, -,line width=6pt}, green, line width=1.1pt},
pot1/.style={preaction={draw=white, -,line width=6pt}, gray, solid, line width=1pt}, 
pot2/.style={preaction={draw=white, -,line width=6pt}, gray, solid, line width=1pt},
inPot1/.style={ out=60,in=120,distance=3cm, min distance=2cm, line width=1.1pt, black!75},
inPot2/.style={ out=-60,in=-120,distance=3cm, min distance=2cm, line width=1.1pt, black!75},
selfloop1/.style={ out=45,in=120,distance=6cm, min distance=4cm, line width=1.1pt, red},
selfloop2/.style={ out=-45,in=-120,distance=6cm, min distance=4cm, line width=1.1pt, blue},
loop1red/.style={ out=45,in=120,distance=6cm, min distance=4cm, line width=1.1pt, red},
loop2red/.style={ out=-45,in=-80,distance=6cm, min distance=4cm, line width=1.1pt, red},
cross line/.style={preaction={draw=white, -,	line width=6pt}}]
\matrix (m) [matrix of math nodes, row sep=3em,
column sep=2em, text height=1.5ex, text depth=0.25ex]
{ \vphantom{a} & \vphantom{b} &  \vphantom{c}  & \vphantom{c}  & & \vphantom{c}\\
 A0  &&  A1  & \vphantom{b} &  A2 && An-1 && An && \\
 \vphantom{a} &  \vphantom{b} & \vphantom{B} &  \vphantom{c}  & \vphantom{C} & \vphantom{d} & \vphantom{D}\\ };
 \path[\pot]
(m-2-1) edge [inPot1,DarkGreen, cross line] node[nodedescr] {$a_1$} (m-2-3) %A to Nx
(m-2-1) edge [inPot2, DarkBlue, cross line] node[nodedescr] {$a_1$} (m-2-3) %A to Ny
(m-2-3) edge [inPot1, DarkGreen, cross line] node[nodedescr] {$a_2$} (m-2-5) %A to Nx
(m-2-3) edge [inPot2, DarkBlue, cross line] node[nodedescr] {$a_2$} (m-2-5) %A to Ny
(m-2-7) edge [inPot1, DarkGreen, cross line] node[nodedescr] {$a_{n}$} (m-2-9) %A to Nx
(m-2-7) edge [inPot2, DarkBlue, cross line] node[nodedescr] {$a_{n}$} (m-2-9); %A to Ny
\path[solid,red!80, line width=6pt]
(m-2-1) edge (m-2-3);
\path[solid,red!80, line width=6pt]
(m-2-3) edge (m-2-5);
\path[dashed,red!90, line width=6pt]
(m-2-5) edge (m-2-7);
\path[solid,red!80, line width=6pt]
(m-2-7) edge (m-2-9);

\end{tikzpicture}
\caption{
    {\bf Network NI$^n$ produces $2^n$ identical cells in $n$ steps:} This network generates $n$ identical divisions to produce $2^n$ identical daughter from one founder cell.    }
      \label{fig:NIn}
\end{figure}

Some networks produce identical cell types.  In this case, the daughter cell networks may or may not generate further multicellular development and structure with possibly multiple cell types (Figures \ref{fig:NI1} and \ref{fig:NIn}).

\subsection{Stem cell networks}

Another basic class of developmental control networks are stem cell networks.  Stem cell networks are nonterminal. Some stem cell networks are unconditionally nonterminal while most are conditionally nonterminal.  We will discuss these in detail when we discuss cancer stem cell networks (see \autoref{sec:NSC}).  Cancer stem cell networks and normal stem cell networks share many of their topological and dynamic properties. Crucially, normal stem cell networks have preconditions for their activation as well as normal cell differentiation patterns.  We will discuss stem cell networks in greater detail in \autoref{sec:NSC}.

\subsection{Orthogonal categories of developmental networks} 

While the networks determine ideal growth rate they also determine cell differentiation and specify organism phenotype in space and time.  Differentiation and organization architectures are orthogonal categories of developmental networks, forming extra dimensions in addition  to the network proliferative architecture.  There are several interrelated properties that can be used to classify developmental networks:

\begin{enumerate} 

\item {\bf Network Architecture:} The network architecture as given by the global interlinking of the developmental control network is one dimension that influences phenotype of developing multicellular systems and organisms. This determines its proliferative capacity,  its developmental dynamics and its complexity.

\item {\bf Differentiation:} Cell differentiation into various cell types is controlled by differentiation control networks that are subsumed under the more global developmental control networks.  Developmental networks can be subclassed according to the kind and number of {\em cell types} they generate and, thereby, implicitly according to the cell differentiation sub-networks they activate.  

\item {\bf Conditionality:}
In addition to an activating site or catcher, a network may have further activation conditions that must hold before some of its nodes are activated and executed.  Networks that contain activating conditions are called {\em conditional networks}.  
%\pagebreak
\item {\bf Cell interactivity:} 
Cell interactivity is closely related to conditionality of networks.    
\begin{enumerate}
\item {\bf Cell communication:} 
Cell to cell signaling protocols and environmental conditions can activate developmental networks including cancer networks. 
\item {\bf Cell physical interactions:}
Cell cohesion and cohesion cell strategies can give a cell the capacity to invade other tissue.  
\end{enumerate}
\item {\bf Regulatory versus mosaic development:} How autonomous is the developmental network from cell communication?
 \item {\bf Stochasticity:} Is the network deterministic or does it have stochastic elements? 

\item {\bf Complexity:} How complex is the network and how complex is the organism it generates? What is the relationship between the complexity of the cell and the complexity of the genome?

\item {\bf Physics:} How dependent is the generated multicellular dynamics and development dependent on the physics of interacting 							cells?

\item {\bf Multi-Cellular System (MCS) spatial and temporal organization:} \\
Another orthogonal category of a network is the {\em MCS-phenotypic organization} (the space-time development and morphology) of a multicellular cancer generated by the network.  

\end{enumerate}

In general, nonterminal and terminal, normal and stem cell, conditional and non-conditional, deterministic and stochastic, communicative and non-communicative networks will be mixed together in one large eclectic, evolutionary network.  
 
\subsection{Development and cancer}

The space-time developmental phenotype of a multicellular system such as cancer is reflected in the cancer network that controls it.  On this view, cancers are pathological multicellular systems controlled by cancer networks that, apart from their deviant architecture, are similar to the networks that control the development of healthy multicellular systems. In other words, cancer is a type of multicellular development similar to normal embryonic development. 

\subsubsection{Cell interactions and cancer}

The effect of cancer control networks will be more or less severe to the extent that cellular interaction can lead to novel outbreaks of cancers spawned by such interactions. For simplicity, we assume that the cancers do not have pathogenic interactions with other cell types unless specifically stated in the example case.  In other words, we assume that cancer is \textit{monoclonal}, i.e., arising from a single cell.  This is usually taken to imply what is actually an additional assumption, namely, that cancer cells cannot influence non-cancerous cells to become cancerous.  However, there are situations of cell communication in which this latter implicit assumption need not hold. 

\subsubsection{Self serving interactions in cancer} 

We also assume that a cancer can engage in self serving interactions in order to gain resources for growth from helper cell systems such as blood vessels.  While these are necessary conditions for the growth of cancers they are tertiary in terms of its conceptual and causal essence even if many of the adverse side effects and spread of cancer in the body depend on such tertiary conditions. 

\section{Classification of cancers by their network architecture}
%was Cancer Network Motifs

We will now go into more detail about the nature of the networks that lie at the origin of all cancers.  Cancer networks can be classified into basic types whose network architecture and topology determines whether the cancers are linear, geometric, or exponential in their proliferative dynamics, as well as determining their general clinical phenotype. Each general class of cancer can be further divided into vast variety of subtypes.  Each cancer subtype may differ diagnostically, clinically and dynamically according to the particular architecture of its regulatory cancer network.  

Each of these subtypes of cancer has been modeled and simulated with software. The mathematical equations that describe particular ideal growth rates of cancer types have been computationally verified. The computationally predicted, static and dynamic phenotypes of these artificial \textit{in silico} cancers show strong correlations with the clinical, static and dynamic phenotypes of \textit{in vivo}, naturally occurring cancers. 

\subsection{Infinite cancer loops}

Fundamental to all cancer is simple repetition of action.  A cell divides and results in at least one daughter cell that is also a cancer cell.  If both daughter cells were normal then the so called cancer cell would  proliferate normally being finitely bounded and all its offspring would be non-cancerous.  Hence, no cancer.  Thus, cancer cells must generate cancer cells if the cancer is to continue.  What must the state of the cancer cell be like to generate another cell that is also cancerous?  

\subsubsection{An example of simple infinite looping}
%was {A simple example of looping program}

You are told to follow the following instructions in their proper order starting at 1 then 2 then 3, etc. And don't stop until you are told to do so by the instructions. 

\begin{enumerate}
\item Open the door.
\item Close the door.
\item Go to instruction 1 and follow it.
\end{enumerate}

Clearly, if you follow the instructions, you will open the door; then you will close the door.  Then you will read instruction 3 and jump to instruction 1, which tells you to open the door, which you do. Next, you read instruction 2, and close the door; next you read instruction 3, and so on. Hence, if you follow the instructions you are in an endless loop and you never stop.  

This sort of program is quite common in computer science. Often it appears in introductory courses when students write a program that "hangs", because it never stops precisely because the student has unknowingly written an infinite loop into the program. However, such loops also have their uses in control programs like operating systems which use endless loops to continuously operate the system.  The computer has the ability for outside input to stop these loops, for which students are quite thankful.  In the case of cancer, each proliferative loop generates new cells and thereby causing endless growth.  Here too there are functional proliferative loops where some cells, such as stem cells, continuously regenerate cells, e.g., skin cells, hair or fingernails.  Hence, what is defined as a cancer will depend on not just on the existence of proliferative loops, but on the nature of the proliferative network and its functional context. 

\subsection{Basic types of cancer networks}
%Fundamental cancer network categories
%was The Developmental Network Architectures of Cancers

We hypothesize that all developmental multicellular processes are controlled by developmental control networks.  We distinguish normal networks from cancerous control networks.  Keep in mind, there are however cases where the control network may not be the distinguishing factor, when, for example, a stem cell gains the capacity to invade other tissues.   These are lower level control strategies of the cell that interact with the developmental control networks.  More on this later. 

According to our network theory, cancer is a regulated developmental multicellular process resulting from the transformation of a normal developmental control network.   All cancers can be grouped into three broad classes according to the architecture of their control networks and their corresponding developmental dynamics. 

\begin{enumerate}

\item {\bf Linear cancers NL } (see~\autoref{sec:NL}) \\
Linear cancer networks generate cancers that proliferate linearly in time.  Linear cancers include 1st order stem cells \autoref{fig:LSC1}. 

\item {\bf Exponential cancers NX}  (see~\autoref{sec:NX})\\ 
Exponential cancer networks generate cancers that proliferate exponentially as a function of time. 

\item {\bf Geometric cancers NG} (see~\autoref{sec:NG})\\
A category of cancer networks lying between linear and exponential cancer networks are the geometric cancer networks whose ideal competence is to grow at a rate in accordance with one of the geometric numbers. We will see that they are related to coefficients of the binomial theorem and Pascal's Triangle (see  \autoref{sec:Pascal}).  Geometric networks are related to the concept of a meta-stem cell, i.e., a stem cell that produces stem cells. 

\item {\bf Cancer stem cell networks} (see \autoref{sec:NSC})\\
Cancer stem cell networks are a subset of geometric cancer networks.  They are closely related to normal stem cell networks. We will describe: 
\begin{enumerate}
\item {\bf Deterministic stem cell networks}  follow a well defined, determined developmental path (see \autoref{sec:NSC1}).
\item {\bf Stochastic stem cell networks} (\autoref{sec:NSSC}) follow developmental paths probabilistically. Depending on their network topology and their developmental probability distribution, stochastic stem cell networks can emulate many of the other cancer networks.  Thus,  some stochastic networks can jump out of their category (e.g., linear or geometric) into another cancer class (e.g., exponential) depending on their developmental topology and probability distribution. 
\end{enumerate}

\item {\bf Reactive signaling networks} react to signals that determine their topology (see \autoref{sec:Rsig}).

\item {\bf Communicating cancer networks} (see \autoref{sec:SigC})\\ %sec:Rsig
Many cancers involve communication with other cells and the environment.  Each of the above cancer networks can combined with or implemented by cell signaling communication protocols resulting in more flexible multicellular development generally and specifically, more flexible cancer networks.  Indeed, many cancers result from mutational transformations of the cell signaling protocols.  

\item {\bf Complex cancer networks in context of developmental networks}\\
All these cancer networks are part of developmental control networks (cenes), that can be combined with other normal and cancer developmental networks (resulting in more complex cenes) to produce complex cancer dynamics together with complex cellular and multicellular, organizational, structural phenotypes.  Moreover, depending on their position in the developmental network hierarchy, and hence, their position and activation in the spatial-temporal development of the organism,  the very same cancer network can exhibit very different cellular and multicellular dynamics and phenotypes (e.g., bilateral cancers, complex teratomas, fetus in fetu). 

\end{enumerate}

\section{Linear cancers}
\label{sec:NL}
The most basic type of cancer network generates linear cancers, e.g., basal cell carcinoma or grade I glioma of the brain.  In a linear cancer the number of cells increases as a linear function of the number of given cells at any given time.  %More on this below.  

{\bf General Phenotype:} Clinically linear cancers are slow growing when compared to geometric or exponential cancers.  With linear cancers no additional cancer cells are generated by the cancer network. In deterministic, non-stochastic linear cancer networks, number of cancer cells in the tumor stays constant.  Each execution of the cancer loop  will result in the generation of new cells, but they are not cancerous in the sense that the cancer network is not active. 

While a linear cancer cell does not produce active cancer cells (other than one daughter cell with the same control state as itself), linear cancer cells produce \emph{potential} cancer cells that have inherited the linear cancer network.  If, by some means,  this network is activated in the passive daughter cell then we have another linear cancer cell.  

Linear cancers have network architectures similar to stem cell networks (see below \autoref{fig:LSC1}). Linear control networks are also common in non-cancerous tissue, examples include warts, finger nails, and hair growth.

%\subsubsection{Network NL\protect\ref{fig:NL} Linear growth,  possibly benign, cancer} 
\subsection{Linear cancer network NL} 

%\vspace*{1cm}
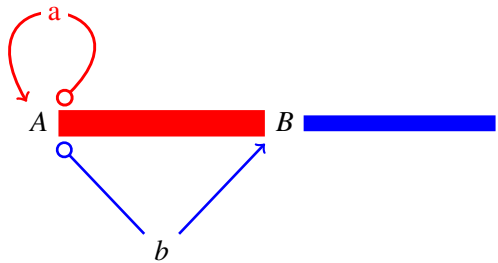
\begin{figure}[H]
\begin{tikzpicture}[style=mystyle]
\matrix (m) [matrix of math nodes, row sep=3em,
column sep=3em, text height=1.5ex, text depth=0.25ex]
{ \vphantom{a} & \vphantom{b}  &  \vphantom{c}  & \vphantom{c}  & \\
 A  &&  B  &&  \vphantom{C}  && \vphantom{D} \\
  \vphantom{a} &  b & \vphantom{b} &  \vphantom{d}  & \\ };
 \path[\poti]
%(m-2-1) edge [pot1]( m-1-2)
%(m-1-2) edge  [cross line, pot1] (m-2-3)
(m-2-1) edge [pot2, blue](m-3-2);
 \path[\pote]
(m-3-2) edge [pot2, blue] (m-2-3);
\path[\pot]
%(m-1-2) edge [pot1 loop, bend left=-80] node[nodedescr] {$ \alpha_1 $} (m-1-1)
(m-2-1) edge [selfloop1, cross line] node[nodedescr] {a} (m-2-1);

\path[solid, red, line width=10pt]
(m-2-1) edge (m-2-3);
\path[solid, blue, line width=6pt]
(m-2-3) edge (m-2-5);
%(m-2-5) edge (m-2-7);
\end{tikzpicture} 
\caption{
    {\bf Network NL: Linear cancer network.} The developmental control network of the parent cell type $A$ self-loops giving a daughter cell of the same cell type $A$ as the parent. The other daughter cell differentiates to type $B$.  The network generates a slow growing cancer containing only one cancer cell $A$.  All the generated cells contain the cancer network.  The difference is that in cells of type $B$ the cancer network that generated them is not active. 
  }
\label{fig:NL}
 \end{figure}

The simplest cancer network NL generates only one cancer cell.  NL controls a cancer that exhibits linear growth and is possibly benign.  The network NL controls  cell division where a cell of type $A$ divides into a cell of type $B$ and a cell of type $A$.  
 
{\bf Phenotype:} This network will result in a tumor whose outward phenotype consists of two main cell types, $A$ and $B$.  There will be only one single cell type $A$ that is cancerous.  The rest will be all cells of type $B$ or, if $B$ is non-terminal, cells derived from type $B$.  The growth rate will depend on the rate of cell division. The complexity (maturity) of the cell type $A$ may influence how quickly it can divide.  More complex cells may take longer to duplicate all their parts making division more complex.  However, whether the cells simple or complex, the growth rate is linear with no increase in tumor cells. Irrespective of growth rate, such tumors are benign in the sense that their source is one or more single cells controlled by a linear cancer network while in all the other cells in the tumor are non-cancerous because the linear cancer network is inactive.  

For example, if the given number of cells is $g$ at time $t$ it will tend to increase by no more than some constant number $c$ to give  $g + c$  after the next round through the cancer loop.  $c$ represents the number of active linear cancer cells in the tumor. The units of time are not necessarily constant units such as seconds, hours, days or months. But they are determined by how long it takes for each of the cancer cells to go through its developmental cancer cycle. Thus, after $n$ rounds of the cancer loop the number of cells will be no more than: 
\begin{equation}
\Cells(n) = n*c + g %\mbox{Cells}(r) = r*c + n
\end{equation}

{\bf Treatment:} Since such tumors have only a few cancer cells that are actually proliferating, the prime objective is to remove these active cancer cells by some method. However, the passive cells generated by the active linear cancer cells still have the mutation, i.e., they have inherited the cancer network even if it is inactive.  Thus, if the network is a stochastic cancer stem cell network and there is a nonzero probability of reactivation then these cancer nets could become active again forming new cancer stem cells (e.g., see the section on linear stochastic caner stem cells \autoref{fig:LSSC}). Hence, depending on the probability of spontaneous reactivation, removal of such inactive tumor cells might still be indicated. Radiation therapy would be relatively contraindicated because of the danger of increasing the likelihood of reactivation of existing passive cancer cells, transforming linear cancer stem cells into exponential cancer stem cells, or generating new cancer cells.    

The network NL results in a benign cancer to the extent that the cells of type $A$ are minimally interactive with other cells leading to no cancerous induction of other cells.  Thus, cell interactivity is an independent component that is relevant in assessing phenotypes of the dynamics of cancer.  Just one network mutation can convert a linear cancer into an aggressive exponential cancer (see \autoref{fig:NX}). 

\subsection{An interactive, conditional linear cancer network NIL}

An interactive or conditional linear cancer network is just like the linear cancer network except that it must be activated by some internal or external signal $\sigma$ with some property $\Phi$.  In this case a passive cell $B$ generated by the network, can revert into a cancer cell of type $A$ given the right signaling context. This may be the context of other cells.  If, for example, the cell $B$ invades other tissue with the appropriate intercellular signaling protocol then $B$ may become cancerous in the new multicellular context. 

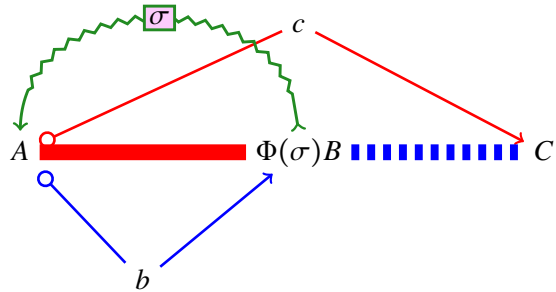
\begin{figure}[H]
\begin{tikzpicture}[style=mystyle]
\matrix (m) [matrix of math nodes, row sep=3em,
column sep=3em, text height=1.5ex, text depth=0.25ex]
{ \vphantom{a} & \vphantom{b}  &  \vphantom{c}  & \vphantom{c} & c  & \\
 \vphantom{C} && A  &&  \Phi(\sigma)B  &&  C  && \vphantom{D} \\
  \vphantom{a} & \vphantom{b} & \vphantom{b} &  b  & \\ };
 \path[\poti]
(m-2-3) edge [pot1, red](m-1-5);%A to C
 \path[\pote]
(m-1-5) edge [pot1, red] (m-2-7); 
 \path[\poti]
(m-2-3) edge [pot2, blue](m-3-4);%A to B
 \path[\pote]
(m-3-4) edge [pot2, blue] (m-2-5);
\path[\sigjump]
 (m-2-5) edge [selfloop1, out=90, in=90, green, receivestyle] node[receivesigstyle] {$\sigma$} (m-2-3);
\path[solid, red, line width=6pt]
(m-2-3) edge (m-2-5);
\path[dashed, blue, line width=6pt]
(m-2-5) edge (m-2-7);
\end{tikzpicture} 
\caption{
    {\bf Network NIL: Interactive linear cancer network.} The developmental control network of the parent cell type $A$ divides into daughter cells $B$ and $C$.  Given a context that sends a signal $\sigma$ with property $\Phi$, then the network generates a slow growing linear cancer.  If the tumor is generated from stem cell $A$ then the tumor contains only one cancer stem cell $A$ which becomes the signal dependent cancer cell $B$ after division.  The rest are terminal cells of type $C$.  Thus, cells of type $B$ have the potential to be cancerous given a signal $\sigma$ with property $\Phi$. 
  }
\label{fig:NIL}
 \end{figure}
 
{\bf Phenotype:} The multicellular phenotype of this cancer network is just like that of the linear network NL except that it contains three cell types $A$, $B$ and $C$ where  $A$ and $B$ transform into each other. Cell $A$ differentiates to $B$ after division. Cell $B$ transforms/dedifferentiates into $A$ given a signal $\sigma$ with property $\Phi$.  Thus, in different contexts $B$ cells may revert to type $A$ cancer cells. This may appear to be stochastic, but is actually determined by some environmental or cellular signal. 

While we will not describe most of them explicitly, all the cancer networks described below can have interactive variants. 
 
\subsection{An interactive linear and stochastic exponential cancer network}

An interactive  linear cancer stochastic exponential network behaves like a linear cancer but turns exponential, with propability $p$, in a signaling context. Given an internal or external signal $\sigma$ with some property $\Phi$, a passive cell $B$ generated by the network from a cell of type $A$, can with probability $p$ revert into that parent cancer cell $A$. Thus, the cellular communicative context can make a passive cell into a cancer cell.  If, for example, the cell $B$ invades other tissue with the appropriate intercellular signaling protocol then $B$ may become cancerous in the new multicellular context. In this network the proliferation turns exponential in a signaling context, but remains linear when no signaling occurs.  

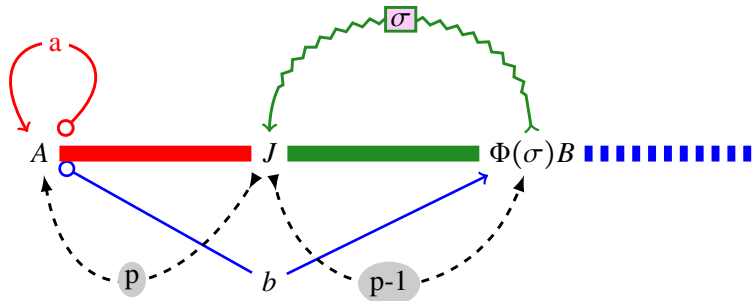
\begin{figure}[H]
\begin{tikzpicture}[style=mystyle]
\matrix (m) [matrix of math nodes, 
row sep=3em,
column sep=3em, 
text height=1.5ex, text depth=0.25ex]
{ \vphantom{a} & \vphantom{b}  &  \vphantom{c}  & \vphantom{c}  & \\
A && J && \Phi(\sigma)B  && \vphantom{b}   &&  \vphantom{C}  && \vphantom{D} \\
  \vphantom{a} &\vphantom{b} & b & & &   & \\ };
 \path[\sigjump]
 (m-2-5) edge [selfloop1, out=90, in=90, green, receivestyle] node[receivesigstyle] {$\sigma$} (m-2-3);
 \path[\poti]
%(m-2-1) edge [pot1]( m-1-2)
%(m-1-2) edge  [cross line, pot1] (m-2-3)
(m-2-1) edge [pot2, blue](m-3-3);
 \path[\pote]
(m-3-3) edge [pot2, blue] (m-2-5);
\path[stochasticPathstyle]
%\colorlet{currentcolor}{green}%to set node color a percent of path color
%(m-1-2) edge [pot1 loop, bend left=-80] node[nodedescr] {$ \alpha_1 $} (m-1-1)
(m-2-3) edge [selfloop1, black, out=-80, in=-110] node[stochasticNodestyle] {p-1} (m-2-5);
\path[stochasticPathstyle]
(m-2-3) edge [selfloop1, out=-120, in=-80, black] node[stochasticNodestyle] {p} (m-2-1);
\path[\pot]
(m-2-1) edge [selfloop1, cross line] node[nodedescr] {a} (m-2-1);
\path[solid, red, line width=6pt]
(m-2-1) edge (m-2-3);
\path[solid, green, line width=6pt]
(m-2-3) edge (m-2-5);
\path[dashed, blue, line width=6pt]
(m-2-5) edge (m-2-7);
%(m-2-5) edge (m-2-7);
\end{tikzpicture} 
\caption{
    {\bf Network NISXL: Interactive linear cancer network.} The developmental control network of the parent cell type $A$ self-loops giving a daughter cell of the same cell type $A$ as the parent. The other daughter cell differentiates to type $B$.  The cell type $B$ has receptors for a signal $\sigma$.  $B$ jumps to state $J$ on receipt of a signal $\sigma$ with property $\Phi$.  Depending on the probability $p$, a cell in state $J$ dedifferentiates either to its grandparent $A$ or its parent $B$.  If it dedifferentiates to $A$ it makes a step in the direction of exponential growth.  Otherwise, it just cycles back to B.   Thus, cells of type $B$ can become cancerous with probability $p$ given a signal $\sigma$ with property $\Phi$.  Interactive cancers are more dangerous since the passive $B$ cells can spontaneously become cancerous in some new multicellular or environmental context.  
  }
\label{fig:NISXL}
 \end{figure}
 
{\bf Phenotype:} The multicellular phenotype of this interactive stochastic cancer network is just like that of a linear cancer network containing only two cell types $A$ and $B$.  However, in different cellular and environmental signaling contexts some of the $B$ cells may stochastically revert to type $A$ cancer cells. Unlike the deterministic communication protocol in the interactive linear network this only activates $A$ with some probability $p$. Hence, even if the right communication context is present,  the lower the probability the lower the chance of generating more cancer cells. As the probability $p$ approaches one and there an ample source of signals $\sigma$ this network approximates an exponential network. As $p$ approaches zero the signals $\sigma$ have no effect and there is no dedifferentiation from $B$ to $A$ making the network linear. When $p = 1/2$ and there is sufficient signaling, half the $B$ cells will dedifferentiate to $A$ cells.  If $B$ is in a signal $\sigma$ free context, the tumor behaves linearly. Once the $B$ cells are in a $\sigma$ signaling context, the tumor behaves exponentially with probability $p$. 

Note, when $B$ loops back to $A$ by way of $J$, the two loops do not form a second order geometric network, because both loops go back to the same parent cell $A$.  

\subsection{An almost invariant transformation of a liner network Network NLJ}

Certain transformations on particular networks leave those networks invariant with respect to the rate of cell growth.  For example, a linear pragmatically equivalent control network to NL contains a control jump that dedifferentiates cell type C back into its parent cell type $A$. 

\begin{figure}[H]
\begin{tikzpicture}[style=mystyle]%, scale=0.4] %scale= \PicSize
\matrix (m) [matrix of math nodes, row sep=3em,
column sep=3em, text height=1.5ex, text depth=0.25ex]
{ \vphantom{a} & \vphantom{b}  &  c  & \vphantom{c}  & \\
 A  &&  B  &&  C  && \vphantom{D} \\
  \vphantom{a} &  b & \vphantom{c} &  \vphantom{d}  & \\ };
 \path[\poti]
%(m-2-1) edge [pot1]( m-1-2)
%(m-1-2) edge  [cross line, pot1] (m-2-3)
(m-2-1) edge [pot1, red](m-1-3);
\path[\pote]
(m-1-3) edge [pot1, red] (m-2-5);
\path[\poti]
(m-2-1) edge [pot2, blue](m-3-2);
\path[\pote]
(m-3-2) edge [pot2, blue] (m-2-3);
\path[right hook->]
%Note, the graphics will be pushed to the right if the angle is bigger e.g., 120 or more
(m-2-5) edge [selfloop1, bend left=-90, dashed] node[nodedescr] {$a$} (m-2-1);
\path[solid, red, line width=6pt]
(m-2-1) edge (m-2-3);
\path[solid, blue, line width=6pt]
(m-2-3) edge (m-2-5);
%(m-2-5) edge (m-2-7);
\end{tikzpicture} 
\caption{
{\bf Network NLJ an almost invariant transform of the linear cancer network NL} 
 An almost equivalent linear cancer with one loop and two cell types but three control states. The dashed arrow represents a jump within the cell cycle and does not involve cell division.  Cell $A$ divides into two daughter cells with control states $B$ and $C$. Daughter cell $C$ jumps to the control state of its parent cell $A$.  The functional result is a linear proliferative network practically equivalent to net  NL~\autoref{fig:NL} above. 
 }
\end{figure}
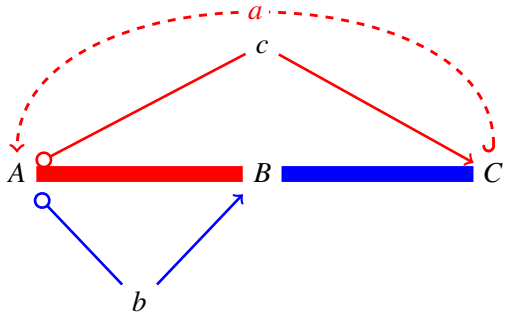

{\bf Phenotype:}  Even though the networks NL and NLJ are almost equivalent, the tumor phenotype of NLJ may show three distinct cell types (A, B and C) and just the two (A and B) of NL. 

%replacing teratoma with multi-cell-type NV Replaced teratoma with mixed cell types
\subsection{Linear mixed cell network  NLM: Linear mixed cell cancer with different cell types and structures} 
%was N2 then replaced NLLT with NLM
\label{sec:NLM}

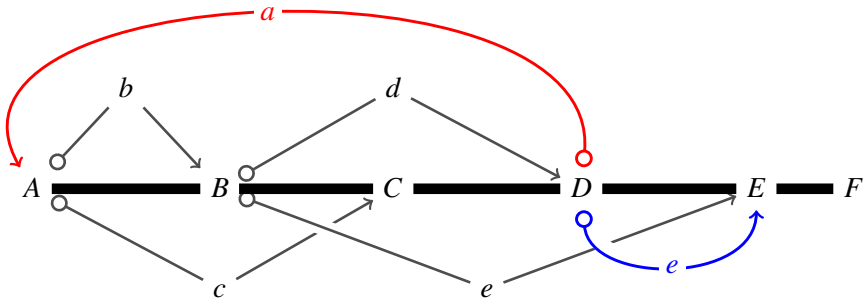
\begin{figure}[H]
\begin{narrow}{-1cm}{-1cm}
\begin{tikzpicture}[style=mystyle]
\pgfpathmoveto{\pgfpointorigin};
\pgftransformxshift{-100.5cm};
%\matrix (m) [matrix of math nodes, fill=red!20, row sep=3em, %Note fill color
\matrix (m) [matrix of math nodes, row sep=2em,
column sep=2em, text height=1.5ex, text depth=0.25ex]
{ \vphantom{a} & b  &  \vphantom{c}  & \vphantom{c}  & d & \vphantom{d} &  \vphantom{d} & \vphantom{e}\\
 A  &&  B  &&  C  && D && E & F\\
  \vphantom{a} &  \vphantom{b} & c &  \vphantom{c}  & \vphantom{b} &  e & \vphantom{e} & \vphantom{e}\\ };
  \path[\poti]
(m-2-1) edge [pot1]( m-1-2);
\path[\pote]
(m-1-2) edge  [cross line, pot1] (m-2-3); %B = 3
  \path[\poti]
(m-2-1) edge [pot2 ](m-3-3);%yshift=-0.5cm
\path[\pote]
(m-3-3) edge [pot2] (m-2-5);%C = 5
  \path[\poti]
(m-2-3) edge  [pot1] (m-1-5);%B = 3 pot1 to D = 7
\path[\pote]
(m-1-5) edge [pot1] (m-2-7);%E = 9
  \path[\poti]
(m-2-3) edge [pot2] (m-3-6);
\path[\pote]
(m-3-6) edge [pot2] (m-2-9);
%  \path[-]
%(m-2-7) edge [pot2] (m-3-8);%D pot2 to E
\path[\pot] 
(m-2-7) edge [inPot1,out=85,in=120,max distance=6cm,red, cross line] node[nodedescr] {$a$} (m-2-1) %D pot1 to A
(m-2-7) edge [inPot2,out=-85,in=-90,distance=3cm, min distance=2cm, blue, cross line] node[nodedescr] {$e$} (m-2-9); %D pot2 to E
\path[solid, line width=4pt]
(m-2-1) edge (m-2-3)
(m-2-3) edge (m-2-5)
(m-2-5) edge (m-2-7)
(m-2-7) edge (m-2-9)
(m-2-9) edge (m-2-10);
\end{tikzpicture}
\end{narrow}
\caption{
{\bf NLM: Linear mixed cell cancer}. The ball-arrow notation $\circleright$ links a parent cell with one of its daughter cells.  It indicates that when the parent cell's execution of the network reaches the control area on DNA that is next to the ball at beginning of the arrow then the parent cell divides and the linked resulting daughter cell activates the control area pointed to by the end of arrow.  In the above network, the cell in control state $A$ divides into daughter cells that differentiate into control states $B$ and $C$.  $C$ is terminal.  $B$ divides into $D$ and $E$.  $E$ is terminal.  A cell in control state $D$ divides into $E$ and $A$.  The link from $D$ to $A$ creates an infinite control loop that, given no inhibiting preconditions, results in endless cellular division.  This sort of single loop architecture occurs both in stem cells and in some cancers (see \autoref{sec:NSC}).  The phenotype of this network is an endlessly growing tumor generating two noncancerous cell types ($C$ and $E$) for each loop, yet there is only one cancer cell that cycles through three cell types ($A$, $B$, and $D$) which are part of the cancer loop. Hence, it is very slow growing tumor with one cancer stem cell (more on this below).
}
\label{fig:NLM}%replaced NLLT with NLM
\end{figure}

This linear cancer network results from a deleterious network transformation of the normal network NN (\autoref{fig:NBMC}) above.   Specifically, it changed the pot1 address of cell control state $D$ to point to $A$ instead of $F$.  The cell state $F$ is now inaccessible by this network.  

% syntax:  \node [options] (name) {label};
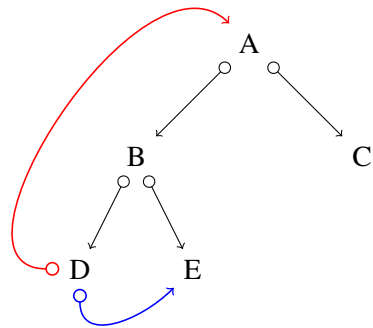
\begin{figure}[H]
\begin{center}
\begin{tikzpicture}[%scale=\PicSize, 
level distance=1.5cm,
  level 1/.style={sibling distance=3cm,\pot},
  level 2/.style={sibling distance=1.5cm,\pot}]
  \node(A) {A}
    child {node(B) {B}
      	child {node(D) {D}
			%child{node(E){F}}
			%child{node(G){}}
		}
      	child {node(E) {E}
      		%child{node(F){F}}
			%child{node(G){}}
		}
    }
    child {node(C) {C}
    		%child {node {lright}}
      	%child {node(t) {rright}}
    };
    \draw[semithick,red, \pot] (D)..controls +(west:2) and +(north west:2) .. (A);
    \draw[semithick,blue, \pot] (D)..controls +(south:1) and +(south west:1) .. (E);
\end{tikzpicture}
\caption{
{\bf NLMtree: Tree view of a Linear mixed cell cancer}.  This is the same network as above, but in a tree view that shows the branching structure of the network.  The network on DNA is the linearized version of this tree graph with loops. The de-differentiating loop in the tree from $D$ to $A$, makes this a never ending cell proliferating network. 
}
\end{center}
\label{fig:NLMtree}%replaced fig:NLLTtree with fig:NLMtree
\end{figure}

\subsection{Linear single teratoma network NLTs}

If we define a {\em teratoma} to be any cancer that generates one or more noncancerous cell types organized in some structure, then we get a whole range of teratomas, from {\em simple teratoma networks} generating just a few non cancerous cell types, such as the mixed cell networks, with no recognizable structure to {\em complex teratoma networks} that generate structure like hair, bone, etc.

{\bf Phenotype:} Under this broad definition of teratoma networks can range from the simple to very complex.  In the traditional usage, teratomas have multicellular structure in addition to diverse cell types.  We can further distinguish bounded teratomas from cancerous teratomas.  A {\em bounded teratoma} results from the activation of a developmental subnetwork that terminates when its cells are differentiated and its possible stem cells are governed by their normal linear control networks.   A bounded teratoma network is thus a natural network in the genome that has been inappropriately activated in time and/or space.  An {\em cancerous teratoma network} is a bounded teratoma network linked inside one or more linear, geometric  or exponential cancer network loops.   For example, the case of a {\em fetus in fetu} (a multicellular system resembling a fetus developing inside the body)  may be  a highly developed teratoma resulting from either a bounded teratoma network or a cancerous teratoma network.  A phenotype of repeating structures would be evidence of a cancerous teratoma network. 

Note, too that in general complex teratomas may contain stem cells. Then if the stem cell loop is linked to the cancerous backwards loop at $D$ then we will get geometric growth where $A$ is in effect a meta-stem cell.

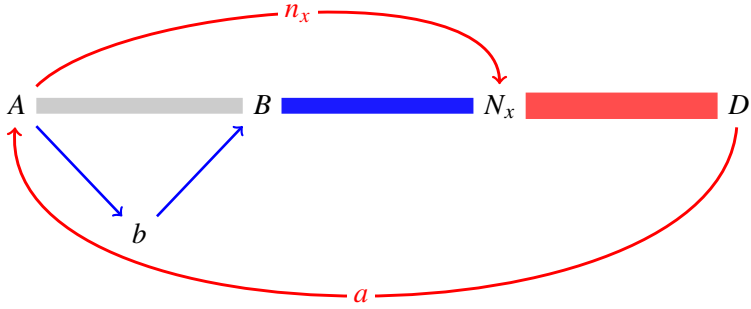
\begin{figure} [H]
\begin{tikzpicture}[style=mystyle]
\matrix (m) [matrix of math nodes, row sep=3em,
column sep=3em, text height=1.5ex, text depth=0.25ex]
{ \vphantom{a} & \vphantom{b} &  \vphantom{c}  & \vphantom{c}  & &&\\
 A  &&  B  &&  N_x && D \\
 \vphantom{a} &  b & \vphantom{B} &  \vphantom{c}  & \vphantom{C} & \vphantom{d} & \vphantom{D}\\ };
\path[->]
%(m-2-1) edge [pot1]( m-1-2)
%(m-1-2) edge  [cross line, pot1] (m-2-5)
(m-2-1) edge [pot2, blue](m-3-2)%to B
(m-3-2) edge [pot2, blue] (m-2-3)
(m-2-1) edge [inPot1, out=45,in=90,red, cross line] node[nodedescr] {$n_x$} (m-2-5) %A to Nx
(m-2-7) edge [loop2red, out=-95,in=-95,distance=6cm, min distance=4cm,cross line] node[nodedescr] {$ a $} (m-2-1); %end of Nx = D to A
%(m-2-3) edge [selfloop2, cross line] node[nodedescr] {$ \alpha_2 $} (m-2-1);
\path[solid,black!20, line width=6pt]
(m-2-1) edge (m-2-3);
\path[solid,blue!90, line width=6pt]
(m-2-3) edge (m-2-5);
\path[solid,red!70, line width=10pt]
(m-2-5) edge (m-2-7);
\end{tikzpicture}
\caption{
    {\bf Network NLTs: Linear single teratoma cancer} where A divides into a cell of type B and a cell controlled by network $N_x$ that generates a structure with possibly multiple cell types before it loops back to A.
  }
  \label{fig:NLTs}
\end{figure}
 
In  the network NLTs in \autoref{fig:NLTs} results from two steps:  First, we amplify the network NL  by appending a more complex network $N_x$ into the genome.  Next, we modify the network link that led from cell type A to point to this new subnetwork while keeping the link to network for daughter cell $B$ the same. In linking the network NL to $N_x$, the cell of type A will divide into an $N_x$-cell with the potency $N_x$ and a differentiated B-cell.  We assume $N_x$ is a semi terminal network that contains no loops except at cell type D.   One of the developmental paths in $N_x$ produces a cell of type D which  dedifferentiates into cell type A again or divides into daughter cells of type A and some other cell type F.  Under the influence of subnetwork $N_x$, the $N_x$-cell goes through a series of differentiations and possible divisions. 

{\bf Remark 1 on NLTs} Assuming synchronous divisions in an ideal, discrete, force free space then NLTs produces $k = N_x(t)$ cells after $t$ rounds of division. Let H(D) be the path in $N_x$ to cell state D. H(D) is the history of network states in $N_x$ that end in state D.  Let there only be one path in $N_x$ leading to state D.  Let |H(D)| be the length of the path H(D). Assume H(D) be a longest path in $N_x$. The $N_x(|H(D)|) = k$ is the total number of cells produced by $N_x$ when it reaches cell state D. 

{\bf Remark 2 on NLTs} We could just count the number of cell divisions in the path H(D).  Or could just count the number of cells $k$ produced by $N_x$ when state D is reached. Assuming D is a longest path then $k$ is the total number of cells produced by $N_x$ when it reaches D. We need a notation for the number of cells produced by a network N when it reaches a cell state X.  If N is terminal, we need a notation for the total number of cells produced by N.  Let $N^*$ be the set of paths in N.  Let $N(X)^*$ or $N^*(X)$ be the set of all paths up to the length of path H(X). We need to be able to talk maximal paths in N and the length |H| of paths  H in N.  We need to talk of a bounded network within an infinite looping network.  

The {\bf clinical phenotype} appears as follows: A cancer controlled by such a network will generally, depending on the complexity of the network $N_x$, will be a more complex tumor with multiple cell types in the $N_x$ cascade together with cells of type $B$.  It may again be benign, with linear non-exponential growth.  Only one cell will be the cancer stem cell.  All the rest will be byproducts. Note however, the growth rate of network $N_x$ may be much slower than a cancer with network NL.  This is because the cell of type A will have to go through multiple differentiations and possibly divisions before it returns to its state $A$.   The salient feature of this type of cancer will be the ratio of $N_x$ generated cells to cells of type $B$.  Let $k$ be the number of cells produced by one loop of $A \rightarrow N_x \rightarrow A$, then the ratio of tumor cells will be $k$ cells of type $A \rightarrow N_x \rightarrow A$ to one of type $B$.  Thus, assuming the network $N_x$ produces $n$ $B$ cells, then a count of $B$ cells in the tumor will show how many loops the cancer has executed.  
  
Let $k$ be the number of cells produced by one round of $N_x$ when it reaches cell state D. Assuming the path to D is a longest path in $N_x$ then multiplying the number $n$  of $B$ cells by $k$ will give the number of cells generated by the subnetwork $N_x$ up to that point.  Since the number of $B$ cells generated by the network NLTs after $n$ loops is just $n$,  the ideal total number Cells$(n)$  of cells generated by the tumor network at loop count $n$ is then given by the following formula:  

%The ideal total number of cells given the count b of cells of type $B$ is: 
%\begin{figure}
\begin{equation}
\mbox{Cells}(n) =  k*n + n
\end{equation}
%\caption{Total number of cells given the count b of cells of type B}
%\end{figure}

Thus, the total number of cells after $n$ loops consists of  $k*n$ cells of type $N_x$ plus $n$ cells of type $B$. 

Note, all equations that give the cell count after a number of rounds of divisions are {\em upper bounds} that indicate the maximum possible number of cells generated given ideal synchronous development without physical constraints.  Given real physics with friction, tension, cell connectivity and other such physical constraints the number of cells generated will in general be less than the upper bound. 

\subsection{A linear double teratoma network NLTd:}

In  a linear double teratoma network a cancer cell A divides into two daughter cells each of which is controlled by a teratoma network. Compare this to the above linear single teratoma network NLTs.  A linear double teratoma still has linear growth because only one of the teratoma subnetworks $N_x$ loops back to A.

\begin{figure}[H]
\begin{tikzpicture}[style=mystyle]
\matrix (m) %\mymatrixmode does not work
[matrix of math nodes, row sep=3em,column sep=3em, text height=1.5ex, text depth=0.25ex]
{ \vphantom{a} & \vphantom{b} &  \vphantom{c}  & \vphantom{c}  & &&\\
 A  &&  N_y  &&  N_x && D \\
 \vphantom{a} &  n_y & \vphantom{B} &  \vphantom{c}  & \vphantom{C} & \vphantom{d} & \vphantom{D}\\ };
 \path[\poti]
(m-2-1) edge [pot2, blue](m-3-2);%to B
 \path[\pote]
(m-3-2) edge [pot2, blue] (m-2-3);
 \path[\pot]
 (m-2-1) edge [inPot1,out=45,in=120,min distance=4cm,red, cross line] node[nodedescr] {$n_x$} (m-2-5) %A to Nx
(m-2-7) edge [inPot2, out=-90,in=-120,distance=6cm,min distance=4cm,cross line, blue] node[nodedescr] {$ a $} (m-2-1); %end of Nx = D to A

\path[solid,black!20, line width=6pt]
(m-2-1) edge (m-2-3);
\path[solid,blue!90, line width=10pt]
(m-2-3) edge (m-2-5);
\path[solid,red!70, line width=10pt]
(m-2-5) edge (m-2-7);
\end{tikzpicture}
\caption{
    {\bf Network NLTd: Linear double teratoma cancer} where A divides into a cell controlled by a bounded teratoma network $N_y$ and a cell controlled by network $N_x$ that generates a structure with possibly multiple cell types including cell type D that loops back to A.
  }
  \label{fig:NLTd}
\end{figure}
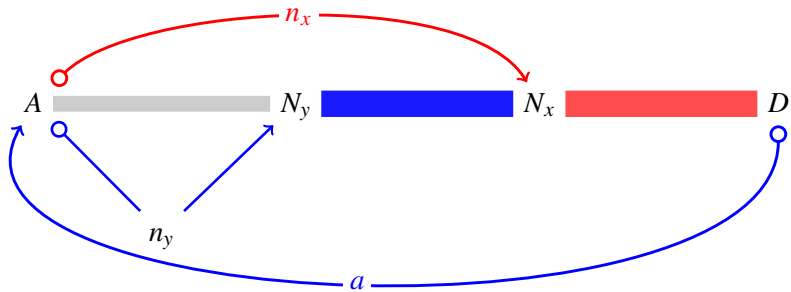

{\bf Phenotype:} A linear double teratoma tumor has all the structures of the linear single teratoma tumor plus the cell types and structures produced by the second teratoma network $N_y$.  Because the cells of the two teratoma networks may interact both physically and via cell signaling, apparently novel structures and cell types may result with the linear double teratoma that are not generated by either of the teratoma networks singly.  Recall, our more general definition of teratoma as any network that produces multiple cell types and/or structures. Hence, double teratomas can range from the very simple to very complex.

\subsection{Multi linear cancer networks NLI$_k$ of identical cells}

A linear $k$-identical cell cancer network NLI$_k$ links an identical cell network NI$_k$ that generates $k$ identical cells with a downstream linear cancer network. The simplest case is where the number of identical linear cancer cells is $k=2$: 

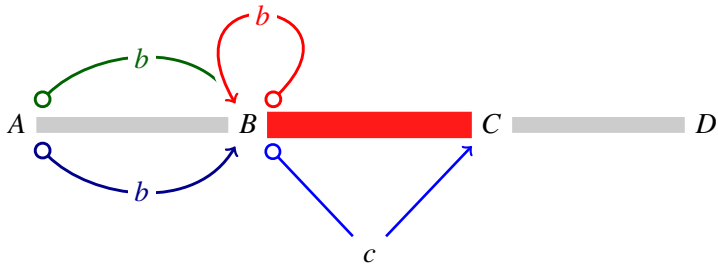
\begin{figure}[H]
\begin{tikzpicture}[style=mystyle]
\matrix (m) [matrix of math nodes, row sep=3em,
column sep=3em, text height=1.5ex, text depth=0.25ex]
{ \vphantom{a} & \vphantom{a}  &  \vphantom{c}  & \vphantom{c}  & \\
 A  &&  B  &&  C  && D \\
  \vphantom{n} & \vphantom{a} & \vphantom{b} &  c  & \\ };
 \path[\pot]
(m-2-1) edge [inPot1, DarkGreen]node[nodedescr] {$ b $}( m-2-3);
 \path[\pot] 
(m-2-1) edge [inPot2, DarkBlue]node[nodedescr] {$ b $}( m-2-3); 
%to A
%(m-1-2) edge  [cross line, pot1] (m-2-3)
 \path[\poti]
(m-2-3) edge [pot2, blue](m-3-4);
 \path[\pote]
(m-3-4) edge [pot2, blue] (m-2-5);
%(m-1-2) edge [pot1 loop, bend left=-80] node[nodedescr] {$ \alpha_1 $} (m-1-1)
 \path[\pot]
(m-2-3) edge [selfloop1, cross line] node[nodedescr] {$ b $} (m-2-3);

\path[solid,black!20, line width=6pt]
(m-2-1) edge (m-2-3);
\path[solid,red!90, line width=10pt]
(m-2-3) edge (m-2-5);
\path[solid,black!20, line width=6pt]
(m-2-5) edge (m-2-7);
\end{tikzpicture}
\caption{
    {\bf Network {\em NLI}$_2$: A multi-linear cancer of identical cells} where the number of starting identical cancer cells $n = 2$. The parent divides into two identical daughter cancer cells of type $B$.  $B$ is controlled by a linear cancer sub-network that generates cell types $B$ and $C$. 
  }
    \label{fig:NLI2}
\end{figure} 

If $2$ is the number of identical cells controlled by the network and $n$ is the number of loops synchronously executed by those cells then in this case the number of cells after $n$ cell divisions is:
\begin{equation}
\Cells(n) = 2 + (2 \times n)
\end{equation}

The \textbf{phenotype} of such a cancer would have identical linearly growing cancers in several places at once. They would have started simultaneously (up to some error range).  

More generally, let NI$_k$ be a network of $k$ identical divisions that generates $2^k$ cells.  Then we can generate a cancer network NLI$_k$ if we link an upstream NI$_k$ to a simple linear cancer of type NL (see \autoref{fig:NL}): 

\begin{figure}[H]
\begin{tikzpicture}[style=mystyle]
\matrix (m) [matrix of math nodes, row sep=3em,
column sep=2em, text height=1.5ex, text depth=0.25ex]
{ \vphantom{a} & \vphantom{b} &  \vphantom{c}  & \vphantom{c}  & & \vphantom{c} &\\
 A_1  &&  A_2  & \vphantom{b} &  A_3 && A_{k-1} && A_k = B && C \\
 \vphantom{a} &  \vphantom{b} & \vphantom{B} &  \vphantom{c}  & \vphantom{C} & \vphantom{d} & \vphantom{D}\\ };
 \path[\pot]

(m-2-1) edge [inPot1,DarkGreen, cross line] node[nodedescr] {$a_2$} (m-2-3) %A to Nx
(m-2-1) edge [inPot2, DarkBlue, cross line] node[nodedescr] {$a_2$} (m-2-3) %A to Ny
(m-2-3) edge [inPot1, DarkGreen , cross line] node[nodedescr] {$a_3$} (m-2-5) %A to Nx
(m-2-3) edge [inPot2,  DarkBlue, cross line] node[nodedescr] {$a_3$} (m-2-5) %A to Ny
(m-2-7) edge [in100Pot1, DarkGreen, cross line] node[nodedescr] {$a_{k}$} (m-2-9) %A to Nx
(m-2-7) edge [in100Pot2, DarkBlue, cross line] node[nodedescr] {$a_{k}$} (m-2-9)%A to Ny

%(m-2-9) edge [selfloop2,blue,  cross line] node[nodedescr] {$ a_1$} (m-2-1) %end of Ny = D to A
(m-2-9) edge [selfloop1, red, cross line] node[nodedescr] {$ k_n$} (m-2-9) % Nx = C to A;
(m-2-9) edge [inPot2, blue, cross line] node[nodedescr] {$c$} (m-2-11); %A to Ny
\path[solid,red!80, line width=6pt]
(m-2-1) edge (m-2-3);
\path[solid,red!80, line width=6pt]
(m-2-3) edge (m-2-5);
\path[dashed,red!90, line width=6pt]
(m-2-5) edge (m-2-7);
\path[solid,red!80, line width=6pt]
(m-2-7) edge (m-2-9);
\path[solid,blue!80, line width=6pt]
(m-2-9) edge (m-2-11);

\end{tikzpicture}
\caption{
    {\bf Network NLI$_k$ Multi linear cancer starting from $2^k$ identical cells:} This network generates $k$ identical divisions to produce $2^k$ identical daughter cancer cells of type $A_k = B$ from one founder cell A$_1$.  The $B$ cells are controlled by a linear cancer sub-network that after $n$ rounds of division, jointly produce $2^k \times n$ cells of type $C$ while retaining a constant $2^{k}$ of linear cancer cells of type B.  Starting form one founder cell, the total number of cells after $n$ rounds of synchronous division is:  $\mbox{Cells}(n, k) = 2^k + 2^{k}\times(n - k + 1) = 2^k\times(n -k +2)$ if $n > k$ and $\mbox{Cells}(n, k) = 2^n$ otherwise.}
    \label{fig:NLIk}
\end{figure}
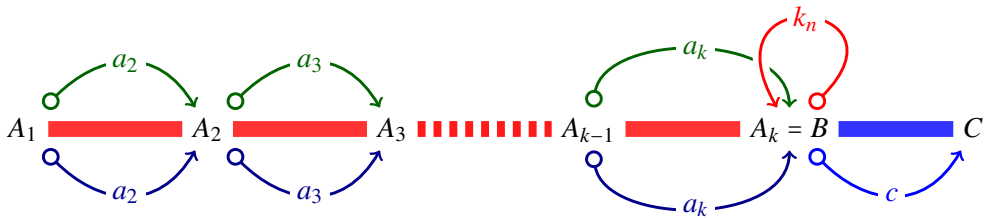 

The below is function describing the ideal rate of growth after $n$ synchronous divisions where $k$ is the number of identical daughter cell divisions nodes in the network in \autoref{fig:NLIk}. 

\begin{eqnarray*}
 \mbox{Cells}(n,k) = \left\{
 \begin{array}{ll} 
 		 2^n  & \mbox{if $n <= k$} \\ 
 		 2^k + 2^{k}\times(n - k + 1) = 2^k\times(n -k +2) & \mbox{for $n > k$}
 						\end{array} 
 				\right.
\end{eqnarray*}

\section{Exponential cancers}
\label{sec:NX}
There are cancers that grow exponentially in that they increase as an exponential function of the number of cells at any given time.  For example, if the number of cells is $k$ at time $t$ of which $c$ are cancer cells, then an exponential cancer will contain at least $c^n$  cells after n units of time.  Now the rate of growth of an exponential process $c^n$ is vastly greater than that of the polynomial $n^c$ or a linear process $n*c$ as will see.  

For example, a cancer that increases by two cells at each time unit will have 2*30 or 60 cells after 30 time units.  Hence, after thirty time units (steps or rounds of cancerous development) we would have a tumor with only 60 cells.  While a cancer that is exponential would double itself with each time unit or step in the cytogenic network.   Hence after 30 time units we would have a tumor with over $2^{30}$ cells which is over a billion cells. Now since it is relative to the length the time unit, if the time unit is one month then after two and a half years our linear cancer is still very small being only 60 cells in size, while our exponential cancer patient has developed a tumor with over one billion cells. If the cancer is fast growing with a cycle of one day, then after one month a patient with the linear cancer will not notice it.  However, a patient with an exponential cancer will have a sizable tumor of some billion cells or if the cancer is fluid, it will have spread very quickly.  An exponential tumor that doubles with every time unit, just one extra day the cancer will a double.  So that after thirty one days our patient will have a tumor of two billion cells.  

All this assumes that there are no physical constraints on tumor growth. In natural systems tremendous pressure is built up in solid tumors that prevent the realization of exponential growth.  Hence, the formulas describe ideal growth without physical constraints. They give an upper bound on the rate of cancer growth.

\subsection{Exponential aggressive  growth cancer network NX} 
%was N2
%\vspace*{1cm}

\begin{figure}[H]
\begin{tikzpicture}[style=mystyle]
\matrix (m) [matrix of math nodes, row sep=3em,
column sep=3em, text height=1.5ex, text depth=0.25ex]
{ \vphantom{a} & \vphantom{a}  &  \vphantom{c}  & \vphantom{c}  & \\
\vphantom{ iNet}  &&  A  &&  B  && C \\
  \vphantom{n} & \vphantom{a} & \vphantom{b} &  \vphantom{c}  & \\ };
 \path[\pot]

(m-2-3) edge [selfloop1, cross line] node[nodedescr] {$a_1$} (m-2-3)
(m-2-3) edge [selfloop2, cross line] node[nodedescr] {$a_2$} (m-2-3);
%\path[solid,black!20, line width=6pt]
(m-2-1) edge (m-2-3);
\path[solid,red!90, line width=10pt]
(m-2-3) edge (m-2-5);
\path[solid,black!20, line width=6pt]
(m-2-5) edge (m-2-7);
\end{tikzpicture}
  \caption{
    {\bf Network NX: Exponential cancer} where A divides into daughter cells that self loop and differentiate to A.
  }
    \label{fig:NX}
\end{figure}
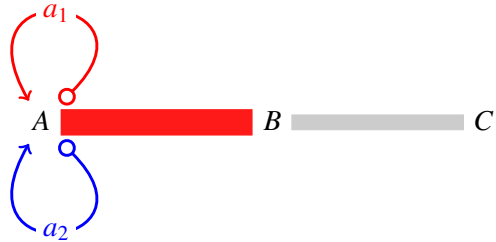

The network NX controls cell division such that a cell of type A divides into two cells both of type $A$.  
The Ideal Total number of cells after n rounds of cell division:
%\begin{figure}[H]
\begin{equation}
\Cells(n) = 2^n
\end{equation}
%\caption{Ideal Total number of cells after r rounds of cell division}
%\end{figure}

{\bf Phenotype} This network will result in a tumor whose growth rate is exponential doubling with every division.  All the cells will be of the same type.  Depending on the connective properties of the cells the tumor will spread quickly if the cells are not easily attached to one another or to other cells types.  If the cells are strongly bound to other cells of type $A$, then we have a tumor that forms a large but quickly growing mass of similar cells.

\subsection{Exponential mixed cell cancer network NXMa} %was NXTa

%  N2.1: A-B,C identical split, B-D,E split, 2D-A loops}

\begin{figure}[H]
\begin{tikzpicture}[style=mystyle]
%%\matrix (m) [matrix of math nodes, fill=red!20, row sep=3em, %Note fill color
\matrix (m) [matrix of math nodes, row sep=3em,
column sep=3em, text height=1.5ex, text depth=0.25ex]
{ \vphantom{a} & b  &  \vphantom{c}  &\vphantom{c}  & d & \vphantom{d} &  \vphantom{d} & \vphantom{e}\\
 A  &&  B  &&  C  && D && E\\
  \vphantom{a} &  \vphantom{b} & c &  \vphantom{c}  & \vphantom{b} & e & \vphantom{e} & \vphantom{e}\\ };
   \path[\poti]
(m-2-1) edge [pot1]( m-1-2);
 \path[\pote]
(m-1-2) edge  [cross line, pot1] (m-2-3); %B = 3
 \path[\poti]
(m-2-1) edge [pot2](m-3-3);
 \path[\pote]
(m-3-3) edge [pot2] (m-2-5);%C = 5
%(m-2-4) edge [pot1] (m-1-5)
 \path[\poti]
(m-2-3) edge  [pot1] (m-1-5);%B = 3 pot1 to D = 7
 \path[\pote]
(m-1-5) edge [pot1](m-2-7);%E = 9
 \path[\poti]
(m-2-3) edge [pot2] (m-3-6);
 \path[\pote]
(m-3-6) edge [pot2] (m-2-9);
 \path[\pot]

(m-2-7) edge [selfloop1, bend left=-90, in=-80, max distance=6cm, min distance=3cm] node[nodedescr] {$ a_1 $} (m-2-1)
(m-2-7) edge [selfloop2, bend left=90, in=90, max distance=6cm, min distance=3cm] node[nodedescr] {$ a_2 $} (m-2-1);
%Note, the graphics will be pushed to the right if the angle is bigger e.g., 120 or more
\path[solid, line width=4pt]
(m-2-1) edge (m-2-3)
(m-2-3) edge (m-2-5)
(m-2-5) edge (m-2-7)
(m-2-7) edge (m-2-9);
\end{tikzpicture}
\caption{
  {\bf Network NXMa: Exponential mixed cell} where a limited number of different cell types in possible repeated patterns are formed.
  }
  \label{fig:NXMa}%replaced fig:NXTa with NXMa
\end{figure}
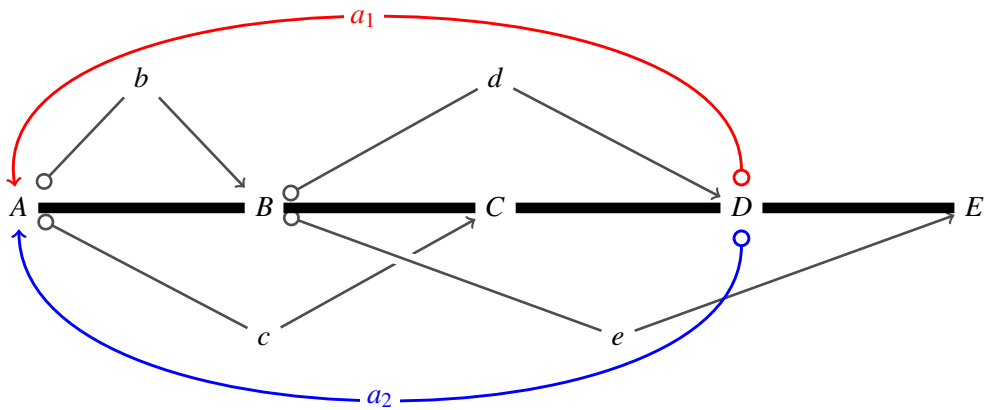

\textbf{Etiology}: The network results from a network transformation of a linear network NLT (\ref{fig:NLT}) that switches a pot pointer.  If this is a genomic cancer network, it is caused by an addressing transformation that in this case switches the pot2 address of the pitcher at area $D$ from $E$ to $A$.  The cell types $C$ and $E$ are non-proliferative while $A$, $B$ and $D$ are part of the cancer cascade. 

\textbf{Phenotype}: This is a mixed cell network since it generates several cell types. The tumor has a limited number cell types $A$, $B$, $C$, $D$ and $E$.  It is highly aggressive, with exponential cell proliferation.  The ideal cell proliferation rate results in $2^{n}$ cells after only $n$ rounds of synchronous cell division.  

\subsubsection{Ideal treatment that would stop exponential cancers} 
%\textbf{Treatment}:  
Ideally one can cure this cancer if one is able to reverse the network transformation that led to the pathological network.  If it is genomic in origin then an addressing inverse transformation  back to the original, presumably normal, developmental pot address would make this a linear developmental network.  

There are two cases depending on if the original linear network was a normal stem cell or a linear cancer: 

\begin{enumerate}
\item \emph{Normal linear stem cell network}:  A method that applies the inverse transformation to all cancer cells will completely stop the exponential cancer. However, then we are still left with an abnormal number ($2^{n}$) of stem cells.  

\item \emph{Linear cancer network}: A second reverse transformation would completely stop cell proliferation resulting in an excess of $2^{n}$ normal cells. 
\end{enumerate}

The above method would be especially valuable for inherited cancer networks, as these could in principle be reversed to normal developmental networks.  

Another method, is simply to transform the pathological address in that is either in the pitcher or the catcher of the pot to a nonfunctional address so that the pathological network link is eliminated in the cancer cells.  The result would be $2^{n}$ cells that no longer have the cancer network.  However, if they contain a still open developmental network responsive to cell signaling, they could continue to grow to they putative multicellular end state, even if this is no longer normal growth in their new context. 

Alternatively, if one had methods to recognize the cells with the pathological network, then drugs might be developed that specifically target cells with the particular cancer network inducing apoptosis (cell death) or otherwise destroying those cells while leaving cells with normal networks untouched. 

\subsection{Exponential single teratoma cancer: Network NXTs}

% N4.  N4 modifies N2}

NXTs modifies NX resulting in a network for an exponential growth cancer with distinct cell types.  Cell A divides into cell N$_x$ and $A$.  The subnetwork N$_x$*A1 for cell type N$_x$ goes through a series of cell differentiations and possible divisions before looping back to cell type $A$.  The dynamic phenotype of this cancer results in exponential growth even if slower than that of NX.  

\begin{figure}[H]
\begin{tikzpicture}[style=mystyle]
\matrix (m) [matrix of math nodes, row sep=3em,
column sep=3em, text height=1.5ex, text depth=0.25ex]
{ \vphantom{a} & \vphantom{b} &  \vphantom{c}  & \vphantom{c}  & &&\\
 A  &&  B  &&  N_x && D \\
 \vphantom{a} &  \vphantom{b} & \vphantom{B} &  \vphantom{c}  & \vphantom{C} & \vphantom{d} & \vphantom{D}\\ };
 \path[\pot]

(m-2-1) edge [inPot,red, cross line] node[nodedescr] {$n_x$} (m-2-5) %A to Nx
(m-2-7) edge [loop2red, cross line] node[nodedescr] {$ a_1 $} (m-2-1) %end of Nx = D to A
(m-2-1) edge [selfloop2, blue, cross line] node[nodedescr] {$ a_2 $} (m-2-1);
\path[solid,black!20, line width=6pt]
(m-2-1) edge (m-2-3);
\path[solid,black!20, line width=6pt]
(m-2-3) edge (m-2-5);
\path[solid,red!70, line width=10pt]
(m-2-5) edge (m-2-7);
\end{tikzpicture}
\caption{
    {\bf Network NXTs: Exponential single teratoma cancer} where A divides into self and network Nx that loops back to A.
  }
  \label{fig:NXTs}
\end{figure}
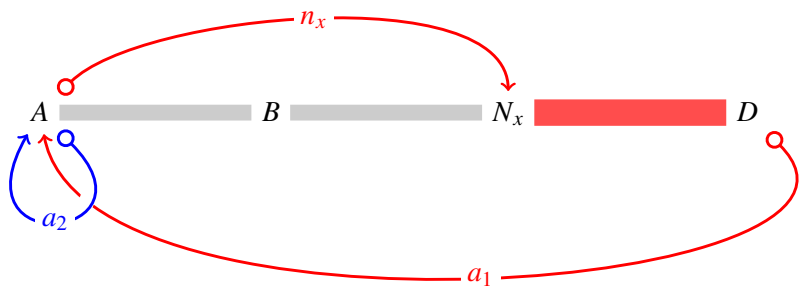

We first observe linear growth of cell types A and N$_x$.  Then the N$_x$ cells form subsystems of cells of various cell types.  These N$_x$ subsystems then generate new cancer stem cells of type $A$.  This leads to gradual but certain exponential growth of the whole tumor and then to exponential growth of each N$_x$ subsystem.  Clinically, then we have a tumor of diverse cell types with possible distinct subsystems with possible morphological structure, yet ultimately while initially slower are still exponential in their growth.  Note, until the N$_x$*A network loops back to $A$, the cancer appears to be a linearly growing cancer of type NL.  Later each of the subsystems induced by N$_x$*A will appear to grow linearly, until sub-pockets of N$_x$*A cells are produced. 

\subsection{Exponential double teratoma cancer network NXTd}

% N5: } 
 
NXTd is a cancer network that modifies NXTs by also making the other A-loop or branch also a subnetwork that first differentiates and possibly divides before looping back to $A$.  Thus, A divides into daughter cells governed by networks Nx and Ny, respectively.  NXTd adds the network Ny to the previous network NXTs which contained only Nx (see  NXTs.)  The NXTd network is a new multicellular subsystem of various cells types and morphology.  Like its partner Nx, Ny ultimately has one or more of its paths loop back to the A cell type. 
 
\begin{figure}[H]
\begin{tikzpicture}[style=mystyle]

\matrix (m) [matrix of math nodes, row sep=3em,
column sep=3em, text height=1.5ex, text depth=0.25ex]
{ \vphantom{a} & \vphantom{b} &  \vphantom{c}  & \vphantom{c}  & &&\\
 A  &&  N_y  & C &  N_x && D \\
 \vphantom{a} &  \vphantom{b} & \vphantom{B} &  \vphantom{c}  & \vphantom{C} & \vphantom{d} & \vphantom{D}\\ };
 \path[\pot]

(m-2-1) edge [inPot2,blue, cross line] node[nodedescr] {$n_y$} (m-2-3) %A to Nx
(m-2-1) edge [inPot1, red, cross line] node[nodedescr] {$n_x$} (m-2-5) %A to Ny
(m-2-7) edge [selfloop1,red,  cross line] node[nodedescr] {$ a_1 $} (m-2-1) %end of Ny = D to A
(m-2-4) edge [selfloop2, blue, cross line] node[nodedescr] {$ a_2 $} (m-2-1);% Nx = C to A;
\path[solid,black!20, line width=6pt]
(m-2-1) edge (m-2-3);
\path[solid,blue!40, line width=10pt]
(m-2-3) edge (m-2-4);
\path[solid,black!40, line width=10pt]
(m-2-4) edge (m-2-5);
\path[solid,red!80, line width=10pt]
(m-2-5) edge (m-2-7);
\end{tikzpicture}
\caption{
    {\bf Network NXTd: Exponential double teratoma cancer} where  A activates networks Nx, Ny, both loop back to A.
  }
 \label{fig:NXTd}
\end{figure}
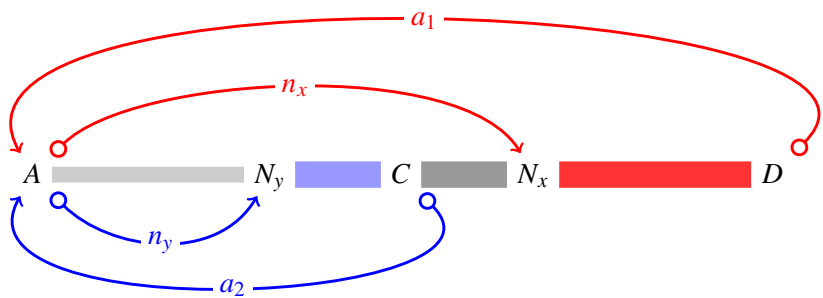

{\bf Phenotype:} Clinically, the phenotype of this cancer starts as a seemingly linearly growing group of variously differentiated cells with perhaps two distinct identifiable developing morphologies.  Then at different points $N_y*C$ and $N_x*D$   will loop back to the A state, causing gradual but exponential growth of the tumor.  The growth may be slower than exponential single teratoma network NXTs, but it will nevertheless be exponential.  

Diagnostically, the key to recognizing these kinds of cancers is the appearance of repeating subsystems with possibly two distinct growth rates and developmental morphological properties. The cancer stem cells of type A will be distributed throughout the tumor.  And if the cancer is fluid or non-connective then the cancer can spread to other cell types that overtly have the same phenotype but are non-cancerous.  Thus the healthy from non healthy cells will be difficult to detect phenomenologically.

\subsection{Hyper cancer networks NXH}

Hyper cancers are exponential cancers with possibly mixed cell types because several identical divisions that lead to possibly distinct differentiation states. The growth rate varies with the number of identical loops and differentiation complexity of the intermediate states. 

\begin{figure}[H]
%\centering
%\subfigure[A 4-node hyper cancer network]{
\subfloat[A 4-node hyper cancer network]{
\includegraphics[scale=0.4]{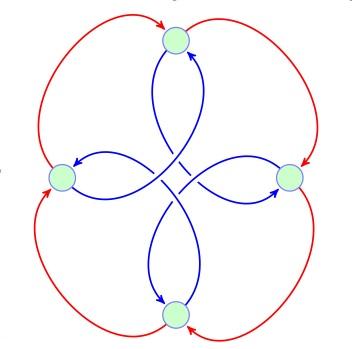}
\label{fig:subfig1} 
}
\hspace{1.0cm}
\subfloat[Many seemingly non-cancerous cell types in the overt phenotype of a tumor]{
\includegraphics[scale=0.4]{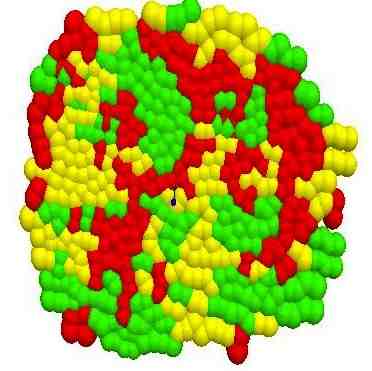}
\label{fig:subfig2}
}
\hspace{1.0cm}
\subfloat[Yet, all cells are cancerous]{
\includegraphics[scale=0.4]{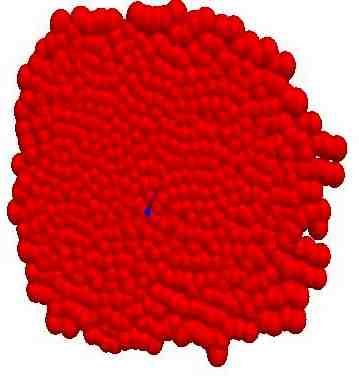}
\label{fig:subfig3}
}
\label{fig:subfigureExample}
\caption{
{\bf Three views of a Hyper Cancer.} In \subref{fig:subfig1} we have a 4 node hyper cancer network view. The overt phenotype of cells differs from their control state. The overt phenotype shown in phenotype view \subref{fig:subfig2} hides the fact that all cells in this tumor are cancerous.  This is evident in the control state view  of the tumor shown in figure \subref{fig:subfig3}. 
}
%and \subref{fig:subfig3}}
\end{figure}

\subsection{Simple Hyper Cancer NXH$^2$}

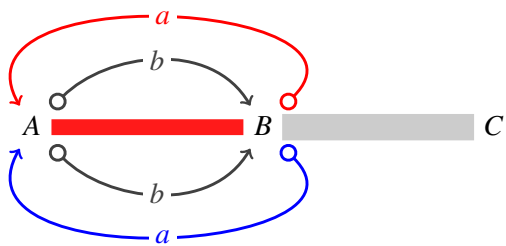
\begin{figure}[H]
\begin{tikzpicture}[style=mystyle]
\matrix (m) [matrix of math nodes, row sep=3em,
column sep=3em, text height=1.5ex, text depth=0.25ex]
{ \vphantom{a} & \vphantom{a}  &  \vphantom{c}  & \vphantom{c}  & \\
 A  &&  B  &&  C  && \vphantom{D} \\
  \vphantom{n} & \vphantom{a} & \vphantom{b} &  \vphantom{c}  & \\ };
 \path[\pot]
(m-2-1) edge [inPot1] node[nodedescr]{$b$}( m-2-3) 
(m-2-1) edge [inPot2] node[nodedescr]{$b$}( m-2-3) 
(m-2-3) edge [selfloop1, cross line] node[nodedescr] {$a$} (m-2-1)

(m-2-3) edge [selfloop2, cross line] node[nodedescr] {$a$} (m-2-1);
\path[solid,red!90, line width=6pt]
(m-2-1) edge (m-2-3);
\path[solid,black!20, line width=10pt]
(m-2-3) edge (m-2-5);
%\path[solid,black!20, line width=6pt]
%(m-2-5) edge (m-2-7);
\end{tikzpicture}
  \caption{
    {\bf Network NXH$^2$ Two node Hyper Cancer:} A simple exponential cancer of identical cells. The parent $A$ divides into two identical daughter cells $B$ such that each daughter differentiate back to the parent control state $A$. Note, this has two control states with the potential of producing two cell types with different differentiation states. 
    }
    \label{fig:NXH2}
\end{figure} 

\subsection{Identical cells with one loop back is still exponential but delayed}

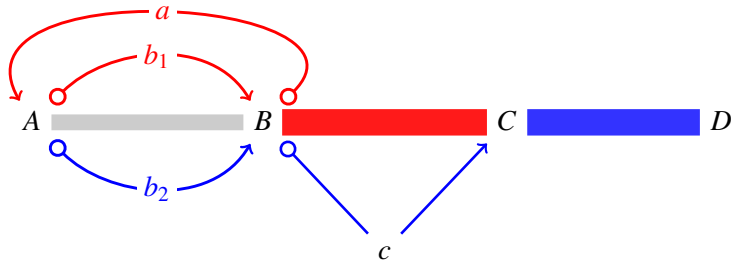
\begin{figure}[H]
\begin{tikzpicture}[style=mystyle]
\matrix (m) [matrix of math nodes, row sep=3em,
column sep=3em, text height=1.5ex, text depth=0.25ex]
{ \vphantom{a} & \vphantom{a}  &  \vphantom{c}  & \vphantom{c}  & \\
 A  &&  B  &&  C  && D \\
  \vphantom{n} & \vphantom{a} & \vphantom{b} &  c  & \\ };
 \path[\pot]
(m-2-1) edge [inPot1, red]node[nodedescr] {$b_1$}( m-2-3) 
(m-2-1) edge [inPot2, blue] node[nodedescr] {$b_2$}( m-2-3) 
(m-2-3) edge [selfloop1, cross line] node[nodedescr] {$ a $} (m-2-1);
%to A
%(m-1-2) edge  [cross line, pot1] (m-2-3)
\path[\poti]
(m-2-3) edge [pot2, blue](m-3-4);
\path[\pote]
(m-3-4) edge [pot2, blue] (m-2-5);

\path[solid,black!20, line width=6pt]
(m-2-1) edge (m-2-3);
\path[solid,red!90, line width=10pt]
(m-2-3) edge (m-2-5);
\path[solid,blue!80, line width=10pt]
(m-2-5) edge (m-2-7);
\end{tikzpicture}
  \caption{
    {\bf Network NXIm Multi-Exponential Cancer:} A simple multi-exponential cancer of identical cells. The parent $A$ divides into two identical daughter cells $B$ such that one daughter differentiate back to the parent control state $A$ and the other differentiates to a cell of type $C$.  
 }
\label{fig:NXIm}
\end{figure} 

The growth is exponential but delayed and hence slower per execution of the cancer loops. The phenotype of this type of cancer is an alternation of build up of cell types $A$ or $B$ while steadily building up of cell type $C$. Cell type $C$ may differentiate and grow into a complex structure. These structures would be repeated in the resulting tissue each developing out of cell of type $C$ if it is linked to a further network.  

\subsection{Invariant transformation of exponential networks}

In terms of growth this cancer is equivalent (up to a delay because of the dedifferentiation jump) in terms of ideal growth rate to the simple exponential network above. Indeed, one can add arbitrarily many identical cell divisions to the network and get an growth rate equivalent networks.  They are all growth rate equivalent to the simplest exponential cancer of simple cell fission growth where daughter cells have the same control state as the parent.  Even though they are growth rate equivalent all these networks may have a different overt phenotype, since each distinct control state as the cells divide may lead to different overt characteristics of the cell. 

There is a distinction between the overt doubling of cells at each ideal synchronous cell division and the number of cells generated after one cancer cycle is completed. A hyper-cancer will produce more cells after a cancer loop cycle, but its growth at each step generates the same number of cells not matter how many identical cell divisions are in the loop. In this latter sense, the cancer networks are equivalent and, thereby, invariant under the transformation. They are not equivalent in terms of phenotype or the number of cells generated over the whole cancer loop cycle. 

\subsection{Exponential cancer network with only one loop NXo}

% N7: : A-B identical split, B-A jump}

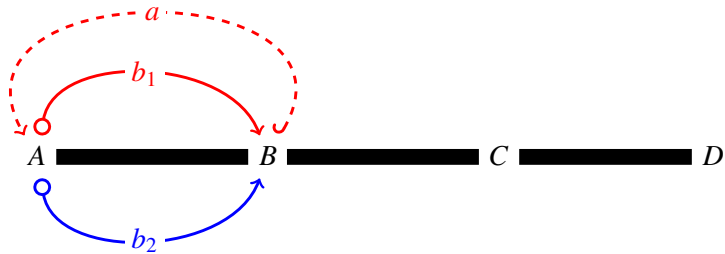
\begin{figure}[H]
\begin{tikzpicture}[style=mystyle]
\matrix (m) [matrix of math nodes, row sep=3em,
column sep=3em, text height=1.5ex, text depth=0.25ex]
{ \vphantom{a} &  \vphantom{b}   &  \vphantom{c}  & \vphantom{c}  & \\
 A  &&  B  &&  C  && D \\
  \vphantom{a} &  \vphantom{b}  & \vphantom{b} &  \vphantom{c}  & \\ };
  \path[\pot]
(m-2-1) edge [inPot1, out= 80,in=110, red]node[nodedescr] {$b_1$}( m-2-3) 
(m-2-1) edge [inPot2, out=-80, in=-110, blue] node[nodedescr] {$b_2$}( m-2-3); 

\path[\jump]
%\path[right hook->]
(m-2-3) edge [selfloop1, out=60, max distance = 6cm, min distance =5cm, dashed, cross line] node[nodedescr] {$ a $} (m-2-1);

\path[solid, line width=6pt]
(m-2-1) edge (m-2-3)
(m-2-3) edge (m-2-5)
(m-2-5) edge (m-2-7);
\end{tikzpicture}
\caption{
  {\bf NXo Exponential cancer with only one loop:} The parent divides into identical daughter cells that differentiate back to $A$. 
  }
  \label{fig:NXo}
\end{figure} 

Interestingly, one can have an exponential cancer with only one self activating loop.  
This can happen when a cell A differentiates into two identical daughter cells and they then 
differentiate back into their parent cell $A$, as illustrated in this example network:  

\subsection{n-Node Hyper Cancer Network NXH$_n$}
% N11: }
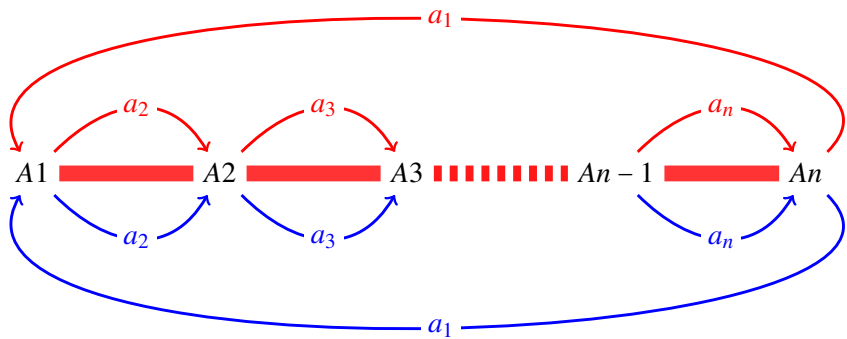
\begin{figure} [H]
\begin{tikzpicture}[style=mystyle]
\matrix (m) [matrix of math nodes, row sep=3em,
column sep=2em, text height=1.5ex, text depth=0.25ex]
{ \vphantom{a} & \vphantom{b} &  \vphantom{c}  & \vphantom{c}  & & \vphantom{c}\\
 A1  &&  A2  & \vphantom{b} &  A3 && An-1 && An && \\
 \vphantom{a} &  \vphantom{b} & \vphantom{B} &  \vphantom{c}  & \vphantom{C} & \vphantom{d} & \vphantom{D}\\ };
\path[->]
(m-2-1) edge [inPot1,red, cross line] node[nodedescr] {$a_2$} (m-2-3) %A to Nx
(m-2-1) edge [inPot2, blue, cross line] node[nodedescr] {$a_2$} (m-2-3) %A to Ny
(m-2-3) edge [inPot1,red, cross line] node[nodedescr] {$a_3$} (m-2-5) %A to Nx
(m-2-3) edge [inPot2, blue, cross line] node[nodedescr] {$a_3$} (m-2-5) %A to Ny
(m-2-7) edge [inPot1,red, cross line] node[nodedescr] {$a_{n}$} (m-2-9) %A to Nx
(m-2-7) edge [inPot2, blue, cross line] node[nodedescr] {$a_{n}$} (m-2-9) %A to Ny

(m-2-9) edge [selfloop2,blue,  cross line] node[nodedescr] {$ a_1$} (m-2-1) %end of Ny = D to A
(m-2-9) edge [selfloop1, red, cross line] node[nodedescr] {$ a_1$} (m-2-1);% Nx = C to A;
\path[solid,red!80, line width=6pt]
(m-2-1) edge (m-2-3);
\path[solid,red!80, line width=6pt]
(m-2-3) edge (m-2-5);
\path[dashed,red!90, line width=6pt]
(m-2-5) edge (m-2-7);
\path[solid,red!80, line width=6pt]
(m-2-7) edge (m-2-9);

\end{tikzpicture}
\caption{
    {\bf Network NXH$^n$ n-Hyper cancer of degree $n$:} This hyper cancer network generates $n$ identical divisions after one loop of the network to produce $2^n$ identical daughter from one founder cell.  At node A$_{n}$ the network generates one more identical cell division as it loops back to its stating point.  The network generates $2^{n^c}$ cells from one founder cell after $c$ cycles of the network loop which is equal to $2^x$ cells after $x$ rounds of synchronous division where $x = n*c$.    
}
\label{fig:NXHn}
\end{figure} 

\begin{equation}
{\mbox Cells}(n, c) = 2^{n^c} = 2^x \mbox{ where } x = n*c
\end{equation}

Here $c$ is used for whole cancer loop cycles,  $n$ is the number of nodes in the network, and $x$ is used for executing all cells by one network control time step. 

\subsection{An almost equivalent Hyper Cancer Network with a jump}

The following network is slower but almost equivalent to the above hyper network. It is slightly slower in growth due to the dedifferentiation step. 
\begin{figure}[H]
\begin{tikzpicture}[style=mystyle]
\matrix (m) [matrix of math nodes, row sep=3em,
column sep=2em, text height=1.5ex, text depth=0.25ex]
{ \vphantom{a} & \vphantom{b} &  \vphantom{c}  & \vphantom{c}  & & \vphantom{c}\\
 A1  &&  A2  & \vphantom{b} &  A3 && An-1 && An && \\
 \vphantom{a} &  \vphantom{b} & \vphantom{B} &  \vphantom{c}  & \vphantom{C} & \vphantom{d} & \vphantom{D}\\ };
 \path[\pot]

(m-2-1) edge [inPot1,red, cross line] node[nodedescr] {$a_2$} (m-2-3) %A to Nx
(m-2-1) edge [inPot2, blue, cross line] node[nodedescr] {$a_2$} (m-2-3) %A to Ny
(m-2-3) edge [inPot1,red, cross line] node[nodedescr] {$a_3$} (m-2-5) %A to Nx
(m-2-3) edge [inPot2, blue, cross line] node[nodedescr] {$a_3$} (m-2-5) %A to Ny
(m-2-7) edge [inPot1,red, cross line] node[nodedescr] {$a_{n}$} (m-2-9) %A to Nx
(m-2-7) edge [inPot2, blue, cross line] node[nodedescr] {$a_{n}$} (m-2-9); %A to Ny

\path[\jump]
%(m-2-9) edge [selfloop2,blue,  cross line] node[nodedescr] {$ a_1$} (m-2-1) %end of Ny = D to A
(m-2-9) edge [jump1, red, max distance =5cm, min distance =4cm, cross line] node[nodedescr] {$ a_1$} (m-2-1);% Nx = C to A;
\path[solid,red!80, line width=6pt]
(m-2-1) edge (m-2-3);
\path[solid,red!80, line width=6pt]
(m-2-3) edge (m-2-5);
\path[dashed,red!90, line width=6pt]
(m-2-5) edge (m-2-7);
\path[solid,red!80, line width=6pt]
(m-2-7) edge (m-2-9);

\end{tikzpicture}
\caption{
    {\bf Network NXHj$^n$ n-Hyper cancer:} This hyper cancer network generates $n$ identical divisions after one cycle of the loop in $n+1$ time steps to produce $2^n$ identical daughter from one founder cell. The dashed arrow represents a jump that does not initiate cell division.  Hence, at node An the network does a jump to the starting point without undergoing cell division. This has the effect of starting the round identical cell divisions again producing an endless series of identical cell divisions.  The network generates $2^{n^c}$ cells after $c$ cycles of the cancer loop from one founder cell.   In $n$ time steps, it produces $2^{n-c}$ cells after $n-c$ rounds of synchronous cell division. 
 }
 \label{fig:NXHjn}
\end{figure}
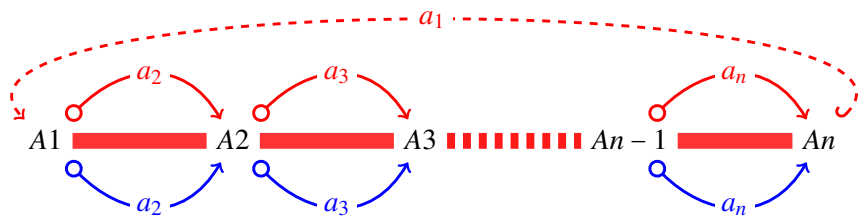 

\subsection{Exponential hyper teratoma cancer NXHTs}

An exponential hyper cancer network is linked to a teratoma subnetwork. 

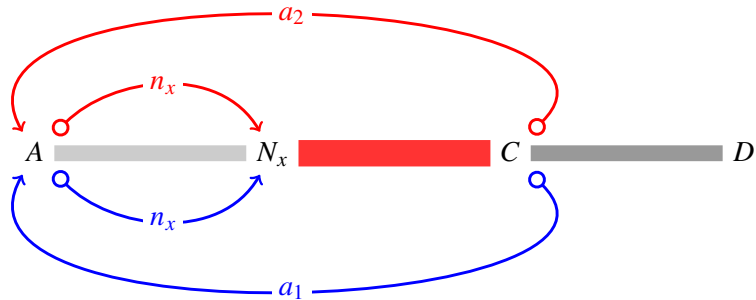
\begin{figure}[H]
\begin{tikzpicture}[style=mystyle]
\matrix (m) [matrix of math nodes, row sep=3em,
column sep=3em, text height=1.5ex, text depth=0.25ex]
{ \vphantom{a} & \vphantom{b} &  \vphantom{c}  & \vphantom{c}  & &&\\
 A  &&  N_x  & \vphantom{b} &  C && D \\
 \vphantom{a} &  \vphantom{b} & \vphantom{B} &  \vphantom{c}  & \vphantom{C} & \vphantom{d} & \vphantom{D}\\ };
 \path[\pot]
(m-2-1) edge [inPot1,red, cross line] node[nodedescr] {$n_x$} (m-2-3) %A to Nx
(m-2-1) edge [inPot2, blue, cross line] node[nodedescr] {$n_x$} (m-2-3) %A to Ny
(m-2-5) edge [selfloop2,blue,  cross line] node[nodedescr] {$ a_1 $} (m-2-1) %end of Ny = D to A
(m-2-5) edge [selfloop1, red, cross line] node[nodedescr] {$ a_2 $} (m-2-1);% Nx = C to A;
\path[solid,black!20, line width=6pt]
(m-2-1) edge (m-2-3);
\path[solid,red!80, line width=10pt]
(m-2-3) edge (m-2-5);
\path[solid,black!40, line width=6pt]
(m-2-5) edge (m-2-7);
\end{tikzpicture}
\caption{
    {\bf Network NXHT: Exponential hyper teratoma cancer network} where identical daughter teratoma networks loop back to A.
  }
  \label{fig:NXHT}
\end{figure}  

NXHT can be viewed as a variation of network NXTd is where subnetwork $N_{x} = N_{y}$.  Thus, $A$ divides into cells of type $N_x$ each controlled by looping network $N_x*A$ where the '*' indicates a link from $N_{x}$ to the network that controls cell type $A$.  Then both will differentiate and possibly divide into a subsystem of some morphology with different cell types.  Ultimately, they loop back to cell type $A$.  Again we have exponential growth.  \autoref{fig:NXHT} is more uniform than \autoref{fig:NXTd}  because it will not have the complexity of having two distinct multicellular subsystems governed by two distinct subnetworks.  The growth rate of the subsystem will depend on the time path through the loop $A \circleright N^{1}_x \circleright N^{2}_x \circleright \ldots \circleright N^{n}_x \circleright C \circleright A$ takes to complete one round.  %$ ^\circ \!{Nx} $

\subsection{Growth rates for Hyper Cancer Networks}
%NXH$_n$ was N11
Let NXH$_n$ be network with a subnetwork Na is linked to Nb.  Let Na be a passive network that causes no cell divisions but only differentiates to Nb. If Nb produces, say 8 identical cells after three rounds of division and then Nb loops back to Na, then for each execution of one loop the network $N8 = Na \otimes Nb$ will produce 8 times as many cells as the previous loop.  For example, after 5 loops it will have produced $8^5$ or 32,768 cells.  After 10 loops, we would have over a billion cells  ($8^{10} = 1,073,741,824$ cells).  While Nb is by itself harmless, when mutated to activate Na, it becomes a very explosive cancer.  If we add just five more rounds, we have $8^{15}$ or over 35 trillion cells. 

Note, this network N8 is similar to  NXHj$^8$ \autoref{fig:NXHjn} and  is equivalent in terms of ideal per loop, growth rate to having a network of type NX \autoref{fig:NX} . However, the latter network requires 30 loops instead of 10 loops to generate $8^{10}$ cells (since $2^{{3}^{10}} = 2^{30}$).  Each loop in networks of type NX doubles the number of cells at each round. And, it would require 45 loops of cell division to generate over 35 trillion cells while NXH$_n$ can do this in 15 loops.   Of course, there is a sleight of hand here since the rounds of division are hidden in the network NXH$_n$.

Our theory predicts that given a subsystem that produces a mass of cells of the same type (with possible multiple cell types during NI$^n$ \autoref{fig:NIn}) by means of several rounds of division, if that system becomes cancerous in just one of those cells, by looping back to the start of the subnetwork controlling its development (to give something like  \autoref{fig:NXHjn}) , then that system will have the clinical phenotype of a highly explosive or hyper cancer.  Indeed, given there are enough resources and the cell physics permits it, then at least $k^n$ cells are produced after n loops where $k$ is the number of identical cells produced by the natural healthy system.  

\subsection{Multi-exponential cancers NiX$_n$}
An added loop transforms a multi-linear cancer into a multi-exponential cancer where a set of two or more identical cells in equivalent control states all proliferate exponentially. 

\begin{figure}[H]
\begin{tikzpicture}[style=mystyle]
\matrix (m) [matrix of math nodes, row sep=3em,
column sep=2em, text height=1.5ex, text depth=0.25ex]
{ \vphantom{a} & \vphantom{b} &  \vphantom{c}  & \vphantom{c}  & & \vphantom{c} &\\
 A0  &&  A1  & \vphantom{b} &  A2 && An-1 && An = B &&  \\
 \vphantom{a} &  \vphantom{b} & \vphantom{B} &  \vphantom{c}  & \vphantom{C} & \vphantom{d} & \vphantom{D}\\ };
 \path[\pot]

(m-2-1) edge [inPot1,DarkGreen, cross line] node[nodedescr] {$a_1$} (m-2-3) %A to Nx
(m-2-1) edge [inPot2, DarkBlue, cross line] node[nodedescr] {$a_1$} (m-2-3) %A to Ny
(m-2-3) edge [inPot1, DarkGreen , cross line] node[nodedescr] {$a_2$} (m-2-5) %A to Nx
(m-2-3) edge [inPot2,  DarkBlue, cross line] node[nodedescr] {$a_2$} (m-2-5) %A to Ny
(m-2-7) edge [in100Pot1, DarkGreen, cross line] node[nodedescr] {$a_{n}$} (m-2-9) %A to Nx
(m-2-7) edge [in100Pot2, DarkBlue, cross line] node[nodedescr] {$a_{n}$} (m-2-9)%A to Ny

(m-2-9) edge [selfloop1, red, cross line] node[nodedescr] {$ a_n$} (m-2-9) % Nx = C to A;
(m-2-9) edge [selfloop2, red, cross line] node[nodedescr] {$ a_n$} (m-2-9);
%(m-2-9) edge [inPot2, blue, cross line] node[nodedescr] {$c$} (m-2-11); %A to Ny
\path[solid,red!80, line width=6pt]
(m-2-1) edge (m-2-3);
\path[solid,red!80, line width=6pt]
(m-2-3) edge (m-2-5);
\path[dashed,red!90, line width=6pt]
(m-2-5) edge (m-2-7);
\path[solid,red!80, line width=6pt]
(m-2-7) edge (m-2-9);

\end{tikzpicture}
\caption{
    {\bf Network NiX$_n$ Multi exponential cancer starting from $2^n$ identical cells:} This network generates $n$ identical divisions to produce $2^n$ identical daughter cancer cells of type $A_n = B$ from one founder cell A$_0$.  The $B$ cells are controlled by an exponential cancer sub-network jointly producing $2^n * 2^k$ cells of type $B$ after $k$ further rounds of division.  Starting form one founder cell, the total number of cells after $r = n+k$ rounds of synchronous division is:  
}
\label{fig:NiXn}     
    %Crit. Need to consistently refer to Cell number for a network N after r rounds of synchronous cell divisions.  Use either:
    % Cells(N, r)  or Cells_N(r) or \Sigma(N,r) or \Sigma_r(N) or \Sigma_N(r)
\end{figure}
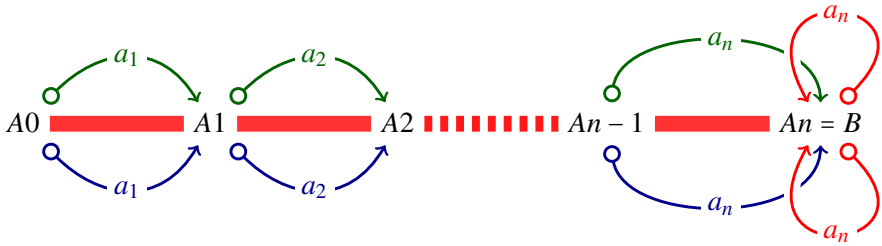 

\begin{equation}
\Cells(n, k) = 2^{n + k}
\end{equation}    

While this is equivalent in terms of ideal growth rate to the simple one node exponential cancer NX above, the tumor may have a different phenotype since each of the control states $A_i$ in NiX$_n^k$ are distinct and may initiate the production different overt cell properties.  Furthermore the growth rates of the cells $C(A_i)$ controlled by $A_i$ may differ from some $C(A_j)$ controlled by $A_j \neq A_i$. Hence, we would get variations in phenotype and growth rates even though ideally with synchronous division the growth rates of NX and NiX$_n^k$ exponentially equivalent since $\Sigma_{n+k}(\mbox{NX}) = \Sigma_{n+k}(\mbox{NiX}_n) = 2^{n + k}$ after $n+k$ rounds of synchronous cell divisions.

An instance where $k = 1$ to give 2 identical cells that continue to grow exponentially. This instance is equivalent to NX since all cell types generated are identical in their control states.  The only difference is that the founder cell $C(A)$ may differ in phenotype from cells of type $C(B)$ whereas all the cells generated by NX are identical to the founder cell. 

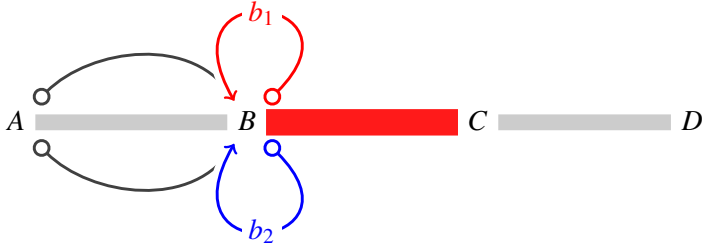
\begin{figure}[H]
\begin{tikzpicture}[style=mystyle]
\matrix (m) [matrix of math nodes, row sep=3em,
column sep=3em, text height=1.5ex, text depth=0.25ex]
{ \vphantom{a} & \vphantom{a}  &  \vphantom{c}  & \vphantom{c}  & \\
 A  &&  B  &&  C  && D \\
  \vphantom{n} & \vphantom{a} & \vphantom{b} &  \vphantom{c}  & \\ };
 \path[\pot]
(m-2-1) edge [inPot1]( m-2-3) 
(m-2-1) edge [inPot2]( m-2-3) 
(m-2-3) edge [selfloop1, cross line] node[nodedescr] {$ b_1 $} (m-2-3)
(m-2-3) edge [selfloop2, cross line] node[nodedescr] {$ b_2 $} (m-2-3);
\path[solid,black!20, line width=6pt]
(m-2-1) edge (m-2-3);
\path[solid,red!90, line width=10pt]
(m-2-3) edge (m-2-5);
\path[solid,black!20, line width=6pt]
(m-2-5) edge (m-2-7);
\end{tikzpicture}
  \caption{
    {\bf Network NiX$_1$ Multi-Exponential Cancer:} The simplest instance of the network NiX$_n$ with $n = 1$. The parent divides into two identical daughter cells that each exponentially generate cells. $\mbox{Cells}(n+1) = 2^{1+ k}$.  $\Sigma_{1+k}(\mbox{NiX}_1 ) = 2^{1+ k}$ after $1+k$ rounds of synchronous cell divisions. 
  }
  \label{fig:NiX1}
\end{figure} 

In this case the number of cells after $n$ cell divisions is:
\begin{equation}
\Cells(n) = c+ (c \times 2^n)
\end{equation}

where $c$ is the number of identical cells controlled by the network and $n$ is the number of loops synchronously executed by those cells.  The phenotype of such a cancer would have identical exponentially growing cancers in $n$ places at once. They would have started simultaneously (up to some error range).

\section{Geometric Cancer Networks}
\label{sec:NG}
%was N8

Having two or more proliferative loops does not necessarily result in an exponential cancer.  However, they can transform into exponential cancers.  For example, using the WHO-grading scheme, astrocytoma grad I, which is slow growing and benign, appears to be a linear cancer of type NG, astrocytoma grade II -> grade III.  Grade III is relatively fast growing compared to grade II, but can be present for several years and then suddenly change to grade IV Glioblastoma Multiforme (GBM). GBM are fast growing and spread quickly.   \nocite{Nakada2011}.

\subsection{1st-Order Geometric Cancer Networks G1}

\label{sec:NG1}
\begin{figure}[H]
\begin{tikzpicture}[style=mystyle]
\matrix (m) [matrix of math nodes, row sep=3em,
column sep=3em, text height=1.5ex, text depth=0.25ex]
{ \vphantom{a} & \vphantom{b}  &  \vphantom{c}  & \vphantom{c}  & \\
 A  &&  B  &&  C  &&  \\
  \vphantom{a} &  b & \vphantom{b} &  \vphantom{c}  & \\ };
 \path[\pot]
(m-2-1) edge [selfloop1, cross line] node[nodedescr] {$ a $} (m-2-1);
\path[\poti]
(m-2-1) edge [pot2](m-3-2);
\path[\pote]
(m-3-2) edge [pot2] (m-2-3);
\path[solid, line width=6pt]
(m-2-1) edge (m-2-3)
(m-2-3) edge (m-2-5);
%(m-2-5) edge (m-2-7);
\end{tikzpicture}
\caption{
    {\bf Network NG$_1$: A 1st-order geometric cancer with one loop:} The cell of type $A$ is like a 1st-order stem cell; it produces terminal cells of some differentiated type $B$ (which we also will refer to as G0 cells). The difference between stem cells and cancer stem cells has to do with their functionality in the system as a whole. 
  }
    \label{fig:NG1}
\end{figure}
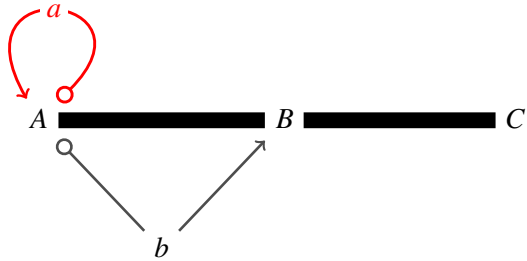 

\begin{figure}[H]
\begin{tikzpicture}[style=mystyle]
\matrix (m) [matrix of math nodes, row sep=3em,
column sep=3em, text height=1.5ex, text depth=0.25ex]
{ \vphantom{a} & \vphantom{b}  &  \vphantom{c}  & \vphantom{c}  & \\
\Phi \rightarrow A  &&  B  &&  \vphantom{C}  && \vphantom{D} \\
  \vphantom{a} &  b & \vphantom{b} &  \vphantom{d}  & \\ };
 \path[\poti]
(m-2-1) edge [pot2, blue](m-3-2);
 \path[\pote]
(m-3-2) edge [pot2, blue] (m-2-3); 
\path[\pot]
(m-2-1) edge [selfloop1, cross line] node[nodedescr] {a} (m-2-1);

\path[solid, red, line width=6pt]
(m-2-1) edge (m-2-3);
\path[solid, blue, line width=6pt]
(m-2-3) edge (m-2-5);
%(m-2-5) edge (m-2-7);
\end{tikzpicture} 

  \caption{
    {\bf Network LSC1: 1st Order (Linear) Stem  Cell Network} The cell of type $A$ is  a regular stem cell that divides conditionally.  If condition $\Phi$ hold, then  $A$ produces cells of type $B$.  One daughter of  $A$ self loops giving a daughter cell of the same cell type $A$ as the parent.  The other daughter cell differentiates to type $B$. 
  }
\label{fig:LSC1}
\end{figure}
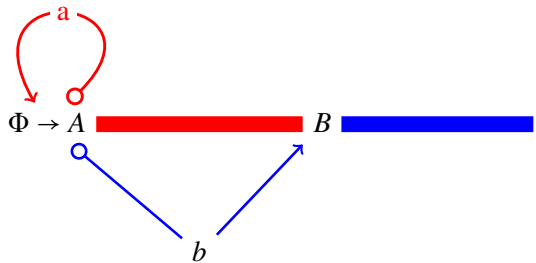

First order geometric cancer networks G1 contain one loop and are the same as the linear cancer networks NL that we studied in \autoref{fig:NL}.  They contain one loop and generate cells linearly, one at a time. The generated cells are terminal cells that do not proliferate.  We will call terminal cells produced by geometric cancer networks G0 cells.  They are related to 1st order stem cell networks (see below \autoref{fig:LSC1}) 

\subsection{2nd-Order Geometric Cancer Networks G2} % was NG$_2$}
\label{sec:NG2}
\begin{figure}[H]
\begin{tikzpicture}[style=mystyle]
\matrix (m) [matrix of math nodes, row sep=3em,
column sep=3em, text height=1.5ex, text depth=0.25ex]
{ \vphantom{a} & \vphantom{b}  &  \vphantom{c}  & \vphantom{c}  & \\
 A  &&  B  &&  C  && D \\
  \vphantom{a} &  b & \vphantom{b} &  c  & \\ };
 \path[\pot]
(m-2-1) edge [selfloop1, cross line] node[nodedescr] {$ a $} (m-2-1);
\path[\poti]
(m-2-1) edge [pot2](m-3-2);
\path[\pote]
(m-3-2) edge [pot2] (m-2-3);
%(m-1-2) edge [pot1 loop, bend left=-80] node[nodedescr] {$ \beta_1 $} (m-1-2)
\path[\pot]
(m-2-3) edge [selfloop1, cross line] node[nodedescr] {$ b $} (m-2-3);
\path[\poti]
(m-2-3) edge [pot2](m-3-4);
\path[\pote]
(m-3-4) edge [pot2] (m-2-5);

\path[solid, line width=6pt]
(m-2-1) edge (m-2-3)
(m-2-3) edge (m-2-5)
(m-2-5) edge (m-2-7);
\end{tikzpicture}
\caption{
    {\bf Network NG$_2$: A 2nd-order geometric cancer with two loops:} Both $A$ and $B$ are linear cancer cells whose joint proliferation potential is given by \autoref{eq:G2}.  The cell of type $B$ is a stem cell; it produces cells of type C. The cell $A$ is a {\em meta-stem cell} that produces stem cells of type $B$.  The difference between stem cells and cancer stem cells has to do with their functionality in the system as a whole. 
  }
    \label{fig:NG2}
\end{figure}
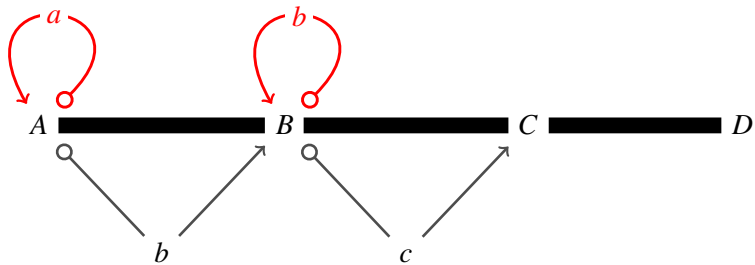  

\subsection{Meta-stem cells}

Geometric networks  of the above type suggest the concept of meta-stem cells (see \autoref{sec:DSC2}).  A {\em linear}  or {\em first order stem cell} is a cell that produces other cells (that are not stem cells) by means of a linear developmental network.  {\em Meta-stem cells} are stem cells that produce stem cells.  Thus a 2nd-order geometric network with two loops contains a meta-stem cell $A$ that produces linear stem cells of type $B$.  A 3rd-order three loop geometric network contains a {\em meta-meta-stem} cell $A$ that produces meta-stem cells $B$ that produce stem cells C that produce cells of type D. And, so on.  Whether a geometric meta-stem cell network is a cancer meta-stem cell network will depend on the properties of the cells and how they function in the organism.  Any stem cell or meta-stem cell network can by mutations become an exponential cancer (see below).  A transformation of linear stem cell network into a meta-stem cell network will make the stem cell proliferate according to \autoref{eq:TriangularNo}  below and be potentially harmful.  Indeed, even the metastatic behavior of cancer stem cells will be seen to be directly related to the properties of their geometric, meta-networks (\autoref{sec:GMetastases}). 

\subsection{Geometric cancer networks NG$_k$ with $k$ loops}
\label{sec:NGk}

\begin{figure}[H]
\begin{tikzpicture}[style=mystyle]
\matrix (m) [matrix of math nodes, row sep=3em,
column sep=3em, text height=1.5ex, text depth=0.25ex]
{ \vphantom{a} & \vphantom{b}  &  \vphantom{c}  & \vphantom{c}  & \\
 A_1  &&  A_2  &&  A_k  && D \\
  \vphantom{a} &  \mbox{$\alpha_2$} & \vphantom{b} &  \vphantom{a3}  &  \vphantom{a3}  & d \\ };
 \path[\pot]
(m-2-1) edge [selfloop1, cross line] node[nodedescr] {$ \alpha_1 $} (m-2-1);
\path[\poti]
(m-2-1) edge [pot2](m-3-2);
\path[\pote]
(m-3-2) edge [pot2] (m-2-3);
\path[\pot]
(m-2-3) edge [selfloop1, cross line] node[nodedescr] {$ \alpha_2 $} (m-2-3);
\path[\poti]
(m-2-5) edge [pot2](m-3-6);
\path[\pote]
(m-3-6) edge [pot2] (m-2-7);

\path[\pot]
(m-2-5) edge [selfloop1, cross line] node[nodedescr] {$ \alpha_k $} (m-2-5);

\path[solid, line width=6pt]
(m-2-1) edge (m-2-3);
\path[dashed, red, line width=8pt]
(m-2-3) edge (m-2-5);s
\path[solid, line width=6pt]
(m-2-5) edge (m-2-7);
\end{tikzpicture}
\caption{
  {\bf Network NGk An k-th Order Geometric Cancer:} A geometric cancer with $k$ loops: Each of the cell types $A_i$, $i = 1 \ldots k$ are linear cancer cells whose joint proliferation potential is give by \autoref{eq:Gk}. Each cell of type $A_i$, $i < k$,  is a higher order meta-stem cell; it produces stem cells of type $A_{i+1}$.  Thus, cell type $A_{k-1}$ is a meta-stem cell that produces stem cells of type $A_k$, etc. The cell type $A_k$ is a stem cell that produces regular non-stem cells of type D.  The difference between stem cells and cancer stem cells has to do with their functionality in the system as a whole. 
  }
  \label{fig:NGk}
\end{figure}
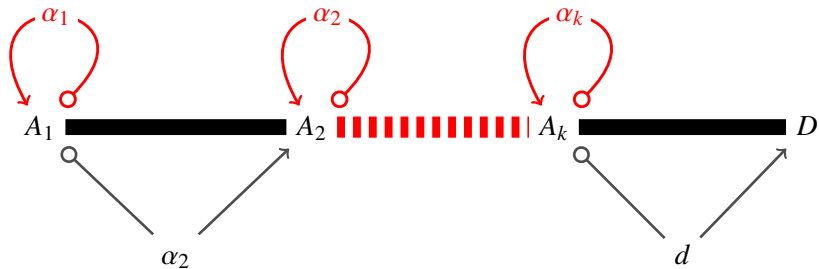 

\subsection{Mathematical properties of geometric cancer networks}
\label{sec:Pascal}
A cancer network of the above type with one or more loops in a linear connected sequence leads has interesting mathematical properties. It turns out that such networks are directly related to Pascal's Triangle, binomial coefficients and geometric numbers. Under the appropriate interpretation, one loop simple produces a line. Two loops produce a triangle. Three loops produce an equilateral pyramid (tetrahedron), four loops a pentalope (a four dimensional pyramid), etc.  In general, an network of $k$ linearly connected loops will produce an $k$-dimensional pyramid. Or viewed numerically, after $n$ synchronous rounds of ideal division, a cancer network with $k$ linearly connected single loops produce the sum of the first $k$ binomial coefficients at level $n$ of Pascal's triangle. 

Recall that by definition:
\begin{equation}
 \left( \begin{array}{c} n \\ k \end{array} \right) =  \frac{n!}{k!(n - k)!}
\end{equation}

Then the number of cells produced by a single cell that is controlled by an cancer network with $k$ linearly connected loops, after undergoing $n$ rounds of synchronous cell division division is given by the following  formula (for  $n > 0$): 
\begin{equation}
{\mbox Cells}(n, k) = \sum_{i = 0}^k \left( \begin{array}{c} n \\ i \end{array} \right) = \sum_{i = 0}^k \frac{n!}{i!(n - i)!}
\label{GGformula}
\end{equation}

The above formula shows the direct relationship between such cancer networks and the binomial coefficients of Pascal's triangle, i.e., the coefficients the binomial theorem. 

Given the following standard definitions: 
\begin{enumerate}

\item Linear number: \begin{equation}
 Lin(n) = 1 + n = \left( \begin{array}{c} n \\ 1 \end{array} \right) \mbox{ when } n \geq 1
\end{equation}

\item Triangular number: 
\begin{equation}
 Tri(n) = \frac{n^2 + n}{2} = \frac{n(n + 1)}{2}  = \left( \begin{array}{c} n \\ 2 \end{array} \right)
 \label{eq:TriangularNo}
\end{equation}

\item Tetrahedral number: 
\begin{equation}
Tet(n) =  \frac{n(n + 1)(n+2)}{6} = \left( \begin{array}{c} n \\ 3 \end{array} \right)
\end{equation}

\item Pentalope number:  
\begin{equation}
Pen(n) = \frac{n(n + 1)(n+2)(n + 3)}{24} = \left( \begin{array}{c} n \\ 4 \end{array} \right)
\end{equation}
\end{enumerate}

Then the number of cells, Cells$(n,k)$, that develop after $n$ rounds of synchronous division obey the sum of the coefficients of Pascal's Triangle where $n$ is the height of the triangle and $k$ is horizontal coordinate corresponding to the number of single connected loops in the regulatory network.  These sums correspond to the volumes of the corresponding $k$-dimensional geometric form in an ideal $k$-dimensional discrete space:  

\begin{enumerate}
\item For one loop we get  a $1$-dimensional structure where its length gives the number of cells: 
\begin{equation}
Cells(n, 1) = Lin(n) = 1 + n
\end{equation}

\item Two loops add the triangular number to give the area of a $2$-dimensional triangle: 
\begin{eqnarray} Cells(n, 2) & = & Lin(n) + \textcolor{red}{Tri(n-1)} \\ \STRUT
& = & 1 + n + \textcolor{red}{\frac{n(n - 1) }{2}}  \\ \STRUT
\label{eq:G2}
\end{eqnarray}

\item For three loops add the tetrahedral number to give the volume of $3$-dimensional pyramid: 
\begin{eqnarray} Cells(n, 3) & = & Lin(n) + Tri(n-1) + \textcolor{red}{Tet(n-2)} \\ \STRUT
& = & 1 + n + \frac{n(n - 1)}{2} + \textcolor{red}{ \frac{n(n-1)(n-2)}{6}} \\ \STRUT
\end{eqnarray}

\item For four loops add the pentalope number to give the volume of a $4$-dimensional pyramid:  
\begin{eqnarray} 
 Cells(n, 4) & = & Lin(n) + Tri(n-1)+ Tet(n-2) + \textcolor{red}{Pen(n-3)} \\ \STRUT
 & = & 1 + n + \frac{n(n - 1)}{2} + \frac{n(n-1)(n-2)}{6} + \textcolor{red}{\frac{n(n-1)(n-2))(n-3)}{24}} \\ \STRUT
 \end{eqnarray}
 
 \item For $k$ loops sum the sequence of numbers through ${\frac{n!}{k!(n - k)!}} $ to give the volume of a $k$-dimensional pyramid:  
\begin{eqnarray} 
 Cells(n, k) & = & Lin(n) + Tri(n-1)+ Tet(n-2) + \ldots + \textcolor{red}{\left( \begin{array}{c} n \\ k \end{array} \right)} \\ \STRUT
 & = & 1 + n + \frac{n(n - 1)}{2} + \frac{n(n-1)(n-2)}{6} + \dots 
 + \textcolor{red}{\frac{n!}{k!(n - k)!}} 
 \label{eq:Gk}
 \end{eqnarray}
 
 \item So, in general we have: 
\begin{eqnarray}
 Cells(n, k) = \left\{ 
 \begin{array}{ll} 
 		1 & \mbox{if $n = 0$} \\ 
 		\sum_{i = 0}^k \frac{n!}{i!(n - i)!} 
 		= \sum_{i = 0}^k \left( \begin{array}{c} n \\ i \end{array} \right) 
 		& \mbox{otherwise}
 \end{array} 
 				\right.
\end{eqnarray}
\label{eq:generalGk}

\end{enumerate}

This view shows the direct relationship between this type of cancer and stem cell network and the geometric numbers. 
What is fascinating is that these ideal networks have a generative competence with such interesting numerical properties.  Their mathematical history spans back from the binomial theorem, to Pascal's triangle, to the Greek's discovery of geometric numbers.

\section{Stem cell networks and cancer stem cells}
\label{sec:NSC}
Normal networks that come closest to cancer networks are stem cell networks.  What distinguishes normal from cancer stem cell networks is not primarily their network architecture but rather when and where a stem cell network is linked and activated, and what further networks the stem cell network activates.  A cancer stem cell network is either formed by mutation in the wrong place in the global developmental network, it is activated inappropriately in the multicellular system, or it activates inappropriate, nonfunctional networks with abnormal and possibly pathological cellular or multicellular phenotypes relative to the overall multicellular system. 

From our theory it follows that there are at least two main kinds of stem cell networks, linear and geometric.  However, linear stem cell networks are just a special case of geometric stem cell networks, namely, 1st order geometric stem cell networks.  These are also fundamental cancer stem cell networks.  A normal linear stem cell produces a cell of a particular type that is either terminal itself or activates a terminal network. A second order geometric stem cell network is a meta-stem cell network that produces linear stem cells.  Thus, linear stem cells produce no further stem cells. Meta stem cells do produce stem cells.  

Since a linear stem cell cannot produce further stem cells there must exist meta-stem cells that produce more than one stem cell. These in turn are produced by yet further upstream embryonic networks.  The original embryonic fertilized egg is controlled by the global embryonic network that is mostly terminal with the exception of its stem cells.  It consists of many subnetworks that may be multiply employed. 

A linear cancer stem cell produces no additional cancer cells.  Interestingly, we will see that the order of a cancer meta-stem cell network controlling a cancer cell determines the possible metastases producible by that cancer cell (see \autoref{sec:GMetastases}). 

A further possibility is that there exist normal {\em exponential stem cell networks} that are activated by cell signaling, in effect a communication network linked with a cytogenic control network to produce cells quickly on demand. However, such networks are be dangerous, leading to a proliferative explosion if something goes wrong with the communication network. 

\subsection{First order stem cells}
\label{sec:NSC1}
A first order stem cell produces no additional stem cells.  Instead a first order stem cell $A$ produces cells $B$  that is in a different control state than its parent cell $A$. The stem cell thus produces one daughter cell that inherits the control state of its parent which is $A$ itself and one daughter cell $B$ that is in a new control state.  Thus stem cells have a self reflexive control system.  Note, that the cell $B$ may still be multi-potent  in that it may be controlled by a network that generates a whole multicellular system. 

First order stem cells have conditional control networks. This means that their activation depends on not just being linked into another network that activates them but that they can only be active if the conditions $\Phi$ are satisfied.  

\subsection{Second order stem cells or meta-stem cells}
\label{sec:DSC2}
Unlike first order stem cells, stem cells or meta-stem cells produce other first order stem cells.  They generate stem cells. Since a first order stem cell cannot generate any further stem cells to produce stem cells we need meta-stem cells or they must be produced by networks of the type NI$_n$ that could produce $2^n$ identical first order stem cells. It is possible that organisms use both strategies. 

Meta stem cells, like first order stem cells,  also have conditional activation in that being linked into another network is not sufficient for their potential activation. Their preconditions $\Theta$ must be satisfied at each pass of the loop to execute the next loop.  

 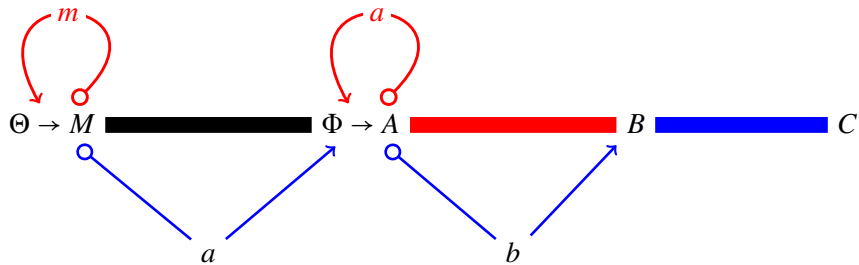
\begin{figure}[H]
\begin{tikzpicture}[style=mystyle]
\matrix (m) [matrix of math nodes, row sep=3em,
column sep=3em, text height=1.5ex, text depth=0.25ex]
{ \vphantom{a} & \vphantom{b}  &  \vphantom{c}  & \vphantom{c}  & \\
 \Theta \rightarrow M &&  \Phi \rightarrow A &&  B  && C \\
  \vphantom{a} &  a & \vphantom{b} &  b  & \\ };
 \path[\pot]
(m-2-1) edge [selfloop1, cross line] node[nodedescr] {$ m $} (m-2-1);
 \path[\poti]
(m-2-1) edge [pot2,blue](m-3-2);
 \path[\pote]
(m-3-2) edge [pot2,blue] (m-2-3); 
\path[\pot]
(m-2-3) edge [selfloop1, cross line] node[nodedescr] {$ a $} (m-2-3);
 \path[\poti]
(m-2-3) edge [pot2,blue](m-3-4);
 \path[\pote]
(m-3-4) edge [pot2,blue] (m-2-5);

\path[solid,  line width=6pt]
(m-2-1) edge (m-2-3);
\path[solid, red, line width=6pt]
(m-2-3) edge (m-2-5);
\path[solid, blue, line width=6pt]
(m-2-5) edge (m-2-7);
\end{tikzpicture}
\caption{
    {\bf Network  DSC2: Second order, meta-stem cell network:} The cell M is a {\em meta-stem cell} that given condition $\Theta$ holds,  produces stem cells of type $A$ which in turn  conditionally produces cells of type $B$. Thus, the cell of type $A$ is  a regular stem cell.  If condition $\Phi$ hold, then  $A$ produces cells of type $B$. 
}
    \label{fig:DSC2}
\end{figure} 

\subsection{Higher order stem cells}
Since stem cells obey the properties of geometric networks, theoretically it is possible to have stem cell networks of arbitrary order $k$, such as 3rd order and 4th order meta-stem cell networks. All such networks can have various preconditions $\Phi_{i}$ attached to the initiation of each the contained stem cell network loops of order $i$.  Without preconditions in an idealized setting without physical constraints, an $k-$th order stem cell network would have the proliferations properties of an $k$-th order geometric network (see \ref{sec:NGk}).

\subsection{Cancer susceptibility of stem cells}
We will see why stem cells are especially susceptible to develop into cancer networks.  First, all that has to happen is that the first order or second order conditional antecedents are mutated to be always true, then the stem cell networks, once activated produces cells indefinitely independent of outside signaling.  Second, any of the links from the other non-proliferative daughter cell differentiation site could loop back to a first or second order loop activation point to produce an exponential cancer. 

\subsection{Normal stem cells and cancer stem cells}

Stem cells have the function of continually producing new cells in tissue that is meant to regenerate. Stem cells have  functional developmental control networks with preexisting loops. This makes the progression to cancer a shorter route, because the linear, proliferating network already exists.  We just need to mutate in one more loop to makes it exponential and then with additional mutations it becomes invasive.  Prostate cancer may be an instance of this process.  

If the stem cell proliferates conditionally, i.e., based on some contextual cellular or environmental condition, then a stem cell cancer can develop by a constant activation of the antecedent condition necessary for the stem cell to divide. This can happen if the activating signal is constant or the receptor system for that signal is mutated to be constantly on.  It may be the case that a stem cell may already be potentially, linearly, geometrically or exponentially, proliferative, but its antecedent conditions may not be satisfied. Hence, if the conditional system remains functional the potential stem cell proliferation whether it be linear or exponential will not show itself.  However, when the antecedent conditions for proliferation are fulfilled or the antecedent testing system is mutated to a constant on state then the stem cell cancer will flower.  An example of this would be a stem cell whose proliferation depends on some signal such as a hormone that, together with a signal transduction pathway, activates the stem cell proliferative loop.  Then if the signal is constantly turned on or the signal transduction mechanism is in a constant "signal received" state, then the stem cell will proliferate either linearly, geometrically or exponentially depending on the structure of its controlling network. 

A further danger with some types of  stem cells is that they may already be functionally, tissue invasive in order to move to the appropriate site, as, for example,  in wound healing. This invasive property would then make a stem cell cancer even more dangerous. 

Stem cells by themselves are not cancerous, but they have the properties of conditional, possibly invasive, linear, geometric or exponential cancers.  The boundary between a cancer and normal cell may not always be clear since in the wrong context a normal stem cell may be cancerous. 

Distinguishing linear cancer cells from stem cells based purely on the network architecture is difficult if not impossible.  Stem cells generate normal cell that are functional in the context of the organism.  Any stem cell that generates such "normal" cells inappropriately is a cancer cell. Thus, the dividing line between linear cancers and stem cells may be indistinct.   Thus, a normal stem cell network becomes cancerous if it is activated inappropriately with respect to the overall functioning of a biological system in the organism, or if the cell differentiation networks the stem network activates generate abnormal cells (e.g., invasive cells, or having  abnormal phenotype) or cells that generate structures inappropriately in the overall system.

\section{Stochastic differentiation in stem cell networks}
\label{sec:NSSC}
It has been argued that the traditional theory of linear stem cells is wrong \cite{Jones2007}.  On the traditional theory of stem cells, stem cells are immortal and constant in number.  However, experimental data appears to show that for some stem cells appear to generate further stem cells.  On a deterministic theory stem cells that produce stem cells are what we have termed meta-stem cells.  However, another model would have stem cells divide and differentiate stochastically either exponentially into two stem cells, linearly into a stem cell and a terminal cell, or into two terminal cells.  The evidence, suggests that epidermal cells divide stochastically according to the distribution 8\% double stem cells, 84\% one stem cell and one terminal cell, and 8\% two terminal cells.  The biological mechanism is not known.  

\begin{figure}[H]
\begin{tikzpicture}[style=mystyle]
\matrix (m) [matrix of math nodes, 
row sep=3em,
column sep=\ColSepThree, %\ColSepTight, \matrixsep, 3em, 
text height=1.5ex, text depth=0.25ex]
{ && \vphantom{b} &&  \vphantom{c}  && \vphantom{c}  & & &\\
 A  && SS &&  L  && TT &&  T && \vphantom{c} \\ %TT\\
  &&  \vphantom{b} && \vphantom{B} &&  \vphantom{c}  && \vphantom{C} & \vphantom{d} & \vphantom{D}\\ };

\path[stochasticPathstyle]
(m-2-1) edge [inPot1, green, cross line] node[stochasticNodestyle] {$p = 0.08$} (m-2-3) %A to SS 3
(m-2-1) edge [in2Pot2,green, cross line] node[stochasticNodestyle] {$p  = 0.08$} (m-2-5) %A to L 5
(m-2-1) edge [inPot2, out=-125, in= -90, distance=6cm, green, cross line] node[stochasticNodestyle] {$p  = 0.84$} (m-2-7); %A to TT 7
\path[\pot]
(m-2-3) edge [inPot1, red, cross line] node[nodedescr] {$l_1$} (m-2-5) %SS to L 5
(m-2-3) edge [inPot2,blue, cross line] node[nodedescr] {$l_2$} (m-2-5) %SS to L

(m-2-5) edge [selfloop1, red,out=45, in=90, cross line] node[nodedescr] {$l$} (m-2-5) %L to L
(m-2-5) edge [selfloop2,blue, in=-80, cross line] node[nodedescr] {$t$} (m-2-9) %L to T

(m-2-7) edge [inPot1, red, cross line] node[nodedescr] {$t$} (m-2-9) %TT 7 to T 9
(m-2-7) edge [inPot2,blue, cross line] node[nodedescr] {$t$} (m-2-9); %TT to T

\path[solid,red!70, line width=6pt]
(m-2-1) edge (m-2-3);
\path[solid,green!60, line width=6pt]
(m-2-3) edge (m-2-5);
\path[solid,blue!40, line width=6pt]
(m-2-5) edge (m-2-7);
\path[solid,black!40, line width=6pt]
(m-2-7) edge (m-2-9);

\end{tikzpicture}
\caption{
    {\bf Network TriSSC: Tripartite stochastic stem cell network that activates one of three developmental networks}. One that produces one meta-stem cell (SS) which, in turn, produces two linear stem cells (L), one that produces one stem cell (L) which generates terminal cells (T) and one that produces a pre-terminal cell (TT) that, in turn, generates two terminal cells (T). Globally, network generates either two terminal cells or a minimum of one and at most two nonterminal stem cells. 
  }
  \label{fig:TriSSC}
\end{figure}
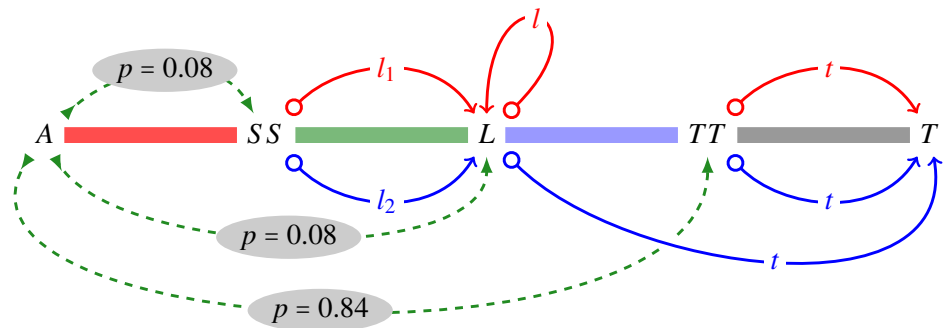

For example, a simple stochastic model might have ternary stochastic differentiation where a cell differentiates by activating, according  to a tripartite probability distribution, one of three developmental networks, namely one that produces a meta-stem cell that generates two stem cells, one that produces one stem cell that generates terminal cells and one that produces a pre-terminal cell that generates two terminal cells. However, such a network would at most produce two stem cells, or one stem cell or no stem cells.   No dedifferentiation to the originating cell $A$ is possible under this network. 

\subsection{Historical background: A network that generates Till's stochastic stem cell model}

One of the first models of stem cells was the stochastic model of hemopoietic cell proliferation  [Till et.al.~\cite{Till1964}].  Till distinguished what he called ``colony forming cells'' that have the capacity to form colonies from differentiated cells without that capacity.  When a colony develops from a single cell only a small number of these cells have ``colony-forming capacity''. The rest are differentiated cells without this capacity.   This he considered a "birth-and-death" process, the generation of colony cells being a birth process and the generation of differentiation a death process. According to the model, a given cell with colony-forming capacity can either divide into two colony forming cells (with probability $p_{2}$) or differentiate into a terminal cell (with probability $p_{0} = 1 - p_{2}$) having no colony forming capacity.  The case of generating mixtures of one colony-forming cells and one differentiated cell was considered a birth process followed by death and was handled by adjusting the probabilities.   
 
Production numbers of blood cells is relatively constant under normal conditions. Under stress or increased demand there is a rapid increase in production of blood cells.  The properties of hemopoiesis (blood cell creation) implies that the production of differentiated cells is precisely controlled.  Hence, Till argues, there must be control mechanisms.  Till then considers if cell proliferation (the numbers of cells produced) is also under precise control or lax control.  He argues for lax control and that the data suggests cell proliferation is stochastic. While a stem cell network was not given by Till, the behavior of the Till model of colony forming cells can be generated by following stochastic stem cell network:

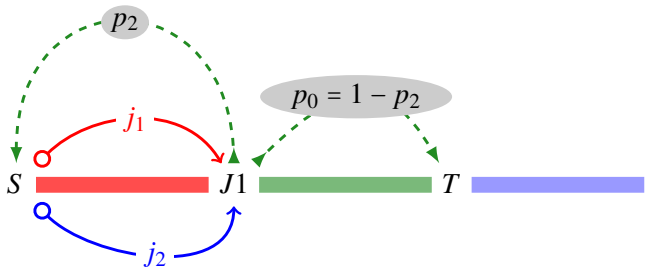
\begin{figure}[H]
\begin{tikzpicture}[style=mystyle]
\matrix (m) [matrix of math nodes, row sep=3em,
column sep=3em, text height=1.5ex, text depth=0.25ex]
{ \vphantom{a} && \vphantom{b} &&  \vphantom{c} \\ %&t_{1} & \vphantom{c}\\ %  & & &\\
 S  && J1 &&  T && \vphantom{b}\\%T1 && T2 &&  \\
 \vphantom{a} &&  \vphantom{b} && \vphantom{B} &&\\ };
 \path[\pot]
(m-2-1) edge [inPot1, red, cross line] node[nodedescr] {$j_1$} (m-2-3) %S to J1 
(m-2-1) edge [inPot2, in=-90, blue, cross line] node[nodedescr] {$j_{2}$} (m-2-3); %S to J1
\path[stochasticPathstyle]
(m-2-3) edge [inPot1, out=90, in=90,distance=5cm, green, cross line] node[stochasticNodestyle] {$p_{2}$} (m-2-1) %J1 to S 
(m-2-3) edge [in2Pot1,in=120,distance=3cm,green, cross line] node[stochasticNodestyle] {$p_{0} = 1 -p_{2}$} (m-2-5); %J1 to T
\path[solid,red!70, line width=6pt]
(m-2-1) edge (m-2-3);
\path[solid,green!60, line width=6pt]
(m-2-3) edge (m-2-5);
\path[solid,blue!40, line width=6pt]
(m-2-5) edge (m-2-7);
\end{tikzpicture}
\caption{
    {\bf Network TillSSC:  Classic stochastic stem cell network}. This network models one of the earliest stochastic stem cell theories (Till~\cite{Till1964}).  A stem cell (S) divides to produce two cell of type J1 that then stochastically change either into a terminal network cell T or ``self renew'' by dedifferentiating to the parent stem cell S type. The cell J1  stochastically either loops back to activate the stem cell S with probability $p_{2}$ or it activates the terminal network cell T with probability $p_{0}= 1- p_{2}$.  The network's behavior approaches a deterministic exponential stem cell network as the probability $p_{2}$  approach $1$.  Stem cells controlled by such a network can spontaneously stop because of the fact that for all points in all possible  paths in the network there is the possibility of reaching the terminal cell state T. The multicellular system controlled by this network consists of two main cell types:  Stem cells (S) that can proliferate and terminal cells (T) with no proliferative potential. 
  }
   \label{fig:TillSSC}
\end{figure}

Under this  network (\autoref{fig:TillSSC}) a stem cell divides into identical daughter cells J1. Each J1 cell  can with probability $p_{2}$  dedifferentiate into a stem cell S or differentiate with probability $p_{0} = 1-p_{2}$ into a terminal cell T.  Hence, unlike Till's original model, this network explicitly handles all the possible outcomes of stem cell division.  Upon stem cell division, the network allows the generation of two stem cells (S,S), as well as the mixed case (S,T) of one stem cell and one differentiated cell, and the case (T,T) of two differentiated cells.  

Till's model  has been criticized because it predicts that stem cells eventually differentiate into terminal cells and, thus, no longer self renew.  However, this depends on the probability distribution.  If $p$ is high then exponential proliferation outruns terminal cell differentiation. 

A central problem with this network is that the key to its behavior is dependent on the value of the probability $p_{2}$.  On the one hand, if the probability $p_{2}$ of stem cell self-renewal is high, it tends toward exponential growth of stem cells. This leads to too many stem cells versus progenitor and terminal, specialized cells. On the other hand, if the self-renewal probability $p_{2}$ is low, it eventually  leads to the elimination of all stem cells. In the latter case, this network does not represent a stem cell with unlimited proliferative potential.  In this network, no dedifferentiation is possible for terminal cells. Hence, the greater the probability $p_{0} = 1-p_{2}$ that the terminal network T is activated, the more likely is the permanent quiescence of the parent stem cell.     

If the probability $p_{2}=1$ we have exponential growth of stem cells.  If $p_{2}=0$ we have differentiation to the terminal cells T.  Hence, this network cannot model the developmental dynamics of deterministic linear networks of the type NL~\autoref{fig:NL}.  However, with the right choice of the probability distribution, it can approximate a relatively constant production of stem cells and terminal cells. 

In their Monte Carlo simulation the probability used was $p_{2} = 0.6$ and $p_{0}=0.4$.  In our Monte Carlo simulation, using their probability distribution, the network proliferation dynamics tends toward a 21\% proportion of stem cells after 52 generations.  
This appears to be much higher than the empirical data for hemopoietic stem cells.  
However, these percentages depend on how many cells are generated by the terminal cell T.  If T is not a terminal cell and instead a progenitor cell controlled by a terminal network T*, and is instead a terminal cell state that does not divide further (as was the case in Till's original model), then the stem cell percentage is lower being 12\% after 20 generations of cell division. We add progenitor cell capacity to T in the network in (\autoref{fig:TillSSCpro}) below. 

Note, in this sort of modeling approach, precise differentiation networks (architectures) are coupled with stochastic activation networks.  The topology of the network architecture sets the boundaries for what stochastic paths are possible at all. 

\begin{figure}[H]
\subfloat[Probability $p_{2}$ = 0.52, gives 4\% stem cells]{
\includegraphics[scale=0.4]{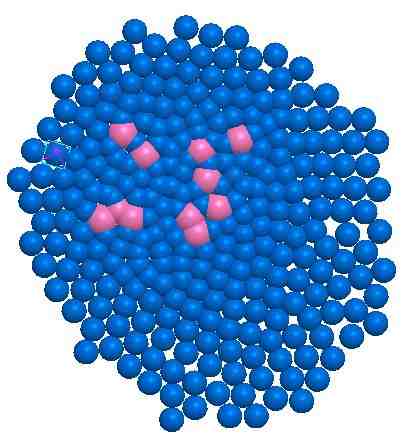}
\label{fig:StocStemSubfig1}
}
\hspace{0.3cm}
\subfloat[When $p_{2}$ = 0.7 gives 36\% stem cells]{
\includegraphics[scale=0.4]{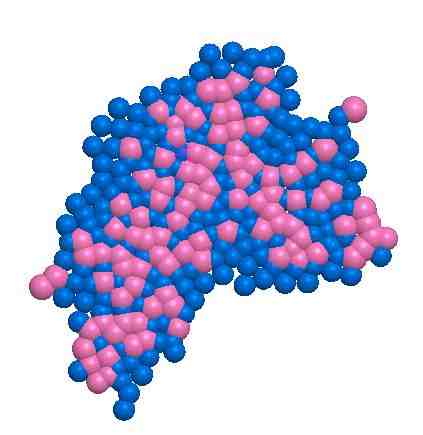}
\label{fig:StocStemSubfig2}
}
\hspace{0.3cm}
\subfloat[Probability $p_{2}$ = 0.9 gives 80\% stem cells]{
\includegraphics[scale=0.4]{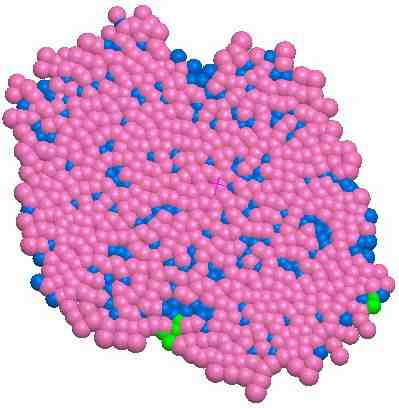}
\label{fig:StocStemSubfig3}
}
\caption{
{\bf Three views of stem cells under different probabilities.} All three tumors are  controlled by same Till network (\autoref{fig:TillSSC} ).  It shows the growth response to changes in the probability $p_{2}$.   As the probability $p_{2}$ is increased, more and more stem cells are generated due to increasing dedifferentiation of the daughter cells to their parent stem cell.  The stochastic exponential network begins to dominate as $p_{2}$ approaches 1. 
}
\label{fig:StocStemProb}
\end{figure}

\subsection{The Till stochastic network extended to generate progenitor cells}
It is very simple to extend Tills original model of hemopoietic (blood) stem cells (S)  [Till et.al.~\cite{Till1964}] to include progenitor networks that control the proliferative capacity of stem cell progeny.  We simply link the terminal cell state to a bounded network T* that in the case below is simply a network that results in one further cell division to produce two distinct daughter cells T1 and T2.  But, T* could be any bounded network.  Thus, in principle, given the right network T*, such progenitor cells could develop into arbitrarily complex structures.  

\begin{figure}[H]
\begin{tikzpicture}[style=mystyle]
\matrix (m) [matrix of math nodes, row sep=3em,
column sep=3em, text height=1.5ex, text depth=0.25ex]
{ \vphantom{a} && \vphantom{b} &&  \vphantom{c}  &t_{1} & \vphantom{c}\\ %  & & &\\
 S  && J1 &&  T*  && T1 && T2 &&  \\
 \vphantom{a} &&  \vphantom{b} && \vphantom{B} && t_{2}  && \vphantom{T}\\}; % & \vphantom{d} & \vphantom{D}\\ };
 \path[\pot]
(m-2-1) edge [inPot1, red, cross line] node[nodedescr] {$j_1$} (m-2-3) %S to J1 3
(m-2-1) edge [inPot2, in=-90, blue, cross line] node[nodedescr] {$j_{2}$} (m-2-3); %S to B 5
%\path[right hook->]
\path[stochasticPathstyle]%[>=latex,>->,dashed]
(m-2-3) edge [inPot1, out=90, in=90,distance=5cm, green, cross line] node[stochasticNodestyle] {$p_{2}$} (m-2-1) %J1 to S 
(m-2-3) edge [in2Pot1,in=120,distance=3cm,green, cross line] node[stochasticNodestyle] {$p_{0} = 1 -p_{2}$} (m-2-5); %J1 to T
 \path[\poti]
(m-2-5) edge [pot2,blue](m-3-7);%T to T2
 \path[\pote]
(m-3-7) edge [pot2,blue] (m-2-9); 
 \path[\poti]
(m-2-5) edge [pot2,red](m-1-6);%T to T1
 \path[\pote]
(m-1-6) edge [pot2,red] (m-2-7);

\path[solid,red!70, line width=6pt]
(m-2-1) edge (m-2-3);
\path[solid,green!60, line width=6pt]
(m-2-3) edge (m-2-5);
\path[solid,blue!40, line width=6pt]
(m-2-5) edge (m-2-7);
\path[solid,black!40, line width=6pt]
(m-2-7) edge (m-2-9);
\end{tikzpicture}
\caption{
    {\bf Network TillSSCpro:  Classic stochastic stem cell network (\autoref{fig:TillSSC}) extended to include progenitor networks}. This network extends one of the earliest stochastic stem cell theories (Till~\cite{Till1964}) to include progenitor cells.  A stem cell (S) divides to produce two cell of type J1 that then either divide into a terminal network cell T* or ``self renew'' by dedifferentiating to the parent stem cell S.  The cell J1 stochastically loops back to activate the stem cell S with probability $p_{2}$ or it activates the terminal network cell T* with probability $p_{0}= 1- p_{2}$.  In Till, T* is a terminal cell type T.  In this example, T* is a terminal network for a progenitor cell that generates two terminal cell types T1 and T2. As with the original model (\autoref{fig:TillSSC}), this network's behavior approaches a deterministic exponential stem cell network as the probability $p_{2}$  approach $1$.  Stem cells controlled by such a network can spontaneously stop because of the fact that for all points in all possible  paths in the network there is the possibility of reaching the terminal cell state T*. The multicellular system controlled by this network consists of three main cell types:  Stem cells (S), progenitor cells (T*) with limited proliferative potential, and terminal cells (T1, T2) with no proliferative potential. 
  }
   \label{fig:TillSSCpro}
\end{figure}
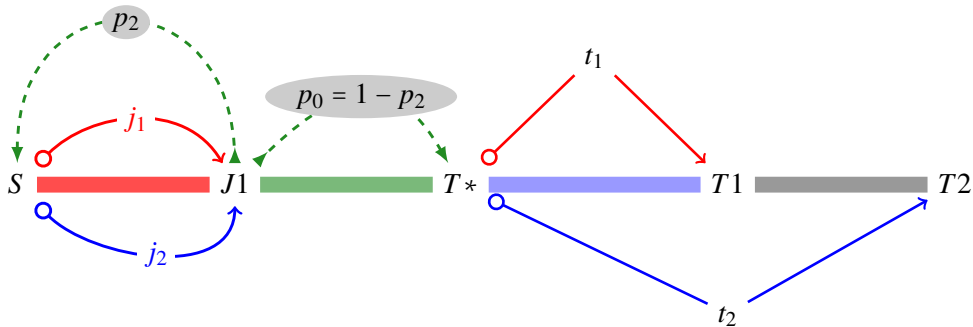

As with the previous network \autoref{fig:TillSSC}, the proliferative potential of stem cells under this network \autoref{fig:TillSSCpro} is dependent on the probability $p_{2}$.  Since no dedifferentiation is possible in this network for terminal cells T1 and T2.  Furthermore, it is not an inherent property of the network topology that stem cells self renew.  However, if the dedifferentiation probability $p_{2}$ of self renewal is high enough then stem cells do self renew and may even dominate the colonies they generate.  

If the probability $p_{2}=1$ we have exponential growth of stem cells, with no progenitor cells T*.  If $p_{2}=0$ we have differentiation to to the terminal progenitor network T* which in this case ends in just two terminal cells T1 and T2.  Hence, this network cannot model the developmental dynamics of deterministic linear networks of the type NL~\autoref{fig:NL}.  However, with the right choice of the probability distribution, it can approximate that a relatively constant ratio of stem cells, progenitor and terminal cells. 

It follows that under the Till model of stem cells we do not need dedifferentiation of terminal cells to maintain stem cell capacity.  All we need is for the 1st order stem cells to be physically loose, mixing in with the terminal cells in the tumor as it grows.  In this case, we would see proliferating stem cells within the tumor in the context of what appear to be only terminal cells.  It may appear as if terminal cells have spontaneously dedifferentiated into stem cells, but this need not be the case since it may be the result of stem cell mixing.  

What could also be happening is that we have a 2nd order geometric stem cell network that is generating 1st order stem cells which due to developmental control in conjunction with physics leads to a mixture of 1st order stem cells with terminal cells. Since the control state of a cell my be hidden by the overt differentiation state of the cell, the 1st order stem cells may be indistinguishable relative to a set of markers from terminal cells.  Stochastically or under particular conditions these 1st order stem cells may then begin to proliferate generating terminal or progenitor cells. 

\subsection{A flexible stochastic network architecture with exponential and linear potential}

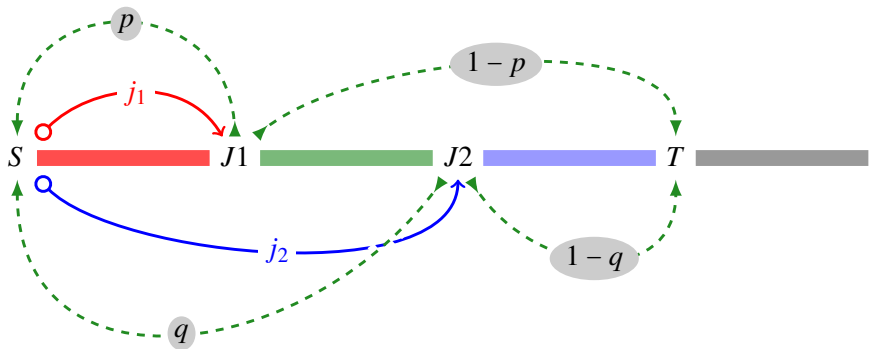
\begin{figure}[H]
\begin{tikzpicture}[style=mystyle]
\matrix (m) [matrix of math nodes, 
row sep=2em,
column sep=3em, 
text height=1.5ex, text depth=0.25ex]
{ \vphantom{a} && \vphantom{b} &&  \vphantom{c}  && \vphantom{c}  & & &\\
 S  && J1 &&  J2  && T && \vphantom{c}  && \vphantom{c} \\ %TT\\
 \vphantom{a} &&  \vphantom{b} && \vphantom{B} &&  \vphantom{c}  && \vphantom{C} & \vphantom{d} & \vphantom{D}\\ };
 \path[\pot]
(m-2-1) edge [inPot1, red, cross line] node[nodedescr] {$j_1$} (m-2-3) %S to J1 5
(m-2-1) edge [inPot2, in=-90, blue, cross line] node[nodedescr] {$j_2$} (m-2-5); %S to 2
 % \path[->]
%\path[right hook->]
\path[stochasticPathstyle]
(m-2-3) edge [inPot1, out=90, in=90,distance=4cm, green, cross line] node[stochasticNodestyle] {$p$} (m-2-1) %J1 to S 1
(m-2-3) edge [in2Pot1,green, cross line] node[stochasticNodestyle] {$1-p$} (m-2-7) %J1 to T 7
(m-2-5) edge [inPot2, out=-125, in= -90, distance=6cm, green, cross line] node[stochasticNodestyle] {$q$} (m-2-1) %J2 to S 
(m-2-5) edge [inPot2, out=-60, in= -90, distance=3cm, green, cross line] node[stochasticNodestyle] {$1-q$} (m-2-7); %J2 to T 
\path[solid,red!70, line width=6pt]
(m-2-1) edge (m-2-3);
\path[solid,green!60, line width=6pt]
(m-2-3) edge (m-2-5);
\path[solid,blue!40, line width=6pt]
(m-2-5) edge (m-2-7);
\path[solid,black!40, line width=6pt]
(m-2-7) edge (m-2-9);
\end{tikzpicture}
\caption{
    {\bf Network XLSSC: Flexible exponential-linear stochastic stem cell network}.  One stem cell (S) divides to produce two cells (J1 and J2) that each stochastically activate either the stem cell itself or a terminal cell.     }
   \label{fig:XLSSC}
\end{figure}

In the network in \autoref{fig:XLSSC} is very flexible. Depending on the probability distribution, the network can range between being exponential, linear, or terminal, as well as every mixture in between.  Furthermore, the network can be deterministic, stochastic or mixture of both.  This flexibility is partly the result of separating out the probability distributions for the behaviors of the two daughter cells.  It shows that the architecture of the network imposes constraints on what kinds of developmental dynamics are in principle possible.  The probability distribution presupposes a network architecture of possible developmental paths.  

The cell S is only a stem cell stochastically and not intrinsically when $p < 1$ and $q < 1$.   When $p = q = 1$ the network is deterministic  exponential. When $p = 1, q = 0$ or $p = 0, q = 1$ the network is deterministic linear, i.e., a deterministic 1st order geometric stem cell network. If $p = q = 0$ the network is terminal.  When $p = 1$  and $q < 1$, or when $q=1$ and $p<1$, then the network is mixed deterministic linear with stochastic exponential tendencies.  

If probabilities $p = q$ then as $p$ and $q$ approach $1$ the network  approaches the behavior of a deterministic exponential network.  However, if $p$ and $q$ are different then this network can simulate a linear stochastic network as well when, for example, $p$ approaches $1$ and $q$ approaches $0$, or vice versa. As the probabilities $p$ and $q$ decrease,  the more frequently the cancer stem cell results in a terminal tumor that does not develop further because it consists only of cells of terminal type T.  This shows that stochastic cancer stem cell networks can in some cases go into spontaneous remission. While this network can also exhibit exponential growth even in a stochastically linear probability distribution, because of the two backward loops, there is a diminishing probability that it remains exponential. Thus, whether this network results in linear or exponential proliferation depends on the probability distribution.  

For the network XLSSC, if the probabilities $p = 1- q$ then the higher the probability of $p$ the more the network approximates a deterministic linear developmental network.  Since, in this case the distribution is anti-symmetric,  the cell population partition of cell types consists of an equal number of exponential stem cells and terminal cells, with the majority of cells being linear stem cells. This corresponds to the observed distribution in epidermal basal stem cells.  If, on the other hand we have a symmetric distribution where $p = q$ then the higher the probability of $p$ the more the network approximates a deterministic exponential network.  Thus the type of cancer network we have depends on the probability distributions over the connecting stochastic links.  

The network XLSSC (\autoref{fig:XLSSC}) has some similarity to the Till model (TillSSC~\autoref{fig:TillSSC}), when $p=q$ (and value of probability $p=p_{2}$ as in the original model by Till~\cite{Till1964} model) and T.  The Till model forces that daughter cells of S are the same. In contrast, this model is more flexible in that, depending on the probability distribution ($p$, $q$), it can model both exponential and linear dynamics of multicellular development. 

\subsection{Linear stochastic stem cell network LSSC}

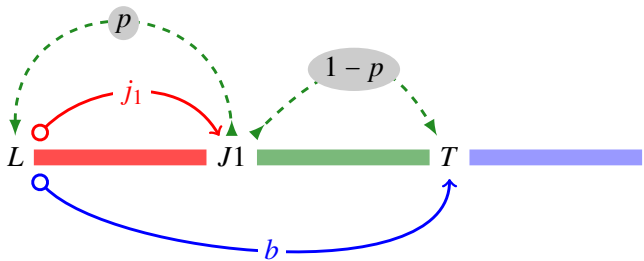
\begin{figure}[H]
\begin{tikzpicture}[style=mystyle]
\matrix (m) [matrix of math nodes, row sep=3em,
column sep=3em, text height=1.5ex, text depth=0.25ex]
{ \vphantom{a} && \vphantom{b} &&  \vphantom{c}  && \vphantom{c}\\ %  & & &\\
 L  && J1 &&  T  && %\vphantom{J2) &&  \vphantom{C) && \vphantom{c} \\ %TT\\
 \vphantom{a} &&  \vphantom{b} && \vphantom{B} &&  \vphantom{c}  && \vphantom{T}\\}; % & \vphantom{d} & \vphantom{D}\\ };
 \path[\pot]
(m-2-1) edge [inPot1, red, cross line] node[nodedescr] {$j_1$} (m-2-3) %S to J1 3
(m-2-1) edge [inPot2, in=-90, blue, cross line] node[nodedescr] {$b$} (m-2-5); %S to B 5
 % \path[->]
%\path[right hook->]
\path[stochasticPathstyle]
(m-2-3) edge [inPot1, out=90, in=90,distance=4cm, green, cross line] node[stochasticNodestyle] {$p$} (m-2-1) %J1 to S 
(m-2-3) edge [in2Pot1,in=120,distance=3cm,green, cross line] node[stochasticNodestyle] {$1-p$} (m-2-5); %J1 to B
\path[solid,red!70, line width=6pt]
(m-2-1) edge (m-2-3);
\path[solid,green!60, line width=6pt]
(m-2-3) edge (m-2-5);
\path[solid,blue!40, line width=6pt]
(m-2-5) edge (m-2-7);
\end{tikzpicture}
\caption{
    {\bf Network LSSC:  A linear stochastic stem cell  network}.  A linear stem cell (L) divides to produce a terminal cell T and cell J1. The cell J1  stochastically loops back to activate the stem cell T or it activates the terminal cell T.  The network's behavior approaches a deterministic linear stem cell network as the probability $p$  approach $1$.  Cancer stem cells controlled by such a network can go into spontaneous  remission because of the fact that for all points in all possible  paths in the network there is the possibility of reaching the terminal cell state T. 
  }
   \label{fig:LSSC}
\end{figure}

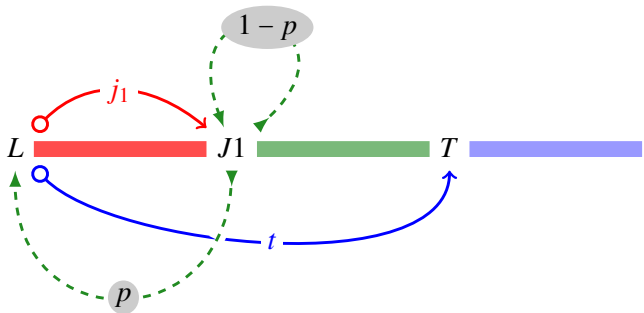
\begin{figure}[H]
\begin{tikzpicture}[style=mystyle]

\matrix (m) [matrix of math nodes, row sep=3em,
column sep=3em, text height=1.5ex, text depth=0.25ex]
{ \vphantom{a} && \vphantom{b} &&  \vphantom{c}  && \vphantom{c}\\ %  & & &\\
 L  && J1 &&  T  && %\vphantom{J2) &&  \vphantom{C) && \vphantom{c} \\ %TT\\
 \vphantom{a} &&  \vphantom{b} && \vphantom{B} &&  \vphantom{c}  && \vphantom{T}\\}; % & \vphantom{d} & \vphantom{D}\\ };
 \path[\pot]
(m-2-1) edge [inPot1, in=140, red, cross line] node[nodedescr] {$j_1$} (m-2-3) %S to J1 3
(m-2-1) edge [inPot2, in=-90, blue, cross line] node[nodedescr] {$t$} (m-2-5); %S to B 5

\path[stochasticPathstyle]
(m-2-3) edge [inPot2, out=-90, in=-90,distance=4.5cm, green, cross line] node[stochasticNodestyle] {$p$} (m-2-1) %J1 to S 
(m-2-3) edge [in2Pot1,in=110,out=40,distance=4.5cm,green, cross line] node[stochasticNodestyle] {$1-p$} (m-2-3); %J1 to B
\path[solid,red!70, line width=6pt]
(m-2-1) edge (m-2-3);
\path[solid,green!60, line width=6pt]
(m-2-3) edge (m-2-5);
\path[solid,blue!40, line width=6pt]
(m-2-5) edge (m-2-7);
\end{tikzpicture}
\caption{
    {\bf Network DLSSC: Deterministic linear stem cell network with stochastic delay}.  A linear stem cell (L) divides to produce a terminal cell T and cell J1. The cell J1  stochastically loops back to activate the stem cell T with probability $p$.  Unlike with a true linear stochastic network like LSS which stochastically self-differentiates to L with probability $p$ or differentiates to  T, this is a deterministic linear stem cell network that always self-differentiates back to L.  The stochasticity only effects the rate at which the cell divides.  The probability $p$ only effects the frequency of looping back to L and, thereby, changes the cycle time cell division.  Unlike LSSC, the linear cancer stem cells controlled by this network never go into spontaneous  remission since the cell endlessly self-renews.  The lower the probability $p$ the longer it takes for the cell to cell renew and divide. 
  }
  \label{fig:DLSSC}
\end{figure}

\subsection{1st-Order Geometric Cancer Networks with open stochastic dedifferentiation}

\label{sec:G1SD}
\begin{figure}[H]
\begin{tikzpicture}[style=mystyle]
\matrix (m) [matrix of math nodes, row sep=3em,
column sep=3em, text height=1.5ex, text depth=0.25ex]
{ \vphantom{a} & \vphantom{b}  &  \vphantom{c}  & \vphantom{c}  & \\
 S  &&  J  &&  T  &&  \\
  \vphantom{a} &   \vphantom{j} & \vphantom{b} &  \vphantom{c}  & \\ };
 \path[\pot]
(m-2-1) edge [selfloop1, cross line] node[nodedescr] {$ a $} (m-2-1)
(m-2-1) edge [inPot2, in=-110, blue, cross line] node[nodedescr] {$j$} (m-2-3); %S to J 
\path[stochasticPathstyle]
(m-2-3) edge [inPot2, out=-80, in=-90,distance=4cm, green, cross line] node[stochasticNodestyle] {$p$} (m-2-1) %J1 to S 
(m-2-3) edge [in2Pot1,in=90,out=80,distance=4cm,green, cross line] node[stochasticNodestyle] {$1-p$} (m-2-5); %J1 to B
\path[solid, line width=6pt]
(m-2-1) edge (m-2-3)
(m-2-3) edge (m-2-5);
%(m-2-5) edge (m-2-7);
\end{tikzpicture}
\caption{
    {\bf Network G1SD: A 1st-order geometric cancer with stochastic dedifferentiation:} In this linear network, a cell of type $S$ is a 1st-order stem cell. It produces progenitor cells of type J that have a stochastic dedifferentiation potential.  A cell J can either dedifferentiate to its parent state S or differentiate into the terminal cell T.  The effect of the stochasticity is seen in the population of progenitor cells J produced by the network.  The lower the probability $p$ of dedifferentiation, the greater the number of terminal cells in the cell population generated by the stem cell S.  
  }
    \label{fig:G1SD}
\end{figure}
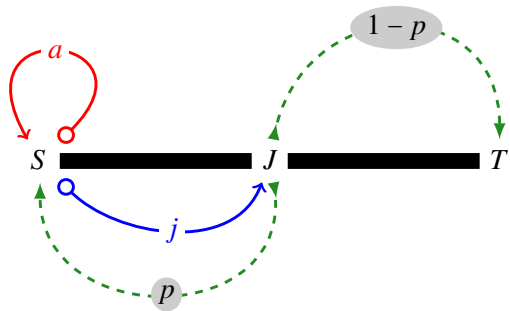 

\subsection{1st-Order Geometric Cancer Networks with closed stochastic dedifferentiation }
In the above network \autoref{fig:G1SD} dedifferentiation can occur once prior to permanent differentiation in a terminal cell state. The effect of the stochasticity can only be seen in the tumor as a whole that is by the cancer stem cell.  In the following network dedifferentiation is a constant possibility for the life of the cell. 

\label{sec:G1SDc}
\begin{figure}[H]
\begin{tikzpicture}[style=mystyle]
\matrix (m) [matrix of math nodes, row sep=3em,
column sep=3em, text height=1.5ex, text depth=0.25ex]
{ \vphantom{a} & \vphantom{b}  &  \vphantom{c}  & \vphantom{c}  & \\
 S  &&  J  &&  T  &&  \\
  \vphantom{a} &   \vphantom{j} & \vphantom{b} &  \vphantom{c}  & \\ };
 \path[\pot]
(m-2-1) edge [selfloop1, cross line] node[nodedescr] {$ a $} (m-2-1)
(m-2-1) edge [inPot2, in=-110, blue, cross line] node[nodedescr] {$j$} (m-2-3); %S to J 
\path[stochasticPathstyle]
(m-2-3) edge [inPot2, out=-80, in=-90,distance=4cm, green, cross line] node[stochasticNodestyle] {$p$} (m-2-1) %J1 to S 
(m-2-3) edge [in2Pot1,in=110,out=40,distance=4.5cm,green, cross line] node[stochasticNodestyle] {$1-p$} (m-2-3); %J1 to B
\path[solid, line width=6pt]
(m-2-1) edge (m-2-3)
(m-2-3) edge (m-2-5);
%(m-2-5) edge (m-2-7);
\end{tikzpicture}
\caption{
    {\bf Network G1SDc: A 1st-order geometric cancer with closed stochastic dedifferentiation}  }
    \label{fig:G1SDc}
\end{figure}
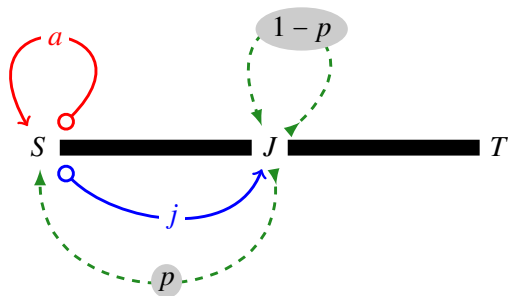 

In this linear network \autoref{fig:G1SDc}, a cell of type $S$ is a 1st-order stem cell. It produces progenitor cells of type J that have a continual stochastic dedifferentiation potential.  A cell J can either dedifferentiate to its parent state S or maintains its present state J.  In this closed stochastic network, a cell J continues to have a dedifferentiation potential for the life of that cell. There is no path to permanent terminal differentiation.  This contrasts with the network \autoref{fig:G1SD} where the cell only has a limited time frame to dedifferentiate prior to its permanent differentiation into a terminal cell T. Thus, all in a population of cells generated by S have the potential to dedifferentiate.  In contrast, in the network \autoref{fig:G1SD} if the probability $p$ is low, most cells in the population generated by S will be terminal cells with no dedifferentiation potential.

\subsection{Closed exponential stem cell network with stochastic delays}
Depending on the probability distribution, the next stochastic network proliferative properties can vary from deterministic terminal, linear, or exponential to stochastic terminal, linear or exponential. 

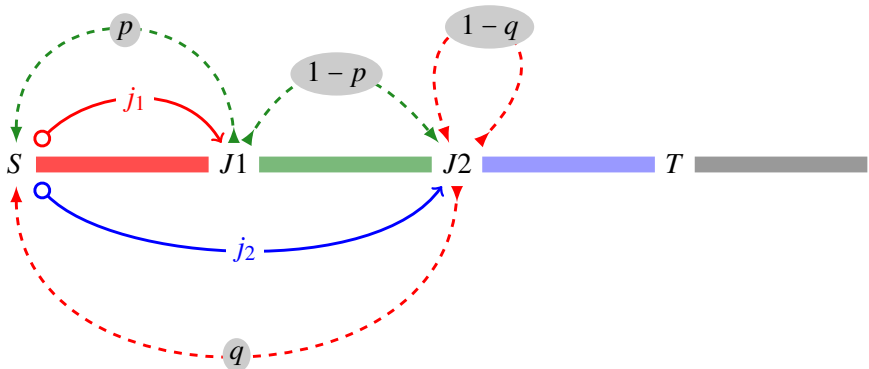
\begin{figure}[H]
\begin{tikzpicture}[style=mystyle]
\matrix (m) [matrix of math nodes, row sep=3em,
column sep=3em, text height=1.5ex, text depth=0.25ex]
{ \vphantom{a} && \vphantom{b} &&  \vphantom{c}  && \vphantom{c}  & & &\\
 S  && J1 &&  J2  && T && \vphantom{c}  && \vphantom{c} \\ %TT\\
 \vphantom{a} &&  \vphantom{b} && \vphantom{B} &&  \vphantom{c}  && \vphantom{C} & \vphantom{d} & \vphantom{D}\\ };
 \path[\pot]
(m-2-1) edge [inPot1, red, cross line] node[nodedescr] {$j_1$} (m-2-3) %S to J1 5
(m-2-1) edge [inPot2, in=-125, blue, cross line] node[nodedescr] {$j_2$} (m-2-5); %S to 2

 % \path[->]
%\path[right hook->]
\path[stochasticPathstyle]

(m-2-3) edge [inPot1, out=90, in=90,distance=4cm, green, cross line] node[stochasticNodestyle] {$p$} (m-2-1) %J1 to S 1
(m-2-3) edge [in2Pot1,in=130,out=60,distance=3cm,green, cross line] node[stochasticNodestyle] {$1-p$} (m-2-5) %J1 to T 7
(m-2-5) edge [inPot2, out=-90, in= -90, distance=6cm, red, cross line] node[stochasticNodestyle] {$q$} (m-2-1) %J2 to S 
(m-2-5) edge [inPot1, out=45, in= 110, distance=5cm, red, cross line] node[stochasticNodestyle] {$1-q$} (m-2-5); %J2 to T 

\path[solid,red!70, line width=6pt]
(m-2-1) edge (m-2-3);
\path[solid,green!60, line width=6pt]
(m-2-3) edge (m-2-5);
\path[solid,blue!40, line width=6pt]
(m-2-5) edge (m-2-7);
\path[solid,black!40, line width=6pt]
(m-2-7) edge (m-2-9);

\end{tikzpicture}
\caption{
    {\bf Network DXSSC modified LSSC: Closed exponential stochastic stem cell  network}.  If the probability $q > 0$ then this network is inherently nonterminal. All stochastic paths lead to potentially nonterminal nodes in the network.  $p = q = 0$ is terminal. $p = q = 1$ is exponential. $p = 1, q = 0$ is linear. $p = 0, q = 1$ is exponential. 
}
 \label{fig:DXSSC}
\end{figure}

Unlike LSSC \autoref{fig:LSSC}, if the probability $q > 0$ in cell state J2 then there are no true, probability independent, terminal cells in DXSSC \autoref{fig:DXSSC}.  True terminal cells have no possible stochastically available paths that lead to a reversal of differentiation to a more dedifferentiated cell type.  Instead, all cell states are stochastic with possible developmental paths that lead to the original founder stem cell S. The cell in state J2 has a probability $q$ of dedifferentiating to S.  The higher the probabilities $q$ and $p$ the more this network mirrors the behavior of a deterministic exponential network.  If either $p$ or $q$ is low while the other high it is more similar to a linear developmental network.  Thus the probability distribution determines the similarity to linear or exponential networks. Even if the probability of  $q$ is very low, as the number of J2 cell increase it becomes more and more likely that one of them will dedifferentiate to S. Thus, the cancer pulls away from linearity with an increasing cell population size.  

If probabilities $q = p = 0$ then the network is deterministic terminal network leading to only one cell division. If $q = 0$ and $p = 1$ we have a deterministic linear stem cell network. However, if the probability $q = 0$ and $p > 0$, then the network is a stochastic linear network where the lower the probability $p$ the greater the chance that proliferation terminates.   If either $p = 1$ or $p = 0$, and if $q = 1$ then the network is a deterministic exponential stem cell network.  If $q > 0$ then no matter how small the nonzero probability $q$ is, proliferation increases with time as more and more cells are created. 

{\bf Remark:} There is another way to escape the maze of exponential proliferation. Cell proliferation can be influenced by the number of times the self-loop at J2 is allowed to repeat.  If it is it is only allowed to repeat only a limited number of times relative then become quiescent, then, depending on the probability $q$, it could result in a terminal differentiation state after some stochastic tries. In other words, if, in parallel with the repetitions of the self-entry loop at J2, there exists a separate simultaneous, parallel process of terminal differentiation by means of some counting mechanism (where the cell at J2 becomes quiescent after some finite number of counting steps), then this network could lead to the ultimate termination of cell proliferation or at most linear proliferation.  The counting mechanism could be dependent on the number of loops executed at J2 or it could be a function of some other temporal variable.  

\subsection{A 1st order geometric/exponential stochastic stem cell network with stochastic progenitor cell dedifferentiation}

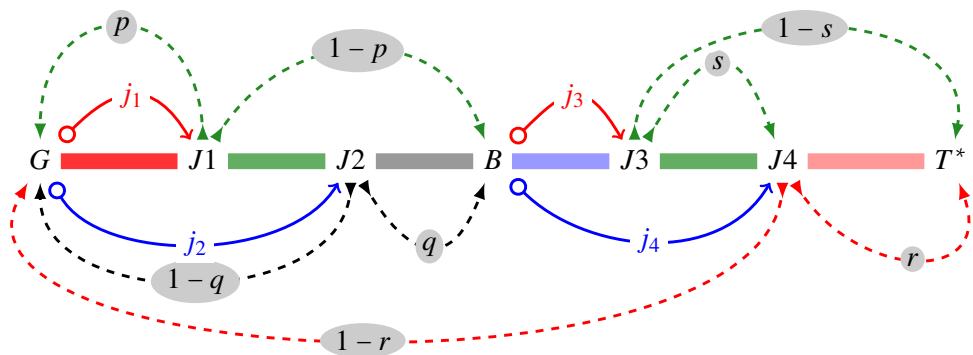
\begin{figure}[H]
% \begin{narrow}{-1.5cm}{-1.5cm}%{-1cm}{-1cm}
\begin{tikzpicture}[style=mystyle]
\matrix (m) [matrix of math nodes, 
row sep=3em,
column sep=\ColSepTight, %\matrixsepless, %\matrixsep, %3em, 
text height=1.5ex, text depth=0.25ex]
{ \vphantom{a} &\vphantom{a}& \vphantom{b} &&  \vphantom{c}  && \vphantom{c}  & & &\\
 G  && J1 && J2 &&  B  && J3 &&  J4 && T^{*} &&\\ %TT\\
 \vphantom{a} &\vphantom{a}&  \vphantom{b} && \vphantom{B} &&  \vphantom{c}  && \vphantom{C} && \vphantom{d} & \vphantom{D}\\ };
 % \path[->]
%\path[right hook->]
\path[stochasticPathstyle]
(m-2-3) edge [inPot1, out=90, in=90,distance=4cm, green, cross line] node[stochasticNodestyle] {$p$} (m-2-1) %J1 to S 
(m-2-3) edge [in2Pot1,out=60,in=110,distance=3.5cm,green, cross line] node[stochasticNodestyle] {$1-p$} (m-2-7) %J1 to B
(m-2-5) edge [in2Pot1,out=-60,in=-110,distance=2.5cm,black, cross line] node[stochasticNodestyle] {$q$} (m-2-7) %J1a to B
(m-2-5) edge [in2Pot2,in=-90,out=-90, distance=3.5cm,black, cross line] node[stochasticNodestyle] {$1-q$} (m-2-1) %J1a to B

(m-2-9) edge [inPot1, out=60, in= 110, distance=3cm, green, cross line] node[stochasticNodestyle] {$s$} (m-2-11) %J3 to J4 
(m-2-9) edge [inPot1, out=90, in= 80, distance=4cm, green, cross line] node[stochasticNodestyle] {$1-s$} (m-2-13) %J2 to C
(m-2-11) edge [inPot2, out=-90, in= -120, distance=6cm, red] node[stochasticNodestyle] {$1-r$} (m-2-1) %J2 to B 
(m-2-11) edge [inPot2, out=-60, in= -70, distance=3cm, red, cross line] node[stochasticNodestyle] {$r$} (m-2-13); %J2 to T 

 \path[\pot]
(m-2-1) edge [inPot1, red, cross line] node[nodedescr] {$j_1$} (m-2-3) %S to J1 3
(m-2-1) edge [inPot2, out=-60, in=-120,distance=2.5cm, blue] node[nodedescr] {$j_{2}$} (m-2-5)%S to J2 5
(m-2-7) edge [inPot1, red, cross line] node[nodedescr] {$j_{3}$} (m-2-9) %S to J2 7
(m-2-7) edge [inPot2, in=-120, blue] node[nodedescr] {$j_{4}$} (m-2-11); %B to  C 9

\path[solid,red!80, line width=6pt]
(m-2-1) edge (m-2-3);
\path[solid,green!70, line width=6pt]
(m-2-3) edge (m-2-5);
\path[solid,black!40, line width=6pt]
(m-2-5) edge (m-2-7);
\path[solid,blue!40, line width=6pt]
(m-2-7) edge (m-2-9);
\path[solid,green!70, line width=6pt]
(m-2-9) edge (m-2-11);
\path[solid,red!40, line width=6pt]
(m-2-11) edge (m-2-13);
%\path[solid,red!80, line width=10pt]
%(m-2-6) edge (m-2-7);
\end{tikzpicture}

\caption{
    {\bf Network G1SSC:   A 1st-order stochastic geometric/exponential stem cell  network with stochastic dedifferentiation}. 
  }
  \label{fig:G1SSC}
%  \end{narrow}
\end{figure}

This stochastic stem cell network (\autoref{fig:G1SSC}) generates progenitor cells B that generate cells J2 that can dedifferentiate to stem cells G or differentiate to a terminal network $T^{*}$. Depending on the probability distribution, this network can exhibit terminal, linear-geometric, and exponential developmental dynamics. If the probability distribution is such that $p=q=r=s=1$ then this is equivalent to a deterministic 1st order geometric stem cell network. A 1st-order stem cell (G) stochastically divides to produce either two progenitor cell B, or or two stem cells G or one progenitor cell B and one stem cell G.   and cell J1. The cells in states J1 or J2  stochastically loop back to activate the stem cell G or to activate the progenitor cell $B$. The progenitor cell $B$ divides into a semi-terminal cell $T^{*}$ by way of J3 or in divides into two cells of type  J4.  J4 either terminates with $T^{*}$ or it dedifferentiates into a stem cell G.  

Its behavior approaches a deterministic 1st-order geometric stem cell network as the probabilities $p$ and $q$ approach $1$.  However, the subnetwork activated by this 1st-order geometric network can still stochastically dedifferentiate into one or two new stem cells G. Cancers controlled by such a network can go into spontaneous  remission because of the fact that for all points in all possible  paths in the network there is the possibility of reaching a terminal cell state.  However, because there is the possibility at J1, J2 and J4 to dedifferentiate to earlier, upstream stem cell states, the relative numbers of stem cells to terminal cells will be higher than in networks that have fewer dedifferentiation pathways. Note, too that the dedifferentiation pathways introduce several possibilities for exponential growth. Even if dedifferentiation probabilities are very small, any increase in these probabilities could have a significant destabilizing influence on the resulting tumor. 

\subsection{A broad spectrum Linear or 2nd order geometric or exponential stochastic stem cell network}

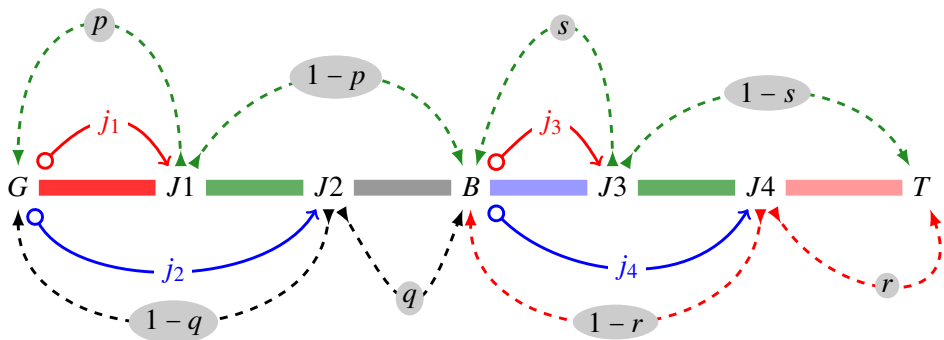
\begin{figure}[H]
% \begin{narrow}{-1.5cm}{-1.5cm}%{-1cm}{-1cm}
\begin{tikzpicture}[style=mystyle]
\matrix (m) [matrix of math nodes, 
row sep=3em,
column sep=\ColSepTight, %\matrixsepless, %\matrixsep, %3em, 
text height=1.5ex, text depth=0.25ex]
{ \vphantom{a} &\vphantom{a}& \vphantom{b} &&  \vphantom{c}  && \vphantom{c}  & & &\\
 G  && J1 && J2 &&  B  && J3 &&  J4 && T &&\\ %TT\\
 \vphantom{a} &\vphantom{a}&  \vphantom{b} && \vphantom{B} &&  \vphantom{c}  && \vphantom{C} && \vphantom{d} & \vphantom{D}\\ };
 % \path[->]
%\path[right hook->]
\path[stochasticPathstyle]
(m-2-3) edge [inPot1, out=90, in=90,distance=5cm, green, cross line] node[stochasticNodestyle] {$p$} (m-2-1) %J1 to S 
(m-2-3) edge [in2Pot1,out=60,in=110,distance=3.5cm,green, cross line] node[stochasticNodestyle] {$1-p$} (m-2-7) %J1 to B
(m-2-5) edge [in2Pot1,out=-60,in=-110,distance=3.5cm,black, cross line] node[stochasticNodestyle] {$q$} (m-2-7) %J1a to B
(m-2-5) edge [in2Pot2,in=-90,out=-90, distance=4cm,black, cross line] node[stochasticNodestyle] {$1-q$} (m-2-1) %J1a to B

(m-2-9) edge [inPot1, out=90, in= 75, distance=5cm, green, cross line] node[stochasticNodestyle] {$s$} (m-2-7) %J2 to S 
(m-2-9) edge [inPot1, out=60, in= 120, distance=3cm, green, cross line] node[stochasticNodestyle] {$1-s$} (m-2-13) %J2 to C
(m-2-11) edge [inPot2, out=-90, in= -90, distance=4cm, red] node[stochasticNodestyle] {$1-r$} (m-2-7) %J2 to B 
(m-2-11) edge [inPot2, out=-60, in= -70, distance=3cm, red, cross line] node[stochasticNodestyle] {$r$} (m-2-13); %J2 to T 

 \path[\pot]
(m-2-1) edge [inPot1, red, cross line] node[nodedescr] {$j_1$} (m-2-3) %S to J1 3
(m-2-1) edge [inPot2, out=-60, in=-120,distance=2.5cm, blue] node[nodedescr] {$j_{2}$} (m-2-5)%S to J2 5
(m-2-7) edge [inPot1, red, cross line] node[nodedescr] {$j_{3}$} (m-2-9) %S to J2 7
(m-2-7) edge [inPot2, in=-120, blue] node[nodedescr] {$j_{4}$} (m-2-11); %B to  C 9

\path[solid,red!80, line width=6pt]
(m-2-1) edge (m-2-3);
\path[solid,green!70, line width=6pt]
(m-2-3) edge (m-2-5);
\path[solid,black!40, line width=6pt]
(m-2-5) edge (m-2-7);
\path[solid,blue!40, line width=6pt]
(m-2-7) edge (m-2-9);
\path[solid,green!70, line width=6pt]
(m-2-9) edge (m-2-11);
\path[solid,red!40, line width=6pt]
(m-2-11) edge (m-2-13);
%\path[solid,red!80, line width=10pt]
%(m-2-6) edge (m-2-7);
\end{tikzpicture}

\caption{
    {\bf Network LG2XSSC:   A stochastic linear, 2nd order geometric or exponential stem cell  network}. If the probability distribution is such that $p=q=r=s=1$ then this is equivalent to a deterministic 2nd order geometric stem cell network. Thus, as these probabilities approach 1 the behavior approaches a 2nd order geometric stem cell network. 
  }
  \label{fig:LG2XSSC}
%  \end{narrow}
\end{figure}

In \autoref{fig:LG2XSSC}, a 2nd-order meta-stem cell (G) divides to produce a 1st-order stem cell $B$ and cell J1. The cell J1  stochastically loops back to activate the meta-stem cell G or it activates the first order linear stem cell $B$. The 1st-order stem cell $B$ divides into a semi-terminal cell C and a cell J2. The cell J2 stochastically activates either its parent stem cell $B$, the semi-terminal cell C or dedifferentiates to G.  The cell C also can stochastically dedifferentiate to any of the previous stem cell fates (G, B) or differentiate into the final terminal cell T. 

The network consists of two linked 1st-order linear stochastic networks. Its behavior approaches a deterministic 2nd-order geometric stem cell network as the probabilities $p$ and $q_1$ approach $1$.  Cancers controlled by such a network can go into spontaneous  remission because of the fact that for all points in all possible  paths in the network there is the possibility of reaching a terminal cell state.  However, because there is the possibility at J2 and C to dedifferentiate to earlier, upstream network states, the relative numbers of stem cells to terminal cells will be higher than in networks that have fewer dedifferentiation pathways. Note, too that the dedifferentiation pathways introduce several possibilities for exponential growth. Even if their probabilities  ($q_3$, $r_2$, $r_3$) are very small, any increase in these probabilities could have a significant destabilizing influence on the resulting tumor.  

The deterministic possible behaviors include: $(p=0, q= 1, s=0, r=1)$ is terminal. $(p=1, q= 0)$ is exponential. $(p=1, q= 1, s=0, r=1)$ is linear.  $(p=1, q= 1, s=1, r=1)$ is 2nd order geometric.  $(p=1, q= 1, s=1, r=0)$ is a linear network linked to an exponential network. 

\subsection{2nd Order Geometric stochastic stem cell  network}

\begin{figure}[H]
\begin{tikzpicture}[style=mystyle]
\matrix (m) [matrix of math nodes, row sep=3em,
column sep=3em, text height=1.5ex, text depth=0.25ex]
{ \vphantom{a} && \vphantom{b} &&  \vphantom{c}  && \vphantom{c}  & & &\\
 G  && J1 &&  B  && J2 &&  C && \vphantom{c} \\ %TT\\
 \vphantom{a} &&  \vphantom{b} && \vphantom{B} &&  \vphantom{c}  && \vphantom{C} & \vphantom{d} & \vphantom{D}\\ };
 \path[\pot]
(m-2-1) edge [inPot1, red, cross line] node[nodedescr] {$j_1$} (m-2-3) %S to J1 3
(m-2-1) edge [inPot2, in=-90, blue, cross line] node[nodedescr] {$b$} (m-2-5) %S to B 5
(m-2-5) edge [inPot1, red, cross line] node[nodedescr] {$j_2$} (m-2-7) %B to J2 7
(m-2-5) edge [inPot2, in=-90, blue, cross line] node[nodedescr] {$c$} (m-2-9); %B to  C 9
 % \path[->]
%\path[right hook->]
\path[stochasticPathstyle]
(m-2-3) edge [inPot1, out=90, in=90,distance=5cm, green, cross line] node[stochasticNodestyle] {$p$} (m-2-1) %J1 to S 
(m-2-3) edge [in2Pot1,in=120,distance=3cm,green, cross line] node[stochasticNodestyle] {$1-p$} (m-2-5) %J1 to B
(m-2-7) edge [inPot1, out=90, in= 75, distance=5cm, green, cross line] node[stochasticNodestyle] {$q$} (m-2-5) %J2 to S 
(m-2-7) edge [inPot1, out=60, in= 90, distance=3cm, green, cross line] node[stochasticNodestyle] {$1-q$} (m-2-9); %J2 to T 
\path[solid,red!80, line width=6pt]
(m-2-1) edge (m-2-3);
\path[solid,green!70, line width=6pt]
(m-2-3) edge (m-2-5);
\path[solid,blue!40, line width=6pt]
(m-2-5) edge (m-2-7);
\path[solid,green!70, line width=6pt]
(m-2-7) edge (m-2-9);
\end{tikzpicture}
\caption{
    {\bf Network G2SSC:  A 2nd order geometric stochastic stem cell  network}.  A 2nd-order meta-stem cell (G) divides to produce a 1st-order stem cell $B$ and cell J1. 
  }
   \label{fig:G2SSC}
\end{figure}
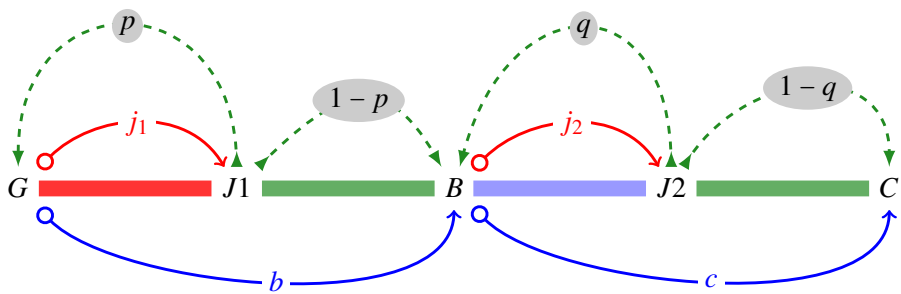

In \autoref{fig:G2SSC} The cell J1  stochastically loops back to activate the meta-stem cell G or it activates the first order linear stem cell $B$. The 1st-order stem cell $B$ divides into a terminal cell C and a cell J2. The cell J2 stochastically activates either the stem cell $B$ or the terminal cell $C$.  The network consists of two linked 1st-order linear stochastic networks. Its behavior approaches a deterministic 2nd-order geometric stem cell network as the probabilities $p$ and $q$ approach $1$.  Cancers controlled by such a network can go into spontaneous  remission because of the fact that for all points in all possible  paths in the network there is the possibility of reaching a terminal cell state. 

\subsection{Geometric stochastic stem cell  network with dedifferentiation}

\begin{figure}[H]
\begin{tikzpicture}[style=mystyle]
\matrix (m) [matrix of math nodes, 
row sep=2.5em,
column sep=\matrixsep, %2.5em, 
text height=1.5ex, text depth=0.25ex]
{ \vphantom{a} && \vphantom{b} &&  \vphantom{c}  && \vphantom{c}  & & &\\
 G  && J1 &&  B  && J2 &&  C && T &&\\ %TT\\
 \vphantom{a} &&  \vphantom{b} && \vphantom{B} &&  \vphantom{c}  && \vphantom{C} && \vphantom{d} & \vphantom{D}\\ };
 \path[\pot]
(m-2-1) edge [inPot1, red, cross line] node[nodedescr] {$j_1$} (m-2-3) %S to J1 3
(m-2-1) edge [inPot2, in=-120, blue, cross line] node[nodedescr] {$b$} (m-2-5) %S to B 5
(m-2-5) edge [inPot1, red, cross line] node[nodedescr] {$j_2$} (m-2-7) %B to J2 7
(m-2-5) edge [inPot2, in=-120, blue, cross line] node[nodedescr] {$c$} (m-2-9); %B to  C 9
 % \path[->]
%\path[right hook->]
\path[stochasticPathstyle]
(m-2-3) edge [inPot1, out=90, in=90,distance=5cm, green, cross line] node[stochasticNodestyle] {$p$} (m-2-1) %J1 to S 
(m-2-3) edge [in2Pot1,in=120,distance=3cm,green, cross line] node[stochasticNodestyle] {$1-p$} (m-2-5) %J1 to B
(m-2-7) edge [inPot1, out=90, in= 75, distance=5cm, green, cross line] node[stochasticNodestyle] {$q_1$} (m-2-5) %J2 to S 
(m-2-7) edge [inPot1, out=60, in= 120, distance=3cm, green, cross line] node[stochasticNodestyle] {$q_2$} (m-2-9) %J2 to C
(m-2-7) edge [inPot2, out=-120, in= -90, distance=4cm, green, cross line] node[stochasticNodestyle] {$q_3$} (m-2-1) %J2 to G 
(m-2-9) edge [inPot1, out=70, in= 110, distance=3cm, black, cross line] node[stochasticNodestyle] {$r_1$} (m-2-11) %J2 to G 
(m-2-9) edge [inPot2, out=-90, in= -90, distance=4cm, black, cross line] node[stochasticNodestyle] {$r_2$} (m-2-5) %J2 to B 
(m-2-9) edge [inPot2, out=-60, in= -120, distance=7cm, black, cross line] node[stochasticNodestyle] {$r_3$} (m-2-1); %J2 to G 
\path[solid,red!80, line width=6pt]
(m-2-1) edge (m-2-3);
\path[solid,green!70, line width=6pt]
(m-2-3) edge (m-2-5);
\path[solid,blue!40, line width=6pt]
(m-2-5) edge (m-2-7);
\path[solid,green!70, line width=6pt]
(m-2-7) edge (m-2-9);\path[solid,black!40, line width=6pt]
(m-2-9) edge (m-2-11);
\end{tikzpicture}
%\end{center}
\caption{
    {\bf Network GSSCdediff:  A geometric stochastic stem cell  network with dedifferentiation}.   }
   \label{fig:GSSCdediff}
\end{figure}
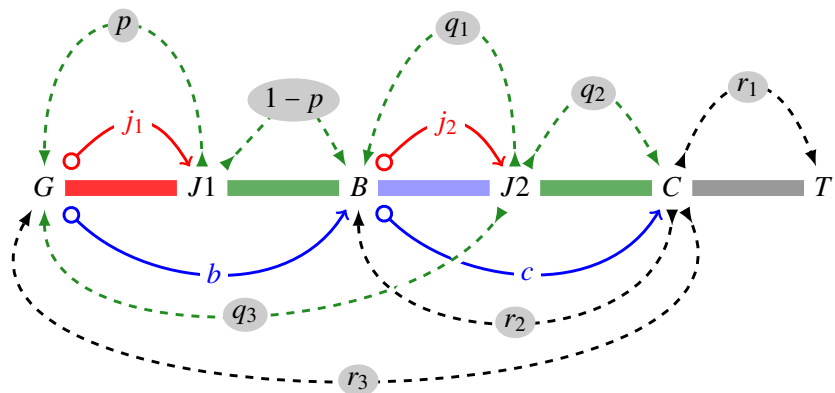

In \autoref{fig:GSSCdediff} let probabilities $q_1 + q_2 +q_3 =  1$ and $r_1+r_2+r_3 = 1$. A 2nd-order meta-stem cell (G) divides to produce a 1st-order stem cell $B$ and cell J1. The cell J1  stochastically loops back to activate the meta-stem cell G or it activates the first order linear stem cell $B$. The 1st-order stem cell $B$ divides into a semi-terminal cell C and a cell J2. The cell J2 stochastically activates either its parent stem cell $B$, the semi-terminal cell C or dedifferentiates to G.  The cell C also can stochastically dedifferentiate to any of the previous stem cell fates (G, B) or differentiate into the final terminal cell T. The network consists of two linked 1st-order linear stochastic networks. Its behavior approaches a deterministic 2nd-order geometric stem cell network as the probabilities $p$ and $q_1$ approach $1$.  Cancers controlled by such a network can go into spontaneous  remission because of the fact that for all points in all possible  paths in the network there is the possibility of reaching a terminal cell state.  However, because there is the possibility at J2 and C to dedifferentiate to earlier, upstream network states, the relative numbers of stem cells to terminal cells will be higher than in networks that have fewer dedifferentiation pathways. Note, too that the dedifferentiation pathways introduce several possibilities for exponential growth. Even if their probabilities  ($q_3$, $r_2$, $r_3$) are very small, any increase in these probabilities could have a significant destabilizing influence on the resulting tumor.

\subsection{Modeling deterministic stem cell networks with stochastic networks}
For stochastic stem cell networks the probability distribution over a network topology determines their dynamic properties.  If the stochastic network allows all possible paths in a set of linear, 2nd-order geometric and exponential networks, then the dynamic behavior of each is approximated as the corresponding embedded links are assigned probabilities that match existence or nonexistence of links in the particular network type,  For example, if a link exists in a network embedded in the stochastic network, then it assigned a probability close to 1 or 1 itself if we want an exact match.  If the link does not exist in the embedded network, it is assigned a probability close to 0 or 0 itself for an exact match.  Hence, the probability distribution distinguishes different network types that are embeddable in the topology or architecture of a given stochastic network. 

\subsection{Transformations of probability distributions change stochastic stem cell behavior}
If we allow the probability distribution over a stochastic network to change with time because of external factors (such as ultra violet radiation) then the same stochastic stem cell network may exhibit various proliferative phenotypes, appearing alternatively as being in remission, linear, geometric or exponential.  Thus, meta-probability functions are at work here that conditionally change the probability distribution. 
  
\section{Conditional cancers}

Some subnetworks are not activated except in special circumstances.  For example the network that controls wound healing is activated if there is injury to tissue.  Now if a cell has a cancerous mutation in its wound healing subnetwork, the cancer will not become evident until some or sufficient surrounding tissue has been injured.  As the wound healing subnetwork is activated the cancer will start.  Repeated injury to, or irritation of tissue has been associated with cancer.  For example, leukoplakia from chewing tobacco and esophageal carcinoma from repeated, chronic episodes of  reflux esophagitis.  It has been speculated that the injury itself causes the cancer.  This may be.  

However, we now have a possible alternative explanation:  repeated injury makes it more likely that a cell with a cancerous mutation in its wound healing subnetwork is found, i.e., induced to initiate its wound healing control subnetwork.  On this account the repeated injury serves more to activate an already existing cancerous subnetwork, rather than being the cause of the mutation in that subnetwork.  Alternatively, the repeated activation of the linear stem cell network regenerates damaged tissue becomes transformed because of error build up resulting from overuse.  The transformation may be to a geometric or exponential cancer.  It may also be a differentiation transformation to an earlier cell type resulting in a new stem cell which would imply geometric or exponential growth with potentially invasive properties.  

An additional aspect of cancerous wound healing subnetworks is that they involve cell dedifferentiation.  If such immature cells are involved in the cancerous subnetwork then the result may be an fast growing, possibly exponential cancer. For example,  if cell differentiates to two identical cells then a single loop can cause an exponential cancer (see \autoref{fig:NXo}).

\section{Reactive communication networks}
  \label{sec:Rsig}
We distinguish \emph{reactive communication networks} from \emph{interactive communication networks}. \emph{Reactive} communication networks respond to input signals with some behavior.  When that behavior leads to sending signals and not just interpreting incoming signals the networks are \emph{interactive}. 

\subsection{A reactive signal based differentiation network architecture with linear and exponential potential}

\begin{figure}[H]
\begin{tikzpicture}[style=mystyle]
\matrix (m) [matrix of math nodes, 
row sep=2em,
column sep=3em, 
text height=1.5ex, text depth=0.25ex]
{ \vphantom{a} && \vphantom{b} &&  \vphantom{c}  && \vphantom{c}  & & &\\
 S  && J1 &&  J2  && T && \vphantom{c}  && \vphantom{c} \\ %TT\\
 \vphantom{a} &&  \vphantom{b} && \vphantom{B} &&  \vphantom{c}  && \vphantom{C} & \vphantom{d} & \vphantom{D}\\ };
 \path[\pot]
(m-2-1) edge [inPot1, red, cross line] node[nodedescr] {$j_1$} (m-2-3) %S to J1 5
(m-2-1) edge [inPot2, in=-90, blue, cross line] node[nodedescr] {$j_2$} (m-2-5); %S to 2
\path[\sigjump]
(m-2-3)[draw=red] edge [inPot1, out=90, in=90,distance=4cm, green, receivestyle, cross line] node[receivesigstyle] {$\sigma_{1}$} (m-2-1) %J1 to S 1
(m-2-3) edge [in2Pot1,green, receivestyle, cross line] node[receivesigstyle] {$\sigma_{2}$} (m-2-7); %J1 to T 7
%\path[>=latex,>->,dotted]
\path[open diamond->]%\sigjump]
(m-2-5) edge [inPot2, out=-125, in= -90, distance=6cm, purple, receivestyle,cross line] node[receivesigstyle] {$\sigma_{3}$} (m-2-1) %J2 to S 
(m-2-5) edge [inPot2, out=-60, in= -90, distance=3cm, purple,  receivestyle,cross line] node[receivesigstyle] {$\sigma_{4}$} (m-2-7); %J2 to T 
\path[solid,red!70, line width=6pt]
(m-2-1) edge (m-2-3);
\path[solid,green!60, line width=6pt]
(m-2-3) edge (m-2-5);
\path[solid,blue!40, line width=6pt]
(m-2-5) edge (m-2-7);
\path[solid,black!40, line width=6pt]
(m-2-7) edge (m-2-9);
\end{tikzpicture}
\caption{
    {\bf Network SigLXSC: Reactive signal based differentiation stem cell network with linear and exponential potential}.  One stem cell (S) divides to produce two cells (J1 and J2).   The red and green squiggly arrows between two cell states indicate cell signaling, e.g., the green arrow $J1 \rightsquigarrow S$ labeled with $\sigma_{1}$ means that on receiving the signal $\sigma_{1}$ a cell in state $J1$ differentiates or jumps to state $S$. On signal $\sigma_{1}$ J1  activates the parent stem cell S. On signal $\sigma_{2}$, J1 differentiates to a terminal cell T.   Similarly, J2 responds to signal $\sigma_{3}$ by dedifferentiating to state S and to $\sigma_{4}$ by activating the terminal network T.  Hence, as long as J1 receives the dedifferentiation signal $\sigma_{1}$ and J2 receives its dedifferentiation signal $\sigma_{3}$ the network will generate exponential cell proliferation. The network has linear proliferative behavior if J1 (or J2) receives the dedifferentiation signal $\sigma_{1}$ ($\sigma_{3}$) while J2 (or J1) receives the terminal signal $\sigma_{4}$ ( or  $\sigma_{2}$). It is a terminal network if both cells J1 and J2 receive their terminal signals $\sigma_{2}$ and $\sigma_{4}$, respectively. The ultimate output of this network whether it has linear, exponential or terminal dynamics is just one terminal cell type T.  If T is the start of a terminal network, it would produce one or more terminal structures each generated by T. If the cells of T interact then the structures T may form composite structures with different properties than T itself even though the whole is terminal. 
  }
   \label{fig:SigLXSC}
\end{figure}
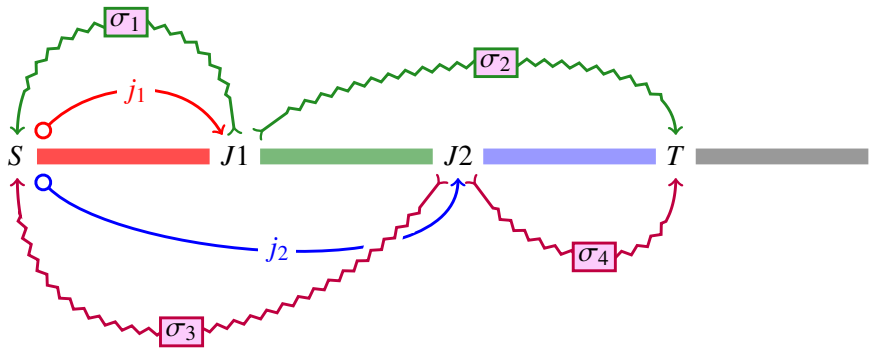

\subsection{A reactive signal based geometric network architecture with 1st and 2nd order geometric potential}

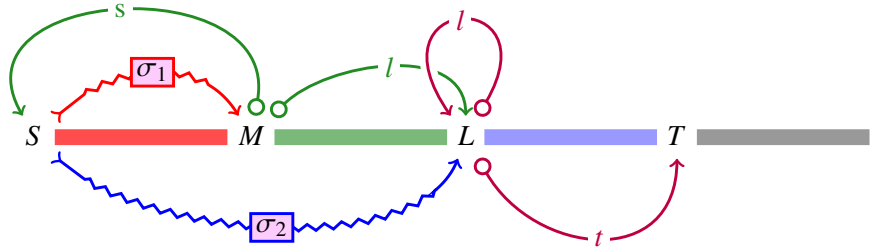
\begin{figure}[H]
\begin{tikzpicture}[style=mystyle]
\matrix (m) [matrix of math nodes, 
row sep=2em,
column sep=3em, 
text height=1.5ex, text depth=0.25ex]
{ \vphantom{a} && \vphantom{b} &&  \vphantom{c}  && \vphantom{c}  & & &\\
 S  && M &&  L  && T && \vphantom{c}  && \vphantom{c} \\ %TT\\
 \vphantom{a} &&  \vphantom{b} && \vphantom{B} &&  \vphantom{c}  && \vphantom{C} & \vphantom{d} & \vphantom{D}\\ };
  \path[\sigjump]
(m-2-1) edge [inPot1, red, receivestyle, cross line] node[receivesigstyle] {$\sigma_{1}$} (m-2-3) %S to J1 5
(m-2-1) edge [inPot2, in=-110, blue, receivestyle, cross line] node[receivesigstyle] {$\sigma_{2}$} (m-2-5); %S to 2

\path[\pot]
(m-2-3) edge [selfloop1, out=80, distance=4cm, green, cross line] node[nodedescr] {s} (m-2-1) %J1 to S 1
(m-2-3) edge [in2Pot1,green, cross line] node[nodedescr] {$l$} (m-2-5); %J1 to T 7
%\path[>=latex,>->,dotted]
\path[\pot]
(m-2-5) edge [selfloop1, out=60, in= 125, purple,cross line] node[nodedescr] {$l$} (m-2-5) %J2 to S 
(m-2-5) edge [inPot2, out=-60, in= -90, distance=3cm, purple,cross line] node[nodedescr] {$t$} (m-2-7); %J2 to T 
\path[solid,red!70, line width=6pt]
(m-2-1) edge (m-2-3);
\path[solid,green!60, line width=6pt]
(m-2-3) edge (m-2-5);
\path[solid,blue!40, line width=6pt]
(m-2-5) edge (m-2-7);
\path[solid,black!40, line width=6pt]
(m-2-7) edge (m-2-9);
\end{tikzpicture}
\caption{
    {\bf Network SigLSC: Reactive signal based geometric stem cell network with linear and meta-stem cell potential}.  On signal $\sigma_{1}$ one stem cell (M) divides to produce two cells (M and L).   M dedifferentiates back to the parent cell S.   On signal $\sigma_{2}$, the linear stem cell network L is activated. Once  L is activated it will continue to produce terminal cells T. Hence, this network is signal controlled. It will continue to produce M and hence S cells as long as it receives the signal $\sigma_{1}$.  S cells will differentiate to L cells on signal $\sigma_{2}$ stopping meta-stem cell production.  The signal $\sigma_{2}$ can thus eliminate all meta-stem cells.  Hence, it may be undesirable for systems that need to preserve their meta-stem cell line. Eliminating the signal  $\sigma_{2}$-path from S to L would insure that the supply of meta-stem cells can  always be increased as long as there are some meta-stem cells or S cells.  The terminal cell type T may be a single terminal cell or a terminal progenitor cell T* that generates a bounded number of further terminal cell types. 
  }
   \label{fig:SigLSC}
\end{figure}

\subsection{A pure reactive signal based geometric network architecture with 1st and 2nd order geometric potential}

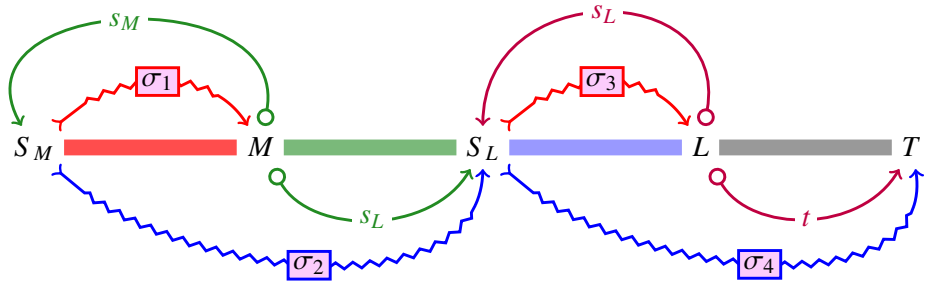
\begin{figure}[H]
\begin{tikzpicture}[style=mystyle]
\matrix (m) [matrix of math nodes, 
row sep=2em,
column sep=3em, 
text height=1.5ex, text depth=0.25ex]
{ \vphantom{a} && \vphantom{b} &&  \vphantom{c}  && \vphantom{c}  & & &\\
 S_{M}  && M &&  S_{L}  && L && T && \vphantom{c} \\ %TT\\
 \vphantom{a} &&  \vphantom{b} && \vphantom{B} &&  \vphantom{c}  && \vphantom{C} & \vphantom{d} & \vphantom{D}\\ };
  \path[\sigjump]
(m-2-1) edge [inPot1, red, receivestyle, cross line] node[receivesigstyle] {$\sigma_{1}$} (m-2-3) %S to J1 5
(m-2-1) edge [inPot2, in=-90, distance=4cm, blue, receivestyle, cross line] node[receivesigstyle] {$\sigma_{2}$} (m-2-5) %S to 2
(m-2-5) edge [inPot1, red, receivestyle, cross line] node[receivesigstyle] {$\sigma_{3}$} (m-2-7) %S to J1 5
(m-2-5) edge [inPot2, in=-80, distance=4cm, blue, receivestyle, cross line] node[receivesigstyle] {$\sigma_{4}$} (m-2-9); %S to 2

\path[\pot]
(m-2-3) edge [selfloop1, out=80, distance=4cm, green, cross line] node[nodedescr] {$s_{M}$} (m-2-1) %J1 to S 1
(m-2-3) edge [inPot2, out=-60, in=-120, green, cross line] node[nodedescr] {$s_{L}$} (m-2-5); %J1 to T 7
%\path[>=latex,>->,dotted]
\path[\pot]
(m-2-7) edge [selfloop1, out=80, in= 90, purple,cross line] node[nodedescr] {$s_{L}$} (m-2-5) %J2 to S 
(m-2-7) edge [inPot2, out=-60, in= -120, purple,cross line] node[nodedescr] {$t$} (m-2-9); %J2 to T 
\path[solid,red!70, line width=6pt]
(m-2-1) edge (m-2-3);
\path[solid,green!60, line width=6pt]
(m-2-3) edge (m-2-5);
\path[solid,blue!40, line width=6pt]
(m-2-5) edge (m-2-7);
\path[solid,black!40, line width=6pt]
(m-2-7) edge (m-2-9);
\end{tikzpicture}
\caption{
    {\bf Network SigG2SC: Pure reactive signal based geometric stem cell network with flexible geometric, linear and terminal potential}.  On signal $\sigma_{1}$, the signal dependent cell $S_{M}$ differentiates to the meta-stem cell (M) which then divides to produce two cells ($S_{M}$ and $S_{L}$).  Both $S_{M}$ and $S_{L}$ are controlled by equivalent signal dependent networks.  Their further behavior depends on the signals they receive. The signal combination $<\sigma_{1}, \sigma_{3}>$ produces a meta-stem cell network with 2nd order geometric potential. The signal combination $<\sigma_{1}, \sigma_{4}>$ and $<\sigma_{2}, \sigma_{3}>$ produce a 1st order stem cell network with linear potential.  The signal combination $<\sigma_{2}, \sigma_{3}>$ causes terminal differentiation stopping all cell production.  This example network shows that different signal combinations can change the architecture of signal based networks leading to distinct stem cell types with different generative potential.  The terminal cell type T may be a single terminal cell or a terminal progenitor cell T* that generates a bounded number of further terminal cell types. 
  }
   \label{fig:SigG2SC}
\end{figure}

\subsection{A hybrid signal stochastic network architecture with exponential and linear potential}

\begin{figure}[H]
\begin{tikzpicture}[style=mystyle]
\matrix (m) [matrix of math nodes, 
row sep=2em,
column sep=3em, 
text height=1.5ex, text depth=0.25ex]
{ \vphantom{a} && \vphantom{b} &&  \vphantom{c}  && \vphantom{c}  & & &\\
 S  && J1 &&  J2  && T && \vphantom{c}  && \vphantom{c} \\ %TT\\
 \vphantom{a} &&  \vphantom{b} && \vphantom{B} &&  \vphantom{c}  && \vphantom{C} & \vphantom{d} & \vphantom{D}\\ };
 \path[\pot]
(m-2-1) edge [inPot1, red, cross line] node[nodedescr] {$j_1$} (m-2-3) %S to J1 5
(m-2-1) edge [inPot2, in=-90, blue, cross line] node[nodedescr] {$j_2$} (m-2-5); %S to 2
 % \path[->]
%\path[right hook->]
\path[stochasticPathstyle]
(m-2-3) edge [inPot1, out=90, in=90,distance=4cm, green, cross line] node[stochasticNodestyle] {$p$} (m-2-1) %J1 to S 1
(m-2-3) edge [in2Pot1,green, cross line] node[stochasticNodestyle] {$1-p$} (m-2-7); %J1 to T 7
%\path[>=latex,>->,dotted]
\path[\sigjump]
(m-2-5) edge [inPot2, out=-125, in= -90,distance=6cm, purple, receivestyle,  cross line] node[receivesigstyle] {$\sigma_{1}$} (m-2-1) %J2 to S 
(m-2-5) edge [inPot2, out=-60, in= -90, distance=3cm, purple,  receivestyle, cross line] node[receivesigstyle] {$\sigma_{2}$} (m-2-7); %J2 to T 
\path[solid,red!70, line width=6pt]
(m-2-1) edge (m-2-3);
\path[solid,green!60, line width=6pt]
(m-2-3) edge (m-2-5);
\path[solid,blue!40, line width=6pt]
(m-2-5) edge (m-2-7);
\path[solid,black!40, line width=6pt]
(m-2-7) edge (m-2-9);
\end{tikzpicture}
\caption{
    {\bf Network SigLXSSC: Hybrid signal stochastic stem cell network.}  One stem cell (S) divides to produce two cells (J1 and $\Sigma$).   J1 stochastically activates either the stem cell itself or a terminal cell.   J2 responds to signal $\sigma_{1}$ by dedifferentiating to state S and to $\sigma_{2}$ by activating the terminal network T.  Hence, as long as a stem cell receives the signal $\sigma_{1}$ it will proliferate exponentially or linearly depending on the probability $p$.  Given the signal $\sigma_{1}$, if the probability $p$ is high the network will proliferate exponentially otherwise if the probablility $p$ is low, the network will proliferate linearly.   Alternatively, if the network S receives the signal $\sigma_{2}$ and the probability $p$ is high it will proliferate linearly, if low, it will eventually terminate in T.   If signal $\sigma_{2}=\overline{\sigma_{1}}$ where $\overline{\sigma_{1}}$ represents the absence or repression of the signal $\sigma_{1}$ then the default behavior of the network without the signal $\sigma_{1}$ is to proliferate linearly if the probability $p$ is high or if the probability $p$ is low, to terminate with T.
 }
   \label{fig:SigLXSSC}
\end{figure}
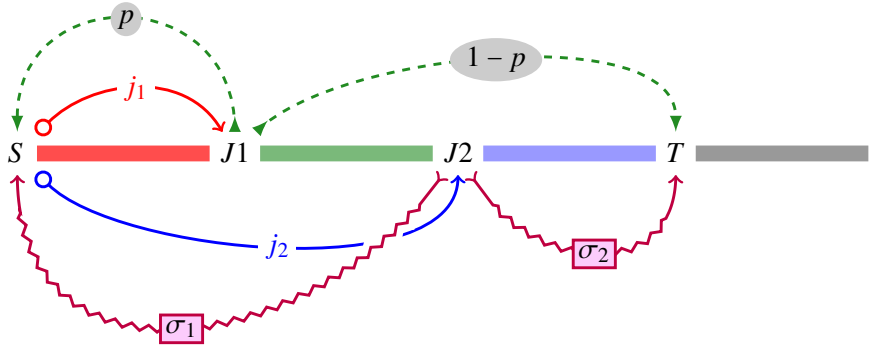 

\section{Communicating cancer networks in systems of cooperating social cells}
\label{sec:SigC}

A type of conditional cancer are {\em social cancers} that depend on cell signaling to be active.  This variety of cancer occurs when the genomic network interacts with receptors, signals and signal transduction pathways.  In that case we can have what might be called social cancers.  For example, a cell $A$ will signal $\alpha$ to cell $B$ and differentiates to A1, $B$ on receiving the signal divides into B1 and B1.  B1 sends a signal $\beta$ back to $A$ and then B1 dedifferentiates to $B$ (its network loops back).  When A1 receives $\beta$ its network loops back  into state $A$.  Now the process repeats with the $A$ cells sending signals $\alpha$ to $B$ cells.  One can see that this process is potentially exponential as long as we have sufficient cells of type $A$ with sufficient signal capacity to continue to activate all the developing $B$ cells.  

With conditional cancers that depend on social communication, the growth rate of the cancer will depend on the accessibility of the signal.  If the receiver of the signal requires direct contact with the sender then even if the cytogenic cell contains a conditional exponential network, since the signal may not be received by those daughter cells the exponential potential may not be realized. Furthermore, if it is a linear network then if the passive daughter cell is interposed between the sender and the receiving cytogenic cell then growth will stop after several divisions when the signal no loner can reach the cytogenic cells.  If later, because of physical pressure or other conditions, the cytogenic cell is again in close enough proximity to the sender then the cancer can start again. 

\subsection{Interactive signaling mono generative networks}
\label{sec:SigC1}
In the figure below we emphasize the communication links.   The network has one cell only signaling and the other cell signaling and conditionally dividing, however it  leaves open whether the cytogenic subnetwork is linear or exponential. The more detailed communication networks and their properties will described in the sections below. 

%was N13, was NSL
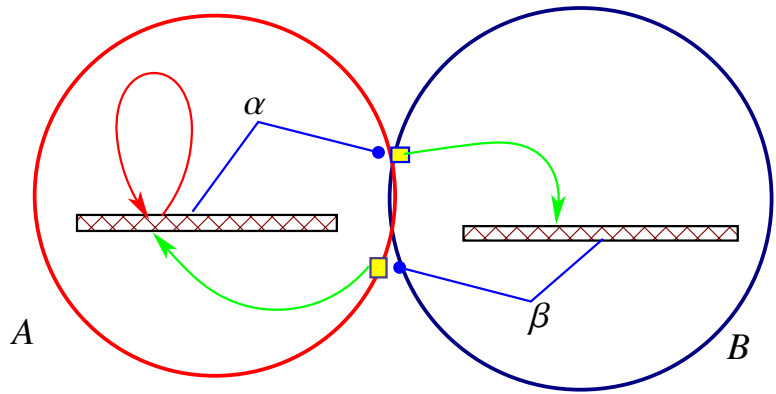
\begin{figure}[H]
\centering
\begin{tikzpicture}%scale= \PicSize, Crit. Scale does not work
[x=1.00mm, y=1.00mm, inner xsep=0pt, inner ysep=0pt, outer xsep=0pt, outer ysep=0pt]
\path[line width=0mm] (51.00,42.86) rectangle +(107.34,54.79);
\definecolor{L}{rgb}{0,0,0.502}
\path[line width= \CellWallThickness, draw=L] (128.68,70.26) circle (25.39mm);
\definecolor{L}{rgb}{1,0,0}
\path[line width= \CellWallThickness, draw=L] (80.06,70.68) circle (23.96mm);
\definecolor{L}{rgb}{0.502,0,0}
\path[line width=0.15mm, draw=L]  (64.12,65.99) -- (61.98,68.12) (66.94,65.99) -- (64.81,68.12) (69.77,65.99) -- (67.64,68.12) (72.60,65.99) -- (70.47,68.12) (75.43,65.99) -- (73.30,68.12) (78.26,65.99) -- (76.13,68.12) (81.09,65.99) -- (78.95,68.12) (83.92,65.99) -- (81.78,68.12) (86.74,65.99) -- (84.61,68.12) (89.57,65.99) -- (87.44,68.12) (92.40,65.99) -- (90.27,68.12) (95.23,65.99) -- (93.10,68.12) (96.27,67.78) -- (95.93,68.12);
\path[line width=0.15mm, draw=L]  (61.73,67.38) -- (62.47,68.12) (63.16,65.99) -- (65.30,68.12) (65.99,65.99) -- (68.12,68.12) (68.82,65.99) -- (70.95,68.12) (71.65,65.99) -- (73.78,68.12) (74.48,65.99) -- (76.61,68.12) (77.31,65.99) -- (79.44,68.12) (80.13,65.99) -- (82.27,68.12) (82.96,65.99) -- (85.09,68.12) (85.79,65.99) -- (87.92,68.12) (88.62,65.99) -- (90.75,68.12) (91.45,65.99) -- (93.58,68.12) (94.28,65.99) -- (96.27,67.98);
\definecolor{L}{rgb}{0,0,0}
\path[line width=0.30mm, draw=L] (61.73,65.99) rectangle +(34.54,2.13);
\definecolor{L}{rgb}{0.502,0,0}
\path[line width=0.15mm, draw=L]  (113.48,64.71) -- (113.11,65.08) (116.31,64.71) -- (114.39,66.63) (119.14,64.71) -- (117.22,66.63) (121.96,64.71) -- (120.05,66.63) (124.79,64.71) -- (122.87,66.63) (127.62,64.71) -- (125.70,66.63) (130.45,64.71) -- (128.53,66.63) (133.28,64.71) -- (131.36,66.63) (136.11,64.71) -- (134.19,66.63) (138.93,64.71) -- (137.02,66.63) (141.76,64.71) -- (139.84,66.63) (144.59,64.71) -- (142.67,66.63) (147.42,64.71) -- (145.50,66.63) (149.57,65.39) -- (148.33,66.63);
\path[line width=0.15mm, draw=L]  (113.11,65.03) -- (114.71,66.63) (115.62,64.71) -- (117.54,66.63) (118.45,64.71) -- (120.37,66.63) (121.28,64.71) -- (123.20,66.63) (124.11,64.71) -- (126.03,66.63) (126.94,64.71) -- (128.86,66.63) (129.77,64.71) -- (131.68,66.63) (132.59,64.71) -- (134.51,66.63) (135.42,64.71) -- (137.34,66.63) (138.25,64.71) -- (140.17,66.63) (141.08,64.71) -- (143.00,66.63) (143.91,64.71) -- (145.83,66.63) (146.74,64.71) -- (148.66,66.63) (149.57,64.71) -- (149.57,64.72);
\definecolor{L}{rgb}{0,0,0}
\path[line width=0.30mm, draw=L] (113.11,64.71) rectangle +(36.46,1.92);
\definecolor{L}{rgb}{0,0,1}
\definecolor{F}{rgb}{1,1,0}
\path[line width=0.30mm, draw=L, fill=F] (103.52,75.16) rectangle +(2.35,1.92);
\definecolor{L}{rgb}{0,1,0}
\path[line width=0.30mm, draw=L] (105.22,76.23) .. controls (108.27,76.71) and (111.33,77.14) .. (114.39,77.51) .. controls (117.44,77.87) and (120.71,78.10) .. (123.13,76.23) .. controls (124.60,75.10) and (125.46,73.37) .. (125.69,71.54) .. controls (125.80,70.69) and (125.78,69.83) .. (125.69,68.98) .. controls (125.64,68.48) and (125.57,67.98) .. (125.48,67.48);
\definecolor{F}{rgb}{0,1,0}
\path[line width=0.30mm, draw=L, fill=F] (125.48,67.48) -- (126.40,70.22) -- (125.65,69.58) -- (125.01,70.33) -- (125.48,67.48) -- cycle;
\definecolor{L}{rgb}{0.282,0.239,0.545}
\definecolor{F}{rgb}{1,1,0}
\path[line width=0.30mm, draw=L, fill=F] (100.75,59.81) rectangle +(2.13,2.56);
\definecolor{L}{rgb}{0,1,0}
\path[line width=0.30mm, draw=L] (100.53,61.30) .. controls (95.60,55.27) and (87.00,53.73) .. (80.28,57.68) .. controls (77.79,59.14) and (75.81,61.28) .. (73.88,63.43) .. controls (73.31,64.07) and (72.74,64.71) .. (72.17,65.35);
\definecolor{F}{rgb}{0,1,0}
\path[line width=0.30mm, draw=L, fill=F] (72.17,65.35) -- (73.52,62.80) -- (73.57,63.78) -- (74.56,63.73) -- (72.17,65.35) -- cycle;
\definecolor{L}{rgb}{0,0,1}
\path[line width=0.30mm, draw=L] (131.66,64.93) -- (122.07,56.61) -- (103.94,61.30);
\definecolor{F}{rgb}{0,0,1}
\path[line width=0.30mm, draw=L, fill=F] (104.62,61.13) circle (0.70mm);
\path[line width=0.30mm, draw=L] (77.08,68.55) -- (85.82,80.49) -- (101.81,76.44);
\path[line width=0.30mm, draw=L, fill=F] (101.81,76.44) circle (0.70mm);
\draw(83.90,81.56) node[anchor=base west]{\fontsize{14.23}{17.07}\selectfont $\alpha$};
\draw(122.07,53.20) node[anchor=base west]{\fontsize{14.23}{17.07}\selectfont $\beta$};
\draw(53.00,51.00) node[anchor=base west]{\fontsize{14.23}{17.07}\selectfont $A$};
\draw(148.00,49.00) node[anchor=base west]{\fontsize{14.23}{17.07}\selectfont $B$};
\definecolor{L}{rgb}{1,0,0}
\path[line width=0.30mm, draw=L] (73.00,68.00) .. controls (75.71,71.40) and (77.13,75.65) .. (77.00,80.00) .. controls (76.89,83.52) and (75.18,86.97) .. (72.00,87.00) .. controls (68.49,87.03) and (66.73,83.02) .. (67.00,79.00) .. controls (67.26,75.03) and (68.65,71.21) .. (71.00,68.00);
\definecolor{F}{rgb}{1,0,0}
\path[line width=0.30mm, draw=L, fill=F] (71.00,68.00) -- (70.15,70.76) -- (69.91,69.80) -- (68.95,70.03) -- (71.00,68.00) -- cycle;
\end{tikzpicture}%
%\fi
\caption{
{\bf A social signaling cancer loop where cell $A$ is cancerous} The cancer is social and conditional in that it depends on a signal from $B$ to divide.  Since the network shows only one loop, it leaves open the potential of the other daughter cell. Hence, it leaves open whether the cytogenic network has linear or exponential or some other potential.
}
\label{fig:SigC1}
\end{figure}

An {\em in vivo} instance of this network type is seen in bone cancer where a signal from the adjacent tissue is required for the other to become cancerous, for example, see (Logothetis~\cite{Logothetis2005}).  

%???This figure is still ambiguous.  

\subsection{Interactive signaling mono-linear network architecture}

The following network controls two cells types A and B that communicate by cell signaling.  The cell A divides only when it receives the signal $\beta$ from cell B.  The cell B only sends its signal after it has received the signal $\alpha$ from cell A. 

 \label{sec:SigL1}
 
\begin{figure}[H]
\begin{tikzpicture}[style=mystyle]
\matrix (m) [matrix of math nodes, 
row sep=3em,
column sep=\ColSepNarrow, %\ColSepTight, 
text height=1.5ex, text depth=0.25ex]
{ \vphantom{a} && \vphantom{b} &&  \vphantom{c}  && \vphantom{c}  &&  \\
 A  && A_{1} &&  A_{2}  &T & B &\vphantom{c}&  B_{1}  & \vphantom{c} \\ %TT\\
 \vphantom{a} &&  \vphantom{b} && \vphantom{B} &&  \vphantom{c}  && \vphantom{C} & \vphantom{d} & \vphantom{D}\\ };
  \path[\sigjump]
(m-2-1) edge [inPot1, green, receivestyle, cross line] node[receivesigstyle] {$\beta$} (m-2-3); %S to J1 5
 \path[\sigjump]
(m-2-7) edge [inPot1, out=90, green, receivestyle, cross line] node[receivesigstyle] {$\alpha$} (m-2-9); 

\path[\pot]
(m-2-5) edge [selfloop2, out=-80, distance=4cm, red, cross line] node[nodedescr] {a} (m-2-1) %J1 to S 1
(m-2-5) edge [in2Pot1,red, cross line] node[nodedescr] {$t$} (m-2-6); %J1 to T 7
\path[\jump, dashed]%Differentiate after send signal
(m-2-3) edge [selfloop2, out=-90, in=-120, distance=2cm, black, cross line] node[nodedescr] {$a_{2}$} (m-2-5)
(m-2-9) edge [selfloop2, out=-60, distance=3cm, black, cross line] node[nodedescr] {b} (m-2-7); 
\path[\sendsig]
%(m-2-5) edge [selfloop1, out=60, in= 125, purple,cross line] node[nodedescr] {$l$} (m-2-5) %J2 to S 
(m-2-3) edge [ in2Pot1, out= 60, in=125, blue,snakesendstyle,cross line] node[sendsigstyle] {$\alpha$} (m-2-7) %A1 sends sig to B
(m-2-9) edge [ in2Pot1, distance=5cm, blue,snakesendstyle,cross line] node[sendsigstyle] {$\beta$} (m-2-1);%B1 sends sig to A

\path[solid,red!70, line width=6pt]
(m-2-1) edge (m-2-3);
\path[solid,green!60, line width=6pt]
(m-2-3) edge (m-2-5);
\path[solid,black!40, line width=6pt]
(m-2-5) edge (m-2-6);
\path[dashed,purple!40, line width=6pt]
(m-2-6) edge (m-2-7);
\path[solid,blue!40, line width=6pt]
(m-2-7) edge (m-2-9);
\path[dashed,purple!40, line width=6pt]
(m-2-9) edge (m-2-10);
\end{tikzpicture}
\caption{
    {\bf Network SigL1: An interactive signal based stem cell network with linear potential.}  Cell A only divides if it receives a signal $\beta$ from cell B.  Cell B only sends a signal if it receives a signal from cell A.  On signal $\beta$ the cell A enters control state A$_{1}$.  Then A$_{1}$  sends a signal $\alpha$ to B. Subsequently A$_{1}$ switches to control state A$_{2}$ whereupon it divides to produce two cells (the dedifferentiated parent cell type A and a terminal cell T).  Hence, one of the daughter cells of  A$_{2}$ dedifferentiates back to the parent cell A.  The cell type B does not divide in response to the signal from A.  Instead, B responds to the signal $\alpha$ by switching to state B$_{1}$ and sending a signal $\beta$ to A.   After that B$_{1}$ dedifferentiates back to B.  The cell A will continue to divide as long the signal loop is maintained. {\bf Notation}:  The blue and green squiggly arrows between two cell states indicate cell signaling. The blue arrow $A_{1} \rightsquigarrow B$ labeled with the triangular $\alpha$ means that a cell in state $A_{1}$ sends a signal $\alpha$ that is received by a cell in state $B$.  The green arrow $B \rightsquigarrow B_{1}$ labeled with the rectangular $\alpha$ means that a cell in state $B$  on receiving the signal $\alpha$ differentiates or jumps to state $B_{1}$. 
  }
   \label{fig:SigL1}
\end{figure}

\subsection{Interactive signaling mono-exponential network with linear growth}

The following network controls two cells types A and B that communicate by cell signaling.  The cell A divides only when it receives the signal $\beta$ from cell B.  The cell B only sends its signal after it has received the signal $\alpha$ from cell A. 

 \label{sec:SigX1}
 
\begin{figure}[H]
\begin{tikzpicture}[style=mystyle]
\matrix (m) [matrix of math nodes, 
row sep=3em,
column sep=\ColSepNarrow, %\ColSepTight, 
text height=1.5ex, text depth=0.25ex]
{ \vphantom{a} && \vphantom{b} &&  \vphantom{c}  && \vphantom{c}  &&  \\
 A  && A_{1} &&  A_{2}  &\vphantom{c} & B &\vphantom{c}&  B_{1}  & \vphantom{c} \\ %TT\\
 \vphantom{a} &&  \vphantom{b} && \vphantom{B} &&  \vphantom{c}  && \vphantom{C} & \vphantom{d} & \vphantom{D}\\ };
  \path[\sigjump]
(m-2-1) edge [inPot1, green, receivestyle, cross line] node[receivesigstyle] {$\beta$} (m-2-3); %S to J1 5
 \path[\sigjump]
(m-2-7) edge [inPot1, out=90, green, receivestyle, cross line] node[receivesigstyle] {$\alpha$} (m-2-9); 
 \path[\sigjump]
(m-2-1) edge [inPot2, black, out=-45, in=-135, receivestyle, cross line] node[receivesigstyle] {$\epsilon$} (m-2-3); %S to J1 5

\path[\pot]
(m-2-5) edge [selfloop2, out=-80, in=-90, distance=4cm, red, cross line] node[nodedescr] {a} (m-2-1) %J1 to S 1
(m-2-5) edge [selfloop1,out=110, distance=5cm, red, cross line] node[nodedescr] {$a$} (m-2-1); %J1 to T 7
\path[\jump, dashed]%Differentiate after send signal
(m-2-3) edge [selfloop2, out=-90, in=-120, distance=2cm, black, cross line] node[nodedescr] {$a_{2}$} (m-2-5)
(m-2-9) edge [selfloop2, out=-60, distance=3cm, black, cross line] node[nodedescr] {b} (m-2-7); 
\path[\sendsig]
%(m-2-5) edge [selfloop1, out=60, in= 125, purple,cross line] node[nodedescr] {$l$} (m-2-5) %J2 to S 
(m-2-3) edge [ in2Pot1, out= 60, in=125, blue,snakesendstyle,cross line] node[sendsigstyle] {$\alpha$} (m-2-7) %A1 sends sig to B
(m-2-9) edge [ in2Pot1, distance=5cm, blue,snakesendstyle,cross line] node[sendsigstyle] {$\beta$} (m-2-1);%B1 sends sig to A

\path[solid,red!70, line width=6pt]
(m-2-1) edge (m-2-3);
\path[solid,green!60, line width=6pt]
(m-2-3) edge (m-2-5);
\path[solid,black!40, line width=6pt]
(m-2-5) edge (m-2-7);
\path[solid,blue!40, line width=6pt]
(m-2-7) edge (m-2-9);
\path[dashed,purple!40, line width=6pt]
(m-2-9) edge (m-2-10);
\end{tikzpicture}
\caption{
    {\bf Network SigX1: Signaling mono exponential cancer network. } Cell A only divides if it receives a signal $\beta$ from cell B.  Cell B only sends a signal if it receives a signal from cell A.  On signal $\beta$ the cell A enters control state A$_{1}$.  Then A$_{1}$  sends a signal $\alpha$ to B. Subsequently A$_{1}$ switches to control state A$_{2}$ whereupon it divides to produce two identical daughter cells (both of the dedifferentiated parent cell type A).  Hence, both daughter cells of  A$_{2}$ dedifferentiate back to the parent cell A.  The cell type B does not divide in response to the signal from A.  Instead, B responds to the signal $\alpha$ by switching to state B$_{1}$ and sending a signal $\beta$ to A.   After that B$_{1}$ dedifferentiates back to B.  The cell A will continue to divide as long the signal loop is maintained.  Prior to any communication the network is in a waiting, passive or deadlocked state. $\epsilon$ is an external signal such as a growth hormone to jump start the network. 
  }
   \label{fig:SigX1}
\end{figure}

%\paragraph{Preconditions} 
\subsubsection{Preconditions}
For this network to be actively cancerous it requires some systemic preconditions:  The existence of at least two cells of type A and B. Furthermore, the cells A and B have to be positioned so that they can receive signals from each other.  Yet even if these conditions are fulfilled, the network may be inactive.  This is results from what is called  \emph{deadlock} where each cell is waiting for a signal from the other before it does anything.  Thus, this network is inactive until an external source sends either the signals $\epsilon$ or $\beta$ to cell A or the signal $\alpha$ to cell B.  The external signal $\epsilon$ might be a growth hormone, for example, that is sent at a certain period of development or the result of environmental factors. Once one of the recognized signals is received, the network starts endless cell proliferation given the above systemic conditions are maintained.  

\subsubsection{Signal differentiation determines linear or exponential growth} 

Interestingly, in this system exponential growth is bounded by the number of B cells.  Even if the cells of type A have the potential to proliferate exponentially, they can only divide if they are close enough to a B cell to receive its signals.  We assume only close (connected) cells can have communication links to each other. Let $m_{b}$ be the number of cells of type B.  Let $c_{a,b}$ be the maximum number of A cells that can connect to a single cell B and, thereby, be close enough to a B cell to send and receive signals. Then the maximum number of cells that can be generated by the above network after $n$ synchronous divisions is given by the following formula:  

\begin{equation}
\mbox{Cells}(n) = n \times c_{a,b} \times m_{b}
\end{equation}
 
Therefore, exponential networks controlled by signaling communication protocols need not exhibit their exponential potential.  Indeed, in this example, the exponential network generates only linear growth. 

\subsubsection{Social cell proliferation as systemic and not cell autonomous}

Under this network only a whole \emph{system} of cells including both types A and B have the potential for multicellular proliferation.   The individual cells alone do not have that potential since each cell type A or B in isolation will not divide. 

\subsubsection{Signal identity changes can transform linear to exponential growth}

A transformation in just one signal type that, for example changes the sender $\alpha$ to send $\beta$ instead, changes the proliferation dynamics of this exponential network to true unhindered exponential growth. For in this case, the A cells signal each other, and possibly themselves, to divide.  Therefore, the signal differentiation is crucial in constraining this exponential network to exhibit only linear growth.  

\subsection{Dual cancer signaling networks}
\label{sec:SigC2}

A the interacting between two cooperative cell types, called partners, is driven by a network where each partner has a separate role driven by a different subnetwork with signaling driving their actions.  In the figure below we emphasize the communication links.   The network has both cells signaling and conditionally dividing, however it  leaves open whether the network is linear or exponential. The more detailed communication networks and their properties will described in the following sections. 
 
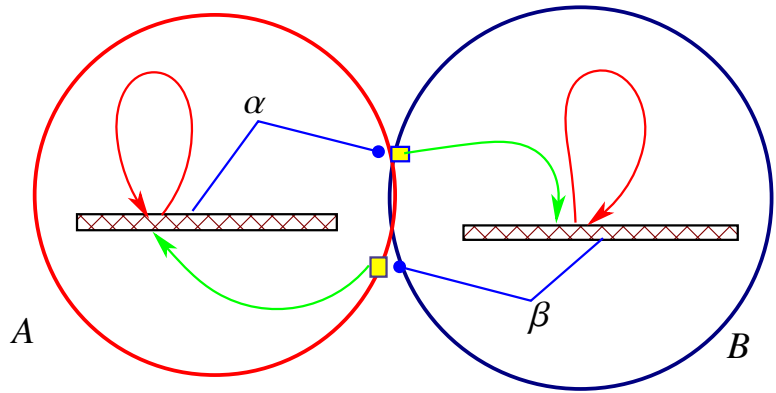
\begin{figure}[H]
\centering
\begin{tikzpicture}%scale= \PicSize, 
[x=1.00mm, y=1.00mm, inner xsep=0pt, inner ysep=0pt, outer xsep=0pt, outer ysep=0pt]
\path[line width=0mm] (51.00,42.86) rectangle +(107.34,54.79);
\definecolor{L}{rgb}{0,0,0.502}
\path[line width= \CellWallThickness, draw=L] (128.68,70.26) circle (25.39mm);
\definecolor{L}{rgb}{1,0,0}
\path[line width= \CellWallThickness, draw=L] (80.06,70.68) circle (23.96mm);
\definecolor{L}{rgb}{0.502,0,0}
\path[line width=0.15mm, draw=L]  (64.12,65.99) -- (61.98,68.12) (66.94,65.99) -- (64.81,68.12) (69.77,65.99) -- (67.64,68.12) (72.60,65.99) -- (70.47,68.12) (75.43,65.99) -- (73.30,68.12) (78.26,65.99) -- (76.13,68.12) (81.09,65.99) -- (78.95,68.12) (83.92,65.99) -- (81.78,68.12) (86.74,65.99) -- (84.61,68.12) (89.57,65.99) -- (87.44,68.12) (92.40,65.99) -- (90.27,68.12) (95.23,65.99) -- (93.10,68.12) (96.27,67.78) -- (95.93,68.12);
\path[line width=0.15mm, draw=L]  (61.73,67.38) -- (62.47,68.12) (63.16,65.99) -- (65.30,68.12) (65.99,65.99) -- (68.12,68.12) (68.82,65.99) -- (70.95,68.12) (71.65,65.99) -- (73.78,68.12) (74.48,65.99) -- (76.61,68.12) (77.31,65.99) -- (79.44,68.12) (80.13,65.99) -- (82.27,68.12) (82.96,65.99) -- (85.09,68.12) (85.79,65.99) -- (87.92,68.12) (88.62,65.99) -- (90.75,68.12) (91.45,65.99) -- (93.58,68.12) (94.28,65.99) -- (96.27,67.98);
\definecolor{L}{rgb}{0,0,0}
\path[line width=0.30mm, draw=L] (61.73,65.99) rectangle +(34.54,2.13);
\definecolor{L}{rgb}{0.502,0,0}
\path[line width=0.15mm, draw=L]  (113.48,64.71) -- (113.11,65.08) (116.31,64.71) -- (114.39,66.63) (119.14,64.71) -- (117.22,66.63) (121.96,64.71) -- (120.05,66.63) (124.79,64.71) -- (122.87,66.63) (127.62,64.71) -- (125.70,66.63) (130.45,64.71) -- (128.53,66.63) (133.28,64.71) -- (131.36,66.63) (136.11,64.71) -- (134.19,66.63) (138.93,64.71) -- (137.02,66.63) (141.76,64.71) -- (139.84,66.63) (144.59,64.71) -- (142.67,66.63) (147.42,64.71) -- (145.50,66.63) (149.57,65.39) -- (148.33,66.63);
\path[line width=0.15mm, draw=L]  (113.11,65.03) -- (114.71,66.63) (115.62,64.71) -- (117.54,66.63) (118.45,64.71) -- (120.37,66.63) (121.28,64.71) -- (123.20,66.63) (124.11,64.71) -- (126.03,66.63) (126.94,64.71) -- (128.86,66.63) (129.77,64.71) -- (131.68,66.63) (132.59,64.71) -- (134.51,66.63) (135.42,64.71) -- (137.34,66.63) (138.25,64.71) -- (140.17,66.63) (141.08,64.71) -- (143.00,66.63) (143.91,64.71) -- (145.83,66.63) (146.74,64.71) -- (148.66,66.63) (149.57,64.71) -- (149.57,64.72);
\definecolor{L}{rgb}{0,0,0}
\path[line width=0.30mm, draw=L] (113.11,64.71) rectangle +(36.46,1.92);
\definecolor{L}{rgb}{0,0,1}
\definecolor{F}{rgb}{1,1,0}
\path[line width=0.30mm, draw=L, fill=F] (103.52,75.16) rectangle +(2.35,1.92);
\definecolor{L}{rgb}{0,1,0}
\path[line width=0.30mm, draw=L] (105.22,76.23) .. controls (108.27,76.71) and (111.33,77.14) .. (114.39,77.51) .. controls (117.44,77.87) and (120.71,78.10) .. (123.13,76.23) .. controls (124.60,75.10) and (125.46,73.37) .. (125.69,71.54) .. controls (125.80,70.69) and (125.78,69.83) .. (125.69,68.98) .. controls (125.64,68.48) and (125.57,67.98) .. (125.48,67.48);
\definecolor{F}{rgb}{0,1,0}
\path[line width=0.30mm, draw=L, fill=F] (125.48,67.48) -- (126.40,70.22) -- (125.65,69.58) -- (125.01,70.33) -- (125.48,67.48) -- cycle;
\definecolor{L}{rgb}{0.282,0.239,0.545}
\definecolor{F}{rgb}{1,1,0}
\path[line width=0.30mm, draw=L, fill=F] (100.75,59.81) rectangle +(2.13,2.56);
\definecolor{L}{rgb}{0,1,0}
\path[line width=0.30mm, draw=L] (100.53,61.30) .. controls (95.60,55.27) and (87.00,53.73) .. (80.28,57.68) .. controls (77.79,59.14) and (75.81,61.28) .. (73.88,63.43) .. controls (73.31,64.07) and (72.74,64.71) .. (72.17,65.35);
\definecolor{F}{rgb}{0,1,0}
\path[line width=0.30mm, draw=L, fill=F] (72.17,65.35) -- (73.52,62.80) -- (73.57,63.78) -- (74.56,63.73) -- (72.17,65.35) -- cycle;
\definecolor{L}{rgb}{0,0,1}
\path[line width=0.30mm, draw=L] (131.66,64.93) -- (122.07,56.61) -- (103.94,61.30);
\definecolor{F}{rgb}{0,0,1}
\path[line width=0.30mm, draw=L, fill=F] (104.62,61.13) circle (0.70mm);
\path[line width=0.30mm, draw=L] (77.08,68.55) -- (85.82,80.49) -- (101.81,76.44);
\path[line width=0.30mm, draw=L, fill=F] (101.81,76.44) circle (0.70mm);
\draw(83.90,81.56) node[anchor=base west]{\fontsize{14.23}{17.07}\selectfont $\alpha$};
\draw(122.07,53.20) node[anchor=base west]{\fontsize{14.23}{17.07}\selectfont $\beta$};
\draw(53.00,51.00) node[anchor=base west]{\fontsize{14.23}{17.07}\selectfont $A$};
\draw(148.00,49.00) node[anchor=base west]{\fontsize{14.23}{17.07}\selectfont $B$};
\definecolor{L}{rgb}{1,0,0}
\path[line width=0.30mm, draw=L] (128.00,67.00) .. controls (127.86,70.35) and (127.53,73.69) .. (127.00,77.00) .. controls (126.34,81.15) and (126.38,85.81) .. (130.00,87.00) .. controls (134.48,88.47) and (138.37,83.08) .. (137.00,77.00) .. controls (136.07,72.88) and (133.55,69.29) .. (130.00,67.00);
\definecolor{F}{rgb}{1,0,0}
\path[line width=0.30mm, draw=L, fill=F] (130.00,67.00) -- (132.61,68.23) -- (131.62,68.33) -- (131.72,69.32) -- (130.00,67.00) -- cycle;
\path[line width=0.30mm, draw=L] (73.00,68.00) .. controls (75.71,71.40) and (77.13,75.65) .. (77.00,80.00) .. controls (76.89,83.52) and (75.18,86.97) .. (72.00,87.00) .. controls (68.49,87.03) and (66.73,83.02) .. (67.00,79.00) .. controls (67.26,75.03) and (68.65,71.21) .. (71.00,68.00);
\path[line width=0.30mm, draw=L, fill=F] (71.00,68.00) -- (70.15,70.76) -- (69.91,69.80) -- (68.95,70.03) -- (71.00,68.00) -- cycle;
\end{tikzpicture}%
%\fi
\caption{
{\bf A dual cancer signaling network} where both communicating partner cells A and B are cancerous. The cancer is social and conditional in that each cell depends on a signal from the other to divide. The rate and extent of growth depends on whether the properties of the cancer subnetworks of the cell types A and B.
}
\label{fig:SigC2}
\end{figure}

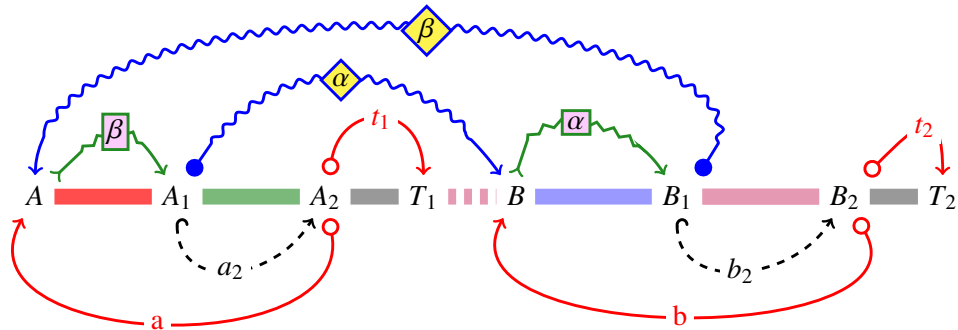
\begin{figure}[H]
\begin{tikzpicture}[style=mystyle]
\matrix (m) [matrix of math nodes, 
row sep=3em,
column sep=\ColSepTight, 
text height=1.5ex, text depth=0.25ex]
{ \vphantom{a} && \vphantom{b} &&  \vphantom{c}  && \vphantom{c}  && \vphantom{c}&&\vphantom{c}&\vphantom{c}&\vphantom{c}& \\
 A  && A_{1} &&  A_{2}  &T_{1}& B &\vphantom{c}& B_{1}  & \vphantom{c}& B_{2} &T_{2}\\ %TT\\
 \vphantom{a} &&  \vphantom{b} && \vphantom{B} &&  \vphantom{c}  && \vphantom{C} & \vphantom{d} & \vphantom{D}\\ };
  \path[\sigjump]
(m-2-1) edge [inPot1, green, receivestyle, cross line] node[receivesigstyle] {$\beta$} (m-2-3); %S to J1 5
 \path[\sigjump]
(m-2-7) edge [inPot1, out=90, green, receivestyle, cross line] node[receivesigstyle] {$\alpha$} (m-2-9); 

\path[\pot]
(m-2-5) edge [selfloop2, out=-80, distance=4cm, red, cross line] node[nodedescr] {a} (m-2-1) %J1 to S 1
(m-2-5) edge [in2Pot1,out=80, red, cross line] node[nodedescr] {$t_{1}$} (m-2-6); %J1 to T 7
\path[\pot]
(m-2-11) edge [in2Pot1, red, in=90, cross line] node[nodedescr] {$t_{2}$} (m-2-12)
(m-2-11) edge [selfloop2, out=-60, distance=4cm, red, cross line] node[nodedescr] {b} (m-2-7); 
\path[\jump, dashed]%Differentiate after send signal
(m-2-3) edge [selfloop2, out=-90, in=-120, distance=2cm, black, cross line] node[nodedescr] {$a_{2}$} (m-2-5)
(m-2-9) edge [selfloop2, out=-90, in=-120,distance=2cm, black, cross line] node[nodedescr] {$b_{2}$} (m-2-11); 
\path[\sendsig] %fill=white,
%(m-2-5) edge [selfloop1, out=60, in= 125, purple,cross line] node[nodedescr] {$l$} (m-2-5) %J2 to S 
(m-2-3) edge [ in2Pot1, out= 60, in=125, distance=4cm, blue,  snakesendstyle,cross line] node[sendsigstyle] {$\alpha$} (m-2-7) %J2 to T 
(m-2-9) edge [ in2Pot1, distance=6cm, blue,snakesendstyle,cross line] node[sendsigstyle] {$\beta$} (m-2-1);
\path[solid,red!70, line width=6pt]
(m-2-1) edge (m-2-3);
\path[solid,green!60, line width=6pt]
(m-2-3) edge (m-2-5);
\path[solid,black!40, line width=6pt]
(m-2-5) edge (m-2-6);
\path[dashed,purple!40, line width=6pt]
(m-2-6) edge (m-2-7);
\path[solid,blue!40, line width=6pt]
(m-2-7) edge (m-2-9);
\path[solid,purple!40, line width=6pt]
(m-2-9) edge (m-2-11);
\path[solid,purple!40, line width=6pt]
(m-2-9) edge (m-2-11);
\path[solid,black!40, line width=6pt]
(m-2-11) edge (m-2-12);
\end{tikzpicture}
\caption{
    {\bf Network SigL2:  A signal based dual linear network with linear potential.} Two cells A and B have equivalent signaling protocols and behavior. Cell A only divides if it receives a signal $\beta$. Upon receiving the signal $\beta$, cell A changes to control state A$_{1}$.  Prior to dividing A$_{1}$ sends a signal $\alpha$ to B.  Then A$_{1}$ changes to control state A$_{2}$ whereupon it divides to produce two cells (the dedifferentiated parent cell type A and a terminal cell type T$_{1}$).  The communication protocol for B is similar to A.  Cell B only divides if it receives a signal $\alpha$.  Upon receiving the signal $\alpha$ cell B differentiates to control state B$_{1}$ which then sends a signal $\beta$ to A.  After that, cell B$_{1}$  switches to control state B$_{2}$.  Then cell B$_{2}$ divides to produce two different cell types (the dedifferentiated parent cell type B and the terminal cell type T$_{2}$).  The cell types A and B will continue to divide as long the signal loop interaction protocol is sustained. 
  }
   \label{fig:SigL2}
\end{figure}

What we have here is a signal loop that is essential for the whole cancer network.  The dual cancer signaling network in  \autoref{fig:SigC2} consists of two subnetworks, and A-subnetwork for cell type A and a B-subnetwork for cell type B. Once activated, the A-subnetwork (e.g., \autoref{fig:SigL2}, \autoref{fig:SigX2} or \autoref{fig:SigXL})  causes a signal $\alpha$ to be sent to cell $B$  followed by activation of subnetwork A$_{2}$ whereupon the cell divides into two daughter cells at least one of which enters the signal receptor state A which enables the cell to receive signals of type $\beta$.  The signaling partner subnetwork of cell type B is activated by a signal transduction cascade initiated by the receipt of  signal $\alpha$.  The B-subnetwork in state B$_{1}$ directs the sending of a signal $\beta$ to cell A.  After sending the signal, the B-subnetwork enters state $B_{2}$ that directs $B$ to divide into two daughter cells, one or both of which loop back, to enter the receptor state B.   

Clinically, we would observe depends on nature of the cytogenic subnetworks, as well as the developing morphology of the tumor in the tissue.  The cells A and B only proliferate if have a neighbor that is a signaling partner.  Thus, the growth of B cells depends on being adjacent to A cells and vice versa.  We would observe the stopping of cell proliferation if either the A or the B cells are removed or some other agent interferes with the signaling loop.  The latter can be effected either by inhibiting the receptor for $\alpha$ or $\beta$, the signal $\alpha$ or $\beta$, or by interfering with either the$\alpha$or $\beta$ signal transduction pathway.  Alternatively, one could interfere with the genome by inhibiting the areas responsible for initiation of signaling or the inhibiting the areas responsible for cellular division.  One can see there are many potential areas where one can break the cancer loop.  In actual in vivo systems one would choose that with minimal risk of side effects.  

We now investigate some of the particular types of dual cancer signaling networks. 

\subsection{A interactive signaling dual linear network architecture}
\label{sec:SigL2}

\subsection{Interactive signaling network with exponential and linear subnetworks}
\label{sec:SigXL}

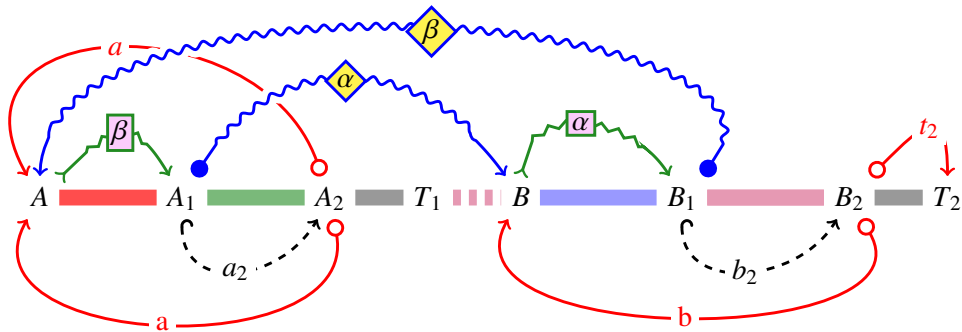
\begin{figure}[H]
\begin{tikzpicture}[style=mystyle]
\matrix (m) [matrix of math nodes, 
row sep=3em,
column sep=\ColSepTight, %\ColSepNarrow, 
text height=1.5ex, text depth=0.25ex]
{ \vphantom{a} && \vphantom{b} &&  \vphantom{c}  && \vphantom{c}  && \vphantom{c}&&\vphantom{c}&\vphantom{c}&\vphantom{c}& \\
 A  && A_{1} &&  A_{2}  &T_{1}& B &\vphantom{c}& B_{1}  & \vphantom{c}& B_{2} &T_{2}\\ %TT\\
 \vphantom{a} &&  \vphantom{b} && \vphantom{B} &&  \vphantom{c}  && \vphantom{C} & \vphantom{d} & \vphantom{D}\\ };
  \path[\sigjump]
(m-2-1) edge [inPot1, green, receivestyle, cross line] node[receivesigstyle] {$\beta$} (m-2-3); %S to J1 5
 \path[\sigjump]
(m-2-7) edge [inPot1, out=90, green, receivestyle, cross line] node[receivesigstyle] {$\alpha$} (m-2-9); 
\path[\pot]
(m-2-5) edge [selfloop2, out=-80, distance=4cm, red, cross line] node[nodedescr] {a} (m-2-1) %J1 to S 1
(m-2-5) edge [selfloop1,out=110, distance=5cm, red, cross line] node[nodedescr] {$a$} (m-2-1); %J1 to T 7
\path[\pot]
(m-2-11) edge [in2Pot1, red, in=90, cross line] node[nodedescr] {$t_{2}$} (m-2-12)
(m-2-11) edge [selfloop2, out=-60, distance=4cm, red, cross line] node[nodedescr] {b} (m-2-7); 
\path[\jump, dashed]%Differentiate after send signal
(m-2-3) edge [selfloop2, out=-90, in=-120, distance=2cm, black, cross line] node[nodedescr] {$a_{2}$} (m-2-5)
(m-2-9) edge [selfloop2, out=-90, in=-120,distance=2cm, black, cross line] node[nodedescr] {$b_{2}$} (m-2-11); 
\path[\sendsig] %fill=white,
(m-2-3) edge [ in2Pot1, out= 60, in=125, distance=4cm, blue,  snakesendstyle, cross line] node[sendsigstyle] {$\alpha$} (m-2-7) 
(m-2-9) edge [ in2Pot1, distance=6cm, blue,snakesendstyle,cross line] node[sendsigstyle] {$\beta$} (m-2-1);
\path[solid,red!70, line width=6pt]
(m-2-1) edge (m-2-3);
\path[solid,green!60, line width=6pt]
(m-2-3) edge (m-2-5);
\path[solid,black!40, line width=6pt]
(m-2-5) edge (m-2-6);
\path[dashed,purple!40, line width=6pt]
(m-2-6) edge (m-2-7);
\path[solid,blue!40, line width=6pt]
(m-2-7) edge (m-2-9);
\path[solid,purple!40, line width=6pt]
(m-2-9) edge (m-2-11);
\path[solid,purple!40, line width=6pt]
(m-2-9) edge (m-2-11);
\path[solid,black!40, line width=6pt]
(m-2-11) edge (m-2-12);
\end{tikzpicture}
\caption{
    {\bf Network SigXL:  A signal based dual network with exponential and linear subnetworks.} Two cells A and B have equivalent signaling protocols but different proliferation behaviors. Cell A only divides if it receives a signal $\beta$. Upon receiving the signal $\beta$, cell A changes to control state A$_{1}$ and sends a signal $\alpha$ to B.  Thereafter A$_{1}$ changes to control state A$_{2}$ and divides to produce two identical daughter cells both of which are of the dedifferentiated parent cell type A. This leads to potential exponential growth of cell type A.   Similarly, cell B only divides if it receives a signal $\alpha$.  Upon receiving the signal $\alpha$ cell B differentiates to control state B$_{1}$ which then sends a signal $\beta$ to A.  After that, cell B$_{1}$  switches to control state B$_{2}$.  However, unlike  A$_{2}$,  B$_{2}$,  divides producing two different cells types -the dedifferentiated parent cell type B and the terminal cell type T$_{2}$.  The cell types A and B will continue to divide as long the signal loop interaction protocol is sustained. 
  }
   \label{fig:SigXL}
\end{figure}

\subsection{A doubly exponential signaling network with linear dynamics}

Interestingly, even a doubly exponential signaling network can have linear proliferative dynamics. If we transform the linear subnetwork in \autoref{fig:SigXL} to an exponential subnetwork or equivalently add an exponential network to the mono exponential signaling network above in \autoref{fig:SigX1} then we get a signaling dual exponential network:

\begin{figure}[H]
\begin{tikzpicture}[style=mystyle]
\matrix (m) [matrix of math nodes, 
row sep=3em,
column sep=\ColSepNarrow, 
text height=1.5ex, text depth=0.25ex]
{ \vphantom{a} && \vphantom{b} &&  \vphantom{c}  && \vphantom{c}  &&  \\
 A  && A_{1} &&  A_{2}  &B && B_{1} &  B_{2}  & \vphantom{c} \\ %TT\\
 \vphantom{a} &&  \vphantom{b} && \vphantom{B} &&  \vphantom{c}  && \vphantom{C} & \vphantom{d} & \vphantom{D}\\ };
  \path[\sigjump]
(m-2-1) edge [inPot1, green, receivestyle, cross line] node[receivesigstyle] {$\beta$} (m-2-3); %S to J1 5
 \path[\sigjump]
(m-2-6) edge [inPot1, out=60, in=135, green, receivestyle, cross line] node[receivesigstyle] {$\alpha$} (m-2-8); 
 \path[\sigjump]
(m-2-1) edge [inPot2, black, out=-45, in=-135, receivestyle, cross line] node[receivesigstyle] {$\epsilon$} (m-2-3); %S to J1 5

\path[\pot]
(m-2-5) edge [selfloop2, out=-80, in=-90, distance=4cm, red, cross line] node[nodedescr] {a} (m-2-1) %J1 to S 1
(m-2-5) edge [selfloop1,out=60, in=90, distance=4.5cm, red, cross line] node[nodedescr] {a} (m-2-1); %J1 to T 7
\path[\pot] %second expo net
(m-2-9) edge [selfloop2, out=-80, in=-90, distance=4cm, red, cross line] node[nodedescr] {b} (m-2-6) %J1 to S 1
(m-2-9) edge [selfloop1,out=80, in=90, distance=4cm, red, cross line] node[nodedescr] {b} (m-2-6); %J1 to T 7
\path[\jump, dashed]%Differentiate after send signal
(m-2-3) edge [selfloop2, out=-90, in=-120, distance=2cm, black, cross line] node[nodedescr] {$a_{2}$} (m-2-5)
(m-2-8) edge [selfloop2, out=-90, distance=2cm, black, cross line] node[nodedescr] {$b_{2}$} (m-2-9); 
\path[\sendsig]
%(m-2-5) edge [selfloop1, out=60, in= 125, purple,cross line] node[nodedescr] {$l$} (m-2-5) %J2 to S 
(m-2-3) edge [ in2Pot1, out= 60, in=125, blue,snakesendstyle,cross line] node[sendsigstyle] {$\alpha$} (m-2-6) %A1 sends sig to B
(m-2-8) edge [ in2Pot1, out=90, in=125, distance=6cm, blue,snakesendstyle,cross line] node[sendsigstyle] {$\beta$} (m-2-1);%B1 sends sig to A

\path[solid,red!70, line width=6pt]
(m-2-1) edge (m-2-3);
\path[solid,green!60, line width=6pt]
(m-2-3) edge (m-2-5);
\path[solid,black!40, line width=6pt]
(m-2-5) edge (m-2-6);
\path[blue, line width=6pt]
(m-2-6) edge (m-2-8);
\path[solid,blue!40, line width=6pt]
(m-2-8) edge (m-2-9);
\end{tikzpicture}
\caption{
    {\bf Network SigX2: Signaling dual exponential cancer network.}  Cell A only divides if it receives a signal $\beta$ from cell B.  Cell B only divides if it receives a signal from cell A.  On signal $\beta$ the cell A enters control state A$_{1}$.  Then A$_{1}$  sends a signal $\alpha$ to B. Subsequently A$_{1}$ switches to control state A$_{2}$ whereupon it divides to produce two identical daughter cells that are both of the dedifferentiated parent cell type A.  Cell B behaves in an equivalent manner in response to the signal $\alpha$ from A. On receipt of $\alpha$ cell B switches to state B$_{1}$ and sends a signal $\beta$ to A. Subsequently B$_{1}$ switches to control state B$_{2}$ whereupon it divides to produce two identical daughter cells that are both of the dedifferentiated parent cell type B.  Both the cells A and B will continue to divide as long their signal loops are maintained.  Prior to any communication the network is in a waiting, passive or deadlocked state. $\epsilon$ is an external signal such as a growth hormone to jump start the network. 
  }
   \label{fig:SigX2}
\end{figure}
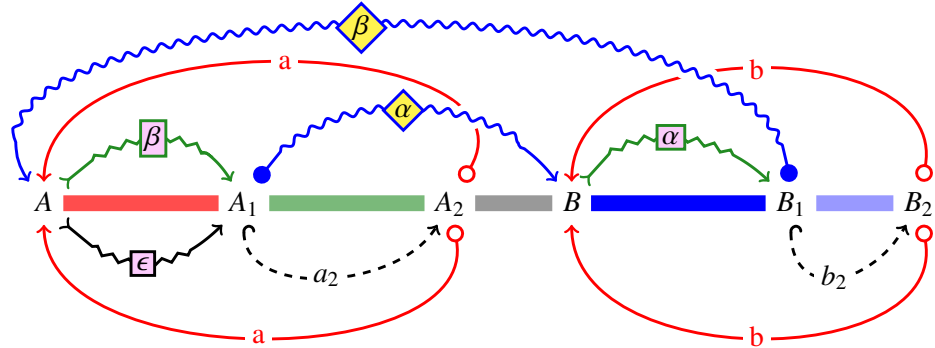

As in the case above (see network \ref{fig:SigX1}), two exponential subnetworks in a signaling network do not necessarily lead to exponential growth.  Instead cell proliferation is limited to those cells of type A and B that are in sufficient contact to maintain their signaling loops.  As the cells proliferate the number of cells in communication contact may change depending on their location in space. Hence, the context created by the multicellular development determines the rate of growth. 

Let $m_{a}$ be the number of cells of type A and $m_{b}$ be the number of cells of type B.  Let $K^{n-1}_{a,b}$ be the total number of cells A and B in communication contact. They are the cells that receive a growth signal from a communication partner cell at time step $n-1$. These are the cells close enough to be in signal contact with a complementary communication partner cell so as to send and receive signals at time step $n-1$. Then the maximum number of cells generated by the above network after $n$ synchronous divisions is given by the following formula:  

\begin{equation}
\mbox{Cells}(n) = 2 \times K^{n-1}_{a,b} 
\end{equation}
 
If for each synchronous round of cell division, all the cells generated by step $n$ are still in contact at time $n+1$ then this network will exhibit   exponential growth. If however only a constant number $K_{a,b}$ of cells remain in contact after each synchronous round of division then we have linear or polynomial growth.  

\subsubsection{Linear development with communicating dual exponential networks}
 \label{sec:SigX2LinGrow}
Therefore, exponential networks controlled by signaling communication protocols need not exhibit their exponential potential.  Indeed, in this example, if the cells in communication contact $K^{i}_{a,b}$ remains constant for all rounds of cell division $i$ then  the exponential network generates only linear growth even if it is double the linear growth of the single exponential network above. 

\begin{figure}[H]
\includegraphics[scale=0.4]{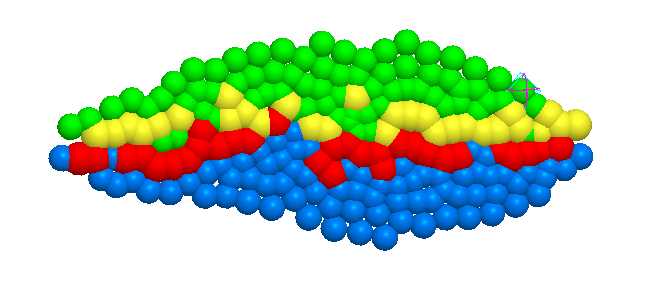}
\caption{
    {\bf Cells generated from a signaling double exponential cancer network.} The yellow and red cells are the only cells that are dividing.  The green and blue cells are signaling partners, while potentially exponential in growth, are mostly passive for lack of a signal from each other.  Hence, while their network (SigX2 in \autoref{fig:SigX2}) is a doubly exponential signaling network, the overall growth is linear or at best first order geometric.  The reason is the lack of communication links to the more distant cells. In consequence, the communication between blue and green cells generates a boundary of red and yellow proliferating cells that in turn produce more blue and green signaling cells. 
  }
   \label{fig:SigX2LinGrow}
\end{figure}

In simulations such as the typical example in (\autoref{fig:SigX2LinGrow}), the exponential networks exhibits only linear or at most 2nd order geometric growth. The reason is that the physics of development is such that there is no full mixture of cells rather there is a boundary or interface between the two cell types where the communicating cells meet and interact.  In the boundary region they can grow.  As cells are formed away from the proliferative communication interface while potentially proliferative are not actually.  

\subsection{Self-signaling cancer networks} %NSsf N14}
\label{SelfSigNets}
Cell communication networks may be self-reflexive. The cell sends a signal to itself that it receives and interprets by activating a cytogenic subnetwork.  The resulting self-induced cell proliferation depends on the properties of the cytogenic subnetwork. In another scenario, a receptor may also get stuck with a constant ``signal received'' message in which case the cell divides without an external signal. 

\begin{figure}[H]
%\centering
\subfloat[A social self signaling cancer loop]{
\begin{tikzpicture}%scale= \PicSize, Crit. scale makes the picture disappear
[x=1.00mm, y=1.00mm, inner xsep=0pt, inner ysep=0pt, outer xsep=0pt, outer ysep=0pt]
\path[line width=0mm] (51.00,44.72) rectangle +(63.70,51.92);
\definecolor{L}{rgb}{1,0,0}
\path[line width= \CellWallThickness, draw=L] (80.06,70.68) circle (23.96mm);% \CellWallThickness 0.80mm
\definecolor{L}{rgb}{0.502,0,0}
\path[line width=0.15mm, draw=L]  (64.12,65.99) -- (61.98,68.12) (66.94,65.99) -- (64.81,68.12) (69.77,65.99) -- (67.64,68.12) (72.60,65.99) -- (70.47,68.12) (75.43,65.99) -- (73.30,68.12) (78.26,65.99) -- (76.13,68.12) (81.09,65.99) -- (78.95,68.12) (83.92,65.99) -- (81.78,68.12) (86.74,65.99) -- (84.61,68.12) (89.57,65.99) -- (87.44,68.12) (92.40,65.99) -- (90.27,68.12) (95.23,65.99) -- (93.10,68.12) (96.27,67.78) -- (95.93,68.12);
\path[line width=0.15mm, draw=L]  (61.73,67.38) -- (62.47,68.12) (63.16,65.99) -- (65.30,68.12) (65.99,65.99) -- (68.12,68.12) (68.82,65.99) -- (70.95,68.12) (71.65,65.99) -- (73.78,68.12) (74.48,65.99) -- (76.61,68.12) (77.31,65.99) -- (79.44,68.12) (80.13,65.99) -- (82.27,68.12) (82.96,65.99) -- (85.09,68.12) (85.79,65.99) -- (87.92,68.12) (88.62,65.99) -- (90.75,68.12) (91.45,65.99) -- (93.58,68.12) (94.28,65.99) -- (96.27,67.98);
\definecolor{L}{rgb}{0,0,0}
\path[line width=0.30mm, draw=L] (61.73,65.99) rectangle +(34.54,2.13);
\definecolor{L}{rgb}{0.282,0.239,0.545}
\definecolor{F}{rgb}{1,1,0}
\path[line width=0.30mm, draw=L, fill=F] (100.75,59.81) rectangle +(2.13,2.56);
\definecolor{L}{rgb}{0,1,0}
\path[line width=0.30mm, draw=L] (100.53,61.30) .. controls (95.60,55.27) and (87.00,53.73) .. (80.28,57.68) .. controls (77.79,59.14) and (75.81,61.28) .. (73.88,63.43) .. controls (73.31,64.07) and (72.74,64.71) .. (72.17,65.35);
\definecolor{F}{rgb}{0,1,0}
\path[line width=0.30mm, draw=L, fill=F] (72.17,65.35) -- (73.52,62.80) -- (73.57,63.78) -- (74.56,63.73) -- (72.17,65.35) -- cycle;
\definecolor{L}{rgb}{0,0,1}
\path[line width=0.30mm, draw=L] (77.08,68.55) -- (85.82,80.49) -- (101.81,76.44);
\definecolor{F}{rgb}{0,0,1}
\path[line width=0.30mm,  draw=L, fill=F] (101.81,76.44) circle (0.70mm);
\draw(83.90,81.56) node[anchor=base west]{\fontsize{14.23}{17.07}\selectfont $\alpha$};
\draw(53.00,51.00) node[anchor=base west]{\fontsize{14.23}{17.07}\selectfont $A$};
\definecolor{L}{rgb}{1,0,0}
\path[line width=0.30mm, draw=L] (73.00,68.00) .. controls (75.71,71.40) and (77.13,75.65) .. (77.00,80.00) .. controls (76.89,83.52) and (75.18,86.97) .. (72.00,87.00) .. controls (68.49,87.03) and (66.73,83.02) .. (67.00,79.00) .. controls (67.26,75.03) and (68.65,71.21) .. (71.00,68.00);
\definecolor{F}{rgb}{1,0,0}
\path[line width=0.30mm, draw=L, fill=F] (71.00,68.00) -- (70.15,70.76) -- (69.91,69.80) -- (68.95,70.03) -- (71.00,68.00) -- cycle;
\definecolor{L}{rgb}{0.294,0,0.51}
\path[line width=0.30mm, draw=L, dash pattern=on 0.30mm off 0.50mm] (103.00,76.00) .. controls (106.31,77.66) and (110.34,76.31) .. (112.00,73.00) .. controls (113.29,70.42) and (112.72,67.35) .. (111.00,65.00) .. controls (109.13,62.45) and (106.16,60.96) .. (103.00,61.00);
\definecolor{F}{rgb}{0.294,0,0.51}
\path[line width=0.30mm, draw=L, fill=F] (103.70,61.05) circle (0.70mm);
\end{tikzpicture}%
\label{fig:SelfSigCell}
}%\subfloat %\caption{A social self signaling cancer loop }
\subfloat[A self-signaling tumor generated by exponential and linear subnetworks]{
\label{fig:SelfSigX1Tumor}
 %\raisebox{-2.1cm}{
\includegraphics[width=0.5\textwidth]{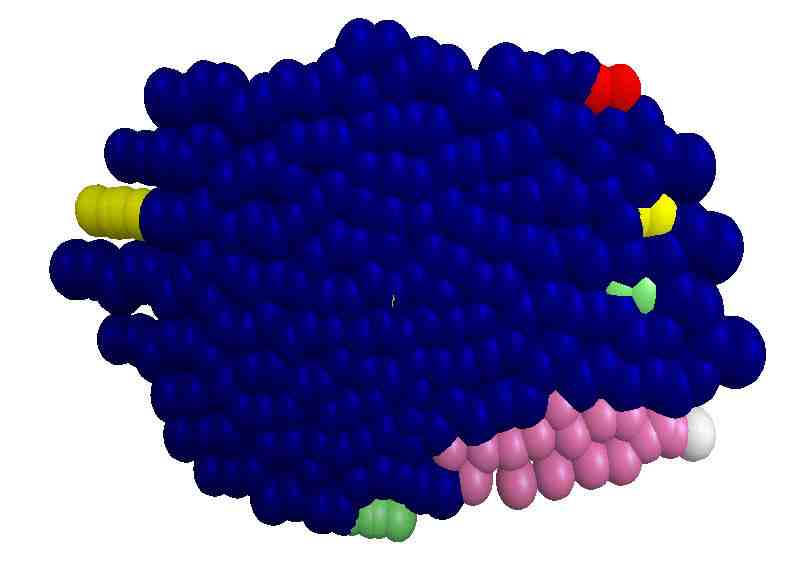} %width=0.28\textwidth
}%\subfloat
\caption{
{\bf A self-signaling cancer cell schemata and its exponential and linear tumor phenotypes.} A cell with self signaling cancer network that induces the cell to send a signal to itself that activates cell division into at least one similar self signaling cells. The growth rate will depend on the properties of the cytogenic subnetwork.  The tumor in \autoref{fig:SelfSigX1Tumor} on the right is the result of two self signaling networks. The cells in dark blue are active tumor cells controlled and generated by of an exponential cytogenic subnetwork. The pink cells are passive, generated by a self-signaling linear network.  
}
\label{fig:SelfSigCellTumor}
\end{figure}
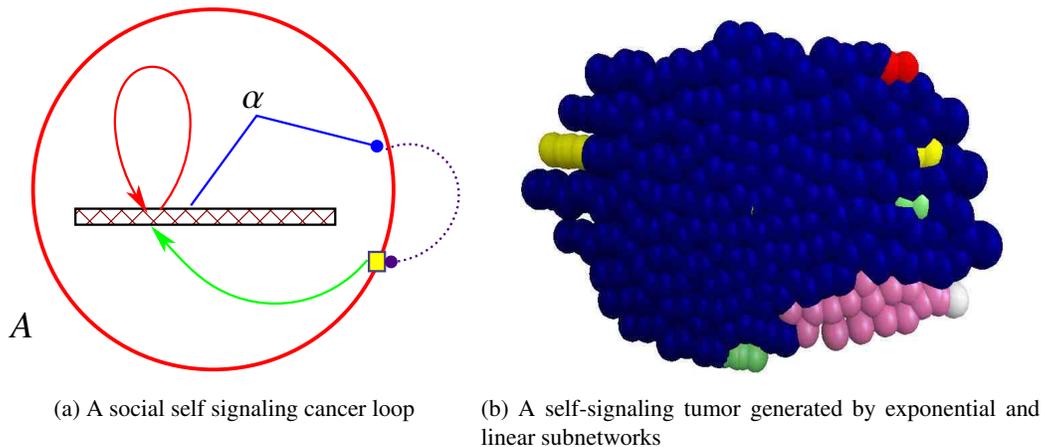

The schematic network illustrated in figure \ref{fig:SelfSigCell} is a variation of signaling cancers (\autoref{fig:SigC1}) where instead of needing a outside signaling partner B,  a cell  A sends a signal $\alpha$ to itself.  If the cytogenic subnetwork is exponential then, on receiving the signal, A divides into two copies of itself, A and A.  These in turn send signals $\alpha$ to themselves and neighbors of the same cell type A, etc.  This results in an infinite loop with a corresponding exponentially proliferating self induced cancer.  The tumor in \autoref{fig:SelfSigX1Tumor} shows the growth of a self-signaling exponential network. 

If the cytogenic subnetwork is linear, then we have a slow growing linear cancer, where the cancer stem cell does not produce more cancer stem cells. Instead it produces cells of some other cell type B.  If these B cells still inherit the cancer receptor for the signal $\alpha$, then they may dedifferentiate to the A cell type and start linear proliferating. 
 
Treatment of signaling cancers would involve agents that attack key control points in the genome network activation signaling loop complex. Another approach would be just to kill off fast growing cells without regard to their control.  That therapy would work for signal autonomous networks like NX networks as well. 

\subsubsection{A self-signaling network architecture with an exponential subnetwork}
\label{sec:Sig}
Clinically, when the cytogenic subnetwork is exponential, then we have a very fast growing and exponentially proliferating cancer.   From a diagnostic perspective, we would observe a relatively uniform cell mass if the cells are cohesive, attaching to cells of their own type.  Else, if the cells are fluid we would have a quickly proliferating cancer permeating all systems of the body.  

\begin{figure}[H]
\centering
\subfloat[Network SelfSigX1]
{\label{fig:SelfSigX1}
\begin{tikzpicture}[style=mystyle]
\matrix (m) [matrix of math nodes, 
row sep=2em,
column sep=2em, %1.8em, 
text height=1.5ex, text depth=0.25ex]
{ \\
A  \pgfmatrixnextcell \pgfmatrixnextcell A_{SR} \pgfmatrixnextcell  \pgfmatrixnextcell   \\ }; 
\path[\pot]
(m-2-1) edge [inPot1, out=90, in=90,blue, cross line] node[nodedescr] {$s$} (m-2-3) %S to J1 5
(m-2-1) edge [inPot2, in=-120, blue, cross line] node[nodedescr] {$s$} (m-2-3); %S to 2
\path[\sigjump]
(m-2-3)[draw=red] edge [inPot1, out=45, in=130,distance=4cm, red, snakesendstyle, cross line] node[sendsigstyle] {$\sigma_{1}$} (m-2-3); %J1 to S 1
\path[open diamond->]%\sigjump]
(m-2-3) edge [inPot2, out=-75, in= -90, distance=3cm, green, receivestyle,cross line] node[receivesigstyle] {$\sigma_{1}$} (m-2-1);  
\path[solid,red!70, line width=6pt]
(m-2-1) edge (m-2-3);
\end{tikzpicture}
}
\subfloat[A state of a tumor controlled by network SelfSigX1]{
\label{fig:SelfSigX1Tumor}
 \raisebox{-2.1cm}{
\includegraphics[width=0.28\textwidth]{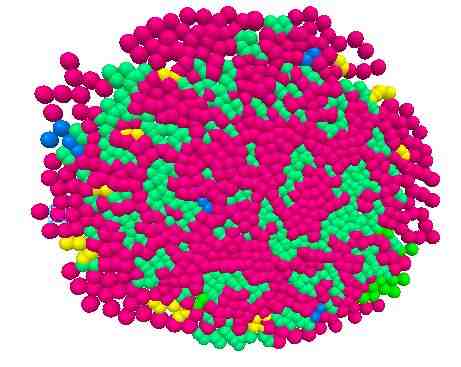}
}% \raisebox
}%\subfloat
\hspace{.1cm}
\subfloat[A advanced stage of a tumor controlled by network SelfSigX1 growing within a bilateral cellular system]{
\label{fig:SelfSigX1BilatCancer}
 \raisebox{-2.1cm}{
\includegraphics[width=0.28\textwidth]{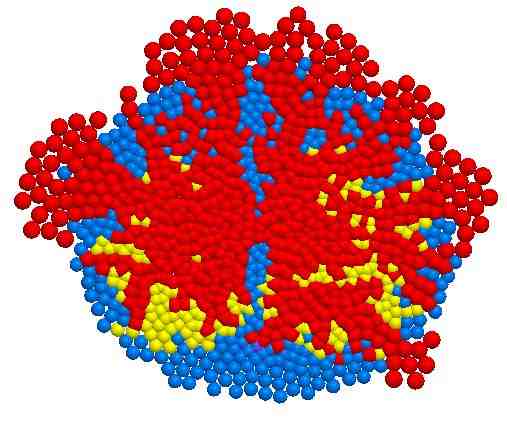}
}% \raisebox
}%\subfloat
\caption{
    {\bf A self-signaling exponential network and its tumor phenotypes} On the left fig. (a) shows a  self reflexive signal based cancer network (SelfSigX1).  The cell A divides and differentiates into two sender-receiver cells of type A$_{SR}$ with receptors for the signal $\sigma_{1}$. In the role of sender cell  A$_{SR}$ sends a signal $\sigma_{1}$ to itself and other neighboring cells.   On receiving a signal $\sigma_{1}$, possibly from itself or from its neighbors, the cell in state A$_{SR}$ dedifferentiates to its parent cell type A and the process starts again.  Fig. (b) shows a stage of a tumor generated by the network and growing independently where dividing cells are in red with communicating cells in yellow and green. Fig. (c) shows an advanced stage of the same type of tumor growing bilaterally in a bilateral multicellular system.  It has breeched the bilateral midline.  Tumor cells are in red. Blue and yellow cells are part of the normal, noncancerous bilateral cellular context. 
  }
   \label{fig:SelfSigX}
\end{figure}
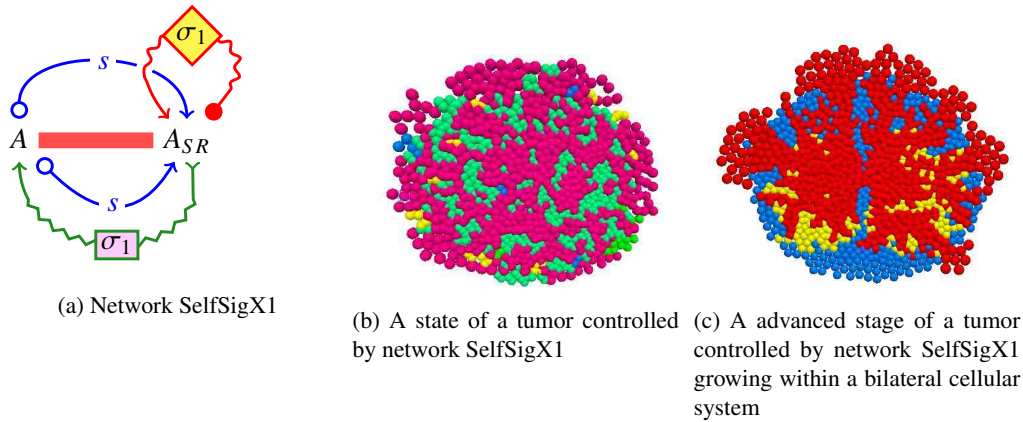

This network outwardly at the cellular level behaves much like the self activating network exponential network NX (see \autoref{fig:NX}).   A differential diagnosis would result if the cancer was restrained by interference with the cell signaling system since that is needed by network \autoref{fig:SigC1} for it to be cancerous.  The distinguishing feature of the NX network would be its resilience and resistance to any form of signaling system therapies, e.g., signaling inhibitors, receptor deactivation, signal-transduction pathway interference. 

\subsubsection{Two self-signaling networks with linear versus exponential subnetworks}
 \label{sec:SelfSigXL}
A linear and an exponential self-signaling network shown below makes the self-signaling versus the cytogenic subnetworks more apparent. The network on the left in the figure below is grows linearly in response to its own signal while the network on the right has exponential potential and once started is equivalent to the network in \autoref{fig:SelfSigX} above.  

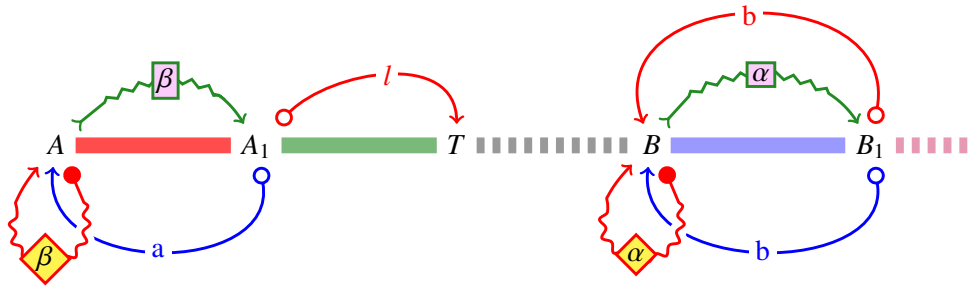
\begin{figure}[H] 
\begin{tikzpicture}[style=mystyle]
\matrix (m) [matrix of math nodes, 
row sep=3em,
column sep=\ColSepNarrow, 
text height=1.5ex, text depth=0.25ex]
{ \vphantom{a} && \vphantom{b} &&  \vphantom{c}  && \vphantom{c}  &&  \\
 A  && A_{1} &&  T  && B &\vphantom{c}& B_{1}  & \vphantom{c}& \vphantom{c} \\ %TT\\
 \vphantom{a} &&  \vphantom{b} && \vphantom{B} &&  \vphantom{c}  && \vphantom{C} & \vphantom{d} & \vphantom{D}\\ };
  \path[\sigjump]
(m-2-1) edge [inPot1, green, receivestyle, cross line] node[receivesigstyle] {$\beta$} (m-2-3); %S to J1 5
 \path[\sigjump]
(m-2-7) edge [inPot1, out=60, green, receivestyle, cross line] node[receivesigstyle] {$\alpha$} (m-2-9); 

\path[\pot]
(m-2-3) edge [in2Pot2, in=-98, out=-80, distance=3cm, blue, cross line] node[nodedescr] {a} (m-2-1) %J1 to S 1
(m-2-3) edge [in2Pot1,red, cross line] node[nodedescr] {$l$} (m-2-5); %J1 to T 7
\path[\pot]
(m-2-9) edge [selfloop1, in=110, out= 80,  red, cross line] node[nodedescr] {b} (m-2-7)
(m-2-9) edge [selfloop2, in=-98, out=-80, distance=3cm, blue, cross line] node[nodedescr] {b} (m-2-7); 
\path[\sendsig]%not needed
%(m-2-5) edge [selfloop1, out=60, in= 125, purple,cross line] node[nodedescr] {$l$} (m-2-5) %J2 to S 
(m-2-7) edge [ selfloop2, out=-60, in=-135, red,snakesendstyle,cross line] node[sendsigstyle] {$\alpha$} (m-2-7) %J2 to T 
(m-2-1) edge [ selfloop2, out=-60, in=-130, red,snakesendstyle,cross line] node[sendsigstyle] {$\beta$} (m-2-1);
%triangle 90 reversed
\path[solid,red!70, line width=6pt]
(m-2-1) edge (m-2-3);
\path[solid,green!60, line width=6pt]
(m-2-3) edge (m-2-5);
\path[dashed,black!40, line width=6pt]
(m-2-5) edge (m-2-7);
\path[solid,blue!40, line width=6pt]
(m-2-7) edge (m-2-9);
\path[dashed,purple!40, line width=6pt]
(m-2-9) edge (m-2-10);
\end{tikzpicture}
\caption[Network SelfSigSC]{
    {\bf Network SelfSigSC: One linear and one exponential self-signaling subnetwork architectures.}  There are two unconnected subnetworks.  In the first subnetwork on the left, the cell type A sends a signal $\beta$ to itself. On receiving the signal $\beta$ it changes to control state A$_{1}$.  It then divides into a dedifferentiated cell of the parent type A and a cell of type T. This subnetwork exhibits linear growth. In the second subnetwork, the cell type B sends a signal $\alpha$ to itself. On receiving the signal $\alpha$ it changes to control state B$_{1}$.  It then divides into two dedifferentiated daughter cells of the parent type A, leading to exponential growth.  
  }
   \label{fig:SelfSigSC}
\end{figure}

\section{Bilateral cancers}

\label{sec:NBilateral}
Some cancers are bilaterally symmetric.  Examples include symmetric breast cancers where the tumor develops in the same place in each breast. Based on computational modeling and simulation, we hypothesize that the mutations that lead to bilateral symmetric cancers must occur very early in development since bilateral symmetry is established very early in the first few divisions of the embryo (Gardner\cite{Gardner2001}, Werner~\cite{Werner2011b}).  If the mutation occurred in a cell after the establishment of bilateral symmetry then the resulting cancer would only occur in one of the body halves.  Most likely, therefore, is that such bilateral cancerous mutations are inherited.  

\begin{figure}[H]
\centering
\subfloat[Top view of two bilateral tumors (red, orange and yellow cells) in a slice of a bilaterally symmetric organism.]{
\includegraphics[scale=0.3]{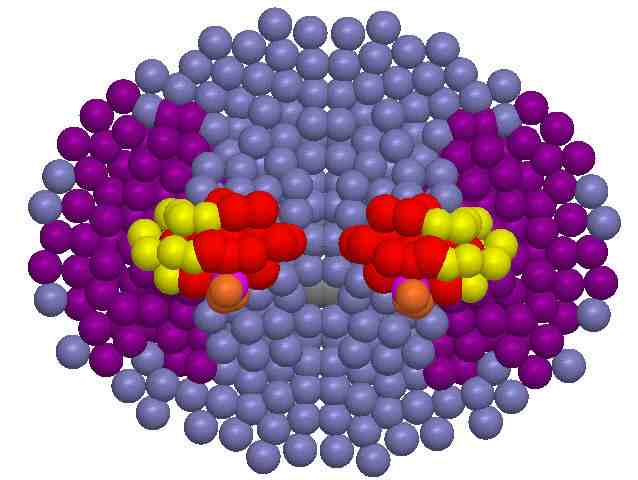}
%{SymSigMulti_Sig2Expo2DSplit_3DTopView.jpg}%{SymSigMulti_Sig2Expo2DSplit_3Dtopview3ortho.jpg} %{symCancer4identical.jpg}
\label{fig:subfig1}
}
\hspace{1cm}
\subfloat[Side view of two bilateral tumors (red, orange and yellow cells) in a slice of a bilaterally symmetric organism]{
\includegraphics[scale=0.3]{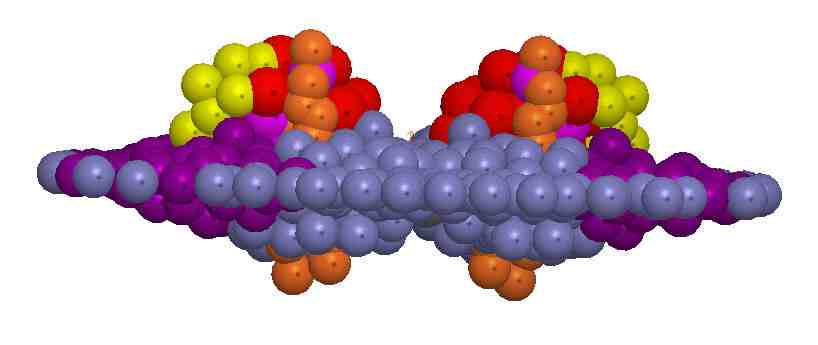}
%{SymSigMulti_Sig2Expo2DSplit_3Dsideview2.jpg}%{SymSigMulti_Sig2Expo2DSplit_3Dsideview.jpg}  % {symCancer4identicalDiff.jpg}
\label{fig:subfig2}
}
%\subfigure[Caption of subfigure 3]{
%\includegraphics[scale=0.5]{HyperIdCancerCellsAllCancerous.jpg}
%\label{fig:subfig3}
%}
\label{fig:subfigureExample}
\caption[Bilateral cancer]{
{\bf Bilateral Cancers:} The top view  of a bilateral cancer is shown on the left in figure \subref{fig:subfig1}.  A side view of the same bilateral cancer in a slice of blue and purple context cells is shown on the right in \subref{fig:subfig2}.   
}
%and \subref{fig:subfig3}}
\end{figure}

Bilateral cancers can be controlled by linear, exponential, geometric cancer networks or signaling cancer networks.  Note, a bilateral symmetric cancer can result from just one mutation.  The very same genome controls the development of each mirror body half.  Thus, the tumor itself is not symmetric, rather the tumors (to the extent that they are not stochastic or context dependent) are mirror images of each other and occupy mirror positions in the respective body half of the person.  

It is extremely unlikely that the very same mutation would occur in the two genomes of two mirror cells whose position and type mirror each other in the two symmetric body halves.  Instead, the best explanation, for symmetrically located cancers that mirror each other in phenotype, is either that they are germline mutations, or that a somatic mutation occurred in the genome of a single cell early in development  prior to the establishment of bilateral symmetry in the embryo, or that the mutation was inherited. 

\subsection{Apparent unilateral cancers generated by submerged bilateral networks}

If tumor development contains stochastic activation points, then Monte Carlo simulations show that in many cases only one of the two potential bilateral cancers will actually continue to grow. The other remains passive because the cancer signaling context did not establish itself.  Thus, the rarity of bilateral cancers need not imply that bilateral cancer networks are as rare.  They may just not show themselves in normal circumstances.  And, some apparently unilateral cancers may in fact be based on stochastic or context dependent bilateral cancer networks.   

\subsection{Temporally displaced bilateral tumors}

It may also happen that the bilaterality of the cancer is temporally displaced and that the complementary tumor starts to grow at some future time.  In this case, apparently distinct tumors are actually caused by the very same underlying bilateral cancer network.  For example, in breast cancer, later tumor development in the second breast may be the result of a stochastic, signal or context dependent bilateral cancer network. 

%SymSigX2Rnd2D.jpg show this with a non-growing unilateral one. 

\section{Vacuous and implicit non growing "cancers"} 
\label{sec:NVacuous}
Interestingly, some cancers do not grow at all either because the loop contains no cytogenic initiation sequence or because because the network is skipped and never activated by the main genomic and epigenetic control networks. 

\subsection{Vacuous non-cytogenic networks with cyclic differentiation} 
%Cyclic but non-cytogenic  (non proliferative) networks}

If we remove the crucial step of cell division we have a loop where the cell just cycles through a series of states endlessly without any cell proliferation.  Such loops can be viewed from two perspectives:  On the one hand, viewed therapeutically, they offer a possible way to convert an explicit cancer into a relatively harmless loop.  On the other hand, viewed diagnostically, such implicit non cytogenic loops, are potentially pre-cancerous in the sense that a mutation that inserts or activates a cell division switch into the loop, transforms the harmless loop into an explicit cancerous loop.  Furthermore, one would expect that some house keeping networks use loops to continually repeat needed cellular processes.  If a copy mutation inserts a cell generating subnetwork into the house keeping cycle then we have a possible cancer.

%\network[NVacCy]
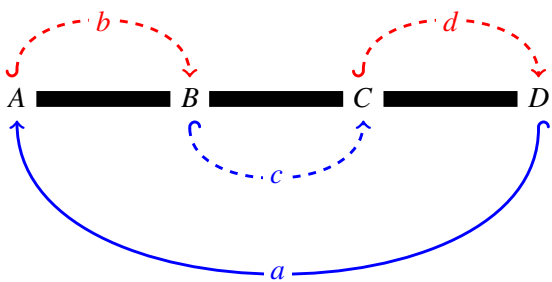
\begin{figure}[H]
\begin{tikzpicture}
%\addtolength{\hoffset}{-0.5cm}
%\addtolength{\textwidth}{1cm}
[scale= \PicSize,
nodedescr/.style={fill=white,inner sep=2.5pt},
description/.style={fill=white,inner sep=2pt},
cancer loop/.style={preaction={draw=white, -,line width=6pt}, line width=1.1pt},
pot1 loop/.style={preaction={draw=white, -,line width=6pt}, red, line width=1.1pt},
pot2 loop/.style={preaction={draw=white, -,line width=6pt}, green, line width=1.1pt},
pot1/.style={preaction={draw=white, -,line width=6pt}, gray, solid, line width=1pt}, 
pot2/.style={preaction={draw=white, -,line width=6pt}, gray, solid, line width=1pt},
selfloop1/.style={ out=45,in=135,distance=6cm, min distance=4cm, line width=1.1pt, red},
selfloop2/.style={ out=-45,in=-135,distance=6cm, min distance=4cm, line width=1.1pt, blue},
jump1/.style={ out=45,in=135,distance=4cm, min distance=2cm, line width=1.1pt, red, dashed},
jump2/.style={ out=-45,in=-135,distance=4cm, min distance=2cm, line width=1.1pt, blue, dashed},
cross line/.style={preaction={draw=white, -,	line width=6pt}}]
%\matrix (m) [matrix of math nodes, fill=red!20, row sep=3em,%Note red fill
\matrix (m) [matrix of math nodes, row sep=2em, 
column sep=2em, text height=1.5ex, text depth=0.25ex]
{ \vphantom{a} & \vphantom{b}  &  \vphantom{c}  & \vphantom{c}  & \vphantom{c} & \vphantom{d} \\
 A  &&  B  &&  C  && D \\
  \vphantom{a} &  \vphantom{b} & \vphantom{b} &  \vphantom{c}  & \vphantom{b} & \vphantom{d} \\ };
\path[right hook->]
(m-2-1) edge [jump1, bend right=-90] node[nodedescr] {$ b $} (m-2-3)
(m-2-3) edge [jump2, bend right=90] node[nodedescr] {$ c $} (m-2-5)
(m-2-5) edge [jump1, bend right=-90] node[nodedescr] {$ d $} (m-2-7)
%Note, the graphics will be pushed to the right if the angle is bigger e.g., 120 or more
(m-2-7) edge [selfloop2, bend left=90] node[nodedescr] {$ a $} (m-2-1);
\path[solid, line width=6pt]
(m-2-1) edge (m-2-3)
(m-2-3) edge (m-2-5)
(m-2-5) edge (m-2-7);
\end{tikzpicture}
\caption{
    {\bf Network NVD: Vacuous non-cytogenic cyclic network with cyclic differentiation}, where $\hookrightarrow$ indicates a non-proliferation jump to a  different differentiation state. The network cycles through different differentiation states without producing any daughter cells. 
  }
\end{figure} 
%was N9

\subsection{Vacuous cancer sub-networks skipped by the developmental controlling network}

Below, we have not just an empty loop but rather a potentially proliferating cycle that would be cancerous if it were not skipped by the dominant activating cascade.  A mutation in $A$ that would lead it to differentiate to $B$ instead of $C$ would then result in an exponential cancer.  Such networks can potentially be activated by cell signaling or by being linked to another network that is linked in to the dominant control network governing development and sustainability. 

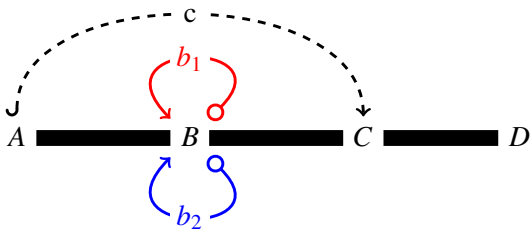
\begin{figure}[H]
\begin{tikzpicture}[scale= \PicSize, 
selfloop1/.style={ out=45,in=135,distance=4cm, min distance=3cm, line width=1.1pt, red},
selfloop2/.style={ out=-45,in=-135,distance=4cm, min distance=3cm, line width=1.1pt, blue},
nodedescr/.style={fill=white,inner sep=2.5pt},
description/.style={fill=white,inner sep=2pt},
cancer loop/.style={preaction={draw=white, -,line width=6pt}, line width=1.1pt},
pot1 loop/.style={preaction={draw=white, -,line width=6pt}, red, line width=1.1pt},
pot2 loop/.style={preaction={draw=white, -,line width=6pt}, blue, line width=1.1pt},
pot1/.style={preaction={draw=white, -,line width=6pt}, gray, solid, line width=1pt}, 
pot2/.style={preaction={draw=white, -,line width=6pt}, gray, solid, line width=1pt},
jumpover/.style={ out=80,in=90,distance=4cm, min distance=3cm, line width=1.1pt, black, dashed},
%cancer loop/.style={line width=2pt},
cross line/.style={preaction={draw=white, -,	line width=6pt}}]
\matrix (m) [matrix of math nodes, row sep=2em,
column sep=2em, text height=1.5ex, text depth=0.25ex]
{ A  &&  B  &&  C  && D \\
  \vphantom{a} & \vphantom{b}  &  \vphantom{c}  & \vphantom{c}  & \\ };
 \path[\pot]
(m-1-3) edge [selfloop1] node[nodedescr] {$ b_1 $}(m-1-3)
(m-1-3) edge [selfloop2] node[nodedescr] {$ b_2 $}(m-1-3);
%(m-1-1) edge [pot2, bend right=-80, distance=5cm, min distance=4cm,green] node[nodedescr] {$ c $}(m-1-5);
\path[right hook->]
(m-1-1)edge [jumpover, bend right=-90] node[nodedescr] {c}(m-1-5);
\path[solid, line width=6pt]
(m-1-1) edge (m-1-3)
(m-1-3) edge (m-1-5)
(m-1-5) edge (m-1-7);
\end{tikzpicture}
\caption{
{\bf Network NVX: A  Vacuous potentially exponential cancer} where the activation path from A to C skips B
}
\end{figure}

%was N10

On the positive side, such networks point to a possible way of curing cancer by converting a cancer network into a vacuous network by skipping it.  This would be achieved by  interfering with any one of the steps in the cancer  cycle such as changing the catching address  that initiates the cancer loop.  
  
\section{Metastases}
In an orthogonal process that adds yet another dimension to cancer networks, cells can break their normal boundaries, enter the blood stream or lymphatic system to travel to new areas of the organism forming potential new cancerous regions called \emph{metastases}.  The underlying cytogenic network controlling the cell is not changed by the cell's change of location.  However, because the cell's context may change, new interaction protocols may change its proliferative behavior by changing its network activation state.  Even the cell's lack of a cellular matrix, may change its control state, causing it to generate its own cellular context.  We consider just a few possible cases of cancer networks in metastatic scenarios.   

\subsection{Metastatic potential of cancer network types}
Each major type of cancer network has different metastatic potential.  A linear cancer is generated by only one stem cell that produces terminal cells or at most terminal progenitor cells.  A 2nd order geometric cancer produces endless 1st order cancer stem cells which in turn produce terminal cells.  Geometric cancers of order 3 produce 2nd order geometric cancer cells and, hence, have greater metastatic potential. Exponential cancers have the greatest metastatic potential for distributed, invasive growth since the daughter cells produce further exponential cancer cells. 
If we add signaling and conditional activation to a given cancer network then, in addition to cell physics, the cellular environment can be both an enabler or a constraint on metastatic potential.  

\subsection{Metastasis via signaling}
\label{sec:SigMetas}

If a cell with an inducible signaling cancer network cell manages to break its boundaries and travel to other parts of the organism, it would not necessarily lead to a cancer developing. First, the invasive cell A cannot endlessly divide without a cooperating communicating partner cell B with the correct complementary signaling protocol.  Second, the cell still needs an appropriate starter signal to activate its cancer network either from a partner B or from an external source such as a growth hormone.  

\begin{figure}[H]
\subfloat[Three metastases in a dual exponential signaling network]{
\includegraphics[scale=0.5]{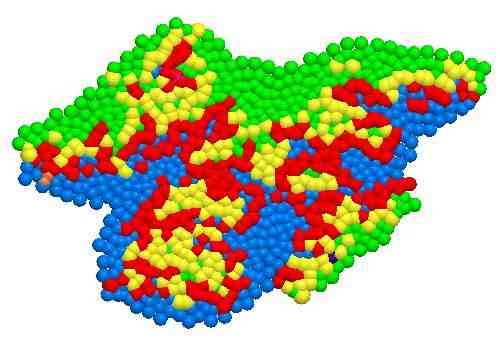}
\label{fig:metsubfig1}
}
\hspace{1.0cm}
\subfloat[Three metastases in a mono exponential signaling network in a context of signal cells]{
\includegraphics[scale=0.4]{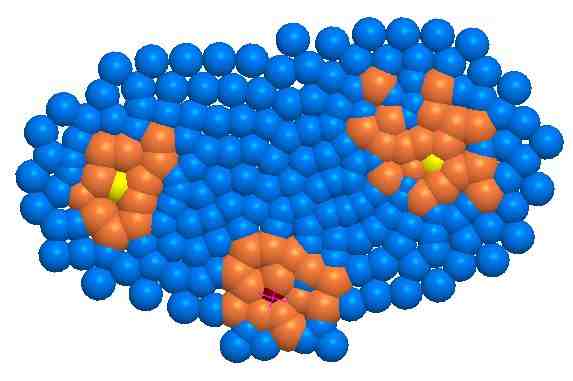}
\label{fig:metsubfig2}
}
\caption{
{\bf Two views of metastasis by different networks.} Figure \subref{fig:metsubfig1} shows the growth response to 3 metastases controlled by dual exponential signaling subnetworks (Network SigX2 \autoref{fig:SigX2} .   The right hand Figure \subref{fig:metsubfig2} shows 3 metastases controlled by the same signaling mono exponential network (Network SigX1 \autoref{fig:SigX1}).   
}
\label{fig:Metastasis}
\end{figure}

The green and blue cells (\autoref{fig:Metastasis} \subref{fig:metsubfig1} above) are communication partners in the control paths of the network but, while they potentially differentiate to exponential yellow and red cells, are passive unless they receive the right signal from each other.  The red and yellow cells divide exponentially but are limited to the differentiation potential of their parent cells, the blue and yellow cells, respectively.  Note the irregular growth.

The orange cells (\autoref{fig:Metastasis}\subref{fig:metsubfig2} divide to form two blue cells. The blue cells are passive receiver cells until they receive a signal from a yellow sender cell.  Once the blue cells receive their signal they differentiate into active dividing orange cells.  Yellow sender cells and receiver blue cells interact by means of a signaling protocol (Network SigX1 \autoref{fig:SigX1}) that regulates the production of the proliferating orange cells. The yellow cells in the center of metastases are purely signaling cells without proliferative potential. The blue cells only differentiate to divide in the context of the yellow signaling cells.  Thus, even though they do not proliferate, the yellow signaling cells are the drivers of this type of cancer.   Here the growth is more regular with fewer distinct differentiation control states.

Thus, \autoref{fig:Metastasis} illustrates different growth characteristics of metastases resulting from two different cancer signaling networks.  Both types of metastases in the two figures illustrates that we can have signal induced metastasis if the appropriate signal reaches any of the invading cancer cells.  Both networks require that an interloper A cell has traveled to the a region or context of cooperating partner cells B. However, the networks have different dynamic growth and differentiation characteristics. 

The the network  (\autoref{fig:Metastasis}\subref{fig:metsubfig2} on the left side initiates metastases when the interloper cell A  or context cells B are activated by an appropriate signal. Yet, even here the doubly exponential network does not produce exponential growth because growth is still dependent on a signaling loop between neighboring cells.  

The second network (\autoref{fig:Metastasis}\subref{fig:metsubfig2}) on the right side, has a more uniform phenotype than the more complex metastases generated by the dual exponential signaling network. 

\subsection{Moving signal-based metastases}

In any signaling cancer network where the sender cells stay constant or not increasing in number, any metastasis would move to a different region if the sender cells move. For example, the mono exponential signaling network shown in \autoref{fig:SigX1}, that controls the yellow signaling cells in subfigure \autoref{fig:metsubfig2}, does not generate any new sender cells.  Hence, without an active other external network that generates further sender cells, the sender cells will not increase in number.  Thus, because the yellow sender cells stay constant in number yet can activate this type of cancer network in a receptive cellular context, we could have the phenomenon of mobile moving cancer areas or moving metastasis that stop in one area but then start anew in another area because the underlying sender cells have moved to new tissue.  

\subsection{Context dependence of signal based metastasis}

Cancer cells controlled by signaling cancer networks cannot form metastases independently without cooperating partner sending cells with a fitting communication protocol. Thus, they require a preexisting context of complementary, passive cancer helper cells.   In an unresponsive context without communicating helper cells, a potentially metastatic cell would lay dormant until a fitting cellular context together with an initiating signal became available. 

Given the context dependence of these kinds of cancers, it is possible that cancer cells with metastasizing potential may be spread more widely through the organism than is apparent from just the presence of explicit cancerous growths.  Many such potentially cancerous cells will just remain inactive in cellular contexts that do not have the partner signaling protocol available.  However, as conditions change, such potential cancer cells may suddenly become active.  

\subsection{Stochastic signal induced metastasis}

Another possible avenue of signal induced metastasis is if there is stochastic signaling where cells with low probability send an proliferation inducing signal to their neighbors.  Yet, another possibility is that these sender cells could form by an occasional stochastic dedifferentiation transforming a passive receiver cell into an active sender cell.   

\subsection{Stopping signal induction of metastases can have side effects} 

One way to prevent metastasis from developing for this kind of cancer network is to inhibit the inducing cell signaling.  However, this may have side effects if the signaling system is used for normal cellular coordination.  

\subsection{A Hierarchy of metastases formed by geometric cancer networks}

Geometric cancer stem cell networks form a hierarchy of interlinked stem cell networks such that an n-th order stem cell network generates a cell controlled by an n-1 order stem cell network. Thus, 3rd order or meta-meta-stem cell networks produce cells controlled by 2nd order, meta-stem cells.  A 2nd order or meta-stem cell generates a cell controlled by a 1st order, linear stem cell network.  At the base of the hierarchy is a terminal network controlling a cell generated by a 1st order, linear stem cell. 

This hierarchy of control in higher order stem cell networks has direct consequences for the types of metastases that can, in principle, be  generated by such higher order geometric networks.  In other words. the metastatic potential of a cancer stem cell is determined by the order of its geometric control network.  As we will see, this the dynamic phenotypic properties of the metastases are a result of their place in the metastatic hierarchy.  This can be used for diagnosis and reverse inference as to the type of stem cell network controlling the metastatic tumor.  

\label{sec:GMetastases}
\begin{figure}[H]
\subfloat[{\bf Terminal metastases from a primary 1st order geometric cancer network.}  ]{
\includegraphics[scale=0.4]{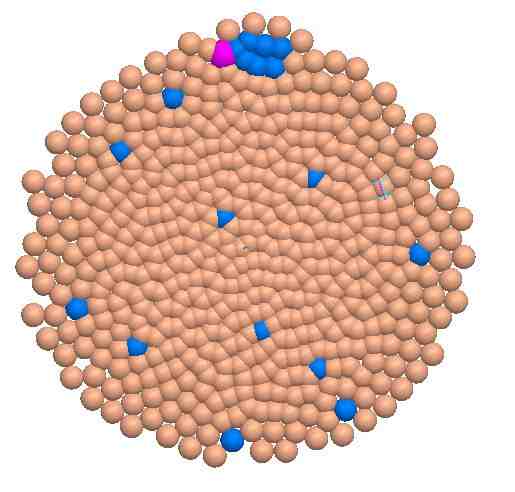}
\label{fig:G1Metsubfig1}
}
\hspace{1.0cm}
\subfloat[{\bf 1st order and terminal metastases from a 2nd order geometric cancer network.}  ]{
\includegraphics[scale=0.4]{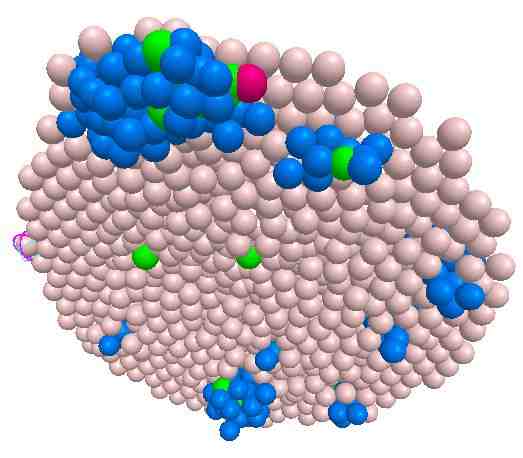}
\label{fig:G2metsubfig2}
}
\caption{
{\bf Two views of metastasis by 1st order and 2nd order geometric networks.}  In Figure \subref{fig:G1Metsubfig1}:  One primary 1st order geometric, linear cancer cell (in red) generates terminal metastatic cells (in blue). But the blue terminal cells are terminal generating no further cells. 
The right hand Figure \subref{fig:G2metsubfig2}: A 2nd order geometric cancer network generates 1st order geometric and terminal metastases.  Against a background of identical cells (in light pink) there is one primary 2nd order geometric cancer cell  (in red) that generates metastases consisting of 1st order geometric cancer cells (in green). These in turn, generate secondary metastatic passive terminal cells (in blue). The blue terminal cells may metastasize, moving to other regions, but they do not generate further cells.  
}
\label{fig:G1G2Metastasis}
\end{figure}

A 2nd order geometric cancer cell generates 1st order linear cancer cells (see \autoref{fig:NG2} and \autoref{fig:DSC2}). If these 1st order linear cancer cells metastasize then we have a phenotype of multiple linear cancer growths in different regions of the body.  For example, one might have multiple slow growing cysts in different regions of the body. In this case, these cysts are metastases that were generated from a single meta-stem cell that generates progenitor cells that generate cysts.  Other examples, might include slowly progressing cancers that move in the lymphatic system.  Here again in may, but need not, be the case that a single meta-cancer stem cell is generating sub-cancers that have the overt phenotype of a uniform cancer progressing in some direction across a region or through the whole body. 

\subsubsection{Secondary and tertiary metastases generated by 3rd order cancer networks}

\begin{figure}[H]
\subfloat[{\bf 1st oder and terminal metastases from a primary 2nd order geometric cancer network.}  ]{
\includegraphics[scale=0.4]{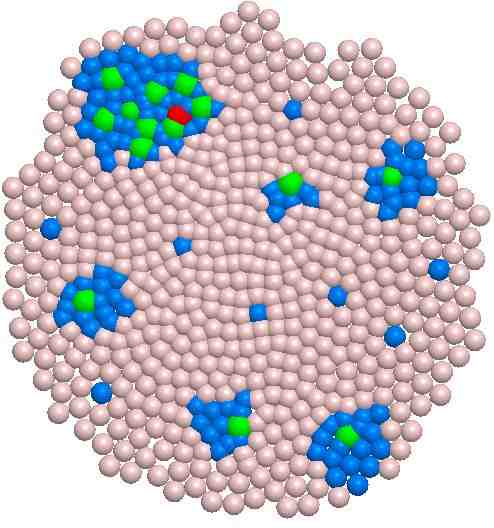}
\label{fig:G2Metsubfig1b}
}
\hspace{0.5cm}%{1.0cm}
\subfloat[{\bf 2nd, 1st order, and terminal metastases from a primary 3rd order geometric cancer network.}  ]{
\includegraphics[scale=0.4]{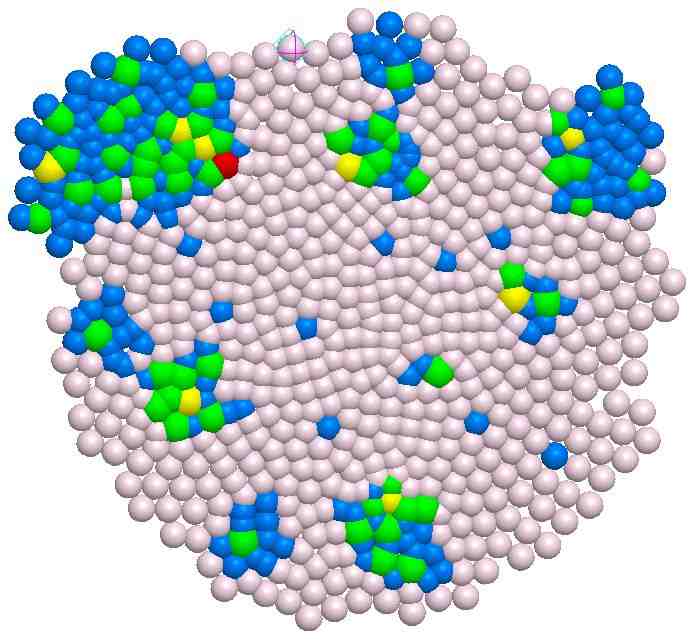}
\label{fig:G3metsubfig2}
}
\caption{
{\bf Comparing metastases by 2nd order and 3rd order geometric cancer networks.}  The {\bf left hand Figure} \subref{fig:G2Metsubfig1b}: A 2nd order geometric cancer network (\autoref{fig:NG2}) generates 1st order geometric and terminal metastases.  Against a background of identical cells (in light pink) there is one primary 2nd order geometric cancer cell  (in red) that generates  metastases consisting of 1st order geometric cancer cells (in green). These in turn, generate secondary metastatic passive terminal cells (in blue). The blue terminal cells may metastasize, moving to other regions, but they do not generate further cells.  
In the {\bf right hand Figure} \subref{fig:G3metsubfig2}: A primary tumor controlled by a 3rd order geometric cancer network generates 2nd order and 1st order geometric and terminal metastases. The primary 3rd order geometric cancer network (G3) generates 2nd order (G2) metastases. These G2 networks in turn generate 1st order (G1) metastases. The 1st order geometric (G1) cancer cells in turn generate terminal (G0) cells that are also potentially metastatic (invasive) but do not generate further cells. 
} 
\label{fig:G2G3Metastasis}
\end{figure}

In general, n-th order geometric cancer networks $G_{n}$ (see \autoref{fig:NGk}) generated cells controlled by $n-1$ order geometric cancer networks $G_{n-1}$.  These $G_{n-1}$ networks, in turn, generate cells controlled by $n-2$ order cancer networks $G_{n-2}$, etc. Therefore, third-order geometric cancer networks $G3$, such as the above simulated network in \autoref{fig:G2G3Metastasis}, generate cells controlled by second-order cancer networks $G2$. In turn, $G2$-networks generate cells controlled by 1st-order geometric networks $G1$.  $G1$-networks are linear cancer networks which generate terminal cells $G0$ that do not proliferate.  

\begin{figure}[H]
\subfloat[{\bf Active cells in 1st oder metastases from a primary 2nd order geometric cancer network.}  ]{
\includegraphics[scale=0.4]{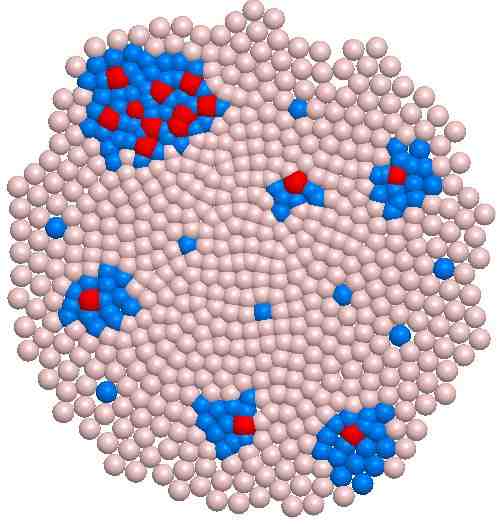}
\label{fig:G2Metsubfig1b}
}
\hspace{0.2cm}
\subfloat[{\bf Active cells in 2nd and 1st order metastases generated by a primary 3rd order geometric cancer network.}  ]{
\includegraphics[scale=0.4]{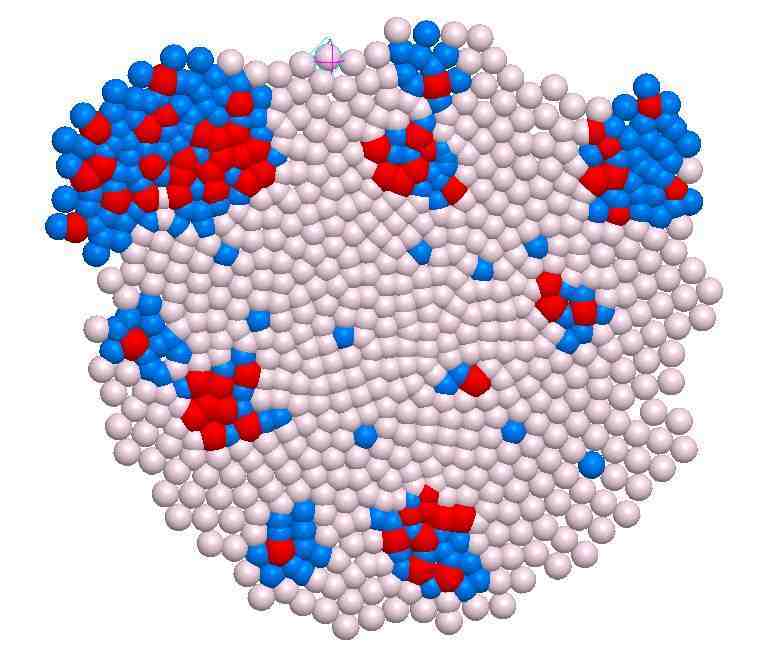}
\label{fig:G3metsubfig2}
}
\caption{
{\bf View of the active proliferating cells in G1, G2 and G3 geometric metastases in \autoref{fig:G2G3Metastasis}}. Cellular proliferation alone does not distinguish higher order from lower order geometric tumors including metastatic tumors. Only terminal cells of type G0 can be distinguished since they do not divide.  Rather the control state and downstream proliferation potential distinguishes geometric networks from each other. 
}
\label{fig:G3MetRedActiveCells}
\end{figure}

\subsubsection{Relating metastatic phenotype with geometric cancer networks}

Hence, since any of the cells generated by a cancer network have the potential to metastasize by moving to other regions of the organism, each primary tumor controlled by a cancer network will have a distinct metastatic phenotype.  Therefore, we can use the phenotype of the metastases to draw inferences about the nature and architecture of the network generating that phenotype.  Moreover, if we find the metastases that together form related patterns as described for the geometric metastases above, it warrants looking for the primary tumor generating the secondary and tertiary metastases.  

\subsubsection{Treatment options for metastases generated by geometric cancer networks}

Treatment would focus on eliminating or transforming cells with the higher level control networks generating the less dangerous lower level networks.  Treatment can be by cell death or by network transformations that block or modify the cancer network.  Different transformations are necessary for different cancer networks.  Once metastases have been created by a higher level network, it is no longer sufficient to block or transform the higher level initiating subnetwork. Instead, each of the cells with lower level cancer subnetworks also have to be transformed or destroyed.  If they are 1st order linear networks, they may be relatively harmless since they only produce cells that no longer proliferate. However, this will depend on their location and effect on the other multicellular contexts in which they reside. Because of cell interactions such as cell signaling, stochastic reactivation of higher level cancer networks such cells, depending on their differentiation state and the cellular context, may still be dangerous. 

\subsection{Stochastic reactivation of terminal cells in geometric networks}

If terminal cells can reactivate a proliferative control state in a network, then the more numerous the number of terminal cells the more likely it is that one or more of them will become an active cancer cell.  Hence, given the possibility of stochastic activation of ancestral cancer networks,  formerly passive metastatic cells can become active tumors. 

\subsection{Metastasizing signal-autonomous exponential cancer networks}  

Metastasizing cells controlled by exponential cancer networks can be curable (e.g., a signal-autonomous mole pregnancy), persistent but manageable if signal dependent on particular partner cells (e.g., \autoref{fig:Metastasis} and \autoref{fig:SigX2LinGrow}), or fatal if resistant to treatment (e.g., resistant self-signaling exponential cancers, \autoref{fig:SelfSigCell} or resistant signal autonomous exponential cancers, \autoref{fig:NX}).  The resistance of a cancer to drugs or other treatment may depend on other factors and molecular pathways and networks somewhat independent and orthogonal to the cytogenic control networks.  The control network theory of cancer offers possible ways to stop resistant cancers by directly attacking the control network topology. 

\begin{figure}[H]
\begin{center}
\includegraphics[scale=0.4]{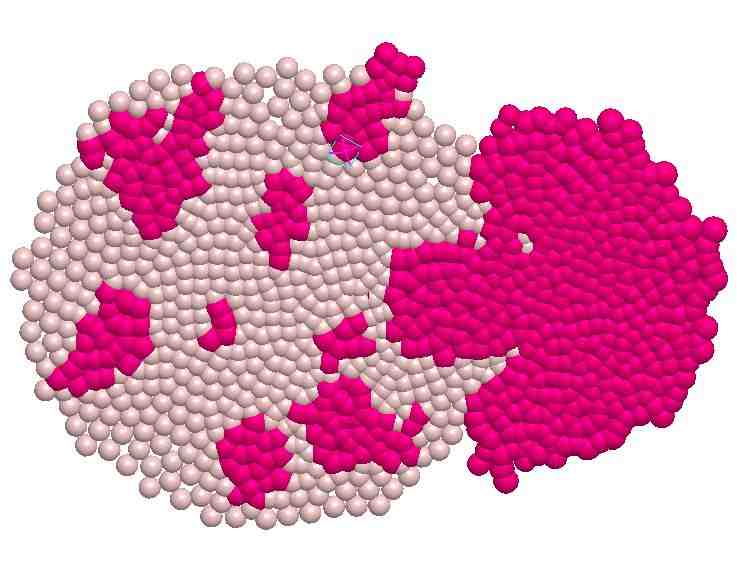}
\caption{
{\bf Metastases of a signal-autonomous exponential cancer network.} All metastases have exponential proliferation potential. Hence, they can generate further metastases with the same exponential potential.  This different from geometric cancer networks that generate cells of a lower proliferative potential.
}
\label{fig:ExpoDet}
\end{center}
\end{figure}

%\section{Cancer cell proliferation in real systems}

\section{Network competence and performance in real continuous physical systems}
%was {Competence and performance}

We distinguish network competence and ideal growth from its actual performance in real space-time.  A  {\em network's competence} is the ideal synchronous growth of cells in a discrete  space with discrete physics with no friction or other properties that would constrain growth.  Network performance involves the dynamic growth of cells in a continuous space governed by the laws of continuous physics. Linguists will recognize the analogy to Chomsky's distinction between linguistic competence as given by an ideal generative grammar and a speaker's actual performance (Chomsky~\cite{Chomsky1965}).  

Similarly, the (ideal) competence or capacity of a generative network exhibits its essential mathematical properties unencumbered by the vagaries physical and temporal contingency. The performance of the network may vary depending on the conditions in a realistic space-time cell physics-in the context of a dynamic physical environment of other growing cells.  The ideal competence of a network shows the ideal, abstract topology of its developmental possibility space (phase space) and the boundaries or limits of its growth potential. The real performance with real continuous physics will in general be less than the ideal competence or capacity.  

This is not to say that the ideal competence is not complex, for the social, communicative and cooperative interactions between cells will also be part of the ideal performance of cells controlled by a proliferative network.  This means we have to distinguish between ideal performance and real performance.  The measures we apply here apply to the ideal performance of cancer control networks.  

\subsection{Ideal developmental networks in real contexts}

%Clearly with such growth rates the patient will soon notice symptoms. 
Bodily resources and cell physics are mitigating factors that limit the growth of cancers.  Hence, our ideal cancers are like the laws of Newtonian mechanics, being limited by friction and other circumstances.  In spite of this it is important to understand the nature of cancer in an idealized developmental environment if we are to decipher cancer networks and thereby potential cancer targets. These and other mitigating factors will be discussed. 

\subsection{Cell physics}

Cell physics becomes important when considering diagnosis of a cancer.  A repeating structure under pressure will not necessarily look the same as the same structure grown in a non-congested environment.  Furthermore, inter-cellular pressures will cause such structures not only to distort, but to break up leading to a confused cell mass of different tissues.  And, this is something that is actually observed in many tumors. 

\subsection{Invasive cancers} 

Cancers can become invasive when the cells change their physical properties or by pressure break out into the vascular system. If they are cancer cells then they are distributed by the blood vessels or lymphatic system to other organs and areas of the body. 

\subsection{Tumor vasculature}

Many interactions between cells appear to from genome autonomous cell strategies that control cell communication and cell actions.  Even though cell strategies are constructed using control information in the genome, once created they may have a life of their own not directly controlled by that genome.   Vascularization is a process that appears to be largely independent of developmental control information in the genome. It is what is referred to as a developmentally regulated rather than a mosaic developmental process. A \emph{regulated development} process is flexible relying on cell communication and cell interaction strategies, while \emph{mosaic development} is inflexible driven without cell signaling.  Vascularization appears to be regulated development being controlled by cell to cell interactions including cell communication strategies.  It results in the formation of blood vessels to feed a growing tumor.  However, even though vascularization requires particular conditions to be initiated, such as anoxia (the lack of oxygen) and even though it may involve genome autonomous cell signaling strategies to spatially and temporally coordinate blood vessel formation, there may still be genomic developmental control networks involved in the process.   These developmental control networks need to be activated for cells to divide and proliferate.   

\subsection{Multiple parallel networks}
Developmental networks or cenes cooperate with the cell's interpretive executive system (IES) to generate cell actions such as cell division and cell communication.  The IES is a complex molecular system that contains lower level networks that interact with the higher level controlling developmental networks.  However, we suggest that the architecture relating these networks is something akin to a subsumption architecture (Brooks~\cite{Brooks1986}) where there is local reactivity in the lower level IES but global control resides in higher level developmental networks.  Many of these networks act in parallel with in the cell.  Within the multicellular system we have vast multi-parallel developmental networks that interact via cell signaling protocols.  The cells are then agents with local strategies implemented by the state of their developmental network and the state of their IES. 

Cancer can be caused by either a transformation in the developmental network alone, or it can result from mutations in the IES or its subsystems.  Thus if a cancer causing mutation affects only cis catchers or trans pots of the genome it is pure developmental cancer. If the cancer results from a mutation in the cell cycle or mutated receptor while the cenome is left intact, it is a cene autonomous cancer.  And, we can have the mixed cases where mutations in the cenome combine with IES mutations to give mixed control cancers.  

\subsection{Relations between gene-based and cene-based networks}
Thus, if the IES is largely implemented by gene-based molecular systems then gene-based cancers may exist that are largely cenome autonomous.  Pure gene-based cancers are conceptually different from cene-based cancers even though their dynamics would still fall under the general architecture and topology of cancer control networks.  Many cancers are a mixture of gene and cene mutations.  For example, gene mutations may increase the invasiveness of a cene based cancer thus contributing to the cancer's metastatic potential. 

Viewed more abstractly where cell's are agents with various strategies, then a cell has a local strategy of behavior and it has a global strategy for multicellular development.  These strategies run in parallel and interact with in the cell. For a multicellular system of cells the gene based local strategies for cell behavior run in parallel and cooperate with the global cene based strategies of development. Each cell may be in a different local and developmental state.  Things can get quite complex and software modeling and simulation is essential for understanding complex multicellular developing systems such as embryo genesis and cancer.  

Both the gene based and the cene based strategies are forms of control. Hence, the dividing line between cenes and genes can be fuzzy. But overall developmental networks (cenes) separate out from local gene based networks. 

\subsection{The addressing architecture and control}
There is a further complication in that catchers and pitchers such as pots may be implemented by more than one addressing architecture. Thus we may have an addressing architecture based on additive but unordered protein transcription factors and we may have more a more precise ordered addressing architecture based on RNA, DNA or RNA in held by proteins such as Argonaut.  Cenes may be implemented by either or any  of these addressing systems.  The local control within IES and interactions between the IES and the cenome may involve both protein and RNA addressing systems (Werner~\cite{Werner2011}). 

\section{Network classification of cancers} %, diagnosis and treatment}
%was {Classifying Cancers, Diagnosis and Treatment}
Using the properties of the above networks types, we can classify cancers according to their phenotype as predicted by their cancer network architecture or by the network itself. A phenotypic classification can then be used in reverse for diagnosis where the cancer network is inferred from the tumor phenotype.  Furthermore, using the network-phenotype mapping, treatment strategies can then be based on a better knowledge of the underlying cancer network. 

We have attempted to provide a theory of all possible cancers giving a kind of periodic table of cancers.  What is striking is the variability of possible cancer networks.  We have a multi-dimensional space of cancer networks.  At the core are the developmental subnetworks that control cell division. These can be mixed and matched with reactivity to context, cell signaling, stochasticity, developmental position (via network linking within the global developmental network that controls the temporal and spatial ontogeny of the organism), and the physics and chemistry of cells. 

The dimensions of the multidimensional space of cancer networks include: Cytogenic capacity or proliferative potential (linear, exponential, geometric), reactive/nonreactive, cell communication, stochasticity, bilaterality, early-mid-late developmental stages depending on where the network is located (including degree of differentiation, complexity of network-MCO, ranging from undifferentiated, totipotent, pluripotent, differentiated, terminal, teratoma, dermoid cyst, fetus in fetu, invasiveness (metastasis, metastatic potential, interactivity with normal tissue, e.g., teratocarcinoma, invasive metastatic teratoma), metabolic control networks, the cell cycle network, cell physics and chemistry.

\subsection{A higher dimensional table of cancer types}

The following table is a summary of all possible dimensions of cancer types and stages.   A cancer can have any or all of the properties of the dimensions listed in the vertical column of the table.  

Together these dimensions of cancer give rise to a large higher-dimensional space, a combinatorial space or manifold of possible cancer network types and their possible implementations.  Each possible cancer has its own phenotype reflecting the cooperative interaction of its dimensions in its formation. The ontogenic phase of a cancer refers to where in the normal developmental network the cancer is situated.  It distinguishes, for example, embryonic, childhood, adolescent, adult and late life cancers.  The proliferation rate depends on the extent of the cancer loops, the degree or complexity of differentiation of the cell types, and the morphological complexity of the developing multicellular structure (e.g., teratomas, fetus in fetu). 

The space of possible cancers is roughly divided into three sections:  The network topology section describes the range of possible topologies of cancer networks.  The network entropy section combines features that make the developmental path the network takes more or less uncertain.  Stochasticity associates probabilities with network links making activation of downstream networks uncertain.  The cell signaling subsection is makes the outcome of a network dependent on what signals the cell sends and receives. Hence, depending on the multicellular context, the path taken in the network can vary. At the same time cell signaling is used to coordinate the actions of cells, making the outcome more robust. The conditional subsection is to distinguish cell signal dependent conditionality and entropy, from external environmental conditions such as temperature that can effect what path is taken in a developmental network.

%???Sum Up of Nets

\begin{table}[H]
\begin{center}
\begin{footnotesize}%0.38\textwidth
\begin{tabular}{|p{0.2\textwidth}|p{0.2\textwidth}|p{0.6\textwidth}|}
\hline
\multicolumn{3}{|c|}{\begin{large}\bf Space of Possible Cancer Networks and their Phenotypic Properties\end{large} \STRUT } \\[1.4ex]
\hline \STRUT
 & \textbf{Dimensions} & \textbf{Effect on Phenotype} \\[1.4ex] \hline 
 \multirow{3}{*}{\textbf{Network Topology}}\STRUT 
 &  \textbf{Linear L}     & Slow growth -one founder cancer stem cell, the rest terminal cells\\ \cline{2-3}\STRUT
  & \textbf{Geometric}  & Dynamics depend on number of loops -meta-stem cells with varying potential \\  \cline{2-3} \STRUT
    & \textbf{Exponential}  & Very fast growth -all cancer cells have equal proliferative potential\\   \hline \STRUT

\multirow{5}{*}{\textbf{Network Entropy}}\STRUT      
& \textbf{Determistic}  & Development is not stochastic with no network entropy and, hence, no developmental entropy resulting from the network topology \\  \cline{2-3} \STRUT
& \textbf{Stochastic}  & Development is probabilistic with developmental entropy \\  \cline{2-3} \STRUT
& \textbf{Signal Autonomous}  & Development is independent of cell signaling with no uncertainty or diversity generated by signals \\ \cline{2-3} \STRUT
& \textbf{Signal Dependent}  & Development is dependent on cell signaling -alternate developmental paths possible \\  \cline{2-3} \STRUT
 &  \textbf{Conditional} & Development is conditional on some property external or internal to the cell.   \\ \hline

\multirow{3}{*}{\textbf{Inter-Network Links}}\STRUT 
& \textbf{Net Locality} \STRUT 	& Locations where the cancer net links into global net  -  Cell differentiation states, complexity and maturity of structure depends on network location via links to and from the global network \\  \cline{2-3} 
&\textbf{Net Extent} \STRUT 	& The length of the range a network covers with its cancer loops  -  Number of cell types and complexity cell  of structure depends on network range and location in the global network \\  \cline{2-3} 
&\textbf{Vacuous Nets} \STRUT  & Unconnected networks not linked to the active global network controlling development - The network phenotype is hidden showing having no effect on development.  But, has the potential to be activated by cis or trans mutations\\ \hline

\multirow{2}{*}{\textbf{Net Complexity}}\STRUT 
&\textbf{Topological} \STRUT 	& Degree of complexity of net topology   -  The complexity of the network determines the possible complexity of the multicellular dynamics and phenotype \\  \cline{2-3} 
& \textbf{Node Complexity}  \STRUT  & Node complexity affects states - The complexity of the nodes has effects on multicellular phenotype orthogonal to network topology   \\  \hline \hline \hline

\textbf{Physics}  \STRUT &  {\bf Cell physical properties orthogonal to cenes} & The connective physical properties of cells and their matrix  - Physics plays an essential role on the phenotypic effect, the space-time development, generated by a developmental network  \\ \hline 	

\textbf{Cell Properties / IES}\STRUT &  {\bf Cell strategies orthogonal to cenes} & Vascularization strategies, invasiveness, acidity, connectedness, flexibility, strength, IES = interpretive-executive system \\ \hline

\end{tabular}
\end{footnotesize}
\end{center}
\caption{Dimensions of the space of all possible cancers}
\label{CancerSpace}
\end{table}

The Inter-Network Links section places a network within the cenome (the global developmental control network of the genome).  If the cancer network is located early in the cenome and its extent is short we get an immature cell mass that may proliferate more quickly than one that occurs later where more complex cell types and structures are developed. Furthermore, the extent of the network determines how many steps are involved in a cancer loop path. The longer the path and the more complex the cell phenotype, the slower the execution of the cancer network.  Vacuous networks are included for the sake of theoretical completeness. Though there may exist such, perhaps archaic networks that may be activated under stressful conditions. 

The Physics section is yet another semi-orthogonal category.  Cell physics makes a substantial contribution to real as opposed to ideal development.  There is an interaction between the activated genes used to structure the cell and the cancer network that in part coordinates the activation of those structural and functional genes. 

The Cell Properties/IES section is another semi-orthogonal dimension: The IES, the interpretive-executive system, that interprets and executes the information in the developmental control network is somewhat autonomous from the network itself.  The very same network can give very different results depending on the interpretation and execution. 

Included in cell properties are its strategies that are somewhat autonomous from its developmental control networks, such as vascularization, invasiveness and other gene based properties of the cell.  Recall, we distinguish the cell differentiation genetic pathways that don't result in multicellular proliferation from from developmental control networks that do result in multicellular genesis. 

\subsection{Request for comments}
Representing this multi-dimensional space of possible cancers graphically is not easy. There are many ways to do it. Each has its advantages and disadvantages. Any suggestions from you, the reader, are welcome.  

One possible and provisional way of making these dimensions more formal is the following notation. 

{\bf Notation:} X$^{n}$ = Exponential hyper cancer of degree $n$, G$^{n}$ = Geometric cancer of order $n$, L = Linear cancer, E = Early, M = Mid, L = Late, F = Fast, ME = Medium, S = Slow, U = Undifferentiated, SD = Semi-differentiated, TT = Terminal, MSC = Meta Stem Cell, SC = Stem Cell, PR = Progenitor cell,  I = Invasive, V = Vascularization, DD = Degree of Differentiation, Y = Yes, N = No.  

\noindent {\bf Dimensional ranges:}  Conditional = Signaling = Stochastic = Bilateral = Vacuous = [Yes, No], Teratoma = [Multiple cell types, Dermoid cyst, fetus in fetu], Ontogenic phase = [E, M, L], Proliferation rate = [S, ME, F], 	Degree of Differentiation = [Immature, Mid, Late, Terminal], Cell types = [U, SD, TT	, MSC, SC, PR], Stage = [I, II, III, IV]. 

\section{Network theory and the underlying cause of cancer}
%was {What causes cancer?}
Many networks can lead to cancer.  One can go on to construct arbitrarily many varieties of cancer networks based on this basic theory.  The key is that cancer may occur when a given genomic network is short-circuited by a backwards loop that activates an earlier site in the network.  This can then lead to the dedifferentiation of the cell to this earlier state.  Then the cell starts developing again according to the network at this earlier activation state until one of the resulting developmental paths inevitably reach the loop point and the process repeats itself, endlessly.  

\subsection{Network theory and gene mutations}

What causes cancer? Perhaps, with cancer, as with developmental biology, the focus on genes (protein coding sequences) has prevented a deeper understanding of the mechanisms underlying cancer.  Based on our theory, we can now give a more detailed answer than merely saying it is caused by a mutation of an oncogene.  In fact, such an explanation is circular since an oncogene is by definition a cancer causing gene.  On the other hand, if we say that cancer is caused by one or more mutations, it does not explain why some mutations of genes cause cancer and most mutations do not cause cancer.  We suspect that for many, if not most, cancers there are no genetic mutations that cause it.  Even for those mutations that are caused directly by genetic mutations, our theory provides an explanation not only why a mutation can cause cancer but precisely under what conditions and how it causes the cancer.   

\subsection{How mutations lead to cancer}

Mutations that lead to cancer transform the developmental control network of a cell.  Developmental networks are based on an addressing system that allows precise control of activation of control states in the cell.  The transformation are the result of a change in the addresses such that looping occurs in the control network.  When these loops are nonfunctional and lead to endless cell proliferation, cancer is the result.  We saw that adding further loops to a linear cancer network can convert it to a geometric or exponential cancer network.  Hence, more mutations can lead to more severe cancer phenotypes. 

\subsection{Viruses and cancer}

Because our theory of developmental evolution postulates that viruses may have be one factor in the build up of the genome and that these repeats were subsequently appropriated as basic control areas linked together into networks, it becomes evident that viruses could change the addressing architecture of genomes and in particular generating specific control networks that are possibly cancerous.  

\subsection{Aneuploidy and network theory}
%was {Aneuploidy and Cancer}
Some argue that too many or too few chromosomes in the nucleus (aneuploidy) is a primary cause of cancer.  It is argued that the probability of cancerous mutations is orders of magnitude too small to explain the high incidence of cancer in human and animal populations.  Instead aneuploidy (chromosomal aberrations in number and structure) is a much more probable phenomenon.  Furthermore, aneuploidy, being associated with a larger than normal cell nucleus, is a common phenotype of aggressive cancers.  The cell then becomes unstable and the normal cell cycle goes awry.

One problem with this hypothesis is that even if true, it provides no explanation as to the regulatory mechanisms that control and generate the cancerous tissue. Instability by itself does not explain the distinct phenotypes seen in different types of cancer.   The regulatory control system still needs to be understood. 
 
Some exponential cancers are not associated with aneuploidy.  Other exponential cancers do satisfy the hypothesis.  For example, in mole pregnancies (hydatidiform mole) which is, in the form of choriocarcinoma, an exponentially proliferating cancer, the female haploid genome is missing in the fertilized egg. 

We have two seemingly contradictory takes on the aneuploidy hypothesis:

First, aneuploidy may the be result of cancer and not its cause. If the cancerous control system and its associated network issues commands too quickly for the cell to execute then chromosomes may be duplicated without the cell physically dividing.  The result is an excess of chromosomes or aneuploidy.  In other words, aneuploidy could result if the control system initiates cell division when the cell is not large enough (does not have the parts or resources) to divide.  When the command to divide is given, chromosomes are duplicated, yet no physical cell division occurs because the cell is not ready.  The control system assumes physical cell division has occurred and initiates post-divisional activation of the genome.  However, since the chromosomes are now partially or totally duplicated, we may get dual activation of both daughter chromosome sets. This may lead to differentiation ambiguities with corresponding phenotypic mixtures.  As a result, we may get mixed phenotypes in the cell that normally would be distributed between the potential but nonexistent daughter cells that were supposed to have been generated by the non-occurring division.  

Thus timing problems can make a mess of the cellular structure and also the resulting control system.  However, there is an alternative explanation for the generation of aneuploidy in a cell.  It can also occur because a mutation (sequence, deletion, or shift) causes the cell division command sequence to be skipped, jumped over or not executed by the control system. The result of such a mutation is the same as with the timing problem above.  We again get duplication of chromosomes without the normally subsequent physical cellular division. 

Whether aneuploidy by itself leads to cancer is independent of the above two possible explanations of the generation of aneuploidy.  The first explanation stated that cancer leads to aneuploidy when the control network is too fast for the cell to keep up with its directives. Then it may partially or fully achieve the first phase of the cycle in copying the chromosomes , however it may not achieve the second phase where it actually duplicates and divides into two daughter cells.  This however, need not cause any additional cancer if no additional loops in a cell proliferating are induced or activated.  We just get more chromosomes not necessarily more cells. In the case of plants chromosome duplications do not usually lead to cancer. 

The second cause of aneuploidy is when a mutation functionally skips the step needed to initiate and complete physical cell division.  Since it executed the step to duplicate the chromosome, the result is aneuploidy.  However, again, by itself this need not lead to cancer. Without further regulatory transformations that result in a regulatory and proliferating loop or activation thereof, such mutations are not the cause of cancer. 

\subsection{Telomeres, aging and network theory}
%was {telomeres and cancer}
%\subsubsection{Cancer}
Cell division has been associated with reduction in the length of telomeres (the ends of eukaryotic chromosomes).  Once the telomeres of the chromosomes are exhausted, the cell stops dividing or cell death occurs.  In some cancers telomeres are not reduced and it is postulated that this leads to endless cellular division.  This fits perfectly into our theory of cancer control.  Telomeres molecularly implement what is known as conditional loop.  The conditional loop is executed as long as some property (exit-property) holds.  In this case, the cell is able to divide as long as the telomeres are of sufficient length.  

In many programming languages there is a directive that is called a  for-loop, the process is executed a set number of times  and then stops (e.g., {\bf for} value $1$ {\bf to} $n = 10$ do something). In the normal case, telomeres give the cell the capacity to exhibit the behavior of a for-loop.  However, if some mechanism prevents the exit-property from ever being true then the loop goes on forever (e.g., in effect making $n$ infinite in our above for-loop example).  

In the case of telomeres, the {\em counting-property} that permits the cell to exit the proliferating loop after a certain number of loops is the finite number of telomeres, plus the {\em reduction condition} that at each cell cycle the length of the telomeres is reduced by one and the {\em exit or stopping condition} for cell division when the length goes below some value.  Since, in some cancer cells the telomeres are not reduced or added again after division, the exit property is never fulfilled because the reduction condition is never realized. Hence we have an endless proliferating loop which results in cancer. 

Note, however that while telomeres may be one form of control over cell division by themselves they do not necessarily cause cancer. If a genome in a cell does not contain and does not participate in a cancer network via cell communication, then even if the telomeres are intact and of sufficient length or numbers, then that  cell may not be cancerous.  Thus the existence of telomeres is not a sufficient condition to generate cancer.  At most their existence may be a necessary condition for the execution of a cancer network.  Therefore, for a cell to be cancerous it needs not only to preserve its telomeres, but most importantly it needs a cancer network that is active and that the cell executes. 
 
\subsubsection{Aging, Cancer Networks and Regenerative Medicine}

Related to this is the problem of aging. If we want to prevent aging we must prevent cells death.  Hence, we need to regenerate cells endlessly.  However, that may lead to endless loops which may be cancerous. To prevent aging we have to create networks that are regenerative without being cancerous.  Therefore, regenerative medicine presupposes a deep understanding of cancer networks and their relation to stem cell networks. 

\section{Curing cancer}
The cure for a particular cancer then lies in either somehow blocking the causative loops, stopping the cell cycle, promoting apotosis (cell death) of the cells controlled by the cancer network, reversing the mutation by an \emph{inverse network transformation} that converts the cancer network into its original natural noncancerous network state.  

Different theories of cancer offer different methods for the treatment and cure of cancer.  Previous attempts to explain cancer have failed to understand the inherent network properties of cancer.  As a result, possible treatment methodologies based on such theories are practically blind.  Therefore, such theories can offer only treatment methods that have inherent, often debilitating, side effects.  They don't eliminate the root cause and instead either kill too many noncancerous cells or not enough cancer cells or both.  In particular, drugs based on gene centered theories of cancer can lead to both overkill and under kill because the gene that supposedly causes or prevents a cancer has functions in normal tissue.  Therefore, activating or deactivating that gene can have significant consequences for normal multicellular developmental processes (e.g., tissue damage, developmental defects, other cancers).  Surgery and radiation are brute force and cannot treat metastases or otherwise inaccessible tumors.  Chemical agents that attack the cell cycle attack all cells that use the cell cycle (e.g., hair loss, intestinal damage, possible new cancers).  

\begin{enumerate}
\item Blocking any point in the cancer loop 
\item In the case of cell signaling cancer networks, blocking cell signaling either by blocking receptors, blocking signal generation, interfering with the signal transduction pathway that leads to activation of the cancer subnetwork
\item Stopping the cell cycle
\item Using genetic properties or markers of cancer cells to kill those cells that express those properties
\item Forcing apotosis (cell death) of the cells controlled by the cancer network
\item Reversing the cancerous network mutations by a series of \emph{inverse network transformations} that convert the cancer network into the original natural noncancerous network
\end{enumerate}

\section{Comparison of  Traditional Theories with our Control Network Theory} 
\subsection{The cell cycle and cancer}

The standard explanation of cancer is that there is uncontrolled growth with a failure of apoptosis (cell death).  Here uncontrolled growth means  that control of the cell cycle is somehow evaded. Prior to division, cells go through several stages with checkpoints before each stage. This is known as the cell cycle.   Cell signaling is assumed to somehow control progression through the cell cycle by way of influencing these checkpoints (Tyson~\cite{Tyson1991,Tyson2002, Tyson2011}).   For example, a signal to progress to cell division may be constantly switched on, leading to endless cell division.  Or, a tumor suppressor gene may be may be deactivated by a signal or mutated leading to uncontrolled growth.  

The problem with these accounts is that they are at best partial explanations of the initiation or control of some particular instances of cancer.   They do not explain the dynamics of cancer development and phenotypes.   They do not explain different growth rates in cancer nor the fact that cancers can generate multiple cell types and go through diverse stages of differentiation.  For example, it does not explain how teratomas differ from mole pregnancies or other differences in the ultrastructure of cancer temporal and spatial phenotypes. 

\subsubsection{Higher level developmental networks control the cell cycle}
The fundamental problem is that the cell cycle view of cancer fails to see that cancer is a developmental process similar to embryonic development. In particular, proliferative cancer networks with their loops are not the same as the cell cycle. Instead, the cell cycle is controlled by developmental control networks.  In other words the cell cycle molecular system is a lower level, cellular control loop that is a sub-network meta-controlled by higher level, developmental epigenetic and genomic control networks.  These higher level developmental networks control the cell cycle by activating or inhibiting the cell cycle at various checkpoints.  

\subsection{Problems with the traditional gene-centered theory of cancer}

Recall our comparison of the gene-centered theory of cancer with our network control theory of cancer in \autoref{CompareGN}.

First, there is no systematic causal relationship between phenotype and genetic mutations. The reason is because the network controlling the activation of genes is left out of the picture. 

Second, a deeper problem is that the theoretical  and explanatory level is too low, omitting the essential factor that causes cancer. For example, a broken transistor in may cause the computer to not work properly. However, this does not explain why it malfunctions the way it does.  Being told a transistor is broken does not in itself have predictive power; we cannot predict what the future the behavior of the computer will be.  A predictive explanation requires that we understand the role of the transistor in the overall architecture of the computer, its control network.   Similarly, knowing a gene or non-coding area DNA is mutated will have limited predictive and explanatory value, unless we understand the role of that gene or area in the overall control architecture of the cell. Phenotypic effect depends on the role of the broken unit in the overall network.   

\subsubsection{Knudson's two hit model}
The idea of Knudson \cite{Knudson1971} is that there may be a predisposition to cancer if one of two allelic genes is non-functional because of a preexisting mutation at birth. This thereby makes the cancer more likely because only one more mutation in the other complementary allele/gene is required to make the gene nonfunctional.  If the gene is a tumor suppressor genes then a cancer may develop if both alleles of the gene are knocked out by mutations. 

We can see there are several assumptions in Knudson's theory of cancer.  

\begin{enumerate}

\item It assumes that genes, and not networks, are the root cause of cancer.  

\item These genes cause cancer if they are not functional because of mutations.  

\item Hence, cancer is the result of the lack of suppression of a process of cell proliferation.  

\item Consequently, all cancers are caused by tumor suppressor genes.  

\end{enumerate}

Knudson's model might be used to try to explain the difference in growth rates of cancers.  If we link the degree of suppression with the number of genes suppressing the cancer, then we might link the growth rate with the number of suppressed genes. If only one gene is suppressing  growth, the rate of growth may still be inhibited. If both alleles of a tumor suppressor genes are disabled then the growth rate might be twice as fast than if only one gene is disabled. 
Hence, if only one tumor suppressor gene is mutated we might get only partial suppression and hence a slow growing tumor. If both genes are suppressed we can have a faster growing tumor. 

However, this would not explain why we get exponential growth, since the suppression of the second gene would, presumably, without other assumptions in the theory, only double the rate of a linear cancer making it twice as fast but not exponential. Thus, Knudson's theory cannot explain why some cancers exhibit linear growth and others are exponential, let alone geometric.  All cancers would be either linear, linear by a factor of two, with no explanation of why some cancers are inherently exponential or geometric.  

This failure of explanatory power is a result of the inherent limitations of atomistic nature of Knudson's model.  It cannot not have the explanatory capacity of the network model we are proposing because there is not enough ontological structure in Knudson's theory.  

Cancer, on Knudson's model,  is seen not as an active developmental process but rather as a result of an insufficiency, a lack of a suppressing component.  As a consequence, Knudson's model cannot account for the differences in phenotype of cancers.  His theory can only be used to explain the existence or nonexistence of a cancer that is suppressed by genes, but it cannot explain the developmental dynamics and phenomenological nature of that cancer. 

\subsubsection{Repressor genes and cancer}
The network model we are proposing explains the underlying reasons why Knudson's two hit model works for some cancers. If a gene is suppressing a cancer network that would otherwise be activated then one suppressor gene is sufficient to prevent the cancer from growing.  If both alleles of a suppressor gene are disabled by mutation then the network is activated and generates the cancer. Different suppressor genes suppress different networks and hence we have different genes associated with different cancers. Yet, note, that the actual formation and dynamics of the particular cancer is the result of the network that is suppressed and not the suppressor gene itself. 

\subsection{How does our theory differ from the classical account of cancer?}

\begin{enumerate}

\item
\textbf{Dynamic and causal properties}
We are able to describe the actual dynamic properties of cancers and how their networks influence these dynamics.  The classical account only describes the etiology of cancer, namely that cancer is caused by mutations in oncogenes.  How oncogenes actually cause cancer is often left unsaid with some exceptions. And those exceptions involve loops which confirm our theory.  In fact, the accounts are consistent with each other.  

\item
\textbf{Targets -safer and unsafe}
A safe target is a cancer target that does not have major developmental or other side effects.  We see that the targets of cancers are points in a network.  However, we have to be careful, since preventing a cancer may also  interfere with a functional network.  Our theory gives an account as to which points in a network may be less subject to side effects if they are interfered with by some therapy. For example, if a cancer occurs in childhood then it may involve networks that are responsible for the further development of the child. In such a case, the target to block the cancer has to be chosen carefully, limited to those that have the least negative developmental consequences. 

\item
\textbf{Typology of cancer}
We have an entirely new view of cancers and their typology.  We classify cancers by their control network architectures and not by individual genes or phenotypes.  However, we also see that phenotype can help us in determining the kind of cancer network that may underlie the phenomenologically and molecular properties of the tumor.
 
%was \textbf{A developmental disease}
\item
\textbf{Cancer is a form of multicellular development}
Cancer is just a type of multicellular process.  The looping conditions cause a normal developmental process to potentially result in pathology. 
We view multicellular development as a multiagent process where cells are active complex agents that have the capacity to perform various actions such as cell division, cell communication and cooperation. A multicellular organism is a vast distributed multiagent system where many of the agents are in different control states.  Control networks are executable by the cell. As we have seen, when control networks have a certain architecture they can lead to cancer.  We saw in the case of stem cells that the dividing line between cancers and normal development may depend on additional differentiation control networks that determine strategic properties such as motility and invasiveness of the cell.  A stem cell network that activates an non-normal cytogenic or differentiation network is usually viewed as being cancerous. 

\item
\textbf{General theory with broad applicability}
We have said nothing of the biochemical details of the loops.  We have left open how the control network architecture is implemented in the cell and its genome.  It may be purely genetic (protein based), it may be a combination of protein and RNA, or it may be pure RNA control with protein helper agents.  That is something for experimentalists to resolve.  However, the overall cellular-genomic control architecture of development and its pathological sibling, cancer, is at a level of abstraction that permits any of these scenarios.  In spite of its generality and abstract conceptual basis, it has predictive and prescriptive power.  It is a paradigm shift that will help lead future research efforts to find the actual basis or biochemical implementation for the network control architecture and topologies responsible for different types of cancers.
  
\item
\textbf{Abstraction has broader explanatory and predictive power}
Thus, we are operating at a higher level of ontology and information that, while dependent on the lower levels for its realization, exhibits regularities and laws that are difficult to deduce and predict using only the ontological and conceptual machinery available the lower levels of the ontological and informational hierarchies we see in living systems.  And, this is at the basis of the utility of investigating such levels of network control.  They help us see patterns and crucial points of control in the cellular system that will help us in the discovery of drug targets as well as the avoidance of the dangerous repercussions of ignorance of such developmental networks.
 
\item
\textbf{A new research paradigm}
As a research paradigm it shows that we will have to pay much closer attention to the control architecture of the cell and the control information in the genome.  Genome semantics is one way of describing the meaning of a genome in the context of the cell as agent that interprets that control information and executes it in the process of development.  If it turns out, as we suspect from other theoretical considerations, that RNA plays a crucial role in combinatorial selection of genomic commands to the cell, then our entire concept of the cell will change.  Genes will be subservient helper agents in executing the commands in the form of non-coding regions in DNA and/or RNA generated from those non-coding regions.  

\end{enumerate}

\subsection{The explanatory power of our theory and diagnostics}

The types of possible cancer on our view can vary as greatly as the complexity of the genome permits.  Given an genomic network one could in principle calculate the combinatorial possibilities or mutations or transformations which would result in cancerous loops.  

\subsection{Cancer growth rates}
The network theory of cancer explains why some cancers are slow growing and others are very fast.  

It distinguishes between linearly, geometric,  and exponentially growing cancers.  Normally, an exponential cancer will soon outpace a geometric or linear cancer.  A geometric cancer will grow more quickly than a linear one.  Therefore, the growth rate may serve as a diagnostic criterion to distinguish linear, geometric and exponential cancers.  However, there may be considerable variation in growth rates within these broad categories for several reasons: 

\begin{enumerate}

\item 
\textbf{Complexity of differentiation} Prior to each cell division, the more complex the cell, the more time each cell cycle takes.  And, therefore, the slower the rate of growth of the cancer be it linear, geometric, or exponential.  In other words, before division can take place, the cellular differentiation pathway may take more time if a cell differentiates from one cell type to another such as from a simple to a more complex cell type.  Furthermore, more complex cells require more steps and energy to grow and divide. 

\item \textbf{Network complexity} The rate of growth of a cancer also depends on the length and complexity of the cancer pathway from the initiation point to the point that the network loops back on itself.   The longer the cancer pathway and the more complex the actions initiated by that pathway, the longer it takes for the cancer cells to loop back to their starting point.   
For nonlinear cancers there will be more than one cancer pathway. If for example, two loops in an exponential cancer each generate a complex structure before they loop back to their initial starting state, then that cancer may execute very slowly.  However, the cancer cells will still double at the completion of both loops. Hence, the growth, though slow in each differentiation step, will be exponential over the loop count. Eventually, it will outpace the fastest growing linear and geometric cancers. 

\item \textbf{Physics and contextual constraints} Cell physics and the multicellular context of the cancer will significantly influence its real versus its ideal growth rate.  

\item \textbf{Cell signaling constraints} For social cancers depending on cell signaling, the growth rate depends on the availability of  communication system in the social context of the cell in the multicellular system.  If the appropriate communication is not there then either the cancer fails to grow if activation of the network requires signaling or the cancer begins to grow if signaling is required to repress the cancer network and thereby preventing its growth.  We saw an example  (\autoref{fig:SigX2LinGrow}) of a doubly exponential cancer network that only grows linearly because of limited available cell signaling between potentially exponential cancer cells.  

\end{enumerate}

Summing up, whether cancers are linear, geometric or exponential, their growth rates will vary.  Our theory predicts that cancers with slow growth rates will normally exhibit complex differentiation states, multiple cell types, and even distinct repeating morphologies whether they be ultra-structure or larger tissue or organ like complexes.  On the other hand, less differentiated cancers that do not differentiate to more complex mature cell types, will grow faster because the network to develop those cells will be shorter and hence the loop will execute more quickly. However, the growth rate of a long cancer loop, in terms of production of cells over time, may be same as a short cancer loop if the differentiation steps in both pathways are similar in complexity.  In general, though, cancer networks with short control loops and simple differentiation steps will be the fastest growing cancers. However, an exponential cancer will eventually outpace geometric and linear cancers if the patient lives long enough. 

\section{What cancer is not}
Our theory differs in fundamental ways from previous theories and models of cancer.  However, it should be said at the outset that despite their limitations, previous models of cancer have intrinsic value.  For reviews and views of major cancer models see \cite{Byrne2009, Anderson2008, Anderson1998, Bizzarri2008, Rejniak2011, Saetzler2011}.  What we are trying to do is to is construct a general unified theory of cancer. The particular models of cancer could then be instantiations, such as special cases of cancer, or be integrated with the general theory by giving more detailed accounts of molecular implementation, mathematical analysis, and physical realization and performance. 

\subsection{Physics-based models: Cancer is not just physics}%: Cancer is not just about the physics of cells
A great deal of insight can be gained by viewing cells as purely physical entities subject to the laws of physics.  Cell behavior is modeled by a set of physical laws. The properties of the cell can then be adjusted by changing the parameters of those laws and/or by changing the boundary conditions under which the cell actions takes place.  Parameters of the cell model can include rate of cell division, cell spatial dimensions, various cell structural parameters such as spring constants that may be used to model cell connections. More and more sophisticated cell models have been developed including models that can give the cell a wide variety of shapes and flexibility (\cite{Rejniak2011,Quaranta2008}).  It is then hypothesized that cell physical parameters correspond to genes or sets of genes.  Changing a physical parameter is then equated with mutations in a gene or set of genes.  A simulated multicellular developing tumor is generated and compared with a natural tumor.  If there is a match then inferences are drawn that the model is confirmed and that particular parameters that generate particular phenotypes in tumors are likely related to mutations in certain genes.  Such a methodology, of relating model parameters to genes and then seeing transforms of model parameters as mutations of related genes, was applied successfully to the study of bacteria \cite{Palsson2006}.  

Applying this methodology to multicellular development has intrinsic challenges.   The physics of a skyscraper is different than the architecture of that skyscraper. The physics does not determine the architecture, nor does the architecture determine the physical laws.  Trying to model multicellular systems with just physics and with no developmental control networks is like trying to build a skyscraper with no architectural plan.  The difference is that developmental control networks are not \emph{descriptions} of an organism or system of cells. Rather, developmental control networks are executable networks or strategies that when interpreted by the cell lead to cellular actions. The combination of all those strategic networks in different states in a vast multicellular system leads to development in the context of physics.  While physical conditions enforce constraints, strategic developmental control networks distributed across the cells in a multicellular system, cooperate to generate ever more complex multicellular systems and ultimately form an organisml. Putting physics in the lead puts the cart before the horse. 

\subsubsection{Physics-based models fail to explain the ontogeny of morphology} 
Many present models of tumor growth focus on cell physics ignoring the underlying developmental control networks.  While the physical interaction between cells and the cell strategies that cells use to respond to and control their interactions with other cells are an essential component in understanding cancer and development of organisms in general, cell physics misses the root cause of cancer.  Some attempt to explain pattern formation in development and cancer purely as the result of the physical interaction between cells \cite{Saetzler2011, Sonnenschein1999} (see below).  Such theories fail to explain the ontogeny of morphology and fine structure of dynamically developing multicellular organisms.  Since cancer is a developmental process such theories fail to explain the ontogeny of structure and function in cancers as well.  

\subsubsection{Physics-based models fail to explain complexity in development}
There are fundamental problems with all such approaches when they are applied to cancer or embryonic development in general.  One problem is the complexity of multicellular life and the information required to generate complex living organisms from a single cell.  Purely local strategies of interaction between cells in a developing multicellular system (whether implemented by simple rules in discrete space-time or by a set of differential equations in continuous space-time), do not contain sufficient control information to generate complex organisms \cite{Werner1996}.  It is one reason why reaction diffusion models of the kind proposed by Turing \cite{Turing1990} fail to be adequate models of development. 

\subsection{Rate-based models: Cancer is not just about rates of cell division}  
Physics-based models are usually conjoined with rate-based models of cancer.  Limitations of rate-based models lie in the  assumptions they make about the nature of cell proliferation.  Almost all previous mathematical and computational models of cancer assume that cancer is caused by differences in the rates of cell proliferation due to uncontrolled cell growth.  While partly true, this is not the cause of cancer, but an effect of the nature of the developmental networks controlling the cancerous cells. Hence, previous models are limited to descriptions of rates of cell proliferation over time using either differential equations or other mathematical formalisms.  Most such models limit the description of cancer cell proliferation to adjusting rate of division parameters (also called cell cycling rates).  These rates of cell proliferation are then either signal induced or by some other deterministic or stochastic condition.  Cell division either terminates in a differentiated cell or cells proliferate by doubling, growing exponentially at a certain rate.   For example, the rate may be a function of another parameter that represents the strength of a diffused signal where the greater signal strength the faster cells proliferate.

These models have some similarity with Till's original stochastic model (\autoref{fig:TillSSC}) \cite{Till1964}. By balancing rates of cell proliferation with rates of terminal cell differentiation or death, have successfully described the growth of spheroid tumors where cells at the center of the tumor die due to lack of oxygen.  

However, like Till's model, these models implicitly presuppose a particular type of developmental control network.  Recall that the Till model has inherent limitations based on the underlying developmental cancer network. It cannot simulate most of the other cancer networks.  So too, rate-based models fail to give an account of all the other types of cancer that do not fit into this scheme.  At best they can attempt to describe other cancers by adjusting the values of the cell cycling rate and or the probability distribution.  As we saw stochastic networks can be very flexible in the kinds of developmental network topologies that they imitate.  Unless we consider the actual architecture of the underlying network, it is questionable how much understanding we get from adjusting probabilities or rate constants. 

Purely rate-based, whether deterministic or stochastic, mathematical models not only fail to grasp the essential unifying features common to all cancers, but also they fail to explain why particular cancers are they way they are.  For example, rate-based theories also fail to explain the hierarchical nature of metastases or why there are teratomas or bilateral cancers.  While they may describe the rate of growth of some tumors they fail to explain how and why the tumor grows at all.   

In sum, rate-based models of cancer fail to provide an explanation of why the tumor grows differently from another tumor.  All they can do is to adjust rate constants or probabilities to get different rates of growth.  They cannot explain why some cancers grow very slowly or why mutations can transform them into suddenly growing exponentially, except by saying the rates of cell proliferation have changed. But, saying that is just circular reasoning: The rate of the tumor growth has changed because the rate of cell proliferation has changed.  

\subsection{Systemic models: Cancer is not the result of self-organization based on physical emergent properties of cells}
Soto and Sonnenschein claim that cells cannot be controlled by a program, a strict recipe or set of instructions: 
\begin{quote}$\ldots$  it is unlikely that each cell can follow and execute each of the encoded rules independently of the crowded environment present in a tissue or organ. Since cells can only sense their local environment, the emergence of tissues can only be driven by rules governed by coordinated interactions with the local environment of each cell. This leads to the conclusion that the dynamic process of tissue formation must mainly be governed by self-organization.\cite{Saetzler2011}
\end{quote}

While multicellular development certainly depends on the local social, communicative and physical context of each each cell, the conclusion that tissue formation is mainly controlled by self organization is not warranted. We have shown that cellular developmental control networks can be conditional, reactive, and interactive. Developmental control networks contain complex executable strategies. Furthermore, each cell can be in a different execution state that activates a distinct developmental subnetwork (cene) in the global developmental network (cenome).  Hence, each cell can have its own unique net-based strategy that interacts with the strategies of other cells -all in the social, communicative and physical context of neighboring cells. 

This is not to say that the physical and social context in which a cell resides is not important.  The physical and social context of other cells influences the how the cell's developmental network interacts and how the network strategy is physically interpreted and realized.  But, self-organization based on physical properties alone without the complex control information in developmental networks is not sufficient to generate particular, complex multicellular structure or morphology. Physical constraints are pleiotropic that, like genes, have a general effect leading to redundant structures.  

It is further argued by Sonnenschein and Soto  \cite{Sonnenschein1999, Saetzler2011} that as multicellular systems develop the physical system changes resulting in the emergence of new properties that then influence the development itself.  We agree. However, the mistake that is made is to conclude that such emergent properties are sufficient for the particular ontogenic development of a species.  

\subsection{Stochastic models: Stochasticity is not a sufficient explanation of cancer dynamics}
We saw that stochasticity in developmental control networks can add great flexibility to the network.  The probability distribution determines many of the proliferative properties of the network. Thus, a general stochastic network like that in \autoref{fig:XLSSC} can emulate many deterministic developmental networks that are topologically distinct.  Moreover, we saw (e.g., \autoref{fig:DLSSC} and \autoref{fig:DXSSC}) that the rate of cancer growth of various cancers can be adjusted by varying probabilities of stochastic delay loops.   

That being said, all stochastic models make implicit assumptions about underlying developmental networks.  Any probability space is based on a foundational possibility space. The topology of the developmental network associated with stochastic probabilities provides this space of possible dynamic paths that a cell can take.  Therefore, stochasticity is not a sufficient condition for cancer development. Neither is it a necessary condition since cancerous developmental stem cell networks may be deterministic.  

\subsubsection{Stochasticity, competence and performance in developmental networks} 
Interestingly, a cancer network can be deterministic, but the performance of the network in real continuous physics may be stochastic.  This is another reason to distinguish between network competence and network performance.  The network acts in the context of a cell with its interpretive-executive system (IES), in the social context of other cells, and in the context of cell and environmental physics.   Each of theses context may by itself be stochastic without the network being stochastic.  Hence, we may have stochastic performance with a network that has a deterministic competence. 

\subsection{Gene-based models: Genes alone are not the cause of cancer}  
Previous so called network theories of cancer focus on interactions between gene products as the cause of cancer.  

\subsubsection{Gene-centered view of development and cancer is a delusion}
Low level protein networks based on genes are a different sort of network than higher level developmental control networks.  While ultimately both kinds of networks are made of molecular components, it is their role, place and function in the organizational, informational and ontological hierarchy of the cell and the multicellular system that distinguishes these networks. The interpretive-executive system (IES) of the cell that gives pragmatic meaning to developmental networks, in part consists of lower level genetically based protein and RNA networks. The point is that, with regard to the multicellular development of complex form and function, be it normal or pathological, it is the higher level control networks (cenes) which play the dominant role.  The gene-centered view of development is the source of the mistake that cancer is caused by genes \cite{Werner2011a,Werner2011b}. 

\subsubsection{Cancer genes: Cancer networks, not cancer genes, cause cancer} 
So called cancer genes are pleiotropic (having multiple functions and different effects throughout the development of an organism) and as such cannot, in general, be the cause of a particular cancer phenotype. While genes are certainly involved in cancer, the gene-based theories of cancer miss the essential nature of cancer.   Such gene based theories fail to see that developmental networks are higher level meta-control networks that control lower level molecular networks such as the cell cycle.  

\subsubsection{Cell cycle models: An apparently uncontrolled cell cycle is side-effect and not the original cause of cancer}  
The view that an uncontrolled cell cycle is the cause of cancer, fits well with the widely held assumption that cancer is uncontrolled cell proliferation.  By focussing on too low a level of organization they miss all together the overarching control networks responsible for not only the development of embryos but also cancer.  The cell cycle is controlled in cancer just as much as it is controlled in normal development.  The cell cycle does its job in cancer just as it does in normal development. It is just activated more frequently. 

\subsection{Agent-based models: Agents controlled the rules of local interaction are not sufficient to explain cancer}
Agent-based systems view cells as agents that interact with other cells \cite{Werner1988, Werner1989}.  However, the complexity of a result is directly related to the complexity of the rules of interaction \cite{Werner1996}. If the rules are purely local, their effect is pervasive leading to redundancies in structure.  This can account for many systemic properties, but agents governed by simple rules of interaction alone, cannot account for structural and morphological complexity in development \cite{Werner2009}. For that more information is necessary. That is why we need all the control information in genomes \cite{Werner2007a}.  While, in principle, we can make a multiagent phenotype as complex as we want by making the control information in the form of rules more and more complex, doing such amounts to constructing developmental control networks of the kind we have introduced in this work.  

\subsection{Hybrid models: Cancer is not explained by hybrid rate-based, genetic models}
Attempts at hybrid theories of cancer composed of rate-based theories of cancer combined with gene-based theories and cell physics still fail as theories of cancer because they fail to provide an explanation for both development of both normal and cancerous multicellular development.  The underlying problem is that multicellular development is not adequately explained by genetic and rate-based theories.  Genes are pleiotropic in their causal roles and cannot account for individual differences in development or cancer \cite{Werner2011b}.  

The basic point is: Because previous models of cancer fail to include the topology of developmental networks, they cannot offer a way forward to curing cancer.  

\subsection{Evolutionary models: Cancer is not just an evolutionary process}  
Neither the development of organisms nor the development of cancer can be explained simply as an evolutionary process. While cancer cells do evolve through a process of mutation, what gives a particular cancer its particular dynamics and morphology is the result of its controlling developmental networks and their transformations.  Our theory explains why particular mutations or transformations of developmental networks change the dynamics and morphology of the developing cancer. In addition, genetic mutations can and do change the physics of cell-cell interactions that make it possible for tumor cells to break free and form metastases, for example.  However, while evolution of the control networks in concert with genetic evolution do result in cancer, the actual behavior of tumors is explained by the resulting higher level developmental control networks together with genetic and epigenetic changes that modify cell behavior and cell physics.  Cells can evolve and adapt to the new multicellular contexts that result from cancer.  In concert with developmental control networks such cellular changes or adaptions can make a cancer more or less severe. What is essential to understand that cancers evolve on at least two levels: The level of cell interaction strategies (both physics and signaling) by changes in genetics and at the level of developmental control networks. 

\subsection{Development as complex social physical system}
The cell is a semi-autonomous agent reacting locally to its social and physical context but globally directed by control information in its developmental networks.  The cell reacts locally to its environment, both internal and external.  A cancer cell has local strategies by which it reacts to stresses such as a lack of oxygen, to pressure, to stretching, to strains, to its neighbors, and to cell signaling.  However, in developing multicellular systems, each cell's local strategies is subsumed by higher level, global control strategies that are in the form of developmental control networks Werner~\cite{Werner2007a}. 

It is similar to a man going on a long hike through the country. Locally he avoids obstacles and may even take detours if the environment is convoluted, but globally he keeps his direction.  His actual path then is the result not just of his global strategy but his local reactive strategies as well.  The combination of theses strategies together with the strategy of his environment leads to a particular path. In social systems these local and global strategies interact with other agents plus the environment making things even more complex (Werner~\cite{Werner1988}).  

Development is complex communicative, social process.  Each cell has local reactivity to impinging circumstance but globally control is the result of developmental control information that is interpreted and executed by that cell.  The more complex and structured the adult organism is the more global control information is required \cite{Werner1996}. Moreover, a multicellular system is a social system of cells that communicate by signals, pushing and pulling and touch (being aware of their neighbors). Thus a developing organism involves a complex time and space dependent interaction of local and global strategies of the cellular agents. At any given point in time and space, a cell finds itself in a possibly new social and physical context that includes other cells and environmental circumstances. It reacts both locally and globally to this context by means of its local and global control networks.  

The attempts to reduce global control to purely local control would have our man getting lost in the woods doing a random walk making the probability of his reaching his goal impossibly small. So it is with the morphogenesis of complex organisms.  Previous models of cancer have focussed on local cell strategies and made attempts to deduce global properties of tumors from local rules or equations of interaction.  This research is essential to understanding the local reactive strategies of cancer cells, but it needs to be supplemented and integrated with global developmental strategies in the form of global developmental control network theory that we have provided here.  To get to foundation of cancer and cure it we need to understand global developmental networks and how they interact with local cell strategies and cellular social and physical contexts. 

\section{Conclusion}  
We have provided a theory that unifies many different types of cancers and their phenotypes by describing the developmental control networks that underly all cancers. 

\subsection{Main Results}
\begin{enumerate}

\item  {\bf  A new paradigm for understanding cancer}.  At the core the of our theory of cancer is a control network based theory of how organisms develop from a single cell.  It it is the basis for all multicellular life.  Cancer is then a special, pathological instance of this more general theory of development. Our theory is in contrast with the dominant gene centered theories of development and cancer.  Cancer, its etiology, dynamics, and development is not explained by just the mutations of genes. 

\item {\bf Curing cancer}. Our theory shows that, in principle, we can stop any cancer.  However, to apply the theory we need new discoveries that relate the theory to its molecular implementation.   In concert, we need to develop new technologies for recognizing and then transforming cancer networks into harmless normal developmental networks.  In spite of these challenges, the theory offers significant new insights into the nature of cancer and, more generally, the development of tissue, organs and organisms. 

\item  {\bf The control networks underlying stem cells}. We generalized the stem cell networks beyond linear networks, to meta-stem cell networks,  and ever higher level geometric networks. We showed the strong relationship between normal stem cell networks and cancer stem cell networks. 

\item  {\bf New categories of stem cells and cancer stem cells}. Stem cells can be understood by way of their control networks, namely the geometric networks.  We showed that geometric networks growth dynamics obey formulas based on the classical mathematics of the coefficients of Pascal's Triangle. 

\item  {\bf Exponential and hyper-exponential cancer networks}.  These networks give us a theory of the extremely fast growing cancers. 

\item  {\bf A novel theory of teratoma tumors.}  Tumors with different tissue, cell types and even structures such as teeth and hair are understood with the control network theory.  We showed how the properties of the teratoma depends on which locations the cancer network is linked to the overall developmental control network of the organism.  This allows the us to distinguish and explain immature teratoma tumors from mature ones.  It also implies new strategies for treatment.  Our theory further explains a very rare and unique type of tumor, the \emph{fetus in fetu} tumors where a mass of tissue resembling an embryo growths inside person's body.  

\item {\bf Rare cases of cancer explained}. Our theory also explains the properties of the rare cases of mole pregnancies that result is a very fast growing cancer. 

\item  {\bf Cancer network theory explains bilaterally symmetric tumors}.  These occur in rare forms of breast cancer.  The properties of the particular tumor will depend on the network's locality and its type of cancer network topology. 

\item {\bf The theory links cancer and normal development}. More generally, our theory implies the properties of a cancer will depend on linking relationships of the cancer network to the global developmental control network of the organism.  

\item  {\bf Stochastic cancer networks}.  Stochastic control networks of development involve probabilities. We showed that stochastic networks can be used to model and approximate all the major non-stochastic, deterministic cancer control networks.  At the same time, it allowed us to model some classical cancer stem cell behavior.

\item  {\bf Cancer signaling networks}. We showed that cell-cell communicative interactions can, under certain conditions, ameliorate the growth potential of even exponential cancer networks.  Yet we also developed models of cancer networks for cancers requiring cell communication with noncancerous cells.  Such is the case in some bone cancers.  Here treatment would require stopping the communication, while in the former case stopping cell signaling might actually exacerbate the exponential cancer. 

\item    {\bf A hierarchy of metastases related to a cancer network hierarchy} We related the properties of cancer metastases to the properties of cancer network of their founding cancer stem cell.  We showed that the dynamics and phenotypes of different types of possible metastases are a direct consequence of their cancer networks.  Interestingly, certain types of metastases are less harmful than others.  This theory also indicates the best way to stop metastatic growth by eliminating or transforming the most dangerous founder cells. 

\item   {\bf New research paradigm}. Any new scientific paradigm should open up previously inconceivable questions and problems. A theory that does this is all the more significant if it shows the directions and methods of how to answer those questions and solve those problems.  The network theory of development and cancer opens up a vast new field of cancer research and gives fundamental insights into how cancer can be cured. 

\end{enumerate}  

\subsection{Open Questions and New Research Problems} 

\begin{enumerate} 

\item How are cancer networks and, more generally, developmental control networks implemented at the multicellular level, the cellular level, the genome level and the molecular level?  

\item We need to integrate our theory of cancer and developmental control networks with the extensive research done on the mathematical modeling on the local physics of cancer cells in chemical, signaling and multicellular contexts.  This will involve integrating genetics and genomics, etc. with {\em cenomics} (the study of cenes and the cenome -the global control networks in the genome). 

\item We need to link the higher level developmental control networks with lower level molecular networks such as the cell cycle and other intracellular signaling networks.   

\item Will we be able to recognize cancer cells just by looking at the control networks implicit in their genomes? 

\item If so, then it calls for new methods to manipulate and transform cancer networks into harmless and normal developmental control networks. 

\item Our cancer theory shows how, in principle, one can stop any cancer in any given cell. How can our theory be used to develop new research strategies for the development of medicines for curing cancer?

\end{enumerate}

\subsection{Beyond Cancer Networks}

Looking beyond cancer networks, our network theory of development has further utility. 

As we saw, the problem of aging and rejuvenation are intimately related to stem cell networks and cancer networks. The control network theory of cancer will help researchers determine the boundary that separates a possible fountain of youth, promised by tissue and organ regeneration, and cancer.  

Our theory points to new strategies and methods for stem cell research and regenerative medicine.  Once the network transformations necessary for tissue and organ generations are understood and made safe by avoiding cancerous network transformations, then we may indeed have a new era of rejuvenating medicine. This is done today by means of surgical operations or exogenous transplantation.  In the future it could be by direct, controlled regrowth of needed cells, tissues and organs using computationally designed, synthetically bio-engineered developmental control networks.  

Our theory of developmental networks make a shift of focus from the single cell to multicellular systems.  It implies that a new era of multicellular biotechnology will be possible.  Multicellular biotechnology will combine computational with synthetic systems biology to design and construct single cells and their genomes that when activated generate designed multicellular systems to perform various functions.   

Finally, more fundamentally, our theory of development has deep scientific consequences for the theory of evolution. It predicts the existence of an as yet hidden developmental control code in genomes, a possibly universal code of life. More this in subsequent publications. 

\section{Materials and methods}
All the stem cell and cancer networks were modeled and simulated with Cellnomica's Software Suite ( http://www.cellnomica.com ).  The normal and cancerous developmental control  networks were constructed with Cellnomica's graphical Network Model Builder.  The multicellular images of tumors were snapshots taken at temporal instances of networks being executed in parallel by cells in a dynamic, simulated multicellular system taking place in continuous space-time with continuous physics. Simulations of networks in action in developing multicellular systems used Cellnomica's Developmental Simulation System.  All the multicellular systems represented in the images in this work were grown from a single cell.  The tumors resulted from mutational transformations of previously designed, non-cancerous developmental networks, again using Cellnomica's software.  

\section{Request for comments}
{\bf Author contact for comments on this paper}: ({\color{blue} eric.werner@oarf.org } Please put the word \underline{Cancer} in the subject line)

This paper is meant to motivate researchers, doctors, clinicians, those in the pharmaceutical and health industries to view cancer from a different perspective -to break the hold of the gene-centered paradigm.  This is consistent with incorporating the tremendous knowledge gained by means of the gene-centered paradigm.  The network control theory is consistent with genetic discoveries.  We only claim that gene-centered paradigms is not sufficient to understand multicellular development and in particular cancer. Hence, we welcome all researchers to participate in a new dialogue on the nature of cancer and how we can cooperate to find a cure for cancer. 

The author welcomes comments.  Please contact the author  especially if: 

\begin{itemize}
\item You have particular examples of cancers that fit or do not seem to fit this theory
\item You have stem cell examples that seem to be higher order stem cells. 
\item You have examples of metastases that seem to fit the metastatic hierarchy hypothesis in this paper. 
\item You have a publication that is not in the references and is relevant. There are thousands of publications on cancer. Please forgive any particularly relevant omissions. 
\item You have ideas for extending the theory and/or linking it with more detailed molecular theories. 
\item You see errors of any kind in the paper. 
\item Any other helpful comments are welcome.

\end{itemize}

\nocite{Werner2003b, Werner2005, Werner2007a, Werner2009,Werner2010}
\nocite{Tyson1991, Tyson2003}
\nocite{Dick2010, Dick2009, Dick2008}
%\nocite{Simons2007a, Simons2007b}
\nocite{Shackleton2010, Shackleton2010a, Shackleton2010b, Shackleton2010c}
\nocite{Gardner1999, Gardner2001, Gardner2006, Gardner2007, Tesar2007}
\nocite{Sonnenschein1999}
\nocite{Bizzarri2008}
%\nocite{*}
 
 For reasons of space only a limited number of references are given.  Let me know if your article is especially relevant and should be included. 

\addcontentsline{toc}{section}{References}% for showing references in table of contents 
\begin{multicols}{2}

\footnotesize %\small %\footnotesize %\tiny
\bibliographystyle{abbrv}%nature}%plain}%for number e.g., [1] citations
\bibliography{BibCancerPaperArXiv} %Refs2,CancerModelingByrne}

\begin{thebibliography}{10}

\bibitem{Anderson1998}
A.~R.~A. Anderson and M.~A.~J. Chaplain.
\newblock Continuous and discrete mathematical models of tumour-induced
  angiogenesis.
\newblock {\em Bull. Math. Biol.}, 60:857--899, 1998.

\bibitem{Anderson2008}
A.~R.~A. Anderson and V.~Quaranta.
\newblock Integrative mathematical oncology.
\newblock {\em Nat Rev Cancer}, 8(3):227--234, Mar. 2008.

\bibitem{Bizzarri2008}
M.~Bizzarri, A.~Cucina, F.~Conti, and F.~D'Anselmi.
\newblock Beyond the oncogene paradigm: Understanding complexity in
  cancerogenesis.
\newblock {\em Acta Biotheoretica}, 56:173--196, 2008.
\newblock 10.1007/s10441-008-9047-8.

\bibitem{Brooks1986}
R.~Brooks.
\newblock A robust layered control system for a mobile robot.
\newblock {\em IEEE Journal of Robotics and Automation}, vol RA-2(No. 1), 1986.

\bibitem{Byrne2009}
H.~Byrne and D.~Drasdo.
\newblock Individual-based and continuum models of growing cell populations: a
  comparison.
\newblock {\em J. Math. Biol.}, 58:657--687, 2009.

\bibitem{Chomsky1965}
N.~Chomsky.
\newblock {\em Aspects of the Theory of Syntax}.
\newblock Cambridge, MA: MIT Press, 1965.

\bibitem{Dick2010}
J.~Dick, R.~H. Begent, and T.~Meyer.
\newblock Sarcoidosis and testicular cancer: A case series and literature
  review.
\newblock {\em Urologic oncology}, 28(4):350--4, 2010.

\bibitem{Dick2008}
J.~E. Dick.
\newblock Stem cell concepts renew cancer research.
\newblock {\em Blood}, 112(13):4793--807, 2008.

\bibitem{Dick2009}
J.~E. Dick.
\newblock Looking ahead in cancer stem cell research.
\newblock {\em Nature biotechnology}, 27(1):44--6, 2009.

\bibitem{Gardner1999}
R.~L. Gardner.
\newblock Polarity in early mammalian development.
\newblock {\em Current opinion in genetics and development}, 9(4):417--21,
  1999.

\bibitem{Gardner2001}
R.~L. Gardner.
\newblock Specification of embryonic axes begins before cleavage in normal
  mouse development.
\newblock {\em Development}, 128(6):839--47, 2001.

\bibitem{Gardner2007}
R.~L. Gardner.
\newblock The axis of polarity of the mouse blastocyst is specified before
  blastulation and independently of the zona pellucida.
\newblock {\em Human reproduction}, 22(3):798--806, 2007.

\bibitem{Gardner2006}
R.~L. Gardner and T.~J. Davies.
\newblock An investigation of the origin and significance of bilateral symmetry
  of the pronuclear zygote in the mouse.
\newblock {\em Human reproduction}, 21(2):492--502, 2006.

\bibitem{Hanahan2000}
D.~Hanahan and R.~Weinberg.
\newblock The hallmarks of cancer,.
\newblock {\em Cell}, 100:57--70, 2000.

\bibitem{Jones2007}
P.~H. Jones, B.~D. Simons, and F.~M. Watt.
\newblock Sic transit gloria: farewell to the epidermal transit amplifying
  cell?
\newblock {\em Cell stem cell}, 1(4):371--81, 2007.

\bibitem{Knudson1971}
A.~G. Knudson.
\newblock Mutation and cancer: statistical study of retinoblastoma.
\newblock {\em Proc. Natl Acad. Sci. USA}, 68:820--823, 1971.

\bibitem{Logothetis2005}
C.~J. Logothetis and S.~H. Lin.
\newblock Osteoblasts in prostate cancer metastasis to bone.
\newblock {\em Nat Rev Cancer}, 5(1):21--8, 2005.

\bibitem{Nakada2011}
M.~Nakada, D.~Kita, T.~Watanabe, Y.~Hayashi, L.~Teng, I.~V. Pyko, and J.-I.
  Hamada.
\newblock Aberrant signaling pathways in glioma.
\newblock {\em Cancers}, 3(3):3242--3278, 2011.

\bibitem{Palsson2006}
B.~Palsson.
\newblock {\em Systems Biology: Properties of Reconstructed Networks}.
\newblock Cambridge University Press, 2006.

\bibitem{Quaranta2008}
V.~Quaranta, K.~A. Rejniak, and A.~R. Anderson.
\newblock Invasion emerges from cancer cell adaptation to competitive
  microenvironments: quantitative predictions from multiscale mathematical
  models.
\newblock {\em Semin. Cancer Biol.}, 18:338--348, 2008.

\bibitem{Rejniak2011}
K.~A. Rejniak and A.~R.~A. Anderson.
\newblock Hybrid models of tumor growth.
\newblock {\em Wiley Interdisciplinary Reviews: Systems Biology and Medicine},
  3(1):115--125, 2011.

\bibitem{Saetzler2011}
K.~Saetzler, C.~Sonnenschein, and A.~Soto.
\newblock Systems biology beyond networks: Generating order from disorder
  through self-organization.
\newblock {\em Seminars in Cancer Biology}, 21(3):165 -- 174, 2011.

\bibitem{Shackleton2010a}
M.~Shackleton.
\newblock Melanoma stem cells--are there devils in the detail?
\newblock {\em Pigment cell \& melanoma research}, 23(5):693--4, 2010.

\bibitem{Shackleton2010b}
M.~Shackleton.
\newblock Moving targets that drive cancer progression.
\newblock {\em The New England journal of medicine}, 363(9):885--6, 2010.

\bibitem{Shackleton2010}
M.~Shackleton.
\newblock Normal stem cells and cancer stem cells: similar and different.
\newblock {\em Seminars in cancer biology}, 20(2):85--92, 2010.

\bibitem{Shackleton2010c}
M.~Shackleton and E.~Quintana.
\newblock Progress in understanding melanoma propagation.
\newblock {\em Molecular oncology}, 4(5):451--7, 2010.

\bibitem{Sonnenschein1999}
A.~Sonnenschein, C. \&~Soto.
\newblock {\em The Society of Cells, Cancer and control of cell proliferation}.
\newblock Bios Scientific Publishers, 1999.

\bibitem{Tesar2007}
P.~J. Tesar, J.~G. Chenoweth, F.~A. Brook, T.~J. Davies, E.~P. Evans, D.~L.
  Mack, R.~L. Gardner, and R.~D. McKay.
\newblock New cell lines from mouse epiblast share defining features with human
  embryonic stem cells.
\newblock {\em Nature}, 448(7150):196--9, 2007.

\bibitem{Till1964}
J.~E. Till, E.~A. McCulloch, and L.~Siminovitch.
\newblock A stochastic model of stem cell proliferation, based on the growth of
  spleen colony-forming cells.
\newblock {\em Proceedings of the National Academy of Sciences of the United
  States of America}, 51:29--36, 1964.

\bibitem{Turing1990}
A.~M. Turing.
\newblock The chemical basis of morphogenesis. 1953.
\newblock {\em Bull Math Biol}, 52(1-2):153--97; discussion 119--52, 1990.

\bibitem{Tyson1991}
J.~J. Tyson.
\newblock Modeling the cell division cycle: cdc2 and cyclin interactions.
\newblock {\em Proc Natl Acad Sci U S A}, 88(16):7328--32, 1991.

\bibitem{Tyson2011}
J.~J. Tyson, W.~T. Baumann, C.~Chen, A.~Verdugo, I.~Tavassoly, Y.~Wang, L.~M.
  Weiner, and R.~Clarke.
\newblock Dynamic modelling of oestrogen signalling and cell fate in breast
  cancer cells.
\newblock {\em Nat Rev Cancer}, 11(7):523--532, July 2011.

\bibitem{Tyson2003}
J.~J. Tyson, K.~C. Chen, and B.~Novak.
\newblock Sniffers, buzzers, toggles and blinkers: dynamics of regulatory and
  signaling pathways in the cell.
\newblock {\em Curr Opin Cell Biol}, 15(2):221--31, 2003.

\bibitem{Tyson2002}
J.~J. Tyson, A.~Csikasz-Nagy, and B.~Novak.
\newblock The dynamics of cell cycle regulation.
\newblock {\em Bioessays}, 24(12):1095--109, 2002.

\bibitem{Werner1988}
E.~Werner.
\newblock Toward a theory of communication and cooperation for multiagent
  planning.
\newblock {\em Theoretical Aspects of Reasoning About Knowledge: Proceedings of
  the Second Conference}, pages 129--143, 1988.

\bibitem{Werner1989}
E.~Werner.
\newblock {\em Cooperating agents: a unified theory of communication and social
  structure}, pages 3--36.
\newblock Morgan Kaufmann Publishers Inc., San Francisco, CA, USA, 1989.

\bibitem{Werner1996}
E.~Werner.
\newblock What ants cannot do.
\newblock In J.~Perram and J.~Mueller, editors, {\em Distributed Software
  Agents and Applications}. Springer Verlag, 1996.

\bibitem{Werner2003b}
E.~Werner.
\newblock In silico multicellular systems biology and minimal genomes.
\newblock {\em DDT}, 8(24):1121--1127, 2003.

\bibitem{Werner2005}
E.~Werner.
\newblock Genome semantics, in silico multicellular systems and the central
  dogma.
\newblock {\em FEBS Letters}, 579(7):1779--1782, 2005.

\bibitem{Werner2007a}
E.~Werner.
\newblock How central is the genome?
\newblock {\em Science}, 317(5839):753--754, 2007.

\bibitem{Werner2009}
E.~Werner.
\newblock Evolutionary embryos.
\newblock {\em Nature.}, 460(7251):35, 2009.

\bibitem{Werner2010}
E.~Werner.
\newblock Meaning in a quantum universe.
\newblock {\em Science}, 329(5992):629--630, 2010.

\bibitem{Werner2011a}
E.~Werner.
\newblock The cenome: On the global architecture of genomes.
\newblock {\em arXiv (forthcoming)}, 2011.

\bibitem{Werner2011b}
E.~Werner.
\newblock Developmental delusions.
\newblock {\em arXiv (forthcoming)}, 2011.
\newblock A critiqud of gene-centered developmental biology.

\bibitem{Werner2011}
E.~Werner.
\newblock On programs and genomes.
\newblock {\em ArXiv e-prints}, Nov. 2011.
\newblock arXiv:1110.5265v1 [q-bio.OT] (pdf at:
  \url{http://arxiv.org/abs/1110.5265}).

\end{thebibliography}
\end{multicols}
\end{document}